\documentclass[preprint,12pt]{elsarticle}

\usepackage{graphicx}

\usepackage{amssymb}

\biboptions{numbers, sort&compress}

\journal{\hspace{-3.5cm} \raisebox{-1mm}{\begin{tikzpicture} \draw [fill = white, white] (0, 0) rectangle (3.5, 0.4); \end{tikzpicture}}}

\usepackage{amsmath}
\usepackage{url}
\usepackage{color}
\usepackage{tensor}
\usepackage{mathrsfs}
\usepackage{nicefrac}
\usepackage{tikz-cd}
\usepackage{tikz}
\usetikzlibrary{decorations.markings}
\usetikzlibrary{matrix,arrows}
\usepackage{array}
\usepackage[linktocpage=true]{hyperref}
\usepackage[latin1]{inputenc}      
\usepackage[T1]{fontenc}    
\usepackage{ae,aecompl}                          
\usepackage{appendix}
\usepackage{amsfonts,latexsym,bbm}
\usepackage{amssymb}
\usepackage{accents}
\usepackage{graphics}
\usepackage{graphicx}
\usepackage{booktabs,longtable}
\usepackage{bm}
\usepackage{dcolumn}

\usepackage{float, placeins}
\usepackage{upgreek}
\usepackage{ulem}
\usepackage{tensor}

\usepackage{tikz}
\usepackage{tikz-cd}
\usepackage{tikz-3dplot}
\usetikzlibrary{decorations.markings}
\usetikzlibrary{matrix,arrows}
\usetikzlibrary{calc}
\usetikzlibrary{shapes}
\usetikzlibrary{positioning}
\usetikzlibrary{fit}

\usetikzlibrary{decorations.pathreplacing}
\makeatletter
\def\underbrace#1{%
  \@ifnextchar_{\tikz@@underbrace{#1}}{\tikz@@underbrace{#1}_{}}}
\def\tikz@@underbrace#1_#2{%
  \tikz[baseline=(a.base)] {\node[inner sep=2] (a) {\(#1\)};
  \draw[thick,line cap=round,decorate,decoration={brace,amplitude=4pt}]
    (a.south east) -- node[pos=0.5,below,inner sep=7pt] {\(\scriptstyle #2\)} (a.south west);}}
\makeatother

\usepackage[linktocpage=true]{hyperref}

\makeatletter
\renewcommand*{\eqref}[1]{%
  \hyperref[{#1}]{\textup{\tagform@{\ref*{#1}}}}%
}
\makeatother

\usepackage{pdflscape}

\renewcommand{\vec}[1]{\boldsymbol{#1}}
\newcommand{\tsr}[1]{\overset\leftrightarrow{#1}}
\newcommand{\ext}{_{\rm ext}}
\newcommand{\tot}{_{\rm tot}}
\newcommand{\ind}{_{\rm ind}}

\newcommand{\h}{\hspace{1pt}}
\newcommand{\mh}{\hspace{-1pt}}
\newcommand{\hh}{\hspace{0.5pt}}
\newcommand{\mhh}{\hspace{-0.5pt}}
\newcommand{\g}{\hspace{1.5pt}}
\renewcommand{\i}{\mathrm i}
\newcommand{\de}{\mathrm d}
\newcommand{\e}{\mathrm e}

\definecolor{oldgray}{gray}{0.4}

\newcommand{\T}{_{\mathrm T}}
\renewcommand{\L}{_{\mathrm L}}

\renewcommand{\mid}{\h | \h}
\newcommand{\ket}[1]{|#1\rangle}

\numberwithin{equation}{section}

\DeclareMathAlphabet{\mathbbmsl}{U}{bbm}{m}{sl}

\DeclareFontFamily{U}{mathx}{\hyphenchar\font45}
\DeclareFontShape{U}{mathx}{m}{n}{
      <5> <6> <7> <8> <9> <10>
      <10.95> <12> <14.4> <17.28> <20.74> <24.88>
      mathx10
      }{}
\DeclareSymbolFont{mathx}{U}{mathx}{m}{n}
\DeclareFontSubstitution{U}{mathx}{m}{n}
\DeclareMathAccent{\widecheck}{0}{mathx}{"71}

\begin{document}

\begin{frontmatter}



\title{Response Theory of the Electron-Phonon Coupling}


\author[freiberg]{R.~Starke}
\ead{starke.ronald@gmail.com}

\author[heidelberg]{G.A.H.~Schober\corref{cor1}}
\ead{giulio.schober@gmail.com}

\cortext[cor1]{Corresponding author.}

\address[freiberg]{Institute for Theoretical Physics, TU Bergakademie Freiberg, Leipziger Stra\ss e 23, \\ 09596 Freiberg, Germany}
\address[heidelberg]{Institute for Theoretical Physics, Heidelberg University, Philosophenweg 19, \\ 69120 Heidelberg, Germany \vspace{-0.8cm}}

\begin{abstract}
This article presents a systematic theoretical enquiry concerning 
the conceptual foundations and the nature
of phonon-mediated electron-electron interactions. Starting from the fundamental many-body Hamiltonian,
we propose a simple scheme to decouple the electrons and nuclei of a crystalline solid via effective interactions.
These effective interactions, which we express in terms of linear response functions, are completely symmetric under the exchange of electrons and nuclei.
Correspondingly, we derive concrete formulae for both the effective electron interaction
mediated by phonons and the effective core interaction mediated by electrons. 
In particular, we rederive from our fundamental ansatz the well-known general expressions of the effective electron-electron interaction
in terms of the elastic Green function and the phonon dispersion relation. We further show that the effective core interaction coincides 
in the instantaneous limit with the dynamical matrix as calculated in electronic structure theory. 
If combined with the Kubo formalism, our general formulae lend themselves to the calculation of effective interactions from first principles. 
By showing the compatibility of our approach with the functional integral formalism, this work also paves the way for the derivation of {\itshape ab initio} initial interactions 
for functional renormalization group applications.
\end{abstract}

\begin{keyword}
electron-phonon coupling, effective interaction, many-body theory, ab initio electronic structure, functional renormalization group \\[6pt]
{\it Cite as:~} ResearchGate pub.~{\bfseries 303747580}; arXiv:1606.00012


\end{keyword}

\end{frontmatter}



\newpage

\ \vspace{-1.6cm}
\tableofcontents

\newpage
\section{Introduction}\label{sec:intro}

Introduced by I.\,E.~Tamm in 1929 \cite{Tamm} and renamed ``phonons'' almost en passant by J.~Frenkel in 1932 \cite[p.~477]{Frenkel},
the ``quanta of sound'' are not as long known as their electromagnetic counterparts, the ``light quanta'' viz. ``photons''.
Nevertheless, the {\it electron-phonon coupling} soon became a classic 
topic in condensed matter physics, and yet it is also one of current interest (see e.g.~\cite{Grimvall} for a classic textbook, and \cite{Giustino} for the most recent review).
By contrast, traditional electronic structure theory \cite{Kaxiras,Kohanoff,Martin} had for the longest time been concerned 
with the quantum dynamics of the electron liquid \cite{Giuliani} interacting through the Coulomb potential in the {\it external}
potential of an ionic lattice. Although the ensuing {\it purely electronic} ab initio theory has delivered quantitative predictions for an 
impressive variety of materials properties such as
the electronic band structure, the fundamental band-gap or the heat of formation (see the review articles \cite{Hafner08,Hafner10}),
it had been clear at least since the development of the Bardeen--Cooper--Schrieffer (BCS) theory of superconductivity 
(see \cite{Froehlich,Froehlich2,Bardeen55,Cooper,Bardeen56,Bardeen57} for 
original articles, \cite{Schrieffer,deGennes,Tinkham,Marsiglio} for textbooks), that 
the electron-phonon coupling may dominate the physical properties of a system.
A logical step to describe this within a purely electronic framework
was the introduction of a {\it phonon-mediated electron-electron} interaction (see e.g.~\cite[Chap.~17]{Bruus}).
In the early stages, however, such phonon-mediated interactions were often ignored in the ab initio electronic structure community,
partly because the usual ans\"{a}tze for this interaction are not parameter free, partly because superconductivity had been out of the
scope of \mbox{electronic structure theory.}

Interest in the first-principles treatment of the electron-phonon coupling
was renewed though, as it became increasingly evident that in many
situations it even influences typical observables {\it within} the scope of electronic
structure theory, such as the band structure, the band gap or the 
optical spectrum (see e.g.~\cite{Draxl01,Thewalt,Ji08,Spitaler,Kresse09,GiustinoPRL,Cannuccia,MonserratNeeds, Ponce,Monserrat,Hasan}). Furthermore,
a density functional theory (DFT) of superconductivity was developed \cite{Gross88,Gross05a,Gross05b}.
Ultimately, it turned out (see e.g.~\cite{Marini,Kawai,Tupitsyn}) that phonon-mediated
electron interactions quite naturally fit into the Green function approach \cite{Abrikosov,Bontsch,LandauStat} 
to electronic structure physics \cite{Aryasetiawan98,Onida,Schindl}, which for independent reasons had already been established as the 
most important post-DFT method (see e.g.~\cite{Shishkin}). Fittingly, it had already been shown by L.~Hedin and S.~Lundqvist in their seminal work~\cite{Hedin69}
that a phonon-mediated interaction can be included in the standard Hedin--Schwinger--Dyson equations. In particular, they have derived a closed
system of phono-electronic Green function equations \cite[Eqs.~(15.19)]{Hedin69}. 
Presently, the electron-phonon coupling and phonon-mediated electron interactions play a key r\^{o}le in electronic structure physics 
for the quantitative description and state-of-the-art numerical simulation of many-body properties and correlation effects in 
realistic material models \cite{Kortus01, Kortus03a, Kortus03b, Kortus06, Kontani, Kortus10, SubediBoeri, Profeta, NomuraArita}. In particular, for recent DFT- and DPFT-related approaches we refer the interested reader to Refs.~\cite{Leeuwen, Gonze95, Baroni01, Ponce14, MariniPonceGonze}. While these works are mainly concerned with the formulation of self-consistent Green function equations for the description of the coupled system of electrons and phonons, the main purpose of the present article is to decouple the system of nuclei and electrons altogether by the introduction of effective interaction Hamiltonians.

Parallel to these developments, the functional renormalization group (fRG) has been developed into a versatile and unbiased method for 
investigating correlated electron systems \cite{SH00, Berges, Metzner}. It requires an {\itshape initial} electron-electron interaction 
to be given as an input, and then allows to predict the {\itshape effective} low-energy 
interaction for the electrons near the Fermi surface. The latter
determines the leading Fermi liquid instabilities and the low-temperature phases of the model under
consideration. Within solid state physics, the fRG has traditionally been applied to the Hubbard model with a 
repulsive \cite{Ho2001a, Ho2001b, Katanin03, KataninPRL, Friederich, Husemann, Giering, Eberlein14, TUfRG} or attractive \cite{Eberlein14b} 
onsite interaction. More recently, its scope has been extended to multiband systems \cite{Metzner, Platt, Schober}. For a realistic material 
description, the band structure and the initial interaction are typically obtained from an {\itshape ab initio} method such as DFT. 
This procedure has been successfully applied to models relevant for the high-temperature superconducting cuprates \cite{Uebelacker, Maier} 
and iron pnictides \cite{DHLee, PlattNJP, Thomale09, Thomale, Platt11, LichtensteinRG}, 
as well as single- and multilayer graphene \cite{Kiesel, Wu, Scherer1, Lang, Pena, Scherer2}. In the context of the fRG, phonon-mediated electron-electron interactions have been employed as initial interactions for studies on superconductivity and its competition with other \mbox{ordering tendencies \cite{Tsai, HSEliashberg, Fu, HonerkampPhon, TamA, TamB, Bakrim, Classen}.}

Astonishingly enough, the electron-phonon coupling has even attracted interest in speculative high-energy physics. 
There, it was recognized that a ``probe'' $\mathrm Dp$-brane in a string theory background 
is ``dual to a quark-like flavor degrees of freedom coupled with a bulk gluon sector'', 
and this ``may be intuitively thought to be analogous to the electron systems (dual to the probe brane) in a phonon bath (dual to the bulk supergravity)'' \cite{RyuHolo}. 
Furthermore, following the early prediction by W.\,G.~Unruh of phonon emission from sonic horizons \cite{Unruh}, 
there have been attempts to realize this effect in Bose--Einstein conden\-{}sates of ultracold atoms \cite{Garay, Larre}, 
and thus to observe the self-amplifying Hawking radiation from a sonic black-hole laser \cite{Steinhauer}. 
It remains to see whether this reasoning will stand the test of time.

In this article, we resume the problem of the electron-phonon coupling on the most elementary level. That means, we take the many-body
Hamiltonian of electrons {\it and} nuclei as our starting point, and thoroughly discuss the logic behind the replacement of this
coupled Hamiltonian by two decoupled, ``effective'' Hamiltonians for electrons and nuclei (or ``phonons'') respectively. 
Thereby, it will turn out that the decoupled Hamiltonians are ultimately obtained by eliminating the coupling 
terms from the fundamental Hamiltonian by means of linear response theory.
In particular, we will thus clarify the connection between {\it effective} and {\it screened} interactions.
Concretely, as a main result of our efforts it will turn out that such effective interactions are in fact screened interactions as a matter of principle,
meaning that they can be expressed in terms of linear response functions. 
Correspondingly, we call the approach expounded in this article the {\it Response Theory of the electron-phonon coupling}.
Moreover, it will be clarified that screened interactions have to be introduced
if a coupled system (as e.g.~electrons and nuclei) is decoupled into formally non-interacting subsystems. The decoupling is then compensated
by the introduction of ``effective interactions'' in the respective decoupled Hamiltonians. 
The crucial point is now that these ``effective interactions'', which are not
present in the original many-body Hamiltonian, precisely correspond to screened interaction kernels, such that each decoupled subsystem screens the fundamental
(i.e.~Coulomb) interaction of the \mbox{respective {\it other} subsystem.}

While the first part of this work investigates this problem from a purely quantum-mechanical point of view, 
the second part treats it within a functional integral approach. In fact, within this latter formalism the notion of effective interactions
is already well established. Correspondingly, we will show that the functional integral formalism, if applied to the system of electrons
and phonons, precisely reproduces the effective interactions introduced in the first part of this article. In particular, 
this also implies that the Response Theory of the electron-phonon coupling is suitable for the {\itshape ab initio} calculation of phonon-mediated initial interactions 
for the subsequent use in fRG calculations. As our most general expressions include all possible effects 
of inhomogeneity and anisotropy, they may even be used for layered materials 
such as the bulk Rashba semiconductor BiTeI (see \cite{Ishizaka, Bahramy, Lee}, and \cite{Yu, Ponosov, WuBiTeI, Akrap} for phonon effects 
in the bismuth tellurohalides). Combining the Response Theory with the multiband 
fRG and mean-field approach proposed recently in Ref.~\cite{Schober} would therefore allow for an unbiased, quantitative analysis 
of phonon-mediated superconductivity in BiTeI and related compounds.

Concretely, this article is organized as follows: The introductory Sec.~\ref{sec_el_nuc_pho} is dedicated to a pedagogical review of the many-body problem
of electrons and nuclei as well as electronic structure theory. This section does not contain new results; instead, it sets the stage, 
fixes the conventions and makes this document completely self-contained.
The Response Theory of the electron-phonon coupling is then introduced in Sec.~\ref{sec_dyndec}.
In the crucial Sec.~\ref{sec_RFA}, we derive the effective interactions from linear response theory applied to the fundamental Hamiltonian of electrons and nuclei.
Next, we show in Sec.~\ref{subsec_screening} that the thus obtained total interaction kernels can be interpreted as screened Coulomb interactions.
The following Sec.~\ref{subsec_total} discusses the somewhat intricate question of how such effective interactions can be included in
a decoupled Hamiltonian. Finally, Sec.~\ref{subsec_InstLim} investigates the problem of frequency-dependent interaction kernels.
After these conceptual considerations, the more practically oriented Sec.~\ref{sec_mediatedInt} investigates the concrete 
form of the effective electron-electron interaction
mediated by phonons. The main goal there lies in the confirmation that the simple decoupling procedure proposed in Sec.~\ref{sec_RFA} 
yields sensible results and recovers standard formulae if taken to suitable limiting cases. In particular, 
we re-express in Sec.~\ref{subsec_EffElInt} the effective phonon-mediated 
electron interaction in terms of the elastic Green function, whereby we ultimately recover a result orginally obtained by Hedin and Lundqvist \cite[Eq.~(16.7)]{Hedin69}. Furthermore, we express this effective interaction in terms of the
phonon dispersion relation and the phonon Green function, and we clarify its relation to second-order perturbation theory and to the Fr\"ohlich Hamiltonian. 
Finally, in Sec.~\ref{sec_JustFI} we prove that the concept of effective phonon-mediated interactions as introduced in this article is in accord with its counterpart 
derived in the functional integral formalism. The final Sec.~\ref{sec_EffNuclInt} of the main text
investigates the nuclear analogon of the effective electron interaction mediated by phonons: {\it the effective core interaction mediated by electrons}.
Concretely, after proving a useful generalization of the Hellmann--Feynman theorem in Sec.~\ref{app_HFT} and reconsidering  the nuclear equilibrium
positions in Sec.~\ref{sec_reconsider}, we will derive the
effective second-order nuclear Hamiltonian in Sec.~\ref{subsec_ExpansionHam}. Finally, we will prove in Sec.~\ref{sec_dyn} a theorem which shows that in the instantaneous limit, this effective Hamiltonian is equivalent to the 
standard phonon Hamiltonian with the dynamical matrix calculated from ab initio electronic structure physics.

The appendices to this article serve various purposes:
First, App.~\ref{app_notconv} fixes the conventions associated with the Fourier transformation in several r\'{e}\-{}gimes.
Next, App.~\ref{sec:quantSF} gives a short introduction to the quantized Schr\"{o}dinger field, and thus makes this article completely self-contained and readable even for the layman in quantum field theory.
The following App.~\ref{app_AspectsResp} presents a short summary of linear response theory,
then goes on with the derivation of the fundamental electromagnetic response tensor of the nuclei in terms of the elastic Green function, and closes with a countercheck of the latter result with the corresponding Kubo  formula.
The penultimate App.~\ref{app_Iso} re-investigates these results in the iso\-{}tropic limit and, in particular, establishes the connection to the treatment of the nuclear subsystem by classical elasticity theory.
Finally, App.~\ref{app_QED} compares the notion of the phonon-mediated electron-electron interaction with its
photon-mediated counterpart in (quantum) electrodynamics, and it compares the Response Theory of the electron-phonon coupling to Feynman--Wheeler electrodynamics.

\section{Electrons, nuclei and phonons} \label{sec_el_nuc_pho}
\subsection{Many-body problem for electrons and nuclei}
\subsubsection{Fundamental Hamiltonian in first quantization}

The fundamental Hamiltonian for a system of $N$ nuclei and $ZN$ electrons interacting through the Coulomb potential can be decomposed as follows 
(see~e.g.~\cite[Eq.~(4.1)]{Grimvall} or \cite[Sec.~2.2.3]{MartinRothen}):
\begin{align}\label{eq_fund}
\hat H & =(\hat T_{\rm n} +\hat V_{\rm n-n}) + (\hat T_{\rm e}+\hat V_{\rm e-e}) + \hat V_{\rm e-n}\,.
\end{align}
Concretely, this means
\begin{equation}
\begin{aligned}
\hat H & = -\frac{\hbar^2}{2M} \h \sum_{k = 1}^{N} \Delta_{\vec y_k} + \frac{Z^2 e^2}{2} \h \sum_{k \not = \ell} v(\hat{\vec y}_k, \hat{\vec y}_\ell) \\[3pt]
& \quad -\frac{\hbar^2}{2m} \h \sum_{i = 1}^{ZN} \Delta_{\vec x_i} + \frac{e^2}{2} \h \sum_{i \not =  j} v(\hat{\vec x}_i, \hat{\vec x}_j) - Ze^2 \h \sum_{k = 1}^N \sum_{i = 1}^{ZN} v(\hat{\vec y}_k, \hat{\vec x}_i) \,.
\end{aligned}
\end{equation}
In other words, $\hat T$ stands for the (nuclear or electronic) kinetic energy operators, 
while $\hat V$ denotes the various two-particle interaction operators determined by the Coulomb interaction kernel,
\begin{equation} \label{coul_ker}
v(\vec x,\vec x') \equiv v(\vec x - \vec x') = \frac{1}{4\pi\varepsilon_0} \h \frac{1}{|\vec x-\vec x'|}\,.
\end{equation}
The electronic and nuclear masses are respectively denoted by $m$ and $M$, while $e$ stands for the elementary charge 
and $Z$ for the atomic number, such that the electronic and nuclear charges are respectively given by $q_{\rm e}=-e$ and $q_{\rm n}=Ze$.
The Hamiltonian $\hat H$ acts on the complex Hilbert space defined by the square-integrable wave functions of the form
\begin{equation} \label{eq_wave function}
\vec \Psi \equiv \vec\Psi_{}(\vec x_1, \sigma_1 \h ; \h \ldots \h ; \h \vec x_{ZN}, \sigma_{ZN} \h ; \, \vec y_1, \tau_1 \h ; \h \ldots \h ; \h \vec y_N, \tau_N)\,,
\end{equation}
where $(\vec x_i, \sigma_i)$ denote the electronic (orbital and spin) coordinates, while $(\vec y_k, \tau_k)$ denote the corresponding nuclear coordinates. 
The wave functions \eqref{eq_wave function} are antisymmetric with respect to the electronic coordinates, 
whereas the symmetry with respect to the nuclear coordinates depends on the nuclear spin. 
In the following, we will suppress the spin indices in the notation, because the fundamental Hamiltonian \eqref{eq_fund}
is diagonal in the spinorial variables, and these can therefore be ignored for many conceptual matters.

In principle, the Hamiltonian \eqref{eq_fund} is completely {\it symmetric} with respect to electrons and nuclei. 
The only differences lie in their respective masses, charges, 
total particle numbers and possibly their spin (although the latter has no bearing on the Hamiltonian \eqref{eq_fund}).
Astonishingly though, these quantitative differences have a profound qualitative effect 
which could hardly be deduced a priori, namely the extremely high degree
of nuclear localization as compared to the oftentimes rather delocalized electronic charge density. 

In this article, we consider this strong nuclear localization as an empirical fact, which we do not strive to justify from the fundamental Hamiltonian.
Instead, we use it as an additional input to the electron-nuclear many-body problem, which 
justifies a completely {\it asymmetric} treatment of electrons and nuclei. This asymmetry does, however, not concern
the way in which the fundamental Hamiltonian \eqref{eq_fund} is decoupled, but the general formalism which we apply
to the electronic and nuclear subsystems:
The electronic subsystem will be treated by quantum field theoretical methods.
This means, we interpret the electronic many-body wave function as a state of 
the {\it quantized Schr\"{o}dinger field} $\hat\psi(\vec x,t)$ (see e.g.~\cite{Huang,Schweber,Thirring} or App.~\ref{sec:quantSF}), 
in terms of which we will rewrite all electronic operators. 
By contrast, we will not introduce a quantized Schr\"{o}dinger field
for the nuclei, although this would be possible in principle. Instead, in the nuclear case we stick to the first-quantized formalism.
Nevertheless, quantum field theory will become necessary for the treatment of the nuclei as soon as we perform the transition from the
nuclear coordinates to a {\it quantized displacement field} (see Sec.~\ref{subsec:dynDF&phon}).

\subsubsection{Second-quantized electrons and first-quantized nuclei}\label{subsec_secQelfistQnuc}

Within the limits of electronic structure theory, the electronic quantum field is given by the quantized Schr\"{o}dinger field (see App.~\ref{sec:quantSF}), 
\begin{equation}
\hat\psi\equiv\hat\psi(\vec x)\,,
\end{equation}
which is an operator whose action on a $(ZN)$-particle {\it electronic} wave-func\-{}tion $\Psi^{ZN}(\vec x_1, \ldots, \vec x_{ZN})$
yields a wave function of $(ZN-1)$ electrons, which is given~explicitly by (cf.~Eq.~\eqref{eq_annBasis})
\begin{equation}\label{eq_fieldOpExpl}
\big(\hat\psi(\vec x) \h \Psi^{ZN} \h \big)(\vec x_2,\ldots,\vec x_{ZN})=\sqrt{ZN} \, \Psi^{ZN}(\vec x,\vec x_2,\ldots,\vec x_{ZN})\,.
\end{equation}
In terms of this field operator, the electronic {\itshape charge density operator} is de-\linebreak fined as \smallskip
\begin{equation}
\hat\rho_{\rm e}(\vec x)= (-e) \h\hh \hat\psi^\dagger(\vec x) \h \hat\psi(\vec x) =(-e) \h \hat n_{\rm e}(\vec x)\,.\label{eq_densOp} \smallskip \vspace{2pt}
\end{equation}
Here, $\hat\psi^\dagger(\vec x)$ denotes the hermitean adjoint of the field operator $\hat\psi(\vec x)$, 
whereas the electronic {\itshape number density operator} is given by (cf.~Eq.~\eqref{eq_numberdens})
\begin{equation}
\hat n_{\rm e}(\vec x)=\hat\psi^\dagger(\vec x) \h \hat\psi(\vec x)\,.
\end{equation}
The restriction of $\hat \rho_{\rm e}(\vec x)$ to a subspace with fixed electron number $(ZN)$ can be equivalently expressed as (see e.g.~\cite[Eq.~(3.120)]{Giuliani})
\begin{equation}
 \hat \rho_{\rm e}(\vec x) = (-e) \, \sum_{i = 1}^{ZN} \delta^3(\vec x - \hat{\vec x}_i) \,.
\end{equation}
Using the spectral decomposition,
\begin{equation}
\delta^3(\vec x-\hat{\vec x})=\int\!\de^3\vec x'\,\delta^3(\vec x-\vec x') \, |\vec x'\rangle\langle\vec x'|\,,
\end{equation}
and the antisymmetry with respect to electronic coordinates,
one verifies that the expectation value of the charge-density in a state $|\Psi^{ZN}\rangle$ with electron number $ZN$,
\begin{equation}
\rho_{\rm e}(\vec x) \equiv \big\langle\Psi^{ZN}\big|\h\hat\rho_{\rm e}(\vec x)\h\big|\Psi^{ZN}\big\rangle
=-e\,\big\langle\Psi^{ZN}\big|\h\hat n_{\rm e}(\vec x)\h\big|\Psi^{ZN}\big\rangle\,,
\end{equation}
yields
\begin{equation} \label{eq_densityExpVal}
\begin{aligned}
\rho_{\rm e}(\vec x) & = (-e)\h ZN\int\!\de^3\vec x_2\ldots\int\!\de^3\vec x_{ZN} \\[3pt]
 & \quad \, \times  (\Psi^{ZN})^*(\vec x,\vec x_2,\ldots,\vec x_{ZN}) \, \Psi^{ZN}(\vec x,\vec x_2,\ldots,\vec x_{ZN})\,.
\end{aligned} \smallskip
\end{equation}
In particular, on account of the normalization condition on the many-body wave function, the charge density integrates to the total charge
\begin{equation}
\int\!\de^3\vec x\,\rho_{\rm e}(\vec x)=(-e) \h ZN\,.
\end{equation}
This is particularly transparent for the evaluation of the density operator in a {\itshape one-particle state} $|\Psi^1\rangle$ giving
\begin{equation}
\rho_{\rm e}(\vec x)=-e \,(\Psi^1)^*(\vec x) \, \Psi^1(\vec x)=-e \h n_{\rm e}(\vec x)\,.
\end{equation}
In fact, the second-quantized form of any {\itshape one-particle operator} (cf.~\cite[Sec.\ 1.4]{Giuliani}) 
is formally identical to its expectation value in a one-particle state \linebreak under the replacements
\begin{align}
\Psi^1(\vec x)&\mapsto\hat\psi(\vec x)\,,\\[3pt]
(\Psi^1)^*(\vec x)&\mapsto\hat\psi^\dagger(\vec x)\,.
\end{align}
For example, the kinetic energy operator can also be expressed as
\begin{equation}
 \hat T_{\rm e} = \int \! \de^3\vec x\,\h \hat\psi^\dagger(\vec x) \mh \left(-\frac{\hbar^2}{2m}\Delta\right) \! \hat\psi(\vec x)\,.\label{eq_kinEnSecQu}
\end{equation}
Similarly, the {\it two-particle} Coulomb interaction operator can be expressed in second quantization as
\begin{equation} \label{sq_Coulomb}
 \hat V_{\rm e-e} =\frac{e^2}{2}\int \! \de^3\vec x \int \! \de^3\vec x'\,\h \hat\psi^\dagger(\vec x)\h \hat\psi^\dagger(\vec x') \h 
v(\vec x,\vec x')\h \hat\psi(\vec x')\h \hat\psi(\vec x)\,.
\end{equation}
In particular, this operator yields automatically zero if applied to any one-particle state. 
For later purposes, we note that the Coulomb interaction operator can also be written as
\begin{equation}
\hat V_{\rm e-e}  = \frac{1}{2} \h \int \! \de^3\vec x \int \! \de^3\vec x'\,\h \hat\rho_{\rm e}(\vec x)\h v(\vec x,\vec x')\h 
 \hat\rho_{\rm e}(\vec x')\,,\label{eq_quantCoul}
\end{equation}
provided that a normal ordering prescription of the field operators (``annihilators to the right'') is understood
when the density operator \eqref{eq_densOp} is plugged into Eq.~\eqref{eq_quantCoul}.

As explained already in the previous subsection, we will stick to the first-quantized formalism in the nuclear case.
Nonetheless, we introduce a nuclear charge operator by the definition
\begin{equation}
\hat \rho_{\rm n}(\vec y)=Ze\h\sum_{\vec n}\delta^3(\vec y-\hat{\vec y}_{\vec n}) \,, \label{eq_fQuantNucDens}
\end{equation}
where $\vec n$ labels the nuclear positions, which we will later assume to form a regular lattice.
Again, one verifies easily that the expectation value of $\hat \rho_{\rm n}(\vec y)$ in a nuclear $N$-particle state $|\Phi^N\rangle$ reads
\begin{equation}
\begin{aligned}
\rho_{\rm n}(\vec y) & = e \, ZN \mh \int\!\de^3\vec y_2\ldots\int\!\de^3\vec y_N \\[5pt]
 & \quad \, \times (\Phi^N)^*(\vec y,\vec y_2,\ldots,\vec y_N)\h (\Phi^N)(\vec y,\vec y_2,\ldots,\vec y_N)\,.
\end{aligned}
\end{equation}
With these definitions, we can write the interaction terms involving nuclear coordinates analogously to the electronic case as
\begin{align}
\hat V_{\rm n-n}&=\frac 1 2 \h \int\! \de^3\vec y \int \! \de^3\vec y'\, \hat\rho_{\rm n}(\vec y) \h\hh v(\vec y,\vec y')\h\hh \hat\rho_{\rm n}(\vec y')\,, \label{eq_Vnn} \\[5pt]
\hat V_{\rm e-n}&=\int \! \de^3\vec y \int \! \de^3\vec y'\, \hat\rho_{\rm e}(\vec y)\h\hh v(\vec y,\vec y')\h\hh \hat\rho_{\rm n}(\vec y')\,. \label{eq_Ven}
\end{align}
We note, however, that self-interaction terms in Eq.~\eqref{eq_Vnn} have to be discarded when the first-quantized expression \eqref{eq_fQuantNucDens} is plugged in.
The same problem does not arise in the second-quantized expression \eqref{eq_quantCoul} for the electrons, where the self-interaction
is removed by the normal ordering prescription.
Anyway, expressing the Coulomb interactions in the form of Eqs.~\eqref{eq_quantCoul}, \eqref{eq_Vnn} and \eqref{eq_Ven}
(rather than the notation used in the fundamental Hamiltonian \eqref{eq_fund}) will prove
crucial for the formulation of effective interactions, because it allows for the decomposition of the densities into reference and fluctuation parts (see Sec.~\ref{sec_RFA}).

Finally, we remark that also {\itshape  external potentials} (which are one-particle operators) can be written in their first- or second-quantized form as
\begin{equation}
\hat V_{\rm ext}
=\int\!\de^3\vec x\,\h v\ext(\vec x)\h\hat n(\vec x)
=\int\!\de^3\vec x\,\h \varphi_{\rm ext}(\vec x) \h \hat\rho(\vec x)\,,\label{eq_extPotSecQu}
\end{equation}
where the potential energy $v_{\rm ext}$ is generally related to the electrostatic potential $\varphi_{\rm ext}$ by
\begin{equation}
v_{\rm ext}(\vec x)=q\,\varphi_{\rm ext}(\vec x)\,, \smallskip
\end{equation}
and where $q$ denotes the respective charge of the particles on which 
the exter\-{}nal potential acts.

\subsection{Static decoupling in electronic structure theory}

The Hamiltonian \eqref{eq_fund} proves completely untractable for a true many-body system, and therefore 
has to be replaced by a simplified Hamiltonian in which electrons and nuclei are decoupled.
In this subsection, we briefly review how the standard electronic structure theory achieves this goal.

\subsubsection{Electronic sector} \label{el_sector}

At the outset, {\it electronic} structure theory treats the nuclei as classical point particles 
located at fixed positions $\vec x_{\vec n0}$\hh. It is an {\it empirical} fact that for many systems, these fixed positions roughly form a regular array, 
the so-\linebreak called {\itshape (direct) Bravais lattice} \cite{Ashcroft,Kittel}. This is  of the form 
\begin{equation}
\vec x_{\vec n0}=n_1\hh\vec a_1+n_2\h \vec a_2+n_3\h \vec a_3\,,\label{eq_bravais}
\end{equation}
where $\vec a_1,\vec a_2,\vec a_3$ are fixed, linearly independent vectors (the {\itshape unit vectors} of the Bravais lattice), and $\vec n=(n_1,n_2,n_3)$ is a triple of integers in the range
\begin{equation}
0\leq n_i<N_i\,,
\end{equation}
where $N_1N_2N_3=N$ is the total number of nuclei. Hence, the {\it crystal lattice} is given by
\begin{equation}
 \Gamma = \big\{\vec x_{\vec n0} \h ; \, \vec n=(n_1,n_2,n_3), \, 0 \leq n_i < N_i \ (i = 1, 2, 3) \big\}\,.
\end{equation}
As this article is mainly about conceptual questions, we restrict ourselves to the case where the basis of the lattice
consists of only one nucleus, and all nuclei are of the same nature and in particular have the same mass $M$ and charge $Ze$.
Electronic structure theory now replaces the electronic part,
\begin{equation}
\hat H_{\rm e}=\hat T_{\rm e}+\hat V_{\rm e-e}+\hat V_{\rm e-n}\,,  
\end{equation}
of the Hamiltonian \eqref{eq_fund}, which still involves nuclear coordinates, by a {\it purely electronic Hamiltonian} of the form
\begin{equation}\label{eq_elHam}
\hat H_{\rm e \hh 0}=\hat T_{\rm e}+\hat V_{\rm e-e}+\hat V^{\rm n}_{\rm ext}\,,
\end{equation}
where
\begin{equation}\label{eq_extPot}
\hat V^{\rm n}_{\rm ext}=\int \! \de^3\vec x\!\int \! \de^3\vec y\,\h\rho_{\rm n0}(\vec y) \h v(\vec y,\vec x) \h \hat \rho_{\rm e}(\vec x)
=\int \! \de^3\vec x\,\h\varphi^{\rm n}_{\rm ext}(\vec x) \h \hat\rho_{\rm e}(\vec x)\,,
\end{equation}
and where
\begin{equation}
\varphi^{\rm n}_{\rm ext}(\vec x)=\int \! \de^3\vec y\,\rho_{\rm n0}(\vec y)\h v(\vec y,\vec x) \smallskip
\end{equation}
is the {\it external} Coulomb potential generated by the nuclei through their classical charge density given by
\begin{equation}\label{eq_nucrefdens}
\rho_{\rm n0}(\vec y)=Ze\sum_{\vec x_{\vec n0} \hh \in \h \Gamma}\delta^3(\vec y-\vec x_{\vec n0}) \smallskip\,.
\end{equation}
Thus, we obtain the explicit expression
\begin{equation} \label{ext_exp}
 \hat V_{\rm ext}^{\rm n} = Ze \sum_{\vec x_{\vec n0} \hh \in \h \Gamma} \h \int\! \de^3 \vec x \,\h v(\vec x_{\vec n0}, \h \vec x) \h \hat \rho_{\rm e}(\vec x) \,.
\end{equation}
Essentially, this whole approximation simply means that the replacement 
\begin{equation}
\hat\rho_{\rm n}(\vec x) \mapsto \rho_{\rm n0}(\vec x)
\end{equation}
is performed in the fundamental interaction term \eqref{eq_Ven}.
In the electronic Hamiltonian \eqref{eq_elHam}, the crystal lattice is therefore a prescribed input assumed to be known in advance.

It is now tempting to say that this external potential has the periodicity of the lattice, 
i.e., for any vector $\vec a$ of the form \eqref{eq_bravais} we have
\begin{equation}
\varphi^{\rm n}\ext(\vec x+\vec a)=\varphi^{\rm n}\ext(\vec x)\,,\label{eq_perpot}
\end{equation}
a relation which is frequently used in solid state physics. However, this is not strictly true, because for any real material sample the nuclei form only a finite lattice. It is plausible though,
that deeply within the {\it bulk} of the solid the lattice periodicity holds, with deviations only occurring in the vicinity of the boundaries.
For the description of {\it bulk properties} (i.e., those properties which do not involve surface or boundary effects and which are hence
independent of the probe geometry),  it is therefore the right idealization to assume Eq.~\eqref{eq_perpot} to hold everywhere.
Mathematically, however, this would require an infinite lattice and hence an infinite number of nuclei, while in the Schr\"{o}dinger equation the particle number
should remain finite. In order to reconcile these apparently contradicting requirements, one regards the lattice integers $n_i$
as equivalence classes (see e.g.~\cite{Salmhofer}),
\begin{equation}
n_i\in\mathbb Z_{N_i}\equiv\mathbb Z\h/ N_i\h \mathbb Z\,,
\end{equation}
which means that one identifies all lattice integers
\begin{equation}
\{ \, \ldots, \, n_i-2N_i, \, n_i-N_i, \, n_i, \, n_i+N_i, \, n_i+2N_i,\, \ldots \, \} \,,
\end{equation}
and requires the Born--von-Karman boundary conditions \cite{Ashcroft,Kittel} for the electronic wave function,
\begin{equation}
\Psi( \h \ldots,\vec x,\ldots \h )=\Psi(\h \ldots,\vec x+N_i\h\vec a_i\hh,\h\ldots \h )\,.
\end{equation}
In the sense of this formalism, the crystal is both finite (particle number) and infinite (no boundaries or surfaces). 
Strictly speaking, the imposition of Born--von-Karman boundary conditions also modifies the fundamental Hamiltonian \eqref{eq_fund}, and hence one has to take the thermodynamic limit in the end of all calculations. This is discussed briefly in App.~\ref{app_bvk}.

\subsubsection{Nuclear sector} \label{nuc_sec}

As the necessity of knowing the correct nuclear equilibrium positions in advance
is a rather dissatisfying state of affairs, electronic structure theory has devised an ingenious way to circumvent this shortcoming.
Instead of the equilibrium positions, one first interprets the $\vec x_{\vec n0} \in \Gamma$ as provisional {\it reference positions}, only to 
consider small deviations from these written as
\begin{equation}
\vec x_{\vec n} = \vec x_{\vec n0}+\vec u_{\vec n} \,, \vspace{-3pt}
\end{equation}
or equivalently, \smallskip
\begin{equation} \label{eq_deviations}
 \vec x_{\vec n}[\vec u] = \vec x_{\vec n0} + \vec u(\vec x_{\vec n0}) \,. \smallskip
\end{equation}
One now investigates the total {\it nuclear potential energy}, which is written in the form
\begin{equation} \label{eq_Vphon}
V_{\rm n}[\vec u] = V_{\rm n-n}[\vec u] + E_{\rm e \hh 0}[\vec u]\,, \smallskip
\end{equation}
as a functional of these deviations. Here, the first term is the internuclear Coulomb interaction potential given explicitly by
\begin{align}
V_{\rm n-n}[\vec u]&=\frac{1}{2} \h Z^2 e^2 \h \sum_{\vec n\neq\vec m}v(\vec x_{\vec n}-\vec x_{\vec m})\\
&=\frac{1}{2} \h Z^2 e^2 \h \sum_{\vec n\neq\vec m}v(\vec x_{\vec n0}-\vec x_{\vec m0}+\vec u_{\vec n}-\vec u_{\vec m})\,,
\end{align}
while the second term,
\begin{equation}
E_{\rm e \hh 0}[\vec u] = \langle\Psi_0[\vec u] \mid \hat H_{\rm e \hh 0}[\vec u] \mid \Psi_0[\vec u]\rangle\,, \label{eq_defpotphon}
\end{equation}
denotes the electronic energy calculated in the external potential of the nuclei with fixed positions as given by Eq.~\eqref{eq_deviations}. 
This means that the electronic Hamiltonian \eqref{eq_elHam} is interpreted as a functional of the deviations,
\begin{equation}
\hat H_{\rm e \hh 0}[\vec u]=\hat T_{\rm e}+\hat V_{\rm e-e}+\hat V^{\rm n}_{\rm ext}[\vec u]\,, \label{eq_elecHam}
\end{equation}
where the operator of the external potential is given by (cf.~Eq.~\eqref{ext_exp})
\begin{equation} \label{eq_elecHam_extpot}
\hat V^{\rm n}_{\rm ext}[\vec u] = Ze \,\sum_{\vec n}\int \! \de^3 \vec x \,\h
v(\vec x_{\vec n}[\vec u] - \vec x) \, \hat \rho_{\rm e}(\vec x) \,.
\end{equation}
Furthermore, $|\Psi_0[\vec u]\rangle$ is determined for each $\vec u$ as the electronic ground state of $\hat H_{\rm e \hh 0}[\vec u]$.
The {\it equilibrium positions} of the nuclei are now defined as those reference positions which minimize the total nuclear potential energy \eqref{eq_Vphon}. In Sec.~\ref{sec_reconsider}, we will show that these equilibrium positions are in fact identical to the classical equilibrium positions defined by the vanishing of the electrostatic forces.
In practice, they can hence be calculated by starting from a guess for the nuclear equilibrium positions and  minimizing the total potential energy with respect to slight variations of these afterwards.
Surprising as it may be at first sight, the purely electronic structure theory
can therefore predict the lattice structure and in particular the lattice constants of solids (see e.g.~\cite{Haas, Harl, Schimka, Lejaeghere}).

\subsection{Dynamical displacement field and phonons}\label{subsec:dynDF&phon}

The above considerations show that nuclear quantities are not completely outside the scope of a purely electronic theory.
While this is still intuitive for {\it static} nuclear properties such as the lattice constant, it is by no means obvious
that electronic structure theory is also capable of incorporating the {\itshape dynamics} of the nuclei. Fortunately though, even this turns out to be feasible.
Here, the nuclear localization is again decisive, as it allows for the introduction of a {\itshape displacement field,}
which we will now review both on the classical and on the quantum level.

\subsubsection{Classical displacement field} \label{cdf}

Precisely as in classical continuum mechanics, the classical nuclear displacement field, 
\begin{equation}
\vec u\equiv\vec u(\vec x_{\vec n0},t)\,,
\end{equation}
is introduced such that the instantaneous nuclear positions are per definitionem given by \cite[Eq.~(1.1)]{Landau65}
\begin{equation}
\vec x_{\vec n}(t)=\vec x_{\vec n0}+\vec u(\vec x_{\vec n0},t)\,. \smallskip
\end{equation}
In other words, $\vec u(\vec x_{\vec n0},t)$ is the momentary elongation of the nucleus with equilibrium position $\vec x_{\vec n0}$ out of that very equilibrium
position. For the dynamics of the displacement field, one writes down a classical equation of motion,
\begin{equation}\label{EoMDF}
M\frac{\partial^2}{\partial t^2}\vec u(\vec x_{\vec n0},t)+\sum_{\vec m}\tsr K(\vec x_{\vec n0},\vec x_{\vec m0}) \h \vec u(\vec x_{\vec m0},t)=0\,, \vspace{-3pt}
\end{equation}
where $\tsr K$ denotes the {\itshape dynamical matrix} (see e.g.~\cite[Chap.~3, Sec.~1]{Grimvall}). Generally, the dynamical matrix is invariant under lattice translations, i.e.
\begin{align} \label{lat_tra}
\tsr K(\vec{x}_{\vec{n}0}+\vec a, \h \vec{x}_{\vec{m}0}+\vec a)&=\tsr K(\vec{x}_{\vec{n}0}, \h \vec{x}_{\vec{m}0})\,
\end{align}
for every lattice vector $\vec a$ (of the form Eq.~\eqref{eq_bravais}), and can therefore be written as \smallskip
\begin{equation}
\tsr K(\vec x_{\vec n0},\h \vec x_{\vec m0}) = \tsr K(\vec x_{\vec n0} - \vec x_{\vec m0}) \,. \smallskip
\end{equation}
This means that the dynamical matrix can be interpreted as a function of one difference lattice vector $\vec r$ only. 
In addition, the dynamical matrix is assumed to be real valued,
\begin{equation} \label{Kre}
 K_{ij}(\vec r_{\vec n0}) = K_{ij}^*(\vec r_{\vec n0}) \,,
\end{equation}
and subject to the symmetry
\begin{equation} \label{Ksy}
K_{ij}(\vec{r}_{\vec{n}0}) = K_{ji}(-\vec{r}_{\vec{n}0})\,.
\end{equation}
As a consequence, the harmonic equation of motion \eqref{EoMDF} for the displacement field 
corresponds to a classical second-order Hamiltonian given by (cf. \cite[Eq.~(1.4.5)]{Ziman2})
\begin{equation}
\begin{aligned}
H_{\rm phon}[\vec u,\vec\pi] & = \frac{1}{2M} \h \sum_{\vec{n}}\vec{\pi}^{\rm T}(\vec{x}_{\vec{n}0},t)
\h \vec{\pi}(\vec{x}_{\vec{n}0},t)\\[3pt]
& \quad \, +\frac{1}{2} \h \sum_{\vec{n}, \h \vec{m}}\vec{u}^{\rm T}(\vec{x}_{\vec{n}0},t) \h 
\tsr{K}(\vec{x}_{\vec{n}0},\vec{x}_{\vec{m}0}) \h \vec{u}(\vec{x}_{\vec{m}0},t)\,,\label{eq_phonoHamClass}
\end{aligned}
\end{equation}
with the conjugate momentum
\begin{equation} \label{eq_conj_momentum}
\vec{\pi}(\vec{x}_{\vec{n}0},t)=M\partial_t\vec{u}(\vec{x}_{\vec{n}0},t)\,.
\end{equation}
Thus, Eq.~\eqref{EoMDF} is equivalent to the Hamilton equations of motion
\begin{align}
\partial_t \hh u_i(\vec x_{\vec n0},t)&=\big\{u_i(\vec x_{\vec n0},t),H_{\rm phon}\big\}=\frac{\delta H_{\rm phon}}{\delta\pi_i(\vec x_{\vec n0},t)}\,,\\[5pt]
\partial_t \hh \pi_i(\vec x_{\vec n0},t)&=\big\{\pi_i(\vec x_{\vec n0},t),H_{\rm phon}\big\}=-\frac{\delta H_{\rm phon}}{\delta u_i(\vec x_{\vec n0},t)}\,.
\end{align}
Here, we have introduced the {\itshape Poisson brackets,} which are defined for functionals of the displacement field and its conjugate momentum as
\begin{align}
 & \{F,G\}(t)= \\[3pt] \nonumber
 & \sum_{\vec{x}_{\vec n0}\hh \in\hh \Gamma} \sum_{i = 1}^3 \left(\frac{\delta F[\vec u,\vec\pi]}{\delta u_i(\vec x_{\vec n0},t)}\h \frac{\delta G[\vec u,\vec\pi]}{\delta \pi_i(\vec x_{\vec n0},t)}-
\frac{\delta G[\vec u,\vec\pi]}{\delta u_i(\vec x_{\vec n0},t)}\h \frac{\delta F[\vec u,\vec\pi]}{\delta \pi_i(\vec x_{\vec n0},t)}\right).
\end{align}
In particular, the fundamental variables obey the Poisson bracket relations
\begin{align}
\big\{u_i(\vec x_{\vec n0},t),\h \pi_j(\vec x_{\vec m0},t)\big\}&=\delta_{\vec{nm}} \h \delta_{ij} \,,\label{eq_Poi1}\\[3pt]
\big\{u_i(\vec x_{\vec n0},t),\h u_j(\vec x_{\vec m0},t)\big\}&=0\,,\label{eq_Poi2}\\[3pt]
\big\{\pi_i(\vec x_{\vec n0},t),\h \pi_j(\vec x_{\vec m0},t)\big\}&=0\,.\label{eq_Poi3}
\end{align}
In Fourier space, these can be written equivalently as
\begin{align}
\big\{u_i(\vec k_{\vec n}, t), \h \pi_j(-\vec k_{\vec m},t)\big\}&=\delta_{\vec{nm}} \h \delta_{ij} \,,\label{eq_Poi1_FT}\\[3pt]
\big\{u_i(\vec k_{\vec n}, t), \h u_j(-\vec k_{\vec m},t)\big\}&=0\,,\label{eq_Poi2_PT}\\[3pt]
\big\{\pi_i(\vec k_{\vec n}, t), \h \pi_j(-\vec k_{\vec m},t)\big\}&=0\,,\label{eq_Poi3_PT}
\end{align}
where $\vec k_{\vec n}, \vec k_{\vec m} \in \Gamma^*$ are dual lattice vectors (see~App.~\ref{app_FT} and \cite[Sec.~2.1]{ED1} for our conventions). We note that the reality of $u_j(\vec x_{\vec m0}, t)$ in real space implies in Fourier space that
\begin{equation}
 u_j(-\vec k_{\vec m}, t)= u_j^*(\vec k_{\vec m}, t) \,,
\end{equation}
and the analogous identity holds for the conjugate momentum. With all these definitions, the Hamiltonian \eqref{eq_phonoHamClass} is of the usual form,
\begin{equation}\label{eq_phonHam}
H_{\rm phon} = T_{\rm phon} + V_{\rm phon}\,,
\end{equation}
and hence the dynamical matrix can be identified with the second-derivative tensor of 
a potential energy $V_{\rm phon} \equiv V_{\rm phon}[\vec u]$ with respect to the displace-\linebreak ment field, i.e.,
\begin{equation}\label{eq_defK}
 K_{ij}(\vec{x}_{\vec{n}0},\vec{x}_{\vec{m}0}) 
 = \left. \frac{\partial^2 V_{\rm phon}[\vec u]}{\partial u_i(\vec x_{\vec n 0}) 
 \h\partial u_j(\vec x_{\vec m 0})} \, \right|_{\vec u \hh = \hh \vec 0}\,. 
\end{equation}
We now turn to the solutions of the equation of motion. First, we rewrite Eq.~\eqref{EoMDF} in Fourier space  as
\begin{equation} \label{evp}
\left(-M \omega^2 \h \tsr 1  + \tsr K(\vec k_{\vec m}) \right) \vec u(\vec k_{\vec m}, \omega) = 0 \,.
\end{equation}
For brevity, we will in the following denote the dual lattice vectors by $\vec k \equiv \vec k_{\vec m} \in \Gamma^*$, 
although it should be kept in mind that at this point they are still discrete. They will become elements of the continuous Brillouin zone later on when we take the 
thermodynamic limit; see App.~\ref{app_thLim}.

The symmetries~\eqref{Kre}--\eqref{Ksy} of the dynamical matrix read in Fourier space
\begin{equation}
 K_{ij}(\vec k) = K_{ij}^*(-\vec k) \label{sym_herm_k} \,,
\end{equation}
and respectively, \smallskip
\begin{equation}
 K_{ij}(\vec k) = K_{ji}(-\vec k) \,. \label{sym_inv_k} \smallskip
\end{equation}
Together, these equations imply the hermiticity of the dynamical matrix,
\begin{equation}
K_{ij}(\vec k) = K^*_{ji}(\vec k) \equiv (K^\dagger)_{ij}(\vec k)\,.
\end{equation}
For each wavevector $\vec k$, Eq.~\eqref{evp} can be interpreted as a hermitean eigenvalue equation. Consequently, for each $\vec k$ 
there are three linearly indepen-\linebreak dent solutions, which we label by the index $\lambda\in\{1, 2, 3\}$. The corresponding eigenfrequencies $\omega_{\vec k\lambda}$ 
are determined by the condition (cf.~\cite[Chap.~4, \S\,7]{Animalu})
\begin{equation}
 \det \mh \left({- M} \hh \omega_{\vec k\lambda}^2 \tsr 1 + \h \tsr K(\vec k) \right) = 0 \,, \smallskip
\end{equation}
and the eigenvectors satisfy
\begin{equation} \label{normalized_ev}
 \tsr K(\vec k) \, \vec e_{\vec k \lambda} = M \omega_{\vec k \lambda}^2 \h \vec e_{\vec k\lambda} \,.
\end{equation}
We assume these eigenvectors to be orthonormal,
\begin{equation}
 \vec e_{\vec k\lambda}^\dagger \h \vec e_{\vec k \lambda'} \h \equiv \h \vec e_{\vec k\lambda}^* \cdot \vec e_{\vec k\lambda'} \h = \h \delta_{\lambda \lambda'} \,,
\end{equation}
and complete,
\begin{equation}
 \sum_{\lambda = 1}^3 \vec e_{\vec k\lambda} \h \vec e_{\vec k\lambda}^\dagger  = \tsr 1 \,. \smallskip
\end{equation}
In particular, assuming that the dynamical matrix $K(\vec k)$ is positive definite, \linebreak there are always {\itshape two} eigenfrequencies $\omega_{\vec k\lambda} > 0$ and $-\omega_{\vec k\lambda} < 0$ for each $\vec k$ \linebreak and $\lambda$ (or an even number if the eigenspaces of the dynamical matrix have further degeneracies). These frequencies differ only by a sign but yield the same eigenvalue of Eq.~\eqref{normalized_ev}.
Therefore, the most general solution of Eq.~\eqref{evp} can be written as a linear superposition,
\begin{equation} \label{mode_inv}
 \vec u(\vec k, \omega) = \sqrt{2\pi} \h c \,  
 \sum_{\lambda = 1}^3 \big( \alpha_{\vec k \lambda} \, \vec e_{\vec k\lambda} \, \delta(\omega - \omega_{\vec k\lambda} )  
 + \beta_{\vec k\lambda} \, \vec e_{\vec k\lambda} \, \delta(\omega + \omega_{\vec k\lambda} ) \big)\,,
\end{equation}
with complex coefficients $\alpha_{\vec k\lambda}$ and $\beta_{\vec k\lambda}$\hh. 
By an inverse Fourier transformation, this is equivalent to
\begin{equation} \label{mode}
 \vec u(\vec x_{\vec n0}, t) = \frac 1 {\sqrt N} \sum_{\vec k, \h \lambda} \left( \alpha_{\vec k\lambda} \h \vec e_{\vec k\lambda} \, \e^{-\i\omega_{\vec k\lambda}t + \i \vec k \cdot \vec x_{\vec n0}} + \beta_{\vec k\lambda} \h \vec e_{\vec k\lambda} \h \e^{\i\omega_{\vec k\lambda}t+\i\vec k\cdot \vec x_{\vec n0}} \h \right) \,,
\end{equation}
and this is the most general solution of the equation of motion \eqref{EoMDF}.
The requirement that the displacement field is real valued, however, implies that the coefficients $\alpha_{\vec k\lambda}$ and $\beta_{\vec k\lambda}$ are not independent of each other. To derive the corresponding condition on them, we first note that by taking the complex conjugate of Eq.~\eqref{normalized_ev} and using Eq.~\eqref{sym_herm_k}, we obtain
\begin{equation}
 \tsr K(-\vec k) \, \vec e^*_{\vec k \lambda} = M \omega_{\vec k \lambda}^2 \h \vec e^*_{\vec k\lambda} \,.
\end{equation}
This shows that $\vec e^*_{\vec k\lambda}$ also solves the equation of motion with the same eigenvalue $\omega_{\vec k\lambda}^2$ but with the inverse wavevector $(-\vec k)$. In particular, we can thus label the solutions with the wavevector $(-\vec k)$ by the same index $\lambda$ as the solutions with  the wavevector $\vec k$, such that generally,
\begin{equation}
 \omega_{-\vec k\lambda} = \omega_{\vec k\lambda} \,, \label{cond_1}  \vspace{-2pt}
\end{equation}
and \smallskip
\begin{equation} 
 \vec e_{-\vec k\lambda} = \vec e_{\vec k\lambda}^* \,. \smallskip \vspace{2pt}\label{cond_2}
\end{equation}
Now, by substituting $\vec k \to -\vec k$ in Eq.~\eqref{mode} and using the above conditions, we get the equivalent expression
\begin{align}
 & \vec u(\vec x_{\vec n0}, t) = \\[3pt] \nonumber
 & \frac 1 {\sqrt N} \sum_{\vec k, \h \lambda} \left( \alpha_{-\vec k\lambda} \h \vec e_{\vec k\lambda}^* \, \e^{-\i\omega_{\vec k\lambda}t - \i \vec k \cdot \vec x_{\vec n0}} + \beta_{-\vec k\lambda} \h \vec e_{\vec k\lambda}^* \h \e^{\i\omega_{\vec k\lambda}t-\i\vec k\cdot \vec x_{\vec n0}} \h \right) \mh .
\end{align}
Taking the complex conjugate further yields
\begin{align}
 & \vec u^*(\vec x_{\vec n0}, t) = \\[3pt] \nonumber
 & \frac 1 {\sqrt N} \sum_{\vec k, \h \lambda} \left( \alpha_{-\vec k\lambda}^* \h \vec e_{\vec k\lambda} \, \e^{\i\omega_{\vec k\lambda}t + \i \vec k \cdot \vec x_{\vec n0}} + \beta_{-\vec k\lambda}^* \h \vec e_{\vec k\lambda} \h \e^{-\i\omega_{\vec k\lambda}t + \i\vec k\cdot \vec x_{\vec n0}} \h \right) \mh.
\end{align}
Comparing this with Eq.~\eqref{mode} shows that the reality condition $\vec u(\vec x_{\vec n0}, t) = \vec u^*(\vec x_{\vec n0}, t)$ is equivalent to the condition on the expansion coefficients,
\begin{equation} \label{reality_ab}
 \alpha_{\vec k\lambda} = \beta_{-\vec k\lambda}^* \,,
\end{equation}
for all $\vec k$ and $\lambda$. With this, we can rewrite Eq.~\eqref{mode} as follows (after substituting \h$\vec k \to -\vec k$ in the second term):
\begin{equation} \label{mode2}
 \vec u(\vec x_{\vec n0}, t) = \frac 1 {\sqrt N} \sum_{\vec k, \h \lambda} \left( \alpha_{\vec k\lambda} \h \vec e_{\vec k\lambda} \, \e^{-\i\omega_{\vec k\lambda}t + \i \vec k \cdot \vec x_{\vec n0}} + \alpha_{\vec k\lambda}^* \h \vec e_{\vec k\lambda}^* \h \e^{\i\omega_{\vec k\lambda}t-\i\vec k\cdot \vec x_{\vec n0}} \right) .
\end{equation}
This expression of the displacement field is manifestly real valued. Finally, we introduce the {\itshape mode expansion coefficients} by
\begin{equation} \label{redef}
 a_{\vec k\lambda} := \sqrt{\frac{2M\omega_{\vec k\lambda}}{\hbar}} \, \alpha_{\vec k\lambda}\,,
\end{equation}
where $\hbar$ is the Planck constant divided by $2\pi$. Then, the general solution of the classical equation of motion \eqref{EoMDF} can be written as
\begin{align} \label{eq_genSol}
 & \vec u(\vec x_{\vec n0}, t) = \\[3pt] \nonumber
 & \frac 1 {\sqrt N} \, \sum_{\vec k, \h \lambda} \sqrt{\frac{\hbar}{2 M \omega_{\vec k\lambda}}} \mh \left( a_{\vec k\lambda} \h \vec e_{\vec k\lambda} \, \e^{-\i\omega_{\vec k\lambda}t + \i \vec k \cdot \vec x_{\vec n0}} + a_{\vec k\lambda}^* \h \vec e_{\vec k\lambda}^* \h \e^{\i\omega_{\vec k\lambda}t-\i\vec k\cdot \vec x_{\vec n0}} \right) .
\end{align}
This is the {\itshape mode expansion} of the classical displacement field. One says that each $(\vec k, \lambda)$ labels a ``phonon'' mode, 
with $\lambda$ distinguishing between the three different {\itshape polarizations} (directions of elongation) of the mode. Correspondingly, 
the $\omega_{\vec k\lambda}$ are called the phonon {\itshape frequencies}, and the complex-valued mode expansion coefficients $a_{\vec k\lambda}$ are 
called the respective phonon {\itshape  amplitudes.} Similarly, by Eq.~\eqref{eq_conj_momentum} the 
conjugate momentum can be expressed as
\begin{align}
 & \vec{\pi}(\vec{x}_{\vec{n}0},t) \label{eq_genSol1} = \\[3pt] \nonumber
 & \frac 1 {\sqrt N} \,\sum_{\vec{k},\h\lambda}\sqrt{\frac{M \hbar \h \omega_{\vec{k}\lambda}}{2}}
\left(-\i a_{\vec{k}\lambda} \h \vec{e}_{\vec{k}\lambda} \h \e^{-{\rm i}\omega_{\vec{k}\lambda}t+{\rm i}\vec{k}\cdot\vec{x}_{\vec{n}0}}
+\i a^*_{\vec{k}\lambda} \h \vec{e}^{*}_{\vec{k}\lambda} \h \e^{{\rm i}\omega_{\vec{k}\lambda}t-{\rm i}\vec{k}\cdot\vec{x}_{\vec{n}0}} \h \right).
\end{align}
Classically, the amplitudes $a_{\vec k\lambda}$ are to be fixed by the {\itshape initial conditions} at time $t = t_0$ on the displacement field,
\begin{align}
\vec u(\vec x_{\vec n0},t_0)&\overset{!}{=}\vec u_0(\vec x_{\vec n0})\,,\\[2pt]
\vec\pi(\vec x_{\vec n0},t_0)&\overset{!}{=}\vec\pi_0(\vec x_{\vec n0})\,.
\end{align}
Concretely, choosing $t_0=0$ for the sake of convenience, the Fourier transforms of the initial conditions can be expressed as
\begin{align}
\vec u_0(\vec k)&=\sum_\lambda \sqrt{\frac{\hbar}{2M\omega_{\vec k\lambda}}}\,(a_{\vec k\lambda}+a^*_{-\vec k\lambda}) \, \vec e_{\vec k\lambda}\,, \label{eq_uini} \\[3pt]
\vec\pi_0(\vec k)&=\sum_\lambda \sqrt{\frac{M\hbar \h \omega_{\vec k\lambda}}{2}}\,(-\i a_{\vec k\lambda}+\i a^*_{-\vec k\lambda}) \, \vec e_{\vec k\lambda}\,. \label{eq_piini}
\end{align}
Conversely, we obtain the following explicit expression for the determination of the mode expansion coefficients,
\begin{align}
 a_{\vec k\lambda} = \frac{1}{\sqrt{2\hbar}}\,\vec e_{\vec k\lambda}^{*} \mh \cdot \left(\sqrt{M\omega_{\vec k\lambda}}\,\vec u_0(\vec k)+
 \frac{\i}{\sqrt{M\omega_{\vec k\lambda}}}\,\vec\pi_0(\vec k)\right).
\end{align}
Furthermore, the classical Hamiltonian \eqref{eq_phonoHamClass} evaluated at \mbox{$t = 0$} can be expressed in terms of the Fourier-transformed initial conditions as
\begin{equation} \label{eq_hphon}
\begin{aligned}
H_{\rm phon}[\vec u,\vec\pi] & = \frac{1}{2M} \h \sum_{\vec{k}}\vec{\pi}^{\dagger}_0(\vec{k})
\h \vec{\pi}_0(\vec{k}) +\frac{1}{2} \h \sum_{\vec{k}}\vec{u}^{\dagger}_0(\vec{k}) \h 
\tsr{K}(\vec{k}) \h \vec{u}_0(\vec{k})\,,
\end{aligned}
\end{equation}
where
\begin{equation}
\vec u_0^\dagger(\vec k) \equiv (\vec u_0^{\rm T})^*(\vec k) = \vec u_0^{\rm T}(-\vec k) \,. \smallskip
\end{equation}
By putting Eqs.~\eqref{eq_uini}--\eqref{eq_piini} into Eq.~\eqref{eq_hphon}, we obtain after a short calculation the {\it classical} \h Hamiltonian in terms of the phonon amplitudes:
\begin{equation}\label{eq_phonHamNormal}
H_{\rm phon}=\sum_{\vec{k}, \h \lambda}\frac{\hbar\hh\omega_{\vec{k}\lambda}}{2}\left(a^*_{\vec{k}\lambda}a_{\vec{k}\lambda}+
a_{-\vec{k}\lambda}a^*_{-\vec{k}\lambda}\right)
= \sum_{\vec{k}, \h \lambda}\hbar\hh\omega_{\vec{k}\lambda} \, a^*_{\vec{k}\lambda} \h a_{\vec{k}\lambda}\,.
\end{equation}
For this concise expression, the normalization of the mode expansion coefficients via Eq.~\eqref{redef} is crucial.
The possibility to express the energy in terms of the expansion coefficients $a_{\vec k\lambda}, \h a^*_{\vec k\lambda}$
has in fact a deeper meaning: On a fundamental level, the general solution \eqref{eq_genSol} has to be written as
\begin{align} \label{eq_genSolrev}
& \vec{u}(\vec{x}_{\vec{n}0},t) \\[3pt] \nonumber
& =\frac{1}{\sqrt N} \h \sum_{\vec{k},\h \lambda} \sqrt{\frac{\hbar}{2 M \omega_{\vec{k}\lambda}}}
\left(a_{\vec{k}\lambda}(t) \, \vec{e}_{\vec{k}\lambda} \, \e^{{\rm i}\vec{k}\cdot\vec{x}_{\vec{n}0}}
+a^*_{\vec{k}\lambda}(t)\, \vec{e}^{*}_{\vec{k}\lambda}\, \e^{-{\rm i}\vec{k}\cdot\vec{x}_{\vec{n}0}} \h \right),
\end{align}
with the time-dependent coefficients
\begin{align}
a_{\vec k\lambda}(t)&=a_{\vec k\lambda} \, \e^{-{\rm i}\omega_{\vec{k}\lambda}t}\,, \label{time_evol_1} \\[3pt]
a^*_{\vec k\lambda}(t)&=a^*_{\vec k\lambda} \, \e^{{\rm i}\omega_{\vec{k}\lambda}t}\,. \label{time_evol_2}
\end{align}
The reason for this is that the general form \eqref{eq_genSolrev} holds for {\it kinetic} reasons, i.e., the displacement field can be written in this way
independently of its equation of motion. On the other hand, the time evolution \eqref{time_evol_1}--\eqref{time_evol_2} of the expansion coefficients is determined by the {\it dynamics} of the system,
i.e.~by the equation of motion. In other words, the expansion coefficients constitute
{\it generalized coordinates} and therefore correspond to the {\it dynamical degrees of freedom}.

It is straightforward to show that the Poisson bracket relations \eqref{eq_Poi1}--\eqref{eq_Poi3} of the fundamental degrees of freedom are equivalent
to the relations
\begin{align}
\big\{a_{\vec k\lambda}(t), \h a^*_{\vec k'\lambda'}(t)\big\}&=({\rm i}\hbar)^{-1}\h \delta_{\vec k \hh \vec k'} \, \delta_{\lambda\lambda'}\,,\label{eq_Poi1a}\\[3pt]
\big\{a^*_{\vec k\lambda}(t), \h a^*_{\vec k'\lambda'}(t)\big\}&=0\,,\label{eq_Poi2a}\\[3pt]
\big\{a_{\vec k\lambda}(t), \h a_{\vec k'\lambda'}(t)\big\}&=0 \label{eq_Poi3a} \,,
\end{align}
of these generalized coordinates. We stress again that all the above considerations are completely classical. 

\subsubsection{Quantized displacement field} \label{sec_quant_disp}

In order to retrieve a full-fledged quantum theory of the nuclear system, one has to quantize the displacement field $\vec u(\vec x_{\vec n0},t)$.
As always, quantization consists in the replacement of the classical variables corresponding to the dynamical degrees of freedom with operators.
This can be done on different levels of sophistication. The most intuitive way is to step back and consider the nuclei as an $N$-particle
system with canonical coordinates $\vec x_{\vec n}$ and $\vec\pi_{\vec n}$, which is quantized by postulating the well-known {\itshape canonical \mbox{commutator relations}},
\begin{align}
\big[\hat{\vec x}_{\vec n},\hat{\vec\pi}_{\vec m}\big]&=\i\hbar\,\delta_{\vec n\vec m}\h\tsr{1}\,,\label{eq_CCR1}\\[3pt]
\big[\hat{\vec x}_{\vec n},\hat{\vec x}_{\vec m}\big]&=0\,,\label{eq_CCR2}\\[3pt]
\big[\hat{\vec\pi}_{\vec n},\hat{\vec\pi}_{\vec m}\big]&=0\,.\label{eq_CCR3}
\end{align}
These replace the classical Poisson bracket relations \eqref{eq_Poi1}--\eqref{eq_Poi3}.
The commutator relations can be realized by the operators
\begin{align}
\vec x_{\vec n} & \mapsto\hat{\vec x}_{\vec n} \,, \\
\vec \pi_{\vec m} & \mapsto\hat{\vec\pi}_{\vec m}=\frac{\hbar}{\i}\frac{\partial}{\partial\vec x_{\vec m}} \,,
\end{align}
acting on $N$-particle nuclear wave functions 
\begin{equation}
\Phi^N \equiv \Phi^N(\vec x_{\vec n_1},\ldots,\vec x_{\vec n_N})\,,
\end{equation}
which possibly also depend on the electronic coordinates. Of course, all of this is just standard quantum mechanics. 
It can be re-interpreted as a quantum field theory though, if we write the nuclear positions in terms of 
the displacement field as in Eq.~\eqref{eq_deviations}. Now, under quantization the equi\-{}librium positions $\vec x_{\vec n0}$ 
remain ordinary numbers, whereas the displacements become operators, i.e., \smallskip
\begin{equation}
\hat{\vec x}_{\vec n}=\vec x_{\vec n0} \h \hat{\mathrm I}+\hat{\vec u}(\vec x_{\vec n0})\,,  \smallskip
\end{equation}
where $\hat{\mathrm I}$ denotes the identity operator.
With this, the canonical commutator relations \eqref{eq_CCR1}--\eqref{eq_CCR3} translate into
\begin{align}
\big[\hh \hat{\vec{u}}(\vec{x}_{\vec{n}0},t), \h M\partial_t\hat{\vec{u}}(\vec{x}_{\vec{m}0},t)\hh\big]&=\i\hbar\,\delta_{\vec n\vec m}\h\tsr{1}\,,\label{eq_CCR11}\\[3pt]
\big[\hh \hat{\vec{u}}(\vec{x}_{\vec{n}0},t), \h \hat{\vec{u}}(\vec{x}_{\vec{m}0},t)\hh\big]&=0\,,\label{eq_CCR22}\\[3pt]
\big[\hh \partial_t\hat{\vec{u}}(\vec{x}_{\vec{n}0},t), \h \partial_t\hat{\vec{u}}(\vec{x}_{\vec{m}0},t)\hh\big]&=0\,.\label{eq_CCR33}
\end{align}
We conclude that {\it standard quantum mechanics for the system of $N$ nuclei is \linebreak 
equivalent to a quantum field theory for the displacement field on a lattice with $N$ points}.
From the quantum mechanical point of view, the quantization implies the transition
from {\it classical trajectories} $\vec x_{\vec n}(t)$ to a time-dependent probability amplitude, \smallskip
\begin{equation}
 \Phi^N_t(\vec x_{\vec n_1},\ldots,\vec x_{\vec n_N})\,, \smallskip
\end{equation}
for the particles to be at the positions $\vec x_{\vec n_1},\ldots,\vec x_{\vec n_N}$ at time $t$. 
By contrast, the same quantization procedure means from the quantum field theoretical point of view the transition from 
the {\it classical displacement field} $\vec u(\vec x_{\vec n0},t)$ to a time-dependent probability amplitude,
\begin{equation}
 \Phi^N_t[\hh\vec u(\vec x_{\vec n0})]\,,
\end{equation}
for the displacement field to be in the classical field configuration $\vec u(\vec x_{\vec n0})$ at the time $t$.
In particular (as is always the case) the term `quantum field theory' denotes none other than the quantum theory of fields, or put differently, quantum 
field theory is related to classical field theory in precisely the same way as quantum mechanics is related to classical mechanics.

We now come to the quantized Hamiltonian of the lattice oscillations.
The decisive fact here is that the field quantization \eqref{eq_CCR11}--\eqref{eq_CCR33} 
is equivalent to the replacement of the classical mode expansion
coefficients $a_{\vec k\lambda}$ and their complex conjugates $a^*_{\vec k\lambda}$ with operators $\hat a_{\vec k\lambda}$ and their hermitean adjoints $\hat a^\dagger_{\vec k\lambda}$ fulfilling the commutation relations
\begin{align}
\big[\hat a_{\vec{k}\lambda}, \h {\hat a}^\dagger_{\vec{k}'\lambda'}\big]&=\h \delta_{\vec{k}\vec{k}'}\,\delta_{\lambda\lambda'}\,, \label{CCRphon1}\\[3pt]
\big[\hat a_{\vec{k}\lambda}, \h \hat a_{\vec{k}'\lambda'}\big]&=0\,, \label{CCRphon2}\\[3pt]
\big[{\hat a}^\dagger_{\vec{k}\lambda}, \h {\hat a}^\dagger_{\vec{k}'\lambda'}\big]&=0\,, \label{CCRphon3}
\end{align}
which replace the classical Poisson bracket relations \eqref{eq_Poi1a}--\eqref{eq_Poi3a}. 
The displacement field \eqref{eq_genSol} now becomes an operator-valued solution,
\begin{align} \label{eq_genSolquant}
& \hat{\vec{u}}(\vec{x}_{\vec{n}0},t)=\\[5pt] \nonumber
& \frac{1}{\sqrt N} \sum_{\vec{k}, \h \lambda}\sqrt{\frac{\hbar}{2 M \omega_{\vec{k}\lambda}}}
\left(\hat a_{\vec{k}\lambda}\h\vec{e}_{\vec{k}\lambda}\, \e^{-{\rm i}\omega_{\vec{k}\lambda}t+{\rm i}\vec{k}\cdot\vec{x}_{\vec{n}0}}
+{\hat a}^\dagger_{\vec{k}\lambda} \h \vec{e}^{*}_{\vec{k}\lambda} \,\e^{{\rm i}\omega_{\vec{k}\lambda}t-{\rm i}\vec{k}\cdot\vec{x}_{\vec{n}0}}\right)\,,
\end{align}
of the equation of motion \eqref{EoMDF}. Correspondingly, the classical Hamiltonian \eqref{eq_phonHamNormal} translates into the Hamiltonian operator
\begin{equation}\label{eq_phonHamNormalquant}
\hat H_{\rm phon}=\sum_{\vec{k},\h\lambda}\hbar\hh\omega_{\vec{k}\lambda} \h {\hat a}^\dagger_{\vec{k}\lambda}\hat a_{\vec{k}\lambda}\,.
\end{equation}
This Hamiltonian is a sum of harmonic oscillator Hamiltonians (one for each mode $(\vec k,\lambda)$),
the energy eigenstates of which can easily be constructed by acting successively with the creation operators
${\hat a}^\dagger_{\vec k\lambda}$ on the quantum field theoretical ``vacuum''. In our case, this ``vacuum'' simply corresponds
to the ground state $|\Phi_0\rangle$ of the $N$-particle nuclear system with the quantized Hamiltonian \eqref{eq_phonoHamClass}. 
A general energy eigenstate then has the form (cf.~Eq.~\eqref{reprper})
\begin{equation}
|n_{1} \h ; \h \ldots \h ;\h n_{M}\rangle =
\frac{1}{\sqrt{n_1!\ldots \h n_M!}} \, ({\hat a}^\dagger_{\vec k_1\lambda_1})^{n_1}
\ldots ( {\hat a}^\dagger_{\vec k_M\lambda_M})^{n_M}\h|\Phi_0\rangle\,,
\end{equation}
with the {\itshape occupation numbers} $n_i \equiv n_{\vec k_i \lambda_i}$\h. Hence, it is interpreted as a state with $(n_1+\ldots+n_M)$ {\itshape phonons} and the total energy
\begin{equation}
\langle n_1; \ldots; n_M \mid \hat H_{\rm phon} \mid n_1; \ldots n_M \rangle=\sum_{i=1}^M n_i \h\hh \hbar\hh\omega_{\vec k_i\lambda_i} \,.
\end{equation}
We conclude that the term `phonon' by no means designates some kind of ``material entity'', let alone a (``quasi'') particle.
Instead, the phonon concept serves to enumerate the energy eigenstates of the free, quantized displacement field. In particular, 
if we encounter a state of $n_{\vec k\lambda}$ phonons, then the energy is necessarily ``localized'' in the mode $(\vec k,\lambda)$,
but it is in no way localized at some alleged positions of the phonons. Similarly, it makes no sense to speak of an electron being
``surrounded'' by a polarization ``cloud'' consisting of phonons, not to mention even more obscure allegories. In this article, we will strictly abstain from such reifications.

\subsubsection{Dynamical matrix in electronic structure theory} \label{sec_dyn_elec}

Thus far, we have obtained a decoupled theory consisting of a quantum field theory for the electrons
(in the form of the quantized Schr\"odinger field $\hat\psi(\vec x,t)$) 
and a quantum field theory for the nuclei (in the form of the quantized displacement field $\hat{\vec u}(\vec x_{\vec n0},t)$).
However, in the quantum theory of the displacement field, the dynamical matrix and its ensuing ``phonon'' frequencies are not specified.

Electronic structure theory now provides for a {\it quantitative} prediction (see e.g.~\cite{Kresse95}) of the phonon frequencies 
and the dynamical matrix by giving an explicit expression for the (yet undetermined) potential energy $V_{\rm phon}$ 
as a functional of the nuclear positions, i.e.~of the displacement field. This expression is given by the sum 
of the nuclear Coulomb interaction energy $V_{\rm n-n}$ and the electronic energy $E_{\rm e \hh 0}$ in the ground state $|\Psi_0\rangle$ (cf.~\cite[Chap.~19.1]{Martin}). 
In other words, in electronic structure theory one identifies 
\begin{equation} \label{eq_identify}
 V_{\rm phon}[\vec u] \equiv V_{\rm n}[\vec u] \,, \medskip
\end{equation}
with the nuclear potential energy $V_{\rm n}[\vec u]$ defined by Eq.~\eqref{eq_Vphon}. 
The dynamical matrix \eqref{eq_defK} is then given explicitly by the second-order derivative
\begin{equation}\label{eq_defK_elstr}
 K_{ij}(\vec{x}_{\vec{n}0},\vec{x}_{\vec{m}0}) 
 = \left. \frac{\partial^2 V_{\rm n}[\vec u]}{\partial u_i(\vec x_{\vec n 0}) 
 \h\partial u_j(\vec x_{\vec m 0})} \, \right|_{\vec u \hh = \hh \vec 0}\,.
\end{equation}
We conclude that within the limits of standard electronic structure theory,
electrons and nuclei decouple in the following sense: in the calculation of the electronic ground state,
the nuclei are assumed to be fixed at \mbox{$\vec x=\vec x_0+\vec u$}, while in the
calculation of the nuclear dynamics, one assumes a fixed Hamiltonian given in terms
of the dynamical matrix \eqref{eq_defK_elstr}.  Thus, the fundamental Hamiltonian \eqref{eq_fund} is replaced by the decoupled Hamiltonians
\begin{equation}\label{eq_elstructappr}
\hat H\mapsto\hat H_{\rm e \hh 0}+\hat H_{\rm phon}\,,
\end{equation}
where the first operator is given by Eq.~\eqref{eq_elHam}, while the second operator
results from Eq.~\eqref{eq_phonoHamClass} by replacing the classical displacement field
with its quantized analogon. The latter operator is (up to normal ordering ambiguities) equivalent to the Hamiltonian \eqref{eq_phonHamNormalquant}.

Unfortunately though, this approach is 
not capable of incorporating the {\it dynamical} effects of the electron-phonon coupling.
The reason for this is that the electronic Hamiltonian \eqref{eq_elHam} is always given in terms of
a {\it fixed} nuclear charge density $\rho_{\rm n0}$\h. The best approximation with a Hamiltonian of this form
is achieved if the input nuclear density corresponds to the exact crystal lattice. 
By contrast, dynamical effects would arise if the {\it dynamical} nuclear displacement field could act back on the electrons. 
We stress that an {\itshape exact} description of such a dynamical electron-phonon coupling is clearly {\it not possible}, simply because it
would lead back to the original, fundamental Hamiltonian \eqref{eq_fund} of electrons and {\it nuclei}.
On the other hand, the ground state of this fundamental Hamiltonian does not
factorize into a product of an electronic and a nuclear wave function, and can therefore
not exactly be described by decoupled electronic and nuclear Hamiltonians as in Eq.~\eqref{eq_elstructappr}. 

\section{Response Theory of effective interactions} \label{sec_dyndec}

The idea of the Response Theory of the electron-nuclear
coupling is to go beyond the standard electronic structure Hamiltonian \eqref{eq_elHam}
by introducing {\itshape effective interactions,} i.e., an effective electron interaction mediated by
phonons {\itshape and} an effective core interaction (i.e.~internuclear interaction) mediated by electrons. 
These effective interactions are introduced {\it in addition} to the ordinary
electron-electron or internuclear Coulomb interactions, respectively, 
and will allow for a {\itshape dynamical} decoupling of electrons 
and nuclei. 

\subsection{Dynamical decoupling of electrons and nuclei} \label{sec_RFA}

In order to derive the effective interactions in their most general form, we
rewrite the non-local but instantaneous, electron-nuclear interaction Hamiltonian as
\begin{align}
\hat V_{\rm e-n}(t)&=\int \! \de^3\vec x\mh\int \! \de^3\vec x'\,
\hat \rho_{\rm e}(\vec x,t)\, v(\vec x,\vec x')\, \hat \rho_{\rm n}(\vec x',t) \label{eq_Hintt} \\[3pt]
&=\int \! \de^3\vec x\mh\int \! \de^3\vec x'\!\int \!  c\,\de t'\,
\hat \rho_{\rm e}(\vec x,t)\, v(\vec x,t;\vec x',t')\, \hat \rho_{\rm n}(\vec x',t')\,,\label{eq_usualCoul}
\end{align}
with the time-dependent, but instantaneous Coulomb interaction kernel
\begin{equation} \label{eq_Coul}
v(\vec x,t;\vec x',t) \equiv v(\vec x - \vec x') \h \delta(c\hh t-c\hh t')=
\frac{1}{4\pi\varepsilon_0} \h \frac{\delta(c\hh t-c\hh t')}{|\vec x-\vec x'|}\,.
\end{equation}
Inspired by \cite[Eq.~(15.2)]{Hedin69}, we decompose the electronic and nuclear densities into a {\itshape reference} part and a {\itshape dynamical} (or {\itshape fluctuation}) part:
\begin{align}
\rho_{\rm e}(\vec x,t)&=\rho_{\rm e \hh 0}(\vec x)+\delta\rho_{\rm e}(\vec x,t)\,, \label{dens_decomp_e} \\[5pt]
\rho_{\rm n}(\vec x,t)&=\rho_{\rm n0}(\vec x)+\delta\rho_{\rm n}(\vec x,t)\,, \label{dens_decomp_n}
\end{align}
where $\rho_{\rm e \hh 0}(\vec x)$ is the electronic ground-state density given by
\begin{equation}
\langle\Psi_0| \h \hat\rho_{\rm e}(\vec x)\h |\Psi_0\rangle=(-e)\h ZN
\int \! \de^3\vec x_2\ldots\de^3\vec x_{ZN}\,(\Psi_0^* \h \Psi_0)(\vec x,\vec x_2,\ldots,\vec x_{ZN})\,,
\end{equation}
corresponding to the electronic Hamiltonian \eqref{eq_elHam}, while the nuclear reference density $\rho_{\rm n0}$ has been defined
in Eq.~\eqref{eq_nucrefdens}. Under quantization, the densities then turn into operators:
\begin{align}
\hat\rho_{\rm e}(\vec x,t)&=\rho_{\rm e \hh 0}(\vec x) \h \hat{\mathrm I}+\delta\hat\rho_{\rm e}(\vec x,t)\,, \label{eq_decom_e} \\[5pt]
\hat\rho_{\rm n}(\vec x,t)&=\rho_{\rm n0}(\vec x) \h \hat{\mathrm I}+\delta\hat\rho_{\rm n}(\vec x,t)\,. \label{eq_decom_n}
\end{align}
Putting these decompositions into the fundamental interaction Hamiltonian \eqref{eq_Hintt} leads in an abridged notation to 
\begin{align} \label{eq_splitting}
\hat V_{\rm e-n} & = \hat\rho_{\rm e} \g v \g \hat\rho_{\rm n} \\[5pt]
&= \hat\rho_{\rm e}\g v\g \rho_{\rm n0}+\rho_{\rm e \hh 0}\g v\g \hat\rho_{\rm n}+
\delta\hat\rho_{\rm e}\g v\g \delta\hat\rho_{\rm n}-\rho_{\rm e \hh 0}\g v\g \rho_{\rm n 0} \\[5pt]
& \equiv \hat V^{\rm n}\ext + \hat V^{\rm e}\ext +\delta^2\hat V_{\rm e-n}-V^0_{\rm e-n}\,. \label{eq_splitting_3}
\end{align}
The first term, \medskip
\begin{equation}
 \hat V_{\rm ext}^{\rm n} = \hat \rho_{\rm e} \h \varphi_{\rm ext}^{\rm n} \quad \textnormal{with} \quad \varphi_{\rm ext}^{\rm n} = v \h \rho_{\rm n0} \,, \medskip
\end{equation}
has already been defined in Eq.~\eqref{eq_extPot} and describes the electron dynamics in the external potential 
of the fixed nuclear charge density $\rho_{\rm n0}$\h. Similarly, the second term,
\begin{equation}
 \hat V_{\rm ext}^{\rm e} = \hat \rho_{\rm n} \h \varphi_{\rm ext}^{\rm e} \quad \textnormal{with} \quad \varphi_{\rm ext}^{\rm e} = v \h \rho_{\rm e \hh 0} \,, \smallskip
\end{equation}
describes the nuclear dynamics in the external potential of the fixed
electronic charge density $\rho_{\rm e \hh 0}$\h. By contrast, the third term, 
\begin{equation}
\delta^2\hat V_{\rm e-n}=\delta\hat\rho_{\rm e}\h v\h \delta\hat\rho_{\rm n}\,, \label{eq_deltaV}
\end{equation}
couples the dynamical parts of the densities. Finally, the last term in Eq. \eqref{eq_splitting_3} is constant and has therefore no bearing on the dynamics. 
Thus, the fundamental Hamiltonian \eqref{eq_fund} can be re-expressed as
\begin{equation}
\hat H=\hat H_{\rm e \hh 0}+\hat H_{\rm n0}+\delta^2\hat V_{\rm e-n}-V^0_{\rm e-n}\label{eq_alternfund}\,,
\end{equation}
where $\hat H_{\rm e \hh 0}$ was defined in Eq.~\eqref{eq_elHam}, while $\hat H_{\rm n0}$ is analogously defined as
\begin{equation} \label{hno}
\hat H_{\rm n0}=\hat T_{\rm n}+\hat V_{\rm n-n}+\hat V^{\rm e}_{\rm ext}\,.
\end{equation}
In particular, at this level of decoupling, the theory is still symmetric under the exchange of electrons and nuclei.
For later purposes, we note that the splitting of the interaction Hamiltonian can alternatively be written as
\begin{align}
\hat V_{\rm e-n} & = \hat\rho_{\rm e} \g v \g \hat\rho_{\rm n} \\[5pt]
& = \delta\hat\rho_{\rm e}\g v\g \rho_{\rm n0}+\rho_{\rm e \hh 0}\g v\g \delta\hat\rho_{\rm n}+
\delta\hat\rho_{\rm e}\g v\g \delta\hat\rho_{\rm n}+\rho_{\rm e \hh 0}\g v\g \rho_{\rm n 0}\\[5pt]
& \equiv \delta\hat V^{\rm n}\ext + \delta\hat V^{\rm e}\ext +\delta^2\hat V_{\rm e-n}+V^0_{\rm e-n}\,,
\end{align}
where this last expression exclusively involves density fluctuation operators.

The purpose of introducing {\itshape effective interactions} is now to replace the electron-nuclear
coupling $\delta^2\hat V_{\rm e-n}$ in the Hamiltonian \eqref{eq_alternfund}
with two interaction terms which involve exclusively electronic or nuclear operators, respectively, thus 
leading to a decoupling of the electronic and nuclear degrees of freedom. To put this idea into practice,
the Response Theory of effective interactions writes the classical interaction term as
\begin{equation}
\delta^2 V_{\rm e-n}=\delta\rho_{\rm e}\h v\h \delta\rho_{\rm n} = \frac{1}{2}\h \delta\rho_{\rm e}\h v\h \delta\rho_{\rm n}+
\frac{1}{2}\h \delta\rho_{\rm e}\h v\h \delta\rho_{\rm n} \,, \label{eq_tobedecoupled}
\end{equation}
and quantizes it, $\delta\rho\mapsto\delta\hat\rho$, after eliminating the nuclear density fluctuation from the
first term and the electronic density fluctuation from the second term. This can be achieved by {\itshape linear response theory}, 
where the density fluctuations in the presence of external potential variations are given in terms of the density response functions as (cf.~App.~\ref{app_Response})
\begin{align}
\delta\rho_{\rm n}(\vec x,t)&=\int \! \de^3\vec x'\!\int \!  c\,\de t' \, \upchi_{\rm n0}(\vec x,t;\vec x',t') \, \delta\varphi_{\rm ext,\h n}(\vec x',t')\,, \label{eq_feddback1} \\[2pt]
\delta\rho_{\rm e}(\vec x,t)&=\int \! \de^3\vec x'\!\int \!  c\,\de t' \, \upchi_{\rm e \hh 0}(\vec x,t;\vec x',t') \, \delta\varphi_{\rm ext,\h e}(\vec x',t')\,.\label{eq_feddback2}
\end{align}
Here, $\upchi_{\rm n0}$ and $\upchi_{\rm e \hh 0}$ refer to nuclear and electronic {\it reference} density response functions, respectively.
To get rid of the external potentials {\it exerted on} the nuclei and electrons, 
we interpret them in turn to be {\it generated by} the electronic or nuclear density fluctuations, respectively. This means, we set
\begin{align}
\delta\varphi_{\rm ext,\h n}&=\delta\varphi\ext^{\rm e}\,,\\[5pt]
\delta\varphi_{\rm ext,\h e}&=\delta\varphi\ext^{\rm n}\,,
\end{align}
where
\begin{align}
\delta\varphi_{\rm ext}^{\rm e}(\vec x',t')&=\int \! \de^3\vec x''\!\int \!  c\,\de t''\, v(\vec x',t';\vec x'',t'') \, \delta\rho_{\rm e}(\vec x'',t'')\,,\label{eq_phi_e_ext}\\[3pt]
\delta\varphi_{\rm ext}^{\rm n}(\vec x',t')&=\int \! \de^3\vec x''\!\int \!  c\,\de t''\, v(\vec x',t';\vec x'',t'')\, \delta\rho_{\rm n}(\vec x'',t'')\,.
\end{align}
In particular, if the nuclear density fluctuation in the last equation is interpreted as being generated by the fluctuations of the nuclei around their equilibrium
positions, then $\delta\varphi_{\rm ext}^{\rm n}$ corresponds to a {\it deformation potential} (see e.g.~\cite[Chap.~1.3]{Mahan}).
With this, putting Eqs.~\eqref{eq_feddback1}--\eqref{eq_feddback2} into Eq.~\eqref{eq_tobedecoupled}, 
we arrive at the expression
\begin{equation}
\delta^2 V_{\rm e-n}=\frac 1 2 \, \delta\rho_{\rm e}\, v^{\rm eff}_{\rm e-e} \, \delta\rho_{\rm e} + \frac 1 2 \, \delta\rho_{\rm n}\, v^{\rm eff}_{\rm n-n} \, \delta\rho_{\rm n} \,,
\end{equation}
where the {\itshape effective electronic or core interaction kernels} are given respectively~by \vspace{-3pt}
\begin{align}
v^{\rm eff}_{\rm e-e}&=v\,\upchi_{\rm n0}\h\hh v\,, \label{eq_veffee} \\[5pt]
v^{\rm eff}_{\rm n-n}&=v\,\upchi_{\rm e \hh 0}\h\hh v\,. \label{eq_veffnn}
\end{align}
Here, we have used again an abridged notation, where \h$v \h\hh \upchi_{0} \hh v$ \h stands for
\begin{align}
& v^{\rm eff}(\vec x,t;\h \vec x',t') = \\[3pt] \nonumber
&  \int \! \de^3\vec y\int \! c\,\de s\int \! \de^3\vec y'\mh \int \! c\,\de s'\,v(\vec x,t; \h \vec y,s) \, \upchi_0(\vec y,s;\h \vec y',s') \, v(\vec y',s'; \h\vec x',t')\,.
\end{align}
The above equations \eqref{eq_veffee}--\eqref{eq_veffnn} constitute a central result of the Response Theory of effective interactions.
In their quantized form, these effective interaction kernels give rise to the interaction Hamiltonians
\begin{align}
\delta^2\hat V_{\rm e-e}^{\rm eff}(t)&=\frac 1 2  \int\!\mh \de^3\vec x\mh\int \!\mh \de^3 \vec x'\!\int \!\mh  c\,\de t'\,\delta\hat\rho_{\rm e}(\vec x,t) 
\, v^{\rm eff}_{\rm e-e}(\vec x,t;\vec x',t')\, \delta\hat\rho_{\rm e}(\vec x',t')\,,\label{eq_effHam_e}\\[5pt]
\delta^2\hat V_{\rm n-n}^{\rm eff}(t)&=\frac 1 2  \int\!\mh \de^3\vec x\mh\int \!\mh \de^3\vec x'\!\int \!\mh  c\,\de t'\,
\delta\hat\rho_{\rm n}(\vec x,t) \, v^{\rm eff}_{\rm n-n}(\vec x,t;\vec x',t')\, \delta\hat\rho_{\rm n}(\vec x',t'). \label{eq_effHam_n}
\end{align}
These expressions are analogous to the usual Coulomb interaction \eqref{eq_usualCoul}, 
but with the Coulomb interaction kernel \eqref{eq_Coul} replaced by the effective interaction kernels
\eqref{eq_veffee} and \eqref{eq_veffnn}, and the density operators replaced by the 
density fluctuation operators defined by Eqs.~\eqref{eq_decom_e} and \eqref{eq_decom_n}. In the electronic case, this general form of a phonon-mediated electron-electron interaction can already be found in the classical article of Hedin and Lundqvist \cite[Eq.~(15.3)]{Hedin69} and in the more recent important work of R.\,van Leeuwen \cite[Eq.~(92)]{Leeuwen}.

In summary, the Response Theory of effective interactions replaces the fundamental Hamiltonian \eqref{eq_fund}
by a decoupled Hamiltonian
\begin{equation}
\hat H\mapsto \hat H^{\rm eff} = \hat H_{\rm e} + \hat H_{\rm n} - V^0_{\rm e-n}\,, 
\end{equation}
where
\begin{align}
 \hat H_{\rm e} & = \h \hat H_{\rm e \hh 0}+\delta^2 \hat V_{\rm e-e}^{\rm eff} \h = \hat T_{\rm e} + \hat V_{\rm e-e}+ \hat V_{\rm ext}^{\rm n} + \delta^2\hat V_{\rm e-e}^{\rm eff}\, , \label{eq_summary_e} \\[5pt]
 \hat H_{\rm n} & = \hat H_{\rm n0}+\delta^2 \hat V_{\rm n-n}^{\rm eff}= \hat T_{\rm n} + \hat V_{\rm n-n}+ \hat V_{\rm ext}^{\rm e} + \delta^2\hat V_{\rm n-n}^{\rm eff}\, . \label{eq_summary_n}
\end{align}
Although by the above effective Hamiltonians the electronic and nuclear
degrees of freedom {\it completely} decouple on the formal level, there is a price to pay, 
namely the enlargement of the input data. While the electronic structure Hamiltonian $\hat H_{\rm e \hh 0}$ 
only assumes the knowledge of a predetermined reference density $\rho_{\rm n0}$\h, 
the effective Hamiltonians \eqref{eq_summary_e} and \eqref{eq_summary_n}
also depend on undetermined {\it reference response functions} $\upchi_{\rm n 0}$ and $\upchi_{\rm e \hh 0}$\h.

\subsection{Total interaction kernels and screening} \label{subsec_screening}

As opposed to the {\itshape effective} interaction kernels, the {\it total} (or ``{\itshape total effective}'') interaction kernels
take into account both the ordinary Coulomb interaction {\itshape and} the effective interactions as derived
in the previous subsection. Hence, for electrons or nuclei they are respectively defined as
\begin{align}
v^{\rm tot}_{\rm e-e}&=v+v\h\hh\upchi_{\rm n0}\h v\,,\label{effint1}\\[5pt]
v^{\rm tot}_{\rm n-n}&=v+v\h\hh\upchi_{\rm e \hh 0}\h v\,.\label{effint2}
\end{align}
Both these total effective interaction kernels have the form of a Hedin-type screened interaction (see e.g.~\cite{Hedin65, Hedin69}). The latter is given by
\begin{equation} \label{eq_HedinW}
W=v+v \h\hh \upchi \h v=v+v \h\hh \widetilde \upchi \h W\,,
\end{equation}
where the {\it proper} density response function $\widetilde\upchi$ is related to the direct density response function $\upchi$ by (see Eq.~\eqref{eq_FD})
\begin{equation} \label{prop_dir}
\upchi=\widetilde\upchi+\widetilde\upchi \h v \h \upchi\,.
\end{equation}
By further introducing the (longitudinal) electronic and nuclear dielectric functions as in Eqs.~\eqref{eq_eps1}--\eqref{eq_eps2}, i.e.,
\begin{align}
 \varepsilon_{\rm n0}^{-1} & = 1 + v \h\hh \upchi_{\rm n0} \,, \\[5pt]
 \varepsilon_{\rm e \hh 0}^{-1} & = 1 + v \h\hh \upchi_{\rm e \hh 0} \,,
\end{align}
the total effective interaction kernels can now be rewritten compactly in the following form:
\begin{align}
v^{\rm tot}_{\rm e-e}&=\varepsilon^{-1}_{\rm n0} \, v\,,\label{eq_effint1}\\[5pt]
v^{\rm tot}_{\rm n-n}&=\varepsilon^{-1}_{\rm e \hh 0} \, v\,.\label{eq_effint2}
\end{align}
Therefore, we arrive at the following simple {\it interpretation of the total inter\-{}action kernels:} the electrons interact effectively with a Coulomb
potential screened {\it by the nuclei}, whereas the nuclei interact effectively with a Coulomb potential screened {\it by the electrons}. 
We remark that the form \eqref{effint1} of the total  electron-electron interaction can already be found in Ref.~\cite[Eqs.~(15.16)]{Hedin69}, 
while the rewritten form \eqref{eq_effint1} can be found in Refs.~\cite[Eq.~(26.24)]{Ashcroft} or \cite[Eq.~(4.20)]{deGennes}.
Furthermore, we remark that these considerations show that the phonon-mediated interaction combines with the usual Coulomb interaction into
a total effective interaction which in fact {\it is} the screened interaction of the electrons. Consequently, there is absolutely no need to further
screen the Coulomb kernel $v$ entering the expression \eqref{eq_effint1} for the effective electron-electron interaction by an ``electronic dielectric constant'', 
as it is usually done in the textbook literature following the ad hoc argument that this ``represent[s] the effect of the other electrons
in screening the interaction between a given pair'' \cite[p.~518]{Ashcroft}. As this would correspond to an {\itshape overscreening} 
of the electron interaction, we abstain from this procedure in this article, and in comparing
our results to the textbook literature we will always \mbox{undo its effect.}

\subsection{Total interaction operators and Hartree potentials} \label{subsec_total}

In this subsection, we consider again the effective interaction Hamiltonians defined by Eqs.~\eqref{eq_effHam_e}--\eqref{eq_effHam_n}. As mentioned above, they differ from the instantaneous Coulomb interaction in {\it two} respects: the interaction kernels are different,
{\it and} the density operators are replaced by the density fluctuation operators. The latter is due to the fact that part of the electron-nuclear interaction is already included in the respective external potentials $\hat V^{\rm n}_{\rm ext}$ and $\hat V^{\rm e}_{\rm ext}$ (see~Eqs.~\eqref{eq_summary_e}--\eqref{eq_summary_n}). 
As a consequence, one cannot directly merge the Coulomb interaction and the effective interaction into a new total interaction Hamiltonian with the interaction kernel~$v^{\rm tot}$.
To overcome this difficulty, we use again the decompositions \eqref{eq_decom_e}--\eqref{eq_decom_n} to 
rewrite the Coulomb interaction operator, e.g.~in the electronic case, as follows:
\begin{align}
\hat V_{\rm e-e} &= \frac 1 2 \h \hat\rho_{\rm e} \h v \h \hat\rho_{\rm e} \\
 & = \rho_{\rm e \hh 0} \h v \h \hat\rho_{\rm e} + \frac 1 2 \h \delta\hat\rho_{\rm e} \h v \h \delta\hat\rho_{\rm e} - \frac 1 2 \h \rho_{\rm e \hh 0} \h v \rho_{\rm e \hh 0} \\[6pt]
 & = \hat V_{\rm H}^{\rm e} + \delta^2 \hat V_{\rm e-e} - V_{\rm e-e}^0  \,.  \label{eq_coulsplitting}
\end{align}
Here, the first term, \smallskip
\begin{equation}
\hat V_{\rm H}^{\rm e}=\int \! \de^3\vec x\,\h \hat\rho_{\rm e}(\vec x) \h \varphi^{\rm e}_{\rm H}(\vec x)\,, \smallskip
\end{equation}
is a Hartree-type operator given in terms of the reference density by 
\begin{equation}
\varphi_{\rm H}^{\rm e}(\vec x) =\int \! \de^3\vec x'\,v(\vec x,\vec x') \h \rho_{\rm e \hh 0}(\vec x')\,.
\end{equation}
The second term in Eq.~\eqref{eq_coulsplitting} involves only density fluctuation operators and can therefore 
be merged with the effective electron interaction into the {\itshape total} interaction operator
\begin{equation}
 \delta^2\hat V_{\rm e-e}^{\rm tot} = \frac 1 2 \, \delta\hat\rho_{\rm e} \, v \, \delta\hat\rho_{\rm e} + 
 \frac 1 2 \, \delta\hat\rho_{\rm e} \, v_{\rm e-e}^{\rm eff} \, \delta\hat\rho_{\rm e} = 
 \frac 1 2 \, \delta\hat\rho_{\rm e} \, v_{\rm e-e}^{\rm tot} \, \delta\hat\rho_{\rm e}\,. \smallskip
\end{equation}
The last term in Eq.~\eqref{eq_coulsplitting} is again constant and has therefore no effect on the dynamics. For later purposes, we note that an alternative splitting of the Coulomb interaction operator is
\begin{equation} \label{eq_CoulPotReform}
 \hat V_{\rm e-e} = \delta \hat V_{\rm H}^{\rm e} + \delta^2 \hat V_{\rm e-e} + V_{\rm e-e}^0 \,,
\end{equation}
where \smallskip
\begin{equation}
\delta\hat V_{\rm H}^{\rm e}=\int \! \de^3\vec x\,\h \delta\hat \rho_{\rm e}(\vec x) \h \varphi^{\rm e}_{\rm H}(\vec x) \,, \smallskip
\end{equation}
and hence this latter decomposition involves only density fluctuation operators. Proceeding analogously in the nuclear case, we can finally rewrite the equations \eqref{eq_summary_e}--\eqref{eq_summary_n} as follows:
\begin{align}
 \hat H_{\rm e} & = \hat T_{\rm e} + \hat V^{\rm n}_{\rm ext} + \hat V_{\rm H}^{\rm e} + \delta^2\hat V_{\rm e-e}^{\rm tot}-V^0_{\rm e-e}\,, \label{eq_totHam_e} \\[5pt]
 \hat H_{\rm n} & = \hat T_{\rm n}  + \hat V^{\rm e}_{\rm ext} + \hat V_{\rm H}^{\rm n} + \delta^2\hat V_{\rm n-n}^{\rm tot}-V^0_{\rm n-n}\,. \label{eq_totHam_n}
\end{align}
In conclusion, we stress that a total effective interaction Hamiltonian can only be introduced if a Hartree contribution
is split off from the Coulomb interaction. In other words, it is not possible to treat the effective interaction na\"{i}vely as a first-quantized interaction kernel.

\subsection{Instantaneous limit and interactions stricto sensu} \label{subsec_InstLim}

As explained in Sec.~\ref{sec_RFA}, the effective interaction kernels will in general not have the simple form of an {\it instantaneous} interaction, i.e.,
\begin{equation}
v^{\rm eff}(\vec x,t;\vec x',t')\neq v^{\rm eff}(\vec x-\vec x') \h \delta(c\h t-c\h t')\,,
\end{equation}
as would be the case for the Coulomb interaction kernel \eqref{eq_Coul}. By consequence, the effective interaction kernels
do not lead to effective Hamiltonians {\itshape stricto sensu}. This can be seen as follows: The purely electronic Hamiltonian~$\hat H_{\rm e \hh 0}$ is given in second quantization as
\begin{equation}\label{eq_HamOrig}
\begin{aligned}
\hat H_{\rm e \hh 0}& =\int\!\de^3\vec x\,\h\hat\psi^\dagger(\vec x)\left(-\frac{\hbar^2}{2m}\Delta+v^{\rm n}\ext(\vec x)\right)\hat\psi(\vec x) \\[3pt]
&\quad \,+\frac{e^2}{2}\int\!\de^3\vec x\!\int\!\de^3\vec x'\,\hat\psi^\dagger(\vec x) \h \hat\psi^\dagger(\vec x') \h v(\vec x,\vec x') \h \hat\psi(\vec x') \h \hat\psi(\vec x)\,.
\end{aligned}
\end{equation}
Defining the time dependence of the field operators in the Heisenberg picture,
\begin{align} \label{H_pic}
\hat\psi(\vec x,t)=\e^{\i t \hat H_{\rm e \hh 0} / \hbar}\,\hat\psi(\vec x)\,\e^{-\i t \hat H_{\rm e \hh 0} / \hbar}\,,
\end{align}
and using that \smallskip
\begin{equation}
\hat H_{\rm e \hh 0} = \e^{\i t \hat H_{\rm e \hh 0}/ \hbar}\,\hat H_{\rm e \hh 0}\,\e^{-\i t \hat H_{\rm e \hh 0} / \hbar} \,, \smallskip
\end{equation}
the electronic Hamiltonian can be written equivalently as
\begin{align}
\hat H_{\rm e \hh 0}&=\int\!\de^3\vec x\,\h\hat\psi^\dagger(\vec x,t)\left(-\frac{\hbar^2}{2m}\Delta+v^{\rm n}\ext(\vec x)\right)\hat\psi(\vec x,t) \label{explicit_He} \\[3pt]
&\quad \! +\frac{e^2}{2}\int\!\de^3\vec x \int\!\de^3\vec x'\!\int \! c\,\de t'\,\hat\psi^\dagger(\vec x,t) \h \hat\psi^\dagger(\vec x',t') \h v(\vec x,t;\vec x',t')\h\hat\psi(\vec x',t')\h \hat\psi(\vec x,t)\,,\nonumber
\end{align}
where $v$ is the instantaneous Coulomb interaction kernel given by Eq.~\eqref{eq_Coul}. Although the time dependence of the field operators $\hat\psi(\vec x,t)$ is determined by the Hamiltonian~$\hat H_{\rm e \hh 0}$ itself,
this equation is still an {\itshape explicit} equation for that very Hamiltonian. The reason for this is, again, that in Eq.~\eqref{explicit_He} all time dependencies from Eq.~\eqref{H_pic} cancel, and thus we simply end up with the original form \eqref{eq_HamOrig} which does not involve any time-dependent operator. This situation changes, however, if the interaction kernel is not instantaneous, 
as in the case of the effective electronic Hamiltonian, $\hat H_{\rm e} = \hat H_{\rm e \hh 0} + \delta^2 \hat V_{\rm e-e}^{\rm eff}$\,. Now, we cannot simply drop all the time dependencies. Instead, the time dependencies of the field operators entering the effective Hamiltonian are themselves to be determined by
a suitable Hamiltonian operator. The corresponding formula for the effective interaction would therefore at best yield an {\it implicit} equation
for the determination of that effective Hamiltonian, since $\hat H_{\rm e}$ would appear in a hidden form also
in the time dependencies of the \mbox{field operators.}

In order to get an effective interaction {\itshape stricto sensu}, one therefore has to take a suitable {\it instantaneous limit},
which is most easily done directly in the interaction kernel.
To put this into practice, we first observe that the effective interaction can be written in terms of its Fourier transform as
\begin{equation}
v^{\rm eff}(\vec x,\vec x';t-t')=\int\frac{\de\omega}{2\pi c}\,\h\e^{-{\rm i}\omega(t-t')}\,v^{\rm eff}(\vec x,\vec x';\omega)\,.
\end{equation}
By approximating the frequency-dependent effective interaction in the integrand by its zero-frequency limit,
\begin{equation}
v^{\rm eff}(\vec x,\vec x';\omega) \h \mapsto \h v^{\rm eff}(\vec x,\vec x';\omega=0) \,, 
\end{equation}
and using the Fourier representation of the Dirac delta distribution,
\begin{equation}
\delta(c\h t-c\h t')=\int\frac{\de\omega}{2\pi c}\,\h\e^{-{\rm i}\omega(t-t')} \,,
\end{equation}
we obtain the concise formula for the instantaneous limit,
\begin{equation}\label{instLim}
v^{\rm eff}(\vec x,\vec x';t-t')\h \mapsto \h v^{\rm eff}(\vec x,\vec x';\omega=0) \, \delta(c \hh t - c\hh t')\,.
\end{equation}
This leads to an interaction Hamiltonian of the form
\begin{equation}
\delta^2\hat V^{\rm eff}(t) = \frac 1 2 \h \int\! \de^3\vec x\!\int \! \de^3\vec x'\,\delta\hat\rho(\vec x, t)\,
v^{\rm eff}(\vec x,\vec x';\omega=0)\,\delta\hat\rho(\vec x', t)\,, \label{eq_intham}
\end{equation}
which in fact is instantaneous. In this approximation, the effective Hamiltonian $\hat H_{\rm e}$ is a true
many-body Hamiltonian and therefore appropriate for phenomenological models, as they are typically used
in the theory of superconductivity (see e.g.~\cite{Fetter, MartinRothen, Tinkham, Schrieffer}).

Before we now come to more detailed investigations of the effective electron and core interactions, 
let us summarize again the fundamental problems of effective interactions in general: 
(i) they require the previous knowledge of reference response functions, 
(ii) they act only on the density fluctuations, not on the densities themselves, and (iii) they are in general frequency dependent. In particular, 
the last two points make it clear that effective interactions must not be employed na\"{i}vely, i.e.~analogously to the Coulomb interaction in a first-quantized Hamiltonian. 
Furthermore, a description in terms of effective interactions is inequivalent to the original coupled problem.

\section{Effective electron interaction mediated by phonons}\label{sec_mediatedInt}

In the last section, we have derived the fundamental equations \eqref{eq_veffee}--\eqref{eq_veffnn} for the effective electron-electron und nucleus-nucleus interactions, respectively. They followed from a heuristic application of linear response theory to the fundamental Hamiltonian of electrons and nuclei.
Quite as the high degree of nuclear localization and the ensuing derivation of the purely electronic structure Hamiltonian,
this decoupling scheme is far from being exact and, in particular, cannot be justified theoretically from first principles by present means.
Instead, it has to be justified by its consequences. Fortunately, the concept of phonon-mediated electron-electron interactions is already
well established in condensed matter physics and hence as such does not have to be justified anew. In this section, we will therefore corroborate
our basic ansatz, Eqs.~\eqref{eq_veffee}--\eqref{eq_veffnn}, by a systematic reproduction of the already existing standard expressions
for the phonon-mediated interaction. Essentially, there are {\it five} such standard expressions for phonon-mediated interactions:
(i) in terms of the elastic Green function, (ii) in terms of the phonon Green function, (iii) in terms of the longitudinal phonon dispersion relation,
(iv) in terms of second-order perturbation theory, and (v) in terms of the Fr\"{o}hlich Hamiltonian. 
Below, we will show that all of these can be reproduced straightforwardly from the simple ansatz \eqref{eq_veffee}.

\subsection{Rederivation of standard expressions}\label{subsec_Rederivation}
\subsubsection{Expression in terms of elastic Green function}\label{subsec_EffElInt}

The fundamental formula \eqref{eq_veffee} for the effective electron interaction mediated by the nuclear density has been derived without any assumptions about the material.
Hence it is generally valid and in particular takes into account all possible effects of inhomogeneity and anisotropy. Any material-specific property affects only the concrete form of the nuclear reference response function. 
However, due to the high degree of localization of the nuclei
around the crystal lattice points, this general formula is not always suitable for {\itshape practical} purposes. 
Taking this fact into account, we will derive in this subsection
a simplified expression for the effective electron interaction
in terms of the {\itshape elastic Green function} (the reason for this name will
become clear in App.~\ref{app_Iso}). The elastic Green function is defined as a $(3\times 3)$ matrix-valued distribution (see \cite[Eq.~(16.8)]{Hedin69}),
\begin{equation}\label{eq_DefPhonoGF}
 {-\i} \hh \hbar \h c \, D_{ij}(\vec x_{\vec n0},t;\h\vec x_{\vec m0},t')=\langle\Phi_0 \h|\h\mathcal
T\hat{u}_i(\vec x_{\vec n0},t)\h\hat{u}_j(\vec x_{\vec m0},t') \h|\hh\Phi_0\rangle\,,
\end{equation}
where $\hat{\vec u} = (\hat u_1, \hat u_2, \hat u_3)^{\rm T}$ denotes the quantized displacement field as defined in Eq.~\eqref{eq_genSolquant}.
Furthermore, $|\Phi_0\rangle$ is the ground state of the nuclear subsystem with the Hamiltonian $\hat H_{\rm phon}$\hh, which is the quantized analogon of Eq.~\eqref{eq_phonoHamClass}.
Finally, the action of the time-ordering operator $\mathcal T$ on the bosonic field operators $\hat u_i$ is given explicitly by
\begin{align}
 & \mathcal T\hat{u}_i(\vec x_{\vec n0},t)\h\hat{u}_j(\vec x_{\vec m0},t') = \\[5pt] \nonumber
 & \varTheta(t- t') \, \hat{u}_i(\vec x_{\vec n0},t)\h\hat{u}_j(\vec x_{\vec m0},t') + \varTheta(t' - t) \, \hat{u}_j(\vec x_{\vec m0},t') \h \hat{u}_i(\vec x_{\vec n0},t) \,,
\end{align}
where $\varTheta$ denotes the Heaviside step function.
The elastic Green function obeys the equation of motion (cf.~\cite[Eq.~(B.23)]{Hedin69})
\begin{equation}\label{eq_EoMPhonoGF}
\big(\big(M\partial_t^2+\tsr{K}\h\big)\tsr{D}\h\big)(\vec{x}_{\vec{n}0},t;\h \vec{x}_{\vec{m}0},t')=
\delta(c \h t-c \h t')\,\delta_{\vec{nm}}\h\tsr{1} \,,
\end{equation}
and is therefore indeed a Green function for the classical displacement field. This can be shown by using the well-known relation
\begin{equation}
\partial_t \h \varTheta(t-t')=\delta(t-t')\,,
\end{equation}
the fact that the quantized displacement field $\hat{\vec u}(\vec x_{\vec n0},t)$ fulfills the classical equation of motion \eqref{EoMDF}, 
the equal-time commutation relations \eqref{eq_CCR11}--\eqref{eq_CCR33}, and the distributional identity \eqref{distr_id}. 

We will now show that due to the high degree of nuclear localization, the nuclear density response
function can be expressed in terms of the elastic Green function. For this purpose, we start from
the classical expression for the density of point particles,
\begin{equation} \label{eq_cldens}
\rho_{\rm n}(\vec{x},t)=Ze \h \sum_{\vec{n}} \delta^3(\vec{x}-\vec{x}_{\vec{n}}(t))\,,
\end{equation}
where for simplicity we have assumed again  that the basis of the lattice consists of only one nucleus.
The first-order deviation from the equilibrium density is then given by
\begin{equation}\label{nucdensdispl}
\delta\rho_{\rm n}(\vec{x},t)=-Ze \h \sum_{\vec{n}, \h i} u_i(\vec{x}_{\vec{n}0},
t) \left( \frac{\partial}{\partial x_i} \, \delta^3(\vec{x}-\vec{x}_{\vec{n}0}) \right)
\end{equation}
in terms of the displacement field.
This function turns into an operator under the replacement $u_i(\vec{x}_{\vec{n}0},t)\mapsto\hat u_i(\vec{x}_{\vec{n}0},t)$. Furthermore, 
the (time-ordered) nuclear density response function is given in a quantum field theoretical setting by 
(see e.g.~\cite[Eq.~(6.42)]{Bruus} or App.~\ref{app_Kubo})
\begin{equation}
\upchi_{\rm n0}(\vec{x},t;\vec{x}',t') = -\frac{\i}{\hbar \h c}\h\hh\langle\Phi_0\mid \mathcal{T}\delta\hat
\rho_{\rm n}(\vec{x},t) \h \delta\hat \rho_{\rm n}(\vec{x}',t')\mid \Phi_0\rangle\,.
\end{equation}
By inserting Eq.~\eqref{nucdensdispl} into this expression and using Eq.~\eqref{eq_DefPhonoGF}, 
the nuclear density response function can be expressed in terms of the elastic Green function as follows (cf.~\cite[Eq.~(11.18)]{SchafWegener}):
\begin{align}
& \upchi_{\rm n0}(\vec x,t;\h \vec x',t') = \label{eq_genConnection}\\[3pt] \nonumber
& {-Z^2e^2}\sum_{\vec{n},\h i, \h \vec{m}, \h j}
\left( \frac{\partial}{\partial x_i} \, \delta^3(\vec{x}-\vec{x}_{\vec{n}0}) \right) D_{ij}(\vec{x}_{\vec{n}0},t;
\vec{x}_{\vec{m}0},t') \left( \frac{\partial}{\partial x'_j} \, \delta^3(\vec{x}' - \vec{x}_{\vec{m}0}) \right).
\end{align}
Putting this result into the effective phonon-mediated interaction given by Eq.~\eqref{eq_veffee} leads after partial integration to the representation
\begin{align}
& v^{\rm eff}_{\rm e-e}(\vec x,t;\h \vec x',t')=\label{eq_effWWphonGreen}\\[3pt] \nonumber
& {-Z^2 e^2} \sum_{\vec n, \h i, \h \vec m, \h j}  \left( \frac{\partial}{\partial x_i} \, v(\vec x - \vec x_{\vec n 0}) \right) D_{ij}(\vec{x}_{\vec{n}
0},t;\vec{x}_{\vec{m}0},t') \left( \frac{\partial}{\partial x'_j} \, v(\vec x_{\vec m 0} - \vec x') \right).
\nonumber
\end{align}
We further introduce the {\itshape Coulomb force kernel,} $\vec{\mathcal E}(\vec r) \equiv \vec{\mathcal E}(\vec x-  \vec x')$, which is defined in terms of the Coulomb interaction kernel by
\begin{equation}\label{eq_Coulforce}
\vec{\mathcal E}(\vec r) = -\nabla v(\vec r)=\frac{1}{4\pi\varepsilon_0} \h
\frac{\vec r}{|\vec r|^3}\,, \smallskip
\end{equation}
and whose Fourier transform is given by
\begin{equation} \label{eq_Coulforce_FT}
\vec{\mathcal E}(\vec k)=\int \! \de^3\vec r\,\h \vec{\mathcal E}(\vec r) \, \e^{-{\rm i}\vec k\cdot\vec r} =-\i \vec k \, v(\vec k) = -\frac{1}{\varepsilon_0} \h \frac{\i\vec k}{|\vec k|^2} \,.
\end{equation}
With this, we can write the result \eqref{eq_effWWphonGreen} more compactly as
\begin{align}
& v^{\rm eff}_{\rm e-e}(\vec x,t;\h \vec x',t') = \label{effelintprop} \\[5pt] \nonumber
& Z^2 e^2 \, \sum_{\vec n,\h \vec m}
\vec{\mathcal E}^{\rm \,T}(\vec x - \vec x_{\vec n0}) \h \tsr D(\vec x_{\vec n0},t;\vec x_{\vec m0},t') \, 
\vec{\mathcal E}(\vec x_{\vec m0} - \vec x')\,.
\end{align}
This formula, which relates the effective electron-electron interaction to the elastic Green function, represents the most general
expression of a phonon-mediated interaction in the linear r\'{e}gime. It includes all possible effects of inhomogeneity and anisotropy, and 
it coincides with the expression given by Hedin and Lundqvist in Ref.~\cite[Eq.~(16.7)]{Hedin69}.

For later purposes, let us also rewrite Eq.~\eqref{effelintprop} in the Fourier domain. While the Coulomb force kernels transform as in Eq.~\eqref{eq_Coulforce_FT}, 
the effective interaction kernel in the Fourier domain is given by
\begin{equation}
v^{\rm eff}_{\rm e-e}(\vec k,\vec k';\omega)=\frac{1}{(2\pi)^3}\int\!\de^3\vec x\int\!\de^3\vec x'\,\e^{-{\rm i}\vec k\cdot\vec x}\,v^{\rm eff}_{\rm e-e}(\vec x,\vec x';\h\omega)\,\e^{{\rm i}\vec k'\cdot\vec x'}\,.
\end{equation}
From this, we conclude that 
\begin{align} \label{ede1}
& v^{\rm eff}_{\rm e-e}(\vec k,\vec k';\omega) = \\[4pt] \nonumber
& -\frac{Z^2e^2}{(2\pi)^3} \h \sum_{\vec n,\h \vec m}\vec{\mathcal E}^{\dagger}(\vec k)\left(\e^{-{\rm i}\vec k\cdot\vec x_{\vec n0}}
\, \tsr D(\vec x_{\vec n0},\vec x_{\vec m0} \h;\h \omega) \, \e^{{\rm i}\vec k'\cdot\vec x_{\vec m0}}\right) \vec{\mathcal E}(\vec k')\,,
\end{align}
where we have used that
\begin{equation}
 \vec{\mathcal E}^*(\vec k) = \vec{\mathcal E}(-\vec k) = -\vec{\mathcal E}(\vec k) \,.
\end{equation}
While Eq.~\eqref{ede1} is valid for arbitrary wavevectors $\vec k, \vec k' \in \mathbb R^3$, the Fourier transformation of the elastic Green function is only defined for dual lattice vectors $\vec k, \vec k' \in \Gamma^*$. In order to simplify the above expression, we therefore have to take the {\itshape thermodynamic limit} as explained in App.~\ref{app_thLim}.
In this limit, any wavevector $\vec k$ can be decomposed as
\begin{equation}
 \vec k = \vec k_0 + \vec G \,,
\end{equation}
where $\vec k_0 \in \mathcal B$ lies in the first Brillouin zone, and $\vec G$ is a reciprocal lattice vector. 
Per definitionem, $\exp(\i\hh\vec G \mh \cdot \mh \vec x_{\vec n0}) = 1$ holds for any direct lattice vector \linebreak $\vec x_{\vec n0}$, and hence 
we can write Eq.~\eqref{ede1} in the thermodynamic limit as
\begin{align}
& v^{\rm eff}_{\rm e-e}(\vec k_0 + \vec G,\vec k'_0 + \vec G';\h\omega) = \\[5pt] \nonumber 
& -\frac{Z^2e^2}{(2\pi)^3}\sum_{\vec n,\h \vec m}\vec{\mathcal E}^{\dagger}(\vec k_0 + \vec G)\left(\e^{-{\rm i}\vec k_0\cdot\vec x_{\vec n0}}
\, \tsr D(\vec x_{\vec n0},\vec x_{\vec m0}\h; \h \omega) \, \e^{{\rm i}\vec k'_0\cdot\vec x_{\vec m0}}\right) \vec{\mathcal E}(\vec k'_0 + \vec G') \,,
\end{align}
which is equivalent to
\begin{align}
& v^{\rm eff}_{\rm e-e}(\vec k_0 + \vec G,\vec k'_0 + \vec G';\h\omega) = \\[5pt] \nonumber
& -\frac{Z^2e^2 N}{V} \, \vec{\mathcal E}^{\dagger}(\vec k_0 + \vec G) \, \tsr D(\vec k_0, \vec k'_0\h; \h \omega) \, \vec{\mathcal E}(\vec k'_0 + \vec G')\,.
\end{align}
Here, the additional factor \h$(2\pi)^3 N / V = |\mathcal B|$ \h in the last line, 
which equals the volume of the Brillouin zone $\mathcal B$, stems 
from the definition of the Fourier transform in the thermodynamic limit (see App.~\ref{app_thLim}, in particular Eqs.~\eqref{ft_dyn_1}--\eqref{ft_dyn_2}). 
Further using the continuum residue of the invariance of the elastic Green function under lattice translations,
\begin{equation}
 \tsr D(\vec k_0, \vec k'_0 \h; \h \omega) = \tsr D(\vec k_0, \omega) \, \delta^3(\vec k_0 - \vec k_0') \,,
\end{equation}
we finally obtain
\begin{equation}
 v_{\rm e-e}^{\rm eff}(\vec k_0 + \vec G, \vec k'_0 + \vec G'; \omega) = ( v_{\rm e-e}^{\rm eff} ){}^{\vec G \vec G'} (\vec k_0, \omega) \, \delta^3(\vec k_0 - \vec k'_0) \,,
\end{equation}
with
\begin{equation}
 ( v_{\rm e-e}^{\rm eff} ){}^{\vec G \vec G'} (\vec k_0, \omega) = -\frac{Z^2 e^2 N}{V} \, \vec{\mathcal E}^{\dagger}(\vec k_0 + \vec G) \, \tsr D(\vec k_0, \omega) \, \vec{\mathcal E}(\vec k_0+\vec G') \,.
\end{equation}
In particular, we see that the effective interaction in Fourier space generally depends on one wavevector $\vec k_0$ in the first Brillouin zone and two reciprocal lattice vectors $\vec G$, $\vec G'$. Usually, however, the effective interaction is assumed to be a function of one Bloch wavevector only (see e.g.~\cite[Eq.~(3.10)]{Tinkham} or \cite[Eq.~(4.7)]{Annett}). In order to recover the usual expressions, we therefore have to neglect all components except for $\vec G = \vec G' = 0$, which results in the simpler expression (valid for $\vec k \equiv \vec k_0 \in \mathcal B$),
\begin{equation} \label{eff_fourier_2_simpl}
 v_{\rm e-e}^{\rm eff}(\vec k, \omega) = -\frac{Z^2 e^2 N}{V} \, \vec{\mathcal E}^{\dagger}(\vec k) \, \tsr D(\vec k, \omega) \, \vec{\mathcal E}(\vec k) \,.
\end{equation}
By the explicit expression \eqref{eq_Coulforce_FT} of the Coulomb force kernel, 
we finally obtain the effective electron-electron interaction kernel in terms of the elastic \linebreak Green function as
\begin{equation}\label{eq_Compare1}
v^{\rm eff}_{\rm e-e}(\vec k,\omega)=-\frac{Z^2e^2 N}{\varepsilon_0^2 \h\hh V} \, \frac{\vec k^{\rm T} \h \tsr D(\vec k, \omega) \h\hh \vec k}{|\vec k|^4} \,.
\end{equation}
In the following, we will derive an even more concrete form of this effective interaction by using an explicit expression of the elastic Green function in terms of the phonon dispersion relation.

\subsubsection{Expression in terms of phonon dispersion relation} \label{subsec_ExprDispRel}

In the preceding subsection, we have derived a general but rather abstract expression for the phonon-mediated interaction.
By contrast, in this subsection we will transform this expression into a form which displays a clearcut analogy to the Coulomb interaction.
In other words, we will show how to complement the Coulomb kernel in the Fourier domain, $v(\vec k) = 1/(\varepsilon_0 \h |\vec k|^2)$, by approximate effective interaction kernels $v^{\rm eff}(\vec k,\omega)$ 
as they are frequently used in tracts on superconductivity (see e.g.~\cite{Fetter, MartinRothen, Tinkham, Schrieffer}). Starting from the elastic Green function, 
it will turn out that this can be done by employing the phonon dispersion relation.

Our aim is to derive a concrete expression for the elastic Green function, which in the wavevector-time domain reads
\begin{equation} \label{dij}
 {-\i} \hh \hbar \h c \, D_{ij}(\vec k, t - t') = \langle \Phi_0 \mid \mathcal T \hat u_i(\vec k, t) \h \hat u_j(-\vec k, t') \mid \Phi_0 \rangle \,,
\end{equation}
with $\vec k  \equiv \vec k_{\vec m} \in \Gamma^*$. For this purpose, we first express the spatial Fourier transform of the displacement field \eqref{eq_genSol} as follows (cf.~\cite[Eq.~(2.129)]{Ziman}):
\begin{equation}\label{eq_reprDisplFourier}
\vec u(\vec k,t)=\sum_\lambda \sqrt{\frac{\hbar}{2M\omega_{\vec k\lambda}}} \h \left(a_{\vec k\lambda} \h \e^{-\i\omega_{\vec k\lambda} t} + a^*_{-\vec k\lambda} \h \e^{\i\omega_{\vec k\lambda} t} \h \right) \vec e_{\vec k\lambda} \,.
\end{equation}
By putting the corresponding quantized field,
\begin{equation}
\hat{\vec u}(\vec k,t)=\sum_\lambda \sqrt{\frac{\hbar}{2M\omega_{\vec k\lambda}}} \h \left(\hat a_{\vec k\lambda} \h \e^{-\i\omega_{\vec k\lambda} t} + \hat a^\dagger_{-\vec k\lambda} \h \e^{\i\omega_{\vec k\lambda} t} \h \right) \vec e_{\vec k\lambda} \,,
\end{equation}
into Eq.~\eqref{dij} and using that in the nuclear ground state,
\begin{equation}
 \langle \Phi_0 \mid \hat a_{\vec k\lambda} \h \hat a^\dagger_{\vec k\lambda'} \mid \Phi_0 \rangle = \delta_{\lambda \lambda'} \,,
\end{equation}
we obtain the following expression for the elastic Green function:
\begin{equation}
 D_{ij}(\vec k, t -t') = \frac{\mathrm i}{c} \, \sum_{\lambda = 1}^3 \frac{\e^{-\i\omega_{\vec k\lambda} \hh |t - t'|}}{2 M \omega_{\vec k \lambda}} \,  [\vec e_{\vec k\lambda}]_i \h [\vec e_{\vec k\lambda}]_j^* \,.
\end{equation}
Here, we have used the symmetries \eqref{cond_1}--\eqref{cond_2} of the unit vectors $\vec e_{\vec k\lambda}$ and the corresponding 
eigenfrequencies $\omega_{\vec k\lambda}$\hh. 
By a Fourier transformation with respect to the time variable (cf.~\cite[Sec.~2.1]{ED1}),
\begin{equation}
 D_{ij}(\vec k, \omega) = \int \! c \, \de \tau \,\h \e^{\i \omega \tau} \h D_{ij}(\vec k, \tau) \,,
\end{equation}
we further obtain the explicit expression
\begin{equation}
D_{ij}(\vec k,\omega)=-\frac{1}{M} \h \sum_{\lambda=1}^3
\frac{1}{\omega^2-\omega_{\vec k\lambda}^2+\i\eta}\, [\vec e_{\vec k\lambda}]_i \, [\vec e_{\vec k\lambda}]_j^* \,, \label{eq_phonFT}
\end{equation}
where the regularization infinitesimal $\i\eta$ ensures that the Green function is time ordered. In the following, we will drop the regularization indicator $\i\eta$ because we are only interested
in the $\vec k$ dependence of the effective interaction.
As shown in App.~\ref{app_thLim}, Eq.~\eqref{repl_dk},
the above expression remains invariant in the thermodynamic limit where $\vec k$ ranges in the first Brillouin zone. By putting this into our result \eqref{eq_Compare1}, 
we obtain
\begin{equation} \label{eq_phonFT_gen}
 v^{\rm eff}_{\rm e-e}(\vec k,\omega)=\frac{Z^2 e^2 N}{\varepsilon_0^2 \h M V} \, 
 \sum_{\lambda=1}^3 \frac{|\vec k \cdot \vec e_{\vec k\lambda}|^2}{|\vec k|^4} \, \frac{1}{\omega^2-\omega_{\vec k\lambda}^2}\,.
\end{equation}
This is the general expression for the phonon-mediated electron interaction 
in terms of the phonon dispersion relation and the phonon polarization vectors. 

We now further assume that the solid is {\itshape isotropic,} such that the dynamical matrix is given by (see App.~\ref{app_ElTensDM})
\begin{equation}
 \tsr K(\vec k)  = K_{\rm L}(\vec k) \, \tsr P_{\rm L}(\vec k) + K_{\rm T}(\vec k) \, \tsr P_{\rm T}(\vec k) \,,
\end{equation}
with the longitudinal and transverse projection operators $P_{\rm L}$ and $P_{\rm T}$ (see \cite[Sec.~2.1]{ED1}). Hence, in this case 
the dynamical matrix is already diagonal and we can directly read off the eigenvalues, which we write as
\begin{align} 
 K_{\rm L}(\vec k) & = M \h \omega_{\vec k {\rm L}}^2 \,,\label{readoff1} \\[5pt]
 K_{\rm T}(\vec k) & = M \h \omega_{\vec k {\rm T}}^2 \,.\label{readoff2}
\end{align}
Correspondingly, there is one ``longitudinal phonon'' mode with an eigenvector $\vec e_{\vec k {\rm L}}$ parallel to $\vec k$ and two degenerate ``transverse phonon'' modes with eigen\-{}vectors $\vec e_{\vec k {\rm T}(1)}, \vec e_{\vec k {\rm T}(2)}$ perpendicular to the wavevector $\vec k$. 
The elastic Green function is therefore given by
\begin{align}
\tsr D(\vec k, \omega) & = D_{\rm L}(\vec k, \omega) \, \tsr P_{\rm L}(\vec k) + D_{\rm T}(\vec k, \omega) \, \tsr P_{\rm T}(\vec k) \\[5pt]
& = -\frac{1}{M} \h \frac{1}{\omega^2 - \omega_{\vec k{\rm L}}^2}\, \tsr P\L(\vec k)
-\frac{1}{M}\h \frac{1}{\omega^2 - \omega_{\vec k{\rm T}}^2} \, \tsr P\T(\vec k) \,. \label{expl_DL}
\end{align}
From Eq.~\eqref{eq_phonFT_gen}, we see  directly that only the longitudinal phonon mode contributes to the effective interaction, thus giving
\begin{equation} \label{eq_blatter}
v^{\rm eff}_{\rm e-e}(\vec{k},\omega)=\frac{Z^2e^2 N}{\varepsilon_0^2 \h M V} \, \frac{1}{|\vec{k}|^2} \,
\frac{1}{\omega^2-\omega^2_{\vec{k}{\rm L}}}\,.
\end{equation}
This expression of the effective electron-electron interaction in terms of the longitudinal phonon dispersion relation $\omega_{\vec k{\rm L}}$ 
coincides with the interaction derived by G.\,Blatter in \cite[Eq.~(11.14)]{Blatter}. 
It is a special case of our more general result \eqref{eq_phonFT_gen}, being valid in the isotropic limit. 
Finally, for the {\itshape total} effective interaction kernel, we obtain from Eqs.~\eqref{coul_ker} and \eqref{eq_blatter} the following expression:
\begin{align}
v^{\rm tot}_{\rm e-e}(\vec{k},\omega) & = v(\vec k) + v_{\rm e-e}^{\rm eff}(\vec k) \\[5pt]
& = \frac{1}{\varepsilon_0|\vec{k}|^2}\left(1+\frac{Z^2e^2 N}
{\varepsilon_0 \hh M V}\h\frac{1}{\omega^2-\omega_{\vec{k}{\rm L}}^2}\right). \label{eq_simplify}
\end{align}
Writing this in the form of Eq.~\eqref{eq_effint1} shows that the term in brackets,
\begin{equation}\label{eq_dielectricFctLimit}
\varepsilon_{\rm n0}^{-1}(\vec k, \omega)=1+\frac{Z^2e^2 N}{\varepsilon_0 \hh M V}\h\frac{1}{\omega^2-\omega_{\vec{k}{\rm L}}^2} \,,
\end{equation}
represents a reference dielectric function of the undamped Lorentz--Drude type (cf.~\cite[Chap.~2.2]{Fox}) with one oscillator for each $\vec{k}$ mode.
This corroborates once more the interpretation that the effective phonon-mediated {\it electron-electron} interaction is simply 
the original electronic Coulomb interaction screened by the {\it nuclei}.

\subsubsection{Expression in terms of phonon Green function} \label{subsec_ExprPhononGF}

In the preceding subsections, we have shown that the Response Theory is consistent with standard expressions 
of the phonon-mediated electron interaction in terms of the elastic Green function. In many treatises though, 
the formula \eqref{eq_Compare1} for the effective interaction is not used. Instead, one uses an apparently
different expression in terms of the {\it phonon Green function} and the {\it electron-phonon coupling strength} (see e.g.~\cite{Bruus, Annett}). 
In this subsection, we will show that this alternative form can as well be reproduced from the Response Theory of the electron-phonon coupling.

To begin with, we rewrite the mode expansion \eqref{eq_reprDisplFourier} of the displacement field in terms of a new set of time-dependent coefficients defined as
\begin{equation} \label{mod_coeff}
b_{\vec k\lambda}(t)= a_{\vec k\lambda} \h \e^{-\i\omega_{\vec k\lambda} t} + a^*_{-\vec k\lambda} \h \e^{\i\omega_{\vec k\lambda} t} \,,
\end{equation}
where $\vec k \equiv \vec k_{\vec m} \in \mathcal B$. By Eq.~\eqref{cond_1}, these coefficients have the property
\begin{equation}\label{eq_property}
b_{\vec k\lambda}(t)=b_{-\vec k\lambda}^*(t)\,.
\end{equation}
The (spatial) Fourier transform of the displacement field can then be written concisely as
\begin{equation} \label{eq_reprDisplFourier_concise}
\vec u(\vec k,t)=\sum_\lambda \sqrt{\frac{\hbar}{2M\omega_{\vec k\lambda}}}\,b_{\vec k\lambda}(t) \, \vec e_{\vec k\lambda}\,.
\end{equation}
Since both $u(\vec k, t)$ and $b_{\vec k\lambda}(t)$ have the same scaling behavior in the thermodynamic limit (see App.~\ref{app_thLim}, Eqs.~\eqref{repl_uk} and \eqref{repl_ak}), the above relation is also valid in the thermodynamic limit, where $\vec k \in \mathcal B$. On the other hand, the Fourier transform of the nuclear density fluctuation \eqref{nucdensdispl} can be expressed in this limit as
\begin{align} 
\delta\rho_{\mathrm n}(\vec k,t) &  = \frac{1}{\sqrt{(2\pi)^3}} \int \! \de^3\vec x \,\h\e^{-{\rm i}\vec k\cdot\vec x} \, \delta\rho_{\rm n}(\vec x,t) \\[5pt]
 & = -\frac{Ze}{\sqrt{(2\pi)^3}}\,  \sum_{\vec n} \i \vec k \cdot \vec  u(\vec x_{\vec n0}, t) \, \e^{-\i\vec k\cdot \vec x_{\vec n0}} \\[5pt]
 & = -Ze \,\sqrt{\frac{N}{V}} \,\h \i\vec k\cdot\vec u(\vec k,t) \label{jene} \\[5pt]
 & = -\i \h Ze \, \sqrt{\frac N V} \, \sum_\lambda \sqrt{\frac{\hbar}{2M\omega_{\vec k\lambda}}} \, (\vec k \cdot \vec e_{\vec k\lambda}) \, b_{\vec k\lambda}(t) \,. \label{gl}
\end{align}
The above relation is valid for all $\vec k \in \mathbb R^3$, provided that the quantities $\omega_{\vec k \lambda}$, $\vec e_{\vec k \lambda}$ and $b_{\vec k\lambda}(t)$ 
are continued to periodic functions on $\mathbb R^3$ (cf.~the discussion in the preceding section). Correspondingly, the electron-phonon coupling can be written as
\begin{align}
 \delta^2 V_{\rm e-n}(t) & = \int \! \de^3 \vec k \,\h \delta\rho_{\rm n}(\vec k,t) \h v(\vec k) \h \delta\rho_{\rm e}(\vec k,t) \label{pre_ElPhonCoup} \\[5pt]
 & = \int \! \de^3\vec k \h \sum_\lambda g_\lambda(\vec k) \h b_{\vec k\lambda}(t) \, \delta\rho_{\rm e}(\vec k,t)\,, \label{eq_ElPhonCoup}
\end{align}
where we have introduced the {\it electron-phonon coupling strength}  as
\begin{equation}\label{eq_CouplingStrength}
g_\lambda(\vec k)=-\i \, \sqrt{\frac{\hbar \h N}{2M \omega_{\vec k\lambda} V }}\,\frac{Ze}{\varepsilon_0}\, \frac{\vec k \cdot \vec e_{\vec k\lambda}}{|\vec k|^2} \,.
\end{equation}
We now define the {\it phonon Green function} in Fourier space as
\begin{equation}
{-\i} \hh \hbar \h c \, \mathscr D_{\lambda\lambda'}(\vec k,t-t')=\langle\Phi_0|\h\mathcal T\h\hat b_{\vec k\lambda}(t)\h \hat b^\dagger_{\vec k\lambda'}(t')\h|\Phi_0\rangle\,,
\end{equation}
where again $\vec k \equiv \vec k_{\vec m} \in \Gamma^*$. The quantized analogon of Eq.~\eqref{mod_coeff} is given by
\begin{equation} \label{eq_property_q}
 \hat b_{\vec k\lambda}(t)= \hat a_{\vec k\lambda} \h \e^{-\i\omega_{\vec k\lambda} t} + \hat a^\dagger_{-\vec k\lambda} \h \e^{\i\omega_{\vec k\lambda} t} \,,
\end{equation}
and has the property that
\begin{equation}
 \hat b^\dagger_{\vec k\lambda}(t) = \hat b_{-\vec k\lambda}(t) \,.
\end{equation}
The phonon Green function is diagonal per constructionem. Thus, following the calculation in the previous subsection, 
we find its explicit expression in terms of the phonon dispersion relation,
\begin{equation} \label{Dphon_expl}
 \mathscr D_{\lambda\lambda'}(\vec k, \omega) = -\frac{2}{\hbar} \, \frac{\omega_{\vec k \lambda}}{\omega^2 - \omega_{\vec k\lambda}^2 + \i \eta} \, \delta_{\lambda \lambda'}\,.
\end{equation}
In particular, we can express the elastic Green function \eqref{eq_phonFT} in terms of the phonon Green function as follows:
\begin{equation}\label{eq_phonoVSelasticGF}
D_{ij}(\vec k,\omega)=\frac{\hbar}{2M} \h \sum_{\lambda = 1}^3 \frac{1}{\omega_{\vec k\lambda}} \, \mathscr D_{\lambda\lambda}(\vec k,\omega) \, [\vec e_{\vec k\lambda}]_i \h [\vec e_{\vec k\lambda}]^*_j\,,
\end{equation}
which squares with \cite[Eq.~(16.9)]{Hedin69}.
By putting this expression into Eq.~\eqref{eq_Compare1} and further using Eq.~\eqref{eq_CouplingStrength}, the phonon-mediated electron interaction  can finally be recast into the following form:
\begin{equation}\label{eq_Compare2}
v^{\rm eff}_{\rm e-e}(\vec k,\omega) = -\sum_{\lambda = 1}^3 |g_{\lambda}(\vec k)|^2 \, \mathscr D_{\lambda\lambda}(\vec k,\omega) \,. \smallskip
\end{equation}
This is the desired expression of the effective electron interaction in terms of the phonon Green function and the electron-phonon coupling strength. 

In the isotropic limit, where the longitudinal and transverse phonon modes decouple, we have by Eq.~\eqref{eq_CouplingStrength},
\begin{equation} \label{gL}
 g_{\rm L}(\vec k) = -\i \, \sqrt{\frac{\hbar \h N}{2M \omega_{\vec k{\rm L}} V}}\,\frac{Ze}{\varepsilon_0|\vec k|} \,,
\end{equation}
while $g_{\rm T}(\vec k) = 0$. Consequently, Eq.~\eqref{eq_Compare2} reduces to the standard expression (see e.g.~\cite[Eq.~(17.16)]{Bruus} or \cite[Eq.~(8.105)]{Coleman}) \smallskip
\begin{equation} \label{sta}
v^{\rm eff}_{\rm e-e}(\vec k,\omega)=-|g_{\rm L}(\vec k)|^2\,\mathscr D_{\rm L}(\vec k,\omega) \h \equiv \h g_{\rm L}^2(\vec k) \,\mathscr D_{\rm L}(\vec k, \omega) \,. \smallskip
\end{equation}
We particularly recognize in this context that this connection can already be found in the profound work \cite[Eq.~(16.11)]{Hedin69}.
Finally, by putting Eq.~\eqref{gL} and the explicit expression for the longitudinal phonon Green function,
\begin{equation} \label{dL}
 \mathscr D_{\rm L}(\vec k, \omega) = -\frac{2}{\hbar} \, \frac{\omega_{\vec k {\rm L}}}{\omega^2 - \omega_{\vec k{\rm L}}^2} \,, \smallskip
\end{equation}
into Eq.~\eqref{sta}, we recover again the result \eqref{eq_blatter} from the preceding subsection. 
In summary, we have re-expressed the effective phonon-mediated interaction in terms of the phonon Green function and the phonon coupling strength, 
and we have shown that the resulting formula \eqref{eq_Compare2} reduces in the isotropic limit to the well-known equation \eqref{sta}.

\subsubsection{Connection to second-order perturbation theory}\label{subsec_2OrdPert}

In this subsection, we rederive yet another well-known expression for the effective interaction, 
which is derived from second-order perturbation theory (see e.g.~\cite[Eq.~(26.38)]{Ashcroft}). The resulting expression is, however, only valid in the instantaneous limit.
Therefore, we restrict ourselves to the nuclear subsystem 
and treat the electronic density $\rho_{\rm e}(\vec x)$ as an external parameter function. 
Let $E_0 \equiv E_{\rm n0}^{(0)}$ be the energy of the  ground state of the nuclear subsystem, 
where we first assume the electron-phonon coupling to be switched off, i.e., \smallskip
\begin{equation}
E_{\rm n0}^{(0)}=\langle\Phi_0 \mid \hat H_{\rm n0} \mid \Phi_0\rangle\,, \smallskip
\end{equation}
with $\hat H_{\rm n0}$ given by Eq.~\eqref{hno}.
We now consider the change of this energy in second-order perturbation theory, where the perturbation is given by the electron-phonon coupling,
\begin{equation} \label{op}
\delta^2 \hat V_{\rm e-n}=\delta\hat \rho_{\rm n} \h v \, \delta \rho_{\rm e}\,.
\end{equation}
Applying standard perturbation theory (see e.g.~\cite[Sec.~11.1]{SchwablQM}), we obtain up to second order in the perturbation,
\begin{align}
E_{\rm n0} & = E_{\rm n0}^{(0)} + E_{\rm n0}^{(1)} + E_{\rm n0}^{(2)} + \h \ldots \\[3pt]
 & \equiv E_0 + \langle\Phi_0\mid \delta^2\hat V_{\rm e-n}\mid \Phi_0\rangle + 
\sum_{s>0}\frac{\langle\Phi_0\mid \delta^2 \hat V_{\rm e-n}\mid \Phi_s\rangle\langle\Phi_s\mid \delta^2\hat V_{\rm e-n}\mid \Phi_0\rangle}{E_0-E_s}\,,\label{eq_PertExpansion}
\end{align}
where we have assumed an orthonormal and complete set of eigenfunctions $\ket{\Phi_s}$, $s \in \mathbb N_0$\hh, of $\hat H_{\rm n0}$ with corresponding eigenenergies $E_{s}$\hh. 
Now, the first-order term in Eq.~\eqref{eq_PertExpansion} vanishes because the expectation value acts only the nuclear degrees of freedom, and
\begin{equation}
 \langle \Phi_0 \mid \delta \hat \rho_{\rm n} \mid \Phi_0 \rangle \h =  \h \langle \Phi_0 \mid \hat \rho_{\rm n} \mid \Phi_0 \rangle - \rho_{\rm n0} \h = \h 0 
\end{equation}
by the definition of $\rho_{\rm n0}$\hh. By contrast, the second-order perturbative correc-\linebreak tion to the energy of the nuclei reads explicitly
\begin{align}
& E_{\rm n0}^{(2)}=\int \! \de^3\vec x\int \! \de^3\vec x'\int \! \de^3\vec y\int \! \de^3\vec y'\\[5pt] \nonumber
&\times \delta\rho_{\rm e}(\vec x) \, v(\vec x,\vec y) \, \bigg( \h \sum_{s>0}
\frac{\langle\Phi_0\mid \delta\hat \rho_{\rm n}(\vec y)\mid \Phi_s\rangle\langle\Phi_s\mid \delta\hat \rho_{\rm n}(\vec y')\mid \Phi_0\rangle}{E_0-E_s}\bigg) \h v(\vec y',\vec x')\, \delta\rho_{\rm e}(\vec x')\,.
\end{align}
By the symmetry of the integrand under exchanging variables ($\vec x \leftrightarrow \vec x'$ and $\vec y \leftrightarrow \vec y'$), only the real part of the term in brackets gives a non-zero contribution. By Eq.~\eqref{eq_statRF}, the latter equals half the  (instantaneous) nuclear  density response function, and hence we can write the whole contribution as
\begin{equation}
\begin{aligned}
E_{\rm n0}^{(2)} & = \frac 1 2 \h \int\!\de^3\vec x\!\int\!\de^3\vec x'\!\int\!\de^3\vec y\!\int\!\de^3\vec y' \\[5pt]
& \quad \, \times \delta\rho_{\rm e}(\vec x) \, v(\vec x,\vec y) \, \upchi_{\rm n0}(\vec y,\vec y') \, v(\vec y',\vec x')\, \delta\rho_{\rm e}(\vec x')\,.
\end{aligned}
\end{equation}
From this, we directly read off the central statement of this subsection: 
the instantaneous limit of the effective electron-electron interaction \eqref{eq_veffee} can be characterized as the functional derivative
\begin{equation}\label{eq_effIntAshcroft1}
v^{\rm eff}_{\rm e-e}(\vec x,\vec x';\h \omega=0)\h=\h\frac{\delta^2 E_{\rm n0}^{(2)}}{\delta\rho_{\rm e}(\vec x) \h \delta\rho_{\rm e}(\vec x')}\,.
\end{equation}
Equivalently, we find in Fourier space
\begin{equation}
E_{\rm n0}^{(2)}= \frac 1 2 \h \int\!\de^3\vec k\!\int\!\de^3\vec k'\,
\delta\rho^*_{\rm e}(\vec k) \, v(\vec k) \, \upchi_{\rm n0}(\vec k,\vec k') \, v(\vec k')\, \delta\rho_{\rm e}(\vec k')\,,
\end{equation}
and hence \smallskip
\begin{equation}\label{eq_effIntAshcroft}
v^{\rm eff}_{\rm e-e}(\vec k,\vec k'; \h \omega = 0) \h 
=\h v(\vec k)\,\upchi_{\rm n0}(\vec k,\vec k')\h\hh v(\vec k')\,. \smallskip
\end{equation}
This formula connects the instantaneous phonon-mediated electron-electron interaction to second-order perturbation theory and reproduces the standard result \cite[Eq.~(26.38)]{Ashcroft}.

\subsubsection{Connection to Fr\"{o}hlich Hamiltonian}

Finally, we come to yet another frequently encountered form of the phonon-mediated electron-electron interaction,
which is derived from the so-called Fr\"{o}hlich Hamiltonian \cite{Froehlich2,Han,TaylorHeinonen} (cf.~also \cite{Wegner,Mielke,Kehrein} 
for a variant of this approach using continuous flow equations).
For the convenience of the reader, we begin by explaining how this Hamiltonian
is obtained from the fundamental Hamiltonian of electrons and nuclei. For this purpose, one first {\it neglects} the Coulomb interaction $V_{\rm e-e}$ of the electrons completely, but includes the external potential $V_{\rm ext}^{\rm n}$ of the nuclei exerted on the electrons. The resulting electronic part of the Fr\"{o}hlich Hamiltonian
is then given by a one-particle operator reading in second quantization,
\begin{equation}
\hat H_{\rm e \hh 0}=\int_V \de^3\vec x\,\h\hat\psi^\dagger(\vec x) \h \bigg({-\frac{\hbar^2}{2m} \h \Delta}+v_{\rm ext}^{\rm n}(\vec x)\bigg) \h \hat\psi(\vec x)\,. \label{free_el} 
\end{equation}
Here, $V$ denotes the volume of the system, for which we assume Born--von-Karman boundary conditions (see App.~\ref{app_bvk}). We further choose a one-particle basis of normalized {\itshape Bloch functions} \cite{Ashcroft,Kittel},
\begin{equation}
 \psi_{n\vec k}(\vec x) = u_{n\vec k}(\vec x) \h \e^{\i\vec k \cdot \vec x} \,,
\end{equation}
indexed by the Bloch momentum $\vec k$ and the band index $n$, where $u_{n\vec k}(\vec x)$ has the periodicity of the crystal lattice. The electronic field operator can then be written as \smallskip
\begin{equation}
\hat\psi(\vec x)=\sum_{\vec k \h \in\h \Gamma^*} \sum_n \psi_{n\vec k}(\vec x) \h \hat c_{n\vec k}\,, \smallskip
\end{equation}
where we denote by $\hat c_{n \vec k}$ and $\hat c^\dagger_{n\vec k}$ the annihilation and creation operators of the Bloch states 
(see \cite[App.~A.4]{Schober}, \cite[App.~B]{Salmhofer}, \cite[Chap.~1]{Nagaosa} and App.~\ref{sec:quantSF} for short introductions).
As these Bloch states diagonalize the one-particle Hamiltonian,
\begin{equation}
 \bigg({-\frac{\hbar^2}{2m} \h \Delta}+v_{\rm ext}^{\rm n}(\vec x)\bigg) \h \psi_{n\vec k}(\vec x) = \varepsilon_{n\vec k} \h \psi_{n \vec k}(\vec x) \,,
\end{equation}
the free electronic Hamiltonian \eqref{free_el} can be written equivalently as
\begin{equation}
\hat H_{\rm e \hh 0}=\sum_{\vec k \h \in\h \Gamma^*} \sum_n \varepsilon_{n\vec k} \h \hat c^\dagger_{n\vec k} \hh \hat c_{n\vec k}\,.
\end{equation}
Next, one drops the band index $n$ and considers the Bloch momentum $\vec k$
as a bona fide Fourier wavevector, meaning that $\vec k \in \mathbb R^3$. Hence, one interprets $\hat c_{\vec k}$ and $\hat c^\dagger_{\vec k}$ 
as annihilation and creation operators of {\itshape plane waves,}
\begin{equation}
 \psi_{\vec k}(\vec x) = \frac{1}{\sqrt V} \, \e^{\i \vec k \cdot \vec x} \,,
\end{equation}
and writes the free electronic Hamiltonian as
\begin{equation}
\hat H_{\rm e \hh 0}=\sum_{\vec k}\varepsilon_{\vec k} \h \hat c^\dagger_{\vec k} \hh \hat c_{\vec k}\,.
\end{equation}
Furthermore, for the nuclei one performs the usual transition to the phononic degrees of freedom and stipulates a harmonic Hamiltonian $\hat H_{\rm phon}$ of the form \eqref{eq_phonHamNormalquant}.
Finally, for the electron-phonon coupling one invokes again the standard ansatz \eqref{pre_ElPhonCoup}--\eqref{eq_ElPhonCoup} where, however,
{\it the electronic density operator is resubsti\-{}tuted for the density fluctuation operator}. 
For this electronic density, we find
\begin{align}
\hat\rho_{\rm e}(\vec x)&=(-e) \h \hat\psi^\dagger(\vec x) \h \hat\psi(\vec x) \\[6pt]
&=-\frac{e}{V}\sum_{\vec k, \h \vec k'}\e^{{\rm i}(\vec k-\vec k')\cdot\vec x} \,\hat c^\dagger_{\vec k'} \hh \hat c_{\vec k} \\[5pt]
&=-\frac{e}{V}\sum_{\vec k, \h \vec q}\e^{{\rm i}\vec q\cdot\vec x} \,\hat c^\dagger_{\vec k-\vec q} \hh \hat c_{\vec k}\,,
\end{align}
and hence, by Fourier transformation (see App.~\ref{app_bvk}, Eq.~\eqref{bvk_Fourier}),
\begin{equation} \label{standard_dens}
\hat\rho_{\rm e}(\vec q)=-\frac{e}{\sqrt{V}}\sum_{\vec k} \hat c^\dagger_{\vec k-\vec q} \hh \hat c_{\vec k}\,.
\end{equation}
For the electron-phonon coupling, one thus obtains
\begin{align}
\hat H_{\rm int} &=\sum_{\vec q, \h \lambda} g_\lambda(\vec q) \h \big(\hat a_{\vec q\lambda} +\hat a_{-\vec q\lambda}^\dagger \big) \h \hat\rho_{\rm e}(\vec q) \label{eq_ElPhonCoup_disc} \\[5pt]
 &=-\frac{e}{\sqrt{V}}\sum_{\vec q, \h \vec k, \h \lambda} g_\lambda(\vec q) \h \big(\hat a_{\vec q\lambda} +\hat a_{-\vec q\lambda}^\dagger \big) \h \hat c^\dagger_{\vec k-\vec q} \h \hat c_{\vec k} \,. \label{eq_ElPhonCoup_res}
\end{align}
We note that the integral in Eq.~\eqref{eq_ElPhonCoup}, which was derived in the thermodynamic limit, is replaced by a discrete sum in Eq.~\eqref{eq_ElPhonCoup_disc}, while all prefactors remain the same. This is in accord with our conventions for the Born--von-Karman boundary conditions as explained in App.~\ref{app_bvk}. Furthermore, the result \eqref{eq_ElPhonCoup_res} concides precisely with \cite[Eq.~(3.41)]{Bruus}. 

Collecting all terms together, the Fr\"{o}hlich Hamiltonian now reads as follows (cf.~\cite[Eq.~(2.15)]{Froehlich2} or \cite[Eqs.~(5.40)--(5.42)]{Han}, \cite[Eq.~(6.1.1)]{TaylorHeinonen}):
\begin{align}
\hat H'&= \hat H_{\rm e \hh 0} + \hat H_{\rm phon} + \hat H_{\rm int} \\[8pt]
& =\h \sum_{\vec k}\varepsilon_{\vec k} \h \hat c^\dagger_{\vec k} \hh \hat c_{\vec k}+
\sum_{\vec{q},\h\lambda}\hbar\hh\omega_{\vec{q}\lambda} \h {\hat a}^\dagger_{\vec{q}\lambda}\hat a_{\vec{q}\lambda} \nonumber \\[3pt]
&\quad \,-\frac{e}{\sqrt V} \, \sum_{\vec q, \h \vec k, \h \lambda} g_\lambda(\vec q) \h \big(\hat a_{\vec q\lambda}+\hat a_{-\vec q\lambda}^\dagger\big) \h \hat c^\dagger_{\vec k-\vec q} \h \hat c_{\vec k} \,.
\end{align}
We summarize again the approximations which lead from the fundamental Hamiltonian of electrons and nuclei to the
Fr\"{o}hlich Hamiltonian: (i) The electrons are treated as being free (see \cite[Eqs.~(2.9)--(2.10)]{Froehlich2}). 
(ii) For the nuclei, the usual second-order Hamiltonian in terms of the displacement field is employed where, however, the dynamical matrix remains unspecified. 
(In the original work \cite{Froehlich2} the index $\lambda$ is also dropped, thus leaving only a single branch of the phonon dispersion.)
(iii) The wavevector $\vec q$ of the phonons is treated as a bona fide Fourier variable in analogy to the wavevector $\vec k$
of the free electrons. In other words, $\vec q$ is not restriced to the first Brillouin zone, and hence the displacement
field is treated in the continuum limit (see \cite[Eq.~(2.2)]{Froehlich2}). (iv) In the electron-phonon coupling,
the displacement field is assumed to couple to the electronic density (as opposed to the density fluctuation).
We note that in H.~Fr\"ohlich's original publication, the displacement field is even assumed to couple directly to the electronic density,
i.e.~without mediation of the Coulomb interaction as in Eq.~\eqref{pre_ElPhonCoup} (see \cite[Eq.~(2.11)]{Froehlich2}).

Already in the original work \cite{Froehlich2}, the Fr\"{o}hlich Hamiltonian served as the starting point to derive an effective electron interaction
mediated by phonons. For this purpose, the ``method of canonical transformation''
(see \cite[p.~229]{Han}, \cite[p.~225]{TaylorHeinonen}) is commonly employed, according to which the original Fr\"{o}hlich Hamiltonian is {\itshape replaced} by a new Hamiltonian written as
\begin{equation}
\hat H_{\rm F}=\e^{-\hat S}\h \hat H' \, \e^{\hat S}\,,
\end{equation}
such that ``the second-order term in the transformed Hamiltonian is identified as the effective electron-electron interaction'' \cite[pp.~229f.]{Han}.
Choosing the undetermined operator $\hat S$ appropriately and neglecting terms which
``obviously do not give rise to electron-electron interactions'', one arrives at an 
``effective electron-electron interaction'' that ``is mediated actually by virtual phonons that do not correspond to real excitations''.
The overall result is a standard two-particle interaction (see \cite[Eq.~(2.4)]{Bardeen57}, \cite[Eq.~(5.49)]{Han}, \cite[Eq.~(6.5.4)]{TaylorHeinonen}, cf.~also \cite[Eq.~(26.37)]{Ashcroft}),
\begin{equation} \label{FroEff}
\hat V^{\rm F}_{\rm e-e}=\frac{e^2}{2V}\sum_{\vec q, \h \vec k, \h \vec k'} v^{\rm F}_{\rm e-e}(\vec q, \h \vec k) \, \hat c^\dagger_{\vec k-\vec q} \h \hat c^\dagger_{\vec k'+\vec q} \h \hat c_{\vec k'} \hh \hat c_{\vec k}\,,
\end{equation}
whose interaction kernel is given by
\begin{equation}\label{eq_FroInt}
v^{\rm F}_{\rm e-e}(\vec q,\vec k)=\sum_\lambda |g_{\lambda}(\vec q)|^2 \h \frac{2\h \hbar\hh\omega_{\vec q \lambda}}{(\varepsilon_{\vec k}-\varepsilon_{\vec k-\vec q})^2-(\hbar\hh\omega_{\vec q \lambda})^2} \,.
\end{equation}
These formulae differ from the ones obtained until now as they (i) apparently follow from a completely different philosophy,
and (ii) yield a different overall result. In the following, we will explain these differences, 
and thereafter show that the above effective interaction can nonetheless be reproduced from the Response Theory proposed in this article.

Regarding the first point, the main {\itshape conceptual} difference is that the effective interaction kernel derived by H.~Fr\"ohlich follows from a ``canonical transformation'' of the original Hamiltonian. Strictly speaking, this can be justified only if the corresponding transformation,
\begin{equation}
 \ket{\Psi} \mapsto \e^{-\hat S} \h \ket{\Psi} \,,
\end{equation}
is also performed on the physical states (cf.~\cite[Sec.~2.2]{Bykov}). Consequently, the transformed Hamiltonian $H_{\rm F}$ as such does not even represent an effective interaction between ordinary {\itshape electrons}. In fact, this last point has been made clear in the original work \cite{Froehlich2} by H.~Fr\"ohlich, who notes that  the ``physical meaning of the quantities $\hat c_{\vec k}$ [\ldots] is different [from the original electronic annihilation operators; notation adapted]''. These operators only satisfy the same anticommutation relations, and hence, ``it is not necessary to use a new notation''. In particular, according to \cite[Eq.~(3.3)]{Froehlich2}, the electronic density can no longer be obtained from these operators by the standard expression \eqref{standard_dens}, but is instead determined by
\begin{equation}
\hat\rho_{\rm e}(\vec q)=\e^{-\hat S} \, \bigg( {-\frac{e}{\sqrt{V}}}\sum_{\vec k} \hat c^\dagger_{\vec k-\vec q} \h \hat c_{\vec k} \bigg) \h\hh \e^{\hat S} \,.
\end{equation}
On the other hand, the standard expression \eqref{standard_dens} now gives the density of ``particles described by plane waves with amplitudes $\hat c_{\vec k}$ [notation adapted] satisfying Fermi statistics. 
These particles are electrons, carrying with them some lattice deformations'' (cf.~remarks in Sec.~\ref{sec_quant_disp}).

Regarding the second point, the main {\itshape formal} difference is that the Fr\"ohlich interaction kernel depends non-trivially on {\itshape two} 
wavevectors (instead of one wavevector and one frequency). Strictly speaking, it is therefore in general neither of the repulsive nor of the attractive type,
because its inverse Fourier transform is not a function of the distance $|\vec r|=|\vec x-\vec x'|$ alone. Note, however, that ``just such an attractive interaction 
is present [\ldots] whenever $|\varepsilon_{\vec k}-\varepsilon_{\vec k-\vec q}| < \hbar\hh \omega_{\vec k}$'' \cite[p.~225]{TaylorHeinonen}.
More precisely, ``the phonon-mediated electron-electron interaction can be attractive for electrons whose energies fall in the shell 
$|\varepsilon - \varepsilon_{\rm Fermi}| < \hbar\hh\omega_{\rm D}$ about the Fermi surface, where $\varepsilon$ is the energy of an electron and $\omega_{\rm D}$ the Debye
frequency'' \cite[p.~229]{Han}. 

For a clearer comparison, recall that in Sec.~\ref{subsec_ExprPhononGF} we have derived an effective frequency-dependent interaction of the standard type with the interaction kernel (see Eqs.~\eqref{Dphon_expl} and \eqref{eq_Compare2})
\begin{equation}\label{eq_Comparison}
v^{\rm eff}_{\rm e-e}(\vec q,\omega) \h \equiv \h  v(\vec q) \h \chi_{\rm n0}(\vec q, \omega) \h v(\vec q) \h = \h \sum_\lambda |g_{\lambda}(\vec q)|^2 \h \frac{2 \h \hbar\hh\omega_{\vec q \lambda}}{(\hbar\hh\omega)^2-(\hbar\hh\omega_{\vec q\lambda})^2} \,.
\end{equation}
Curiously, this result coincides exactly with the one derived from the Fr\"{o}hlich Hamiltonian, Eq.~\eqref{eq_FroInt}, provided that the frequency is identified with the energy differences determined by the free electron dispersion relation. In particular, this means that we can write the Fr\"ohlich interaction as
\begin{align}
 v^{\rm F}_{\rm e-e}(\vec q,\vec k) & = v(\vec q) \h\hh \chi_{\rm n0}\big(\vec q,(\varepsilon_{\vec k} - \varepsilon_{\vec k - \vec q})/\hbar \hh \big) \hh v(\vec q) \\[5pt]
 & \equiv v^{\rm eff}_{\rm e-e}\big(\vec q, \h (\varepsilon_{\vec k} - \varepsilon_{\vec k - \vec q})/\hbar \hh \big)\,,
\end{align}
in accord with our fundamental ansatz \eqref{eq_veffee}. On the other hand, performing in our result \eqref{eq_Comparison} the instantaneous limit as explained in Sec.~\ref{subsec_InstLim}
would yield a standard interaction of the density-density type reading in second quantization, \smallskip
\begin{equation}
\hat V^{\rm eff}_{\rm e-e}=\frac{e^2}{2V} \sum_{\vec q, \h \vec k, \h \vec k'} v^{\rm eff}_{\rm e-e}(\vec q) \, \hat c^\dagger_{\vec k-\vec q} \h \hat c^\dagger_{\vec k'+\vec q} \h \hat c_{\vec k'} \h \hat c_{\vec k}\,. \smallskip
\end{equation}
The kernel of this interaction is different from the one employed in the Fr\"{o}hlich approach, Eq. \eqref{FroEff}, and in particular depends on only one wavevector.

Despite the aforementioned conceptual and formal differences, we will now show that the Fr\"ohlich interaction \eqref{eq_FroInt} can 
nonetheless also be reproduced from the Response Theory of the electron-phonon coupling. 
In order to do this, we start again from the fundamental expression \eqref{eq_effHam_e} for the effective electron-electron interaction, i.e.,
\begin{equation} \label{eq_effHam_e_densities}
 \hat V_{\rm e-e}^{\rm eff}(t) =\frac 1 2 \h \int_V\!\mh \de^3\vec x\mh\int_V \!\mh \de^3 \vec x'\!\int \!\mh  c\,\de t'\,\hat\rho_{\rm e}(\vec x,t) 
\, v^{\rm eff}_{\rm e-e}(\vec x - \vec x', \h t - t')\, \hat\rho_{\rm e}(\vec x',t')\,.
\end{equation}
Here, we have assumed a translation-invariant interaction kernel and---in accordance with the approximation used 
by H.~Fr\"ohlich---resubstituted the densities for the density fluctuations. By Fourier transforming the densities and the interaction kernel 
with respect to the spatial variables, and partially also with respect to the time variables, Eq.~\eqref{eq_effHam_e_densities} is equivalent to
\begin{equation}
 \hat V_{\rm e-e}^{\rm eff}(t) =\frac 1 {2 \hh \sqrt{2\pi}} \h \sum_{\vec q}  \int \frac{\de \omega}{c} \, \e^{-\i\omega t}\h \hat\rho_{\rm e}(-\vec q,t) 
\, v^{\rm eff}_{\rm e-e}(\vec q, \omega)\, \hat\rho_{\rm e}(\vec q, \omega)\,.
\end{equation}
As explained already in Sec.~\ref{subsec_InstLim}, such a formal interaction entails the problem that one must get rid of the time dependencies (or frequency dependencies in Fourier space)
in order to get an interaction operator {\it stricto sensu}. As we will now show, in the case of the Fr\"{o}hlich approach 
one has to evaluate the effective interaction $V^{\rm eff}_{\rm e-e}(t)$
at time $t=0$, such that
\begin{equation}
 \hat V_{\rm e-e}^{\rm eff}(t = 0) =\frac 1 {2 \hh \sqrt{2\pi}} \h \sum_{\vec q}  \int \frac{\de \omega}{c} \, \hat\rho_{\rm e}(-\vec q, \h t = 0) 
\, v^{\rm eff}_{\rm e-e}(\vec q, \omega)\, \hat\rho_{\rm e}(\vec q, \omega)\,. \label{eq_GenInstLim}
\end{equation}
Furthermore, one has to make an ansatz for the frequency dependence of the second density operator, where---again in accordance with H. Fr\"ohlich's approach---the simplest possibility is a free-electron approximation. In this approximation, the time-dependent density reads (see Eq.~\eqref{standard_dens})
\begin{align}
 \hat \rho_{\rm e}(\vec q, t) & = -\frac{e}{\sqrt V}\sum_{\vec k} \hat c^\dagger_{\vec k-\vec q}(t) \h \hat c_{\vec k}(t) \\[5pt]
 & = -\frac{e}{\sqrt V}\sum_{\vec k} \e^{-\i t (\varepsilon_{\vec k} - \varepsilon_{\vec k - \vec q})/\hbar} \,\hat c^\dagger_{\vec k-\vec q} \h \hat c_{\vec k} \,,
\end{align}
and hence, by Fourier transformation, we find for the frequency-dependent density operator the expression
\begin{equation}
\hat\rho_{\rm e}(\vec q,\omega)
=-e \,\hh \sqrt{\frac{2\pi}{V}} \h \sum_{\vec k}  c \, \delta\big(\omega-(\varepsilon_{\vec k} - \varepsilon_{\vec k - \vec q})/\hbar\big) \h \hat c^\dagger_{\vec k-\vec q} \h \hat c_{\vec k} \,.
\end{equation}
By putting this into Eq.~\eqref{eq_GenInstLim}, we obtain
\begin{equation}
\hat V^{\rm eff}_{\rm e-e}(t = 0) = \frac{e^2}{2V} \sum_{\vec q, \h \vec k, \h \vec k'} v^{\rm eff}_{\rm e-e}\big(\vec q, \h \omega = (\varepsilon_{\vec k} - \varepsilon_{\vec k - \vec q})/\hbar \hh \big) \h \hat c^\dagger_{\vec k - \vec q} \hh \hat c^\dagger_{\vec k' + \vec q} \hh \hat c_{\vec k'} \h \hat c_{\vec k} \,,
\end{equation}
where, as always, the operators have been brought into normal order after plugging in the expression for the density.
Comparing the above result with the direct instantaneous limit derived in Sec.~\ref{subsec_InstLim}, we see precisely the effect of including the frequency dependence in the density operator: instead
of neglecting the frequency in the effective interaction kernel, it is now evaluated at the energy difference of the one-electron levels. This corresponds exactly 
to the difference between Eqs.~\eqref{eq_FroInt} and \eqref{eq_Comparison}, which we wanted to explain within the Response Theory of the electron-phonon coupling.

In summary, we have shown that the effective electron-electron interaction as originally derived from the Fr\"{o}hlich Hamiltonian
can also be reproduced straightforwardly from the Response Theory proposed in this article.

\subsection{Justification from functional integral approach}\label{sec_JustFI}

In the preceding subsections, we have shown that the simple and straightforward ansatz \eqref{eq_veffee} reproduces all
standard expressions for the effective phonon-mediated interaction. In particular, we have rederived the effective electron-electron interaction
in terms of (i) the elastic Green function, (ii) the phonon dispersion relation, (iii) the phonon Green function, (iv) second-order perturbation theory, and
(v) the Fr\"ohlich Hamiltonian.
Satisfying as these results may be from a heuristic vantage point, they still have the drawback of lacking a more rigorous justification
from established physical principles. The goal of this closing subsection therefore precisely lies in the physical underpinning of the ad hoc procedure
used in Sec.~\ref{sec_RFA}. For this purpose, we will take recourse to quantum field theory, where the concept
of effective interactions is fortunately well established. In particular, this statement applies to the functional integral approach, where such effective
interactions for a given subset of degrees of freedom are derived by {\itshape integrating out} the remaining degrees freedom. We will show that this well-established notion of effective interactions, if applied properly to the system of electrons and phonons, exactly reproduces the Response Theory
originally obtained by purely heuristic considerations.

\subsubsection{Displacement field in the continuum limit} \label{disp_cont}

In order to apply the machinery of both classical and quantum field theory to the elastic solid formed by the nuclei, we consider in this subsection the displacement field in the thermodynamic {\itshape and} continuum limit (see Apps.~\ref{app_thLim}--\ref{app_comblim}). Thus, we consider
\begin{equation}
 \vec u \equiv \vec u(\vec x, t) \,,
\end{equation}
with $\vec x$ ranging in the three-dimensional continuous space $\mathbb R^3$. 
This formulation will simplify several formulae, although in principle, (quantum) field theory can be formulated for a finite crystal as well.
Classical field theory, for its part, is best formulated in terms of a classical action,
\begin{equation} \label{Sn}
S_{\rm n0}=\int \! \de t\,L_{\rm n0}(t)\,.
\end{equation}
In the case of an elastic solid, we assume the Lagrangrean to be given by the difference of the kinetic and the potential energy,
\begin{equation}
L_{\rm n0}(t) \equiv L_{\rm n0}[\vec u(t), \partial_t\vec u(t)]=T_{\rm n}[\partial_t\vec u(t)]-V_{\rm n}[\vec u(t)]\,,\label{eq_classactionDF}
\end{equation}
which is analogous to point-particle mechanics (see e.g.~\cite[Chap.~2]{Goldstein}). Correspondingly, 
the kinetic energy functional is given by (cf.~\cite[Sec.~11.1]{FungTong})
\begin{equation}
	T_{\rm n}[\partial_t\vec u]=\frac{\rho_0}{2}\int \! \de^3\vec x\,\h(\partial_t\vec u)(\vec x,t)\mh\cdot\mh(\partial_t\vec u)(\vec x,t)\,, \label{kin_func}
\end{equation}
which is the direct continuum generalization of the kinetic energy of a finite set of point particles. In particular,
\begin{equation}
 \rho_0 = \frac{NM}{V} \label{mass_dens}
\end{equation}
denotes the reference {\itshape mass} density, which is related to the reference {\itshape charge} density of the nuclear elastic solid by
\begin{equation}\label{eq_refChargeDens}
\rho_{\rm n0} \h \equiv\h \frac{N Ze}{V} \h =\h \frac{Ze}{M} \h \rho_0\,.
\end{equation}
The dynamics of the system is determined by the extremization of the action,
\begin{equation} \label{extr}
	\frac{\de S_{\rm n0}}{\de\vec u(\vec x,t)}\overset{!}{=}0\,,
\end{equation}
where the {\itshape total functional derivative} is defined as
\begin{equation}
\frac{\de}{\de u_i(\vec x,t)} \equiv \frac{\delta}{\delta u_i(\vec x,t)}+ \h \sum_j \int \! \de^3 \vec x' \int\! c \,\de t' \,\h \frac{\delta(\partial_{t'} u_j(\vec x',t'))}{\delta u_i(\vec x,t)} \h \frac{\delta}{\delta(\partial_{t'}  u_j(\vec x',t'))}\,. \label{func_der}
\end{equation}
By using the trivial identities
\begin{align}
\frac{\delta(\partial_{t'} u_j(\vec x',t'))}{\delta u_i(\vec x,t)}
 & = \frac{\delta}{\delta u_i(\vec x, t)} \int \! c \, \de s \,\h \delta(c \hh t' - c \hh s) \, (\partial_s u_j)(\vec x', s) \\[5pt]
 & = \frac{\delta}{\delta u_i(\vec x, t)} \int \! c \, \de s \, \frac 1 c \h (\partial_{t'} \delta)(t' - s) \, u_j(\vec x', s) \\[5pt]
 & = \delta_{ij} \, \delta^3(\vec x - \vec x') \, \frac 1 c \h (\partial_{t'} \delta)(t'-t) \,,
\end{align}
putting this result into Eq.~\eqref{func_der} and performing a partial integration, we see that
\begin{equation}
 \frac{\de}{\de u_i(\vec x,t)} \equiv \frac{\delta}{\delta u_i(\vec x,t)} - \frac{\partial}{\partial t} \h \frac{\delta}{\delta(\partial_t u_i(\vec x,t))} \,. \smallskip  \vspace{2pt}
\end{equation}
Therefore, Eq.~\eqref{extr} is indeed equivalent to the  Euler--Lagrange equations (cf.~Ref.~\cite[Eq.~(11.1.1)]{WeinbergQM}),
\begin{equation}
  \frac{\delta L_{\rm n0}(t)}{\delta \vec u(\vec x,t)} - \frac{\partial}{\partial t} \h \frac{\delta L_{\rm n0}(t)}{\delta(\partial_t\vec u(\vec x,t))} \h= \h 0\,.
\end{equation}
These are the direct continuum analoga of the Euler--Lagrange equations for point particles.
By the explicit expressions \eqref{eq_classactionDF}--\eqref{kin_func}, they lead to the equation of motion for the displacement field,
\begin{equation} \label{cont_eom}
	\rho_0 \, \frac{\partial^2}{\partial t^2} \h \vec{u}(\vec x,t)=-\frac{\delta V_{\rm n}[\vec u]}{\delta\vec u(\vec x,t)}\,,
\end{equation}
which is the continuum version of the Newtonian equation of motion,
\begin{equation}
M \h \frac{\partial^2}{\partial t^2} \h \vec{x}_i(t)=-\frac{\partial}{\partial\vec x_i} \h V(\vec x_1,\ldots,\vec x_N) \h \bigg|_{\h\vec x_j \hh =\h\vec x_j(t) \, \forall j}\,,
\end{equation}
for $N$ point-particles with mass $M$ interacting through a potential $V$.

In accordance with the microscopic Hamiltonian \eqref{eq_phonoHamClass}, we now write the potential energy in Eq.~\eqref{eq_classactionDF} as
\begin{equation} \label{pot_lag}
 V_{\rm n}[\vec u] = \frac 1 2 \h \int \! \de^3 \vec x \int \! \de^3 \vec x'\, \vec u^{\rm T}(\vec x, t) \h\hh \tsr K(\vec x-\vec x') \, \vec u(\vec x', t) \,,
\end{equation}
or equivalently, in Fourier space,
\begin{equation}\label{eq_potEnContLim}
 V_{\rm n}[\vec u] = \frac 1 2 \h \int \! \de^3 \vec k \,\h  \vec u^{\dagger}(\vec k, t) \h\hh \tsr K(\vec k) \, \vec u(\vec k, t) \,.
\end{equation}
Here, $K(\vec x - \vec x')$ is the dynamical matrix in the thermodynamic and continuum limit.
The Euler--Lagrange equations then lead to the most general linear, homogeneous, Newtonian equation of motion with a time-independent wave-operator,
\begin{equation}
	\rho_0 \h \frac{\partial^2}{\partial t^2} \h \vec u(\vec x,t)+\int \! \de^3\vec x'\,\tsr K(\vec x-\vec x') \,\vec u(\vec x',t)=0\,,\label{eq_EoMcontlimit}
\end{equation}
which can be rewritten in Fourier space as
\begin{equation}
\left(-\rho_0 \, \omega^2 \h \tsr 1 +\tsr K(\vec k)\right)  \vec u(\vec k,\omega)=0\,.
\end{equation}
These are the direct continuum analoga of Eqs.~\eqref{EoMDF} and \eqref{evp}.

Up to now, linear terms have been neglected in the Lagrangean \eqref{eq_classactionDF}.
This was motivated by the expansion of the displacement field around the nuclear equilibrium positions, where the restoring forces vanish (see Sec.~\ref{sec_reconsider}).
However, this logic does not apply in the presence of {\it external} forces on the solid. These can be included through the linear term \cite[Eq.~(1.52)]{Kleinert892},
\begin{equation} \label{lag_int}
L_{\rm int}(t)= \int\!\de^3\vec x\,\h\vec u(\vec x,t)\cdot\vec f(\vec x,t)\,,
\end{equation}
where $\vec f(\vec x, t)$ represents the spatially and temporally varying external force field. 
By adding this term to the Lagrangean \eqref{eq_classactionDF}, we are led to the inhomogeneous equation of motion
\begin{equation}\label{eq_EoMisotropInhomo}
\rho_0 \h \frac{\partial^2}{\partial t^2} \h \vec u(\vec x,t)+\int\!\de^3\vec x'\,\tsr K(\vec x-\vec x') \, \vec u(\vec x',t)=\vec f(\vec x,t)\,.
\end{equation}
The {\it elastic Green function} of this classical equation of motion is now introduced by the defining equation (cf.~\cite[Sec.~1.5]{Kleinert892})
\begin{equation}
\rho_0 \h \frac{\partial^2}{\partial t^2} \h \tsr D(\vec x,t;\vec x',t')+\int\!\de^3\vec y\,\tsr K(\vec x-\vec y) \h \tsr D(\vec y,t;\vec x',t')=
\tsr 1 \h \delta^4(x-x')\,,\label{eq_GFdef}
\end{equation}
where the four-dimensional Dirac delta reads explictly
\begin{equation}
\delta^4(x-x')=\delta^3(\vec x-\vec x') \, \delta(c \h t-c \h t') \,.
\end{equation}
The formula \eqref{eq_GFdef} is the continuum analogon of Eq.~\eqref{eq_EoMPhonoGF}. Given an elastic Green function satisfying this equation, a particular solution of the equation of motion \eqref{eq_EoMisotropInhomo} can be written as
\begin{equation} \label{uresp}
\vec u(\vec x,t)=\int \! \de^3\vec x'\mh\int \! c \, \de t'\,\tsr D(\vec x,t;\vec x',t') \h \vec f(\vec x',t')\,. 
\end{equation}
The elastic Green function itself is most easily constructed in Fourier space, where its defining equation reads
\begin{equation}\label{eq_EoMFourier}
\left(-\rho_0 \h\hh \omega^2\h\tsr 1+\tsr K(\vec k)\right) \mh \tsr D(\vec k,\omega)=\tsr 1\,.
\end{equation}
Now, precisely as in the discrete case (see the discussion in Sec.~\ref{cdf}), the dynamical matrix in Fourier space is hermitean, and we are hence entitled to 
introduce $\vec k$-dependent polarization vectors,
\begin{equation}
\tsr K(\vec k) \h \vec e_{\vec k\lambda}=\rho_0\, \omega^2_{\vec k\lambda}\h \vec e_{\vec k\lambda}\,.
\end{equation}
In terms of these, the Green function can be written as
\begin{equation} \label{D_concr}
\tsr D(\vec k,\omega)=-\frac{1}{\rho_0} \h \sum_{\lambda=1}^3 \frac{\vec e_{\vec k\lambda} \h \vec e^\dagger_{\vec k\lambda}}{\omega^2-\omega^2_{\vec k\lambda}}\,,
\end{equation}
which is the continuum analogon of Eq.~\eqref{eq_phonFT}. In real space, the elastic Green function is then given by
\begin{equation}
\tsr D(\vec x-\vec x',t-t')=-\frac{1}{\rho_0}\h\sum_{\lambda=1}^3\h \int\mh\frac{\de^3\vec k}{(2\pi)^3}\int\mh\frac{\de\omega}{2\pi c}\,
\frac{\e^{{\rm i}\vec k\cdot(\vec x-\vec x')-{\rm i}\omega(t-t')}}
{\omega^2-\omega^2_{\vec k\lambda}}\,\vec e_{\vec k\lambda}\, \vec e^\dagger_{\vec k\lambda}\,.
\end{equation}
As always, this expression defines a whole class of Green functions due to the undefined integration over the poles.
In order to get a concrete solution, one has to apply a suitable regularization procedure. Concretely, $\omega \mapsto \omega + \i\eta$ \h or $\omega\mapsto\omega-\i\eta$ \h
defines the {\it retarded} or the {\it advanced} Green function, respectively, while $\omega\mapsto\omega+{\rm sign}(\omega)\i\eta$ \h yields the {\it time-ordered} Green function.

Finally, to employ {\itshape quantum} field theoretical methods, one has perform the transition from a classical to a quantized displacement field,
\begin{equation}
\vec u(\vec x,t)\mapsto\hat{\vec u}(\vec x,t)\,.
\end{equation}
This is done by postulating the canonical commutator relations,
\begin{align}
\big[\h\hat{\vec{u}}(\vec x,t), \h \rho_0 \h \partial_t\hat{\vec{u}}(\vec x',t)\h\big]&=\i\hbar\,\delta^3(\vec x-\vec x')\tsr 1\,,\label{CCR1}\\[3pt]
\big[\h\hat{\vec{u}}(\vec x,t),\h \hat{\vec{u}}(\vec x',t)\h\big]&=0\,, \\[3pt]
\big[\h\partial_t\hat{\vec{u}}(\vec x,t), \h \partial_t\hat{\vec{u}}(\vec x',t)\h\big]&=0\,,
\end{align}
which are the continuum analoga of Eqs.~\eqref{eq_CCR11}--\eqref{eq_CCR33}. They turn the classical displacement field into an operator field, 
while all other classical relations remain the same. For example, the quantized displacement field is an 
operator-valued solution of the classical equation of motion \eqref{eq_EoMcontlimit}, and the canonical momentum translates into the operator given by
\begin{equation}
\hat{\vec{\pi}}(\vec x,t)=\rho_0 \h \partial_t \h \hat{\vec u}(\vec x,t)\,. \label{eq_CanMomTimeDer}
\end{equation}
Hence, we can re-express Eq.~\eqref{CCR1} as
\begin{equation}
\big[\h\hat{\vec{u}}(\vec x,t),\h \hat{\vec{\pi}}(\vec x',t)\h\big]=\i\hbar\,\delta^3(\vec x-\vec x')\tsr 1\,,\label{eq_CCR}
\end{equation}
which is the continuum analogon of the canonical commutator for point particles.
Within the quantum field theoretical setting, we define the {\itshape time-ordered} elastic Green function as follows (cf.~Eq.~\eqref{eq_DefPhonoGF} in the discrete case):
\begin{equation}\label{eq_QFTGreenFct}
{-\i} \hh \hbar \h c \, D_{ij}(\vec x,t;\vec x',t')=\langle\Phi_0 \mid \mathcal T\hat{u}_i(\vec x,t) \h \hat{u}_j(\vec x',t') \mid\Phi_0\rangle\,,
\end{equation}
where $\mathcal T$ is the time-ordering operator, and the expectation value \mbox{$\langle\Phi_0|\cdot|\Phi_0\rangle$} refers to a time-independent reference state
(typically, the ground state or a thermodynamic ensemble). Contrary to the quantized displacement field, the Green function is not operator valued
but an ordinary complex-valued function (or distribution, to be more precise). In fact, one can show that---quite as its name suggest---this quantum field theoretical Green function is a Green function in the sense of Eq.~\eqref{eq_GFdef} (cf.~the remarks at the beginning of Sec.~\ref{subsec_EffElInt}).

\subsubsection{Classical Lagrangean and functional integral} \label{fi_nucl}

We now come to the derivation of the effective interaction by quantum field theoretical methods. We first give a very brief introduction
to the functional integral calculus as applied to the displacement field in the continuum limit, and then go on to rederive
the effective phonon-mediated electron-electron interaction within this formalism. After that, we will show 
that this phonon-mediated interaction can be derived even more directly by a classical elimination procedure. 
The derivations in this section are completely analogous to the corresponding derivations of the 
photon-mediated interaction in quantum electrodynamics as explained in App.~\ref{rev_qed}.

\bigskip \noindent
{\itshape Elastic Green function in the functional integral formalism.}---In the quantum field theoretical setting, 
the time-ordered elastic Green function can be obtained from the functional integral with sources (see e.g.~\cite{Salmhofer, Altland}),
\begin{equation} \label{ZJ}
Z[\vec J] =\int\! \mathcal D\vec u\, \exp \mh \bigg( \frac{1}{\hbar} \h S_{\rm n0}[\vec u] + \frac{1}{\hbar \h c} \int \! \de^4 x \,\h\vec u(x) \cdot\vec J(x)\bigg)\,,
\end{equation}
where $S_{\rm n0}[\vec u]$ denotes the classical action \eqref{Sn} for the displacement field, as a second-order functional derivative, 
\begin{equation} \label{fder}
\frac{1}{\hbar \h c} \, D_{k\ell}(x, x')= \frac{1}{Z_0} \h \frac{\delta^2 Z[\vec J]}{\delta J_k(x) \h \delta J_\ell(x')} \h \bigg|_{\vec J \h \equiv \h 0}\,,
\end{equation}
where $Z_0 = Z[\vec J \equiv 0]$. The proof of this representation is based on the evalua-\linebreak tion of Gaussian functional integrals, which will also be needed later in the rederivation of the effective phonon-mediated interaction from the functional integral formalism. For the convenience of the reader, we therefore provide a brief heuristic proof of the above representation of the Green function. We first rewrite the kinetic term of the Lagrangean \eqref{eq_classactionDF} as
\begin{equation}
T_{\mathrm n} = \frac{\rho_0}{2} \int \! \de^3 \vec x \int \! \de^3 \vec x' \mh \int \! \de t' \, (\partial_t \vec u)^{\rm T}(\vec x, t) \, \delta^3(\vec x - \vec x') \h \delta(t - t') \, (\partial_{t'} \vec u)(\vec x', t') \,,
\end{equation}
which by partial integration is equivalent to
\begin{equation}
T_{\mathrm n} = -\frac{\rho_0}{2} \int \! \de^3 \vec x \int \! \de^3 \vec x' \mh \int \! \de t' \, \vec u^{\rm T}(\vec x, t) \, \delta^3(\vec x - \vec x') \h (\partial_t^2  \h \delta)(t - t') \, \vec u(\vec x', t') \,.
\end{equation}
Similarly, the potential term \eqref{pot_lag} can be written as
\begin{equation}
 V_{\mathrm n} = \frac 1 2 \h \int \! \de^3 \vec x \int \! \de^3 \vec x' \mh \int \! \de t' \, \vec u^{\rm T}(\vec x, t) \, \tsr K(\vec x - \vec x') \h \delta(t - t') \, \vec u(\vec x', t') \,.
\end{equation}
The action \eqref{Sn}, which is the time integral of the Lagrangean, can therefore be expressed as
\begin{equation} \label{free_n}
 S_{\mathrm n0} = -\frac 1 {2 \h  c} \h \int \! \de^4 x \mh \int \! \de^4 x' \, \vec u^{\rm T}(x) \, \tsr D{}^{-1}(x, x') \, \vec u(x') \,,
\end{equation}
where
\begin{equation}
 \tsr D{}^{-1}(x, x')  = \rho_0 \, \frac{\partial^2}{\partial t^2} \, \delta^3(\vec x - \vec x') \h \delta(c \h t - c \h t') \h \tsr 1 + \tsr K(\vec x - \vec x') \, \delta(c \h t - c \h t')
\end{equation}
is the inverse of the elastic Green function defined by Eq.~\eqref{eq_GFdef}. This is in fact a general relation (see e.g.~\cite[p.~216]{Ryder}): 
the classical Green function is the inverse of the operator in the free (bilinear) part of the classical Lagrangean. In particular, 
we can now write the functional integral \eqref{ZJ} formally as \smallskip
\begin{equation} \label{zw_3}
 Z[\vec J] = \int \! \mathcal D \vec u \, \exp \mh \bigg( \frac{1}{\hbar \h c} \h \bigg( {-\frac 1 2} \h \vec u^{\rm T} \h \tsr D{}^{-1} \h \vec u + \vec u^{\rm T} \vec J \bigg) \bigg). \smallskip
\end{equation}
This is a Gaussian functional integral (see e.g.~\cite{Salmhofer}), which can be evaluated explicitly as follows: by completing the square in the exponent, we have
\begin{equation} \label{form_val}
\begin{aligned}
 Z[\vec J] & = \exp \mh \bigg( \frac 1 {2 \h \hbar \h c} \, \vec J^{\rm T} \h \tsr D \h \vec J \bigg) \\[5pt]
 & \quad \, \times \int \! \mathcal D \vec u \, \exp \mh \bigg( {-\frac 1 {2 \hh \hbar \h c} } \, (\vec u - \tsr D \h \vec J)^{\rm T} \, \tsr D{}^{-1} \h (\vec u - \tsr D \h \vec J) \bigg)  \,.
\end{aligned}
\end{equation}
Note that for this formula to be valid, it is necessary that the Green function is symmetric, $D^{\rm T} = D$, or more explicitly,
\begin{equation} \label{sym_D}
 D_{ij}(x, x') = D_{ji}(x', x) \,.
\end{equation}
This property holds indeed for time-ordered Green functions, to which the functional integral formalism applies (as opposed to, say, retarded Green functions).
Coming back to Eq.~\eqref{form_val}, we can now substitute
\begin{equation}
 \vec u - \tsr D \h \vec J \mapsto \vec u
\end{equation}
to obtain the concise result for the Gaussian integral,
\begin{equation} \label{zw_4}
 Z[\vec J] = Z_0 \h \exp \mh \bigg( \frac 1 {2 \h \hbar \h c} \, \vec J^{\rm T} \h \tsr D \h \vec J \bigg) \,.
\end{equation}
From this, we can further evaluate the functional derivatives in Eq.~\eqref{fder} as follows: by expanding the exponential,
\begin{equation}
 \exp \mh \bigg( \frac 1 {2 \hh \hbar \h c} \, \vec J^{\rm T} \h \tsr D \h \vec J \bigg) = 1 + \frac 1 {2 \h \hbar \h c} \,\vec J^{\rm T} \h \tsr D \h \vec J + \mathcal O(\vec J^4) \,,
\end{equation}
and using that only the quadratic term gives a non-vanishing contribution to the second derivative evaluated at $\vec J \equiv 0$, we obtain
\begin{align}
\frac{1}{Z_0} \h \frac{\delta^2 Z[\vec J]}{\delta J_i(x) \h \delta J_j(x')} \h \bigg|_{\vec J \h \equiv \h 0} & = \frac{\delta^2}{\delta J_i(x) \h \delta J_j(x')} \h \exp \mh \bigg( \frac 1 {2 \h \hbar \h c} \, \vec J^{\rm T} \h \tsr D \h \vec J \bigg) \bigg|_{\vec J \h \equiv\h  0} \\[5pt]
& = \frac{\delta^2}{\delta J_i(x) \h \delta J_j(x')} \, \bigg( \frac 1 {2 \h \hbar \h c} \, \vec J^{\rm T} \h \tsr D \h \vec J \bigg) \\[8pt]
& = \frac 1 {2 \h \hbar \h c} \, \big( D_{ij}(x, x') + D_{ji}(x', x) \big) \\[8pt]
& = \frac{1}{\hbar \h c} \, D_{ij}(x, x') \,,
\end{align}
which was the assertion. Note that in the last step, we have used again the symmetry \eqref{sym_D} of the elastic Green function. 

\bigskip \noindent
{\itshape Phonon-mediated interaction in the functional integral formalism.}---In order to derive an effective electron interaction in the functional integral forma\-lism, we need a field theoretical Lagrangean 
for the electrons coupled to the displacement field. Formally, we write this Lagrangean as
\begin{equation} \label{coup_lag}
 L[\vec u, \psi] = L_{\rm n0}[\vec u] + L_{\rm int}[\vec u, \psi] + L_{\rm e \hh 0}[\psi] \,.
\end{equation}
Here, $L_{\rm n0}[\vec u]$ and $L_{\rm e\hh 0}[\psi]$ denote the purely nuclear or
electronic parts of the classical action, respectively.  Concretely, $L_{\rm n0}[\vec u]$ is given by Eq.~\eqref{eq_classactionDF}, 
while $L_{\rm e \hh 0}[\psi]$ may refer to the classical Schr\"{o}dinger field or, in principle, even to the classical Dirac field (see App.~\ref{sec_review}).
In any case, the interaction Lagrangean $L_{\rm int}[\vec u, \psi]$ is given by the negative of the classical Coulomb interaction energy,
\begin{align} 
	L_{\rm int}(t) & \equiv -\hh\delta^2 V_{\rm e-n}(t) \\[5pt]
 &  =-\mh\int \! \de^3\vec x\!\int \! \de^3\vec x'\,\delta\rho_{\rm n}(\vec x,t) \, v(\vec x-\vec x') \, \delta\rho_{\rm e}(\vec x',t)\,, \label{lint}
\end{align}
As we want to interpret all terms involving nuclear coordinates as functionals of the displacement field, we write the
nuclear charge density as the continuum version of Eq.~\eqref{nucdensdispl}:
\begin{align}
 \delta \rho_{\rm n}(\vec x, t) & = -\rho_{\rm n0} \int \! \de^3 \vec x' \, \vec u(\vec x', t) \cdot (\nabla \delta^3)(\vec x - \vec x') \\[5pt]
 & = -\rho_{\rm n0} \, (\nabla \cdot \vec u)(\vec x, t) \,, \label{eq_densContVers}
\end{align}
which is a well-known formula \cite[Eq.~(45.3)]{Fetter}. By putting this into Eq. \eqref{lint} and performing a partial integration, 
the interaction Lagrangean can be expressed in terms of the displacement field as
\begin{equation}
L_{\rm int}(t)=-\rho_{\rm n0}\int \! \de^3\vec x\!\int\!\de^3\vec x'\,\vec u(\vec x,t)\cdot (\nabla v)(\vec x -\vec x') \, \delta\rho_{\rm e}(\vec x',t)\,.
\end{equation}
This interaction Lagrangean is of the standard form \eqref{lag_int} provided that we identify the force density $\vec f$ with
\begin{align} \label{to_vgl}
	\vec f(\vec x,t)&=-\rho_{\rm n0} \int\!\de^3\vec x'\, (\nabla v) (\vec x - \vec x') \, \delta\rho_{\rm e}(\vec x',t)\,.
\end{align}
For later purposes, we note that in terms of the electrostatic force kernel \eqref{eq_Coulforce} the force density can be written even more concisely as
\begin{equation} \label{conc}
\vec f(\vec x,t)=\rho_{\rm n0}\int \! \de^3\vec x'\,\vec{\mathcal E}(\vec x-\vec x') \, \delta\rho_{\rm e}(\vec x',t)\,.
\end{equation}
We now come back to the action of the coupled nuclear-electronic system, which is given by the time integral of the Lagrangean \eqref{coup_lag}, i.e.,
\begin{equation}
 S[\vec u, \psi] \h = \h \int \! \de t \, L(t) \h = \h S_{\rm n0}[\vec u] + S_{\rm int}[\vec u, \psi] + S_{\rm e \hh 0}[\psi] \,.
\end{equation}
From the above form of the interaction Lagrangean, it follows that the interaction term to the classical action is given by
\begin{equation}
 S_{\rm int}[\vec u, \psi] = \frac 1 c \h \int \! \de^4 x \,\h \vec u(x) \cdot \vec f(x) \,.
\end{equation}
Combining this with Eq.~\eqref{free_n} for the nuclear part, we can write the total action formally as
\begin{equation} \label{S_form}
 S[\vec u, \psi] = -\frac 1 {2 \h c} \, \vec u^{\rm T} \h \tsr D{}^{-1} \h \vec u + \frac 1 c \, \vec u^{\rm T} \vec f + S_{\rm e \hh 0}[\psi] \,.
\end{equation}
Now, within the functional integral approach, the {\itshape effective action} $S_{\rm eff}[\psi]$ for the electronic subsystem is defined by {\itshape integrating out} the nuclear degrees of freedom (see e.g.~\cite{Feldman, Tsai, Weinberg, Altland}), i.e., through the relation
\begin{equation} \label{def_eff_int}
 \exp \mh \bigg( \frac{1}{\hbar} \h S_{\rm eff}[\psi] \bigg) = \frac{1}{Z_0} \h \int \! \mathcal D \vec u \,\h \exp \mh \bigg( \frac{1}{\hbar} \h S[\vec u, \psi] \bigg) \,.
\end{equation}
By putting Eq.~\eqref{S_form} into this equation and performing the Gaussian functional integral as in Eqs.~\eqref{zw_3}--\eqref{zw_4}, we obtain immediately
\begin{equation}
 S_{\rm eff}[\psi] \h = \h S_{\rm e \hh 0}[\psi] + \frac 1 {2 \h c} \, \vec f^{\rm T} \h \tsr D \h \vec f \h \equiv \h S_{\rm e \hh 0}[\psi] + S_{\rm e-e}^{\rm eff}[\psi] \,.
\end{equation}
Hence, it follows that the effective electron-electron interaction is given by
\begin{equation}
 S_{\rm e-e}^{\rm eff} = \int \! \de t \, L_{\rm e-e}^{\rm eff}(t) \,,
\end{equation}
with
\begin{equation}
 L^{\rm eff}_{\rm e-e}(t)=\frac 1 2 \h \int \! \de^3\vec x\int\! \de^3\vec x'\mh\int\! c \, \de t'\,\vec f^{\rm T}(\vec x,t) \, \tsr D(\vec x,t;\vec x',t')\,\vec f(\vec x',t')\,. \label{eq_fv}
\end{equation}
Here, the electronic field variables $\psi(x)$ are hidden in the force vector fields. In fact, rewriting the effective interaction Lagrangean
in terms of the electronic density fluctuations by means of Eq.~\eqref{conc} yields
\begin{equation}
 L^{\rm eff}_{\rm e-e}(t)=-\frac 1 2\int \! \de^3\vec x\int\! \de^3\vec x'\mh\int\! c \, \de t'\, \delta\rho_{\rm e}(\vec x,t) \, v^{\rm eff}_{\rm e-e}(\vec x,t;\vec x',t') \, \delta\rho_{\rm e}(\vec x',t')\,,
\end{equation}
where the effective interaction kernel is given by
\begin{equation} \label{veffcont}
 v_{\rm e-e}^{\rm eff}(\vec x, t; \vec x', t') = \rho_{\rm n0}^2 \int \! \de^3 \vec y \int \! \de^3 \vec y' \, \vec{\mathcal E}^{\rm T}(\vec x - \vec y) \, \tsr D(\vec y, t; \vec y', t')  \, \vec{\mathcal E}(\vec y' - \vec x') \,.
\end{equation}
By Fourier transformation, this is equivalent to
\begin{equation} \label{eq_fundEq_equiv}
v^{\rm eff}_{\rm e-e}(\vec k,\omega)= \rho_{\rm n0}^2 \, \vec{\mathcal E}^{\dagger}(\vec k) \h \tsr D(\vec k,\omega) \, \vec{\mathcal E}(\vec k)\,.
\end{equation}
This effective interaction kernel is the continuum analogon of Eq.~\eqref{eff_fourier_2_simpl}.
Finally, with the representation of the nuclear cur\-{}rent response function through the elastic Green function, Eq.~\eqref{eq_densRespFuncGF}, 
and the relation \eqref{eq_Coulforce_FT} between the Coulomb force kernel $\vec {\mathcal E}(\vec k)$ and the Coulomb interaction kernel $v(\vec k)$,
the expression for the effective electron-electron interaction reverts to
\begin{equation}\label{eq_fundEq}
v^{\rm eff}_{\rm e-e}(\vec k,\omega)=v(\vec k) \, \upchi_{\rm n0}(\vec k,\omega) \, v(\vec k)\,,
\end{equation}
which agrees indeed exactly with our central result \eqref{eq_veffee}.

\bigskip \noindent
{\itshape Classical rederivation of phonon-mediated interaction.}---Finally, we give a field theoretical interpretation of the above findings. In fact,
the result \eqref{eq_fv} for the effective electron-electron interaction can be recovered even more transparently by considerations
which exclusively rely on classical Lagrangean field theory. For that purpose, we consider again the Lagrangean of the displacement field in the form
\begin{equation}\label{eq_LagDPF}
	L_{\rm n}[\vec u, \vec f \h ]\equiv  L_{\rm n0}[\vec u]+ L_{\rm int}[\vec u]
	=-\frac{1}{2}\h\vec u^{\rm T}\h\tsr D{}^{-1}\h\vec u+\vec u^{\rm T} \vec f\,.
\end{equation}
This leads via the Euler--Lagrange equations to the equation of motion 
\begin{equation}
	\tsr D{}^{-1} \h \vec u=\vec f\,.
\end{equation}
With this formula, the displacement field $\vec u$ can be formally eliminated from the Lagrangean. This leaves us with the purely electronic Lagrangean,
\begin{equation}
	L_{\rm eff}[\psi] \equiv L_{\rm e \hh 0}[\psi]+ L^{\rm eff}_{\rm e-e}\h[\vec f]\,,
\end{equation}
where the interaction Lagrangean is obtained by expressing the displacement field through the Green function,
\begin{equation}
	L^{\rm eff}_{\rm e-e}\h[\vec f\h ]\h \equiv\h L_{\rm n}[\vec u(\vec f), \h \vec f \h ]= L_{\rm n}[\h\tsr D\vec f, \h\vec f \h] \,.
\end{equation}
Explicitly, we then retrieve directly the expression 
\begin{align}
L^{\rm eff}_{\rm e-e}\h[\vec f]
& = -\frac 1 2 \h (\tsr D \vec f)^{\rm T} \h\hh \tsr D{}^{-1} \h (\tsr D \vec f) + (\tsr D \vec f)^{\rm T}  \vec f \\[5pt]
& =\frac{1}{2}\,\vec f^{\rm \, T}\h\tsr D\h\hh\vec f\label{eq_cont1}
\end{align}
for the effective interaction Lagrangean, which coincides exactly with the result from the functional integral approach. 
Note that in Eq.~\eqref{eq_cont1} we have used again the symmetry $D^{\rm T} = D$ of the elastic Green function.
Thus, we have shown that on a formal level the effective electron-electron interac-\linebreak tion is simply obtained from the coupled Lagrangean \eqref{eq_LagDPF} by eliminating the dynamical displacement
field by means of its response function (i.e.~the elastic Green function).

\section{Effective core interaction mediated by electrons}\label{sec_EffNuclInt}

In this final section, we investigate in detail the effective electron-mediated core interaction.
In stark contrast to its phonon-mediated counterpart, this core interaction does not constitute a very widespread concept, to the extent
that one may even doubt its actual usefulness. To counter this skepticism, we will show
in this section that our simple ansatz \eqref{eq_veffee} for the electron-mediated interaction is in fact already common practice, although its
fundamental form has apparently gone unnoticed so far. Concretely, we will show that the standard treatment
of the dynamical matrix in electronic structure theory precisely corresponds to the introduction of the
effective electron-mediated core interaction in the instantaneous limit.

\subsection{Second-order Hellmann--Feynman theorem}\label{app_HFT}

We begin this section by proving a useful lemma, which will be employed later and which is a 
generalization of the well-known Hellmann--Feynman theorem (see e.g.~\cite[pp.~56--59]{Martin}). A slight variant of this generalization can also be found in \cite[Eqs.~(10.34)--(10.35)]{Jensen}.

\bigskip\noindent
{\bfseries Lemma.} Consider a Hamiltonian $\hat H(\lambda)$ which depends smoothly on a set of real parameters $\lambda = (\lambda_j)_{j = 1, \ldots, J}$ and which has an orthonormal basis of eigenvectors $\{ \ket{\Psi_s(\lambda)}, \, s \in \mathbb N_0 \}$ 
with corresponding eigenvalues $E_s(\lambda)$, i.e.,
\begin{equation}
 \hat H(\lambda) \, \ket{\Psi_s(\lambda)} = E_s(\lambda) \, \ket{\Psi_s(\lambda)} \,.
\end{equation}
We assume for simplicity that these eigenvalues are non-degenerate.
Then the following identities hold:
\begin{equation}
 \partial_j \h \langle \Psi_s \mid \hat H \mid \Psi_s \rangle  = \langle \Psi_s \mid \partial_j \hat H \mid \Psi_s \rangle \,, \label{eq_hf_1}
\end{equation}
and
\begin{equation}
 \begin{aligned}
 \partial_i \h \partial_j \h \langle \Psi_s \mid \hat H \mid \Psi_s \rangle & = \langle \Psi_s \mid \partial_i \hh \partial_j \hat H \mid \Psi_s \rangle \\[2pt]
 & \quad + 2 \, \mathfrak{Re}\bigg( \h \sum_{r \not = s} \frac{\langle \Psi_s \mid \partial_i \hat H \mid \Psi_r \rangle \langle \Psi_r \mid \partial_j \hat H \mid \Psi_s \rangle}{E_s - E_r} \h \bigg) \,, \label{eq_hf_2}
\end{aligned}
\end{equation}
where we have abbreviated $\partial_i = \partial/\partial \lambda_i$ and suppressed the $\lambda$ dependencies, and where $\mathfrak{Re}$ denotes the real part.

\bigskip
\noindent
{\bfseries Proof.} Equation~\eqref{eq_hf_1} is the (first-order) Hellmann--Feynman theorem and therefore does not need to be proven here.
For the second derivative, we obtain
\begin{align}
 & \partial_i \h \partial_j \h \langle \Psi_s \mid \hat H \mid \Psi_s \rangle = \partial_j \h \langle \Psi_s \mid \partial_i \hat H \mid \Psi_s \rangle \\[5pt]
 & = \langle \Psi_s \mid \partial_i \hh \partial_j \hat H \mid \Psi_s \rangle + \langle \partial_j \Psi_s \mid \partial_i \hat H \mid \Psi_s \rangle + \langle \Psi_s \mid \partial_i \hat H \mid \partial_j \Psi_s \rangle \\[5pt]
 & = \langle \Psi_s \mid \partial_i \hh \partial_j \hat H \mid \Psi_s \rangle + 2 \, \mathfrak{Re} \h \langle \Psi_s \mid \partial_i \hat H \mid \partial_j \Psi_s \rangle \,. \label{zw1}
\end{align}
In the last step we have used that the operator $\hat H(\lambda)$ is hermitean for any~$\lambda$, and consequently $(\partial_i \hat H)(\lambda)$ is hermitean as well. 
To evaluate the derivative of the vector $\ket{\Psi_s(\lambda)}$, we write
\begin{equation}
 \hat H(\lambda + \de \lambda_j) = \hat H(\lambda) + \sum_j \de \lambda_j \h (\partial_j \hat H)(\lambda) \,,
\end{equation}
and regard the second term as a perturbation to $\hat H(\lambda)$. By the usual Rayleigh-Schr\"{o}dinger 
perturbation theory \cite[Sec.~11.1]{SchwablQM}, this yields
\begin{align}
 & \ket{\Psi_s(\lambda + \de \lambda_j)} \nonumber \\[5pt]
 & = \h \ket{\Psi_s(\lambda)} + \sum_j \de \lambda_j \sum_{r \not = s} \ket{\Psi_r(\lambda)} \frac{\langle \Psi_r(\lambda) \mid \partial_j \hat H(\lambda) \mid \Psi_s(\lambda) \rangle}{E_s - E_r} \\[5pt]
 & \equiv \h \ket{\Psi_s(\lambda)} + \sum_j \de \lambda_j \h \ket{\partial_j \Psi_s(\lambda)} \,.
\end{align}
Putting this result into Eq.~\eqref{zw1} shows the assertion. $\Box$

\subsection{Nuclear equilibrium positions}\label{sec_reconsider}

The connection \eqref{nucdensdispl} between the displacement field and the density fluctuation as well as
the expression \eqref{eq_CoulPotReform} for the Coulomb interaction
in terms of the fluctuation interaction and the Hartree potential allow for 
a rederivation of the nuclear equilibrium positions, which will become important later
and is hence to be explained in this subsection.

The equilibrium positions as calculated in electronic structure theory (see Sec.~\ref{nuc_sec}) are defined such that they minimize 
the total potential energy $V_{\rm n}$ of the nuclei. In terms of the deviations $\vec u_{\vec n}$, this implies the necessary condition
\begin{equation}\label{eq_classEquCond}
\left.\frac{\delta V_{\rm n}[\vec u]}{\delta\vec u_{\vec n}} \, \right|_{\vec u = \vec 0} =0\,,
\end{equation}
where $V_{\rm n}[\vec u]  = V_{\rm n-n}[\vec u] + E_{\rm e \hh 0}[\vec u]$ is given by Eqs.~\eqref{eq_Vphon}--\eqref{eq_elecHam_extpot}. 
To evaluate \linebreak this condition, we first rewrite the internuclear Coulomb interaction in terms of a fluctuation interaction and a Hartree potential analogous to Eq.~\eqref{eq_CoulPotReform}:
\begin{equation} \label{zw_1}
V_{\rm n-n}=\delta V^{\rm n}_{\rm H}+\delta^2 V_{\rm n-n}+V^0_{\rm n-n}\,,
\end{equation}
where \smallskip
\begin{equation}
 \delta V^{\rm n}_{\rm H} = \int \! \de^3 \vec x \,\h \delta \rho_{\rm n}(\vec x) \h \varphi_{\rm H}^{\rm n}(\vec x) \,, \smallskip
\end{equation}
and the Hartree potential is given by
\begin{equation}
 \varphi_{\rm H}^{\rm n}(\vec x) = \int \! \de^3 \vec x' \, v(\vec x - \vec x') \h \rho_{\rm n0}(\vec x') \,.
\end{equation}
We further use the expression \eqref{nucdensdispl} for the nuclear density fluctuation in terms of the displacement field, which can be written equivalently as
\begin{equation}
 \delta \rho_{\rm n}(\vec x) = -Z e \h \sum_{\vec m} \vec u_{\vec m} \mh \cdot (\nabla \delta^3)(\vec x - \vec x_{\vec m0}) \,.
\end{equation}
Now, only the first term on the right hand side of Eq.~\eqref{zw_1} enters the equi\-{}librium condition \eqref{eq_classEquCond}, because the second term is of order two, and the third term is of order zero in the displacement field. Thus, we obtain
\begin{align}
 \left. \frac{\delta V_{\rm n-n}[\vec u]}{\delta\vec u_{\vec n}} \h \right|_{\vec u = \vec 0} & = -Ze \, \frac{\delta}{\delta\vec u_{\vec n}} \int \! \de^3 \vec x \,\h \varphi_{\rm H}^{\rm n}(\vec x) \h \sum_{\vec m} \vec u_{\vec m} \mh \cdot  (\nabla \delta^3)(\vec x - \vec x_{\vec m0}) \\[2pt]
 & = -Ze \int \! \de^3 \vec x \,\h \varphi_{\rm H}^{\rm n}(\vec x) \, (\nabla\delta^3)(\vec x - \vec x_{\vec n0}) \\[8pt]
 & = Ze \, (\nabla \varphi_{\rm H}^{\rm n})(\vec x_{\vec n0}) \,,
\end{align}
where we have used again partial integration. Next, we calculate the derivative of the electronic contribution to $V_{\rm n}[\vec u]$, using the Hellmann--Feynman theorem, as follows:
\begin{align}
\frac{\delta E_{\rm e \hh 0}[\vec u]}{\delta\vec u_{\vec n}} & =
\frac{\delta}{\delta\vec u_{\vec n}} \, \big\langle\Psi_0[\vec u] \h \big| \h \hat H_{\rm e \hh 0}[\vec u] \h \big| \hh \Psi_0[\vec u] \h \big\rangle \\[3pt]
& =\bigg\langle\mh\Psi_0[\vec u] \g \bigg| \, \frac{\delta\hat H_{\rm e \hh 0}[\vec u]}{\delta\vec u_{\vec n}} \, \bigg| \h \Psi_0[\vec u]\bigg\rangle\,.
\end{align}
The electronic Hamiltonian $\hat H_{\rm e \hh 0}$ as defined by Eqs.~\eqref{eq_elecHam}--\eqref{eq_elecHam_extpot} depends on the displacement field only through the external potential. This yields
\begin{align}
 \frac{\delta\hat H_{\rm e \hh 0}[\vec u]}{\delta\vec u_{\vec n}} \, \bigg|_{\vec u = \vec 0} & = Ze \, \frac{\delta}{\delta \vec u_{\vec n}} \, \sum_{\vec m} \int \! \de^3 \vec x \,\h v(\vec x - \vec x_{\vec m0} -\vec u_{\vec m}) \, \hat \rho_{\rm e}(\vec x) \, \bigg|_{\vec u = \vec 0} \\[5pt]
 & = -Z e \int \! \de^3 \vec x \,\h (\nabla v)(\vec x - \vec x_{\vec n0}) \, \hat \rho_{\rm e}(\vec x) \,,
\end{align}
and consequently,
\begin{align}
\left. \frac{\delta E_{\rm e \hh 0}[\vec u]}{\delta\vec u_{\vec n}} \right|_{\vec u = \vec 0} & = -Z e \int \! \de^3 \vec x \,\h (\nabla v)(\vec x - \vec x_{\vec n0}) \, \rho_{\rm e \hh 0}(\vec x)  \\[6pt]
 & = Z e \, (\nabla \varphi_{\rm ext}^{\rm e})(\vec x_{\vec n0}) \,,
\end{align}
where $\rho_{\rm e\hh 0}(\vec x) = \langle \Psi_0 \mid \hat \rho_{\rm e}(\vec x) \mid \Psi_0\rangle$, and the external potential is given by
\begin{equation}
 \varphi_{\rm ext}^{\rm e}(\vec x) = \int \! \de^3 \vec x' \, v(\vec x - \vec x') \, \rho_{\rm e \hh 0}(\vec x') \,.
\end{equation}
In summary, we have thus shown that the equilibrium condition \eqref{eq_classEquCond} is equivalent to the equation
\begin{equation}
\vec E^{\rm n}_{\rm H}(\vec x_{\vec n0})+\vec E^{\rm e}\ext(\vec x_{\vec n0})=0\,,\label{eq_equilibriumCond} \smallskip
\end{equation}
where the {\itshape Hartree} and the {\itshape external} electric field are respectively defined by
\begin{align}
\vec E^{\rm n}_{\rm H}(\vec x)=-\nabla\varphi^{\rm n}_{\rm H}(\vec x)&=
-\nabla \h \bigg(\int\!\de^3\vec x'\,v(\vec x-\vec x') \h \rho_{\rm n0}(\vec x')\bigg)\,,\label{eq_HartreeForce}\\[5pt]
\vec E^{\rm e}_{\rm ext}(\vec x)=-\nabla\varphi^{\rm e}_{\rm ext}(\vec x)&=
-\nabla \h \bigg(\int\!\de^3\vec x'\,v(\vec x-\vec x') \h \rho_{\rm e \hh 0}(\vec x')\bigg)\,. \label{eq_extForce}
\end{align}
These are none other than the Coulomb force fields generated respectively by the nuclear and the electronic reference charge densities. 
We therefore obtain the highly intuitive and satisfying result that the equilibrium positions as calculated in electronic structure theory are identical to the classical
equilibrium positions defined by the vanishing of the electrostatic forces exerted on the nuclei by the electrons and by the nuclei themselves.

\subsection{Expansion of the core Hamiltonian} \label{subsec_ExpansionHam}

Quite as in the electronic case, the fundamental expression \eqref{effint2} for the total effective core interaction 
may not be useful in practice due to the high degree of the nuclear localization.
Fortunately, however, we may use again Eq.~\eqref{eq_cldens} to express the effective interaction in terms of the displacement field.
For this purpose, we start from the effective nuclear Hamiltonian given by Eqs.~\eqref{eq_summary_n} and \eqref{eq_totHam_n}, i.e.,
\begin{align}
H_{\rm n} & = T_{\rm n} + V_{\rm n-n} + V_{\rm ext}^{\rm e} + \delta^2 V^{\rm eff}_{\rm n-n} \label{eq_nuclHam_1} \\[6pt]
& = T_{\rm n} + V_{\rm ext}^{\rm e} + V_{\rm H}^{\rm n} + \delta^2 V_{\rm n-n}^{\rm tot}-V_{\rm n-n}^0\,, \label{eq_nuclHam}
\end{align}
and perform a second-order expansion in terms of the displacement field. We use again the decomposition $\rho_{\rm n} = \rho_{\rm n0} + \delta \rho_{\rm n}$, 
but with the second-order expansion of the density fluctuation,
\begin{equation}
\begin{aligned}
 \delta \rho_{\rm n}(\vec x,t) & = -Ze \h \sum_{\vec n, \h i} u_i(\vec x_{\vec n0},t) \left( \frac{\partial}{\partial x_i} \, \delta^3(\vec x - \vec x_{\vec n0}) \right) \\[3pt]
 & \quad +\frac{1}{2} \h Ze \sum_{\vec n, \h i, \h j} u_i(\vec x_{\vec n0},t) \left( \frac{\partial^2}{\partial x_i \h \partial x_j} \, \delta^3(\vec x - \vec x_{\vec n0}) \right) u_j(\vec x_{\vec n0},t) \,.
\end{aligned}
\end{equation}
For the term $V_{\rm ext}^{\rm e} = \rho_{\rm n} \h \varphi_{\rm ext}^{\rm e} = \rho_{\rm n0} \h\hh \varphi_{\rm ext}^{\rm e} + \delta V_{\rm ext}^{\rm e}$\h, we then find up to second order in the displacement field,
\begin{equation}
\begin{aligned}
\delta V_{\rm ext}^{\rm e}[\vec u] & = Ze \h \sum_{\vec n, \h i} u_i(\vec x_{\vec n 0}, t) \left.\frac{\partial \varphi_{\rm ext}^{\rm e}(\vec x)}{\partial x_i} \right|_{\vec x \hh = \hh \vec x_{\vec n 0}} \\[5pt]
& \quad + \frac 1 2 \h Z e \sum_{\vec n, \h i, \h j} u_i(\vec x_{\vec n 0}, t) \, u_j(\vec x_{\vec n 0}, t) \left. \frac{\partial^2 \varphi_{\rm ext}^{\rm e}(\vec x)}{\partial x_i \h \partial x_j} \right|_{\vec x \hh = \hh \vec x_{\vec n 0}} .
\end{aligned}
\end{equation}
The analogous expansion holds for the term $V_{\rm H}^{\rm n}  = \rho_{\rm n} \h \varphi_{\rm H}^{\rm n}$\h. Furthermore, for the term $\delta^2 V_{\rm n-n}^{\rm tot} = \delta \rho_{\rm n} \h v_{\rm n-n}^{\rm tot} \h \delta \rho_{\rm n}$ (cf.~the explicit expression \eqref{eq_effHam_n}), we find
\begin{align}
 & \delta^2 V^{\rm tot}_{\rm n-n}[\vec u] = \\[3pt] \nonumber
 & \frac{1}{2} \h Z^2 e^2 \sum_{\vec n, \h i, \h \vec m, \h j} \h \int \! c\, \de t' \, u_i(\vec x_{\vec n 0}, t) \h u_j(\vec x_{\vec m 0}, t') \left. \frac{\partial^2 v^{\rm tot}_{\rm n-n}(\vec x, t; \vec x',t')}{\partial x_i \h \partial x'_j} 
 \right|_{\substack{ \vec x \, = \, \vec x_{\vec n 0} \\ \vec x' =\,\vec x_{\vec m 0}}}.
\end{align}
Finally, the kinetic energy is already a second-order expression in the displacement field given by
\begin{equation}
T_{\rm n}[\vec u]=\frac{M}{2}\sum_{\vec n}|\dot{\vec u}(\vec x_{\vec n0},t)|^2\,.
\end{equation}
Thus, up to second order in the displacement field (and neglecting constant terms), we obtain from Eq.~\eqref{eq_nuclHam} the effective nuclear Hamiltonian,
\begin{align}
 & H^{(2)}_{\rm n}[\vec u] = \frac{M}{2}\sum_{\vec n, \h i}\dot u_i(\vec x_{\vec n0},t) \, \dot u_i(\vec x_{\vec n0},t) \,  \label{eq_2orderHam} \\[3pt]
   &+ Ze \h \sum_{\vec n, \h i} u_i(\vec x_{\vec n 0}, t) \left.\frac{\partial (\varphi_{\rm ext}^{\rm e} + \varphi_{\rm H}^{\rm n})(\vec x)}{\partial x_i} \right|_{\vec x \hh = \hh \vec x_{\vec n 0}} \nonumber \\[5pt]
   & + \frac 1 2 \h Z e \sum_{\vec n, \h i, \h j} u_i(\vec x_{\vec n 0}, t) \h u_j(\vec x_{\vec n 0}, t) \left. \frac{\partial^2 (\varphi_{\rm ext}^{\rm e} + \varphi_{\rm H}^{\rm n})(\vec x)}{\partial x_i \h \partial x_j} \right|_{\vec x \hh = \hh \vec x_{\vec n 0}} \nonumber \\[5pt]
   & + \frac{1}{2} \h Z^2 e^2 \sum_{\vec n, \h i, \h \vec m, \h j} \h \int \! c\,\de t' \, u_i(\vec x_{\vec n 0}, t) \h u_j(\vec x_{\vec m 0}, t') \left. \frac{\partial^2 v^{\rm tot}_{\rm n-n}(\vec x, \vec x'; t - t')}{\partial x_i \h \partial x'_j} 
 \right|_{\substack{ \vec x \, = \, \vec x_{\vec n 0} \\ \vec x' =\,\vec x_{\vec m 0}}}. \nonumber
\end{align}
Now, since $\vec x_{\vec n0}$ are the nuclear equilibrium positions, the $\vec u$-linear contribution to this Hamiltonian has to vanish (see Sec.~\ref{sec_reconsider}, Eq.~\eqref{eq_equilibriumCond}).
Hence, taking the instantaneous limit of Eq.~\eqref{eq_2orderHam}
leads to a true, second-order Hamiltonian for the displacement field.

\subsection{Connection to the dynamical matrix} \label{sec_dyn}

In the preceding subsection, we have derived a second-order effective Hamiltonian for the nuclear displacement field.
On the other hand, such a Hamiltonian is already known from the treatment of the quantized displacement field in electronic structure theory 
(see Sec.~\ref{subsec:dynDF&phon}, Eq.~\eqref{eq_phonoHamClass} with the dynamical matrix given by Eq.~\eqref{eq_defK_elstr}). 
The question therefore arises: how are these two Hamiltonians related? To answer this question, 
we first rewrite Eq.~\eqref{eq_2orderHam} (disregarding the linear terms) as
\begin{align} 
 H^{(2)}_{\rm n}[\vec u] & = T_{\rm n}[\vec u] + ( V_{\rm n-n} + V_{\rm ext}^{\rm e} + \delta^2 V_{\rm n-n}^{\rm eff} \h ){}^{(2)}[\vec u] \\[5pt]
 & \equiv T_{\rm n}[\vec u] + V^{(2)}_{\rm n-n}[\vec u] + \frac 1 2 \h \vec u^{\rm T} \tsr M \h \vec u \,. \label{eq_nucHamil}
\end{align}
Here, $V^{(2)}_{\rm n-n}$ denotes second-order contribution to the internuclear Coulomb interaction, and we have used the shorthand notation
\begin{equation}
 \vec u^{\rm T} \h \tsr M \h \vec u = 
 \sum_{\vec n, \h i, \h \vec m, \h j} \h \int \! c \, \de t' \, u_i(\vec x_{\vec n 0}, t) \, M_{ij}(\vec x_{\vec n 0}, \vec x_{\vec m 0} \h ; t - t') \, u_j(\vec x_{\vec m 0}, t') \,, \label{eq_nucHam} 
\end{equation}
with the matrix $M$ given explicitly by
\begin{equation}
\begin{aligned}
 M_{ij}(\vec x_{\vec n 0}, \vec x_{\vec m 0} \h ; t - t') & = \left. Z e \, \delta_{\vec n \vec m} \, \delta(c \h t - c\h t') \, \frac{\partial^2 \varphi_{\rm ext}^{\rm e}(\vec x)}{\partial x_i \h \partial x_j} \, \right|_{\vec x \hh = \hh \vec x_{\vec n 0}} \\[5pt]
 & \quad \, + Z^2 e^2 \left. \frac{\partial^2 v^{\rm eff}_{\rm n-n}(\vec x, \vec x'; t - t')}{\partial x_i \h \partial x'_j} \, \right|_{\substack{ \vec x \, = \, \vec x_{\vec n 0} \\ \vec x' =\,\vec x_{\vec m 0}}} \,. \label{eq_defM}
\end{aligned}
\end{equation}
Note that we have used the splitting \eqref{eq_nuclHam_1} instead of \eqref{eq_nuclHam}, 
and hence the Hartree potential $\varphi_{\rm H}^{\rm n}$ does not appear in this expression. We will now prove the following astonishing fact: 

\bigskip\noindent
{\bfseries Theorem.}
The instantaneous limit of the total effective core Hamiltonian \eqref{eq_nucHamil} coincides with the standard phonon Hamiltonian $H_{\rm phon}$ 
defined by Eqs.~\eqref{eq_phonoHamClass}, \eqref{eq_defK_elstr} and \eqref{eq_Vphon}--\eqref{eq_elecHam_extpot}, i.e., we have the identity
\begin{equation}
H^{(2)}_{\rm n}[\vec u]\h(\omega=0) = H_{\rm phon}[\vec u] \,.\label{assertion}
\end{equation}
In particular, this implies that the dynamical matrix from electronic structure theory (see Sec.~\ref{sec_dyn_elec})
already incorporates the effective electron-mediated core interaction as derived in this article.

\bigskip\noindent
{\bfseries Proof.}
By comparing Eqs.~\eqref{eq_nucHamil} and \eqref{eq_phonoHamClass}, we first note that both terms contain the kinetic contribution $T_{\rm n}[\vec u]$, and hence
our assertion is in fact equivalent to
\begin{equation} \label{zw_2}
 V_{\rm n-n}^{(2)}[\vec u] + \frac 1 2 \h \vec u^{\rm T} \h \tsr M(\omega = 0) \, \vec u \h = \h \frac 1 2 \h \vec u^{\rm T} \h \tsr K \, \vec u \,. \smallskip
\end{equation}
Furthermore, by Eqs.~\eqref{eq_defK_elstr} and \eqref{eq_Vphon}, we can write
\begin{equation}
 \frac 1 2 \h \vec u^{\rm T} \h \tsr K \h \vec u \h = \h V_{\rm n}^{(2)}[\vec u] \h = \h V_{\rm n-n}^{(2)}[\vec u] + E_{\rm e \hh 0}^{(2)}[\vec u] \,.
\end{equation}
Hence, both sides of Eq.~\eqref{zw_2} also contain the second-order contribution to the internuclear Coulomb interaction, 
and it only remains to show that
\begin{equation}
 \frac 1 2 \h \vec u^{\rm T} \h \tsr M(\omega = 0) \, \vec u \h = \h E^{(2)}_{\rm e \hh 0}[\vec u] \,,
\end{equation}
or equivalently,
\begin{equation} \label{eq_matM}
 M_{ij}(\vec x_{\vec n0}, \vec x_{\vec m0} \h ; \h \omega = 0) = \left. \frac{\partial^2 E_{\rm e \hh 0}[\vec u]}{\partial u_i(\vec x_{\vec n 0}) \h \partial u_j(\vec x_{\vec m 0})} \h \right|_{\vec u = \vec 0}\,. 
\end{equation}
To prove this equation, we employ the second-order Hellmann--Feynman theorem as stated in Sec.~\ref{app_HFT}: 
using the defining equation \eqref{eq_defpotphon}  for $E_{\rm e \hh 0}$\hh, we obtain
\begin{align}
& \frac{\partial^2 E_{\rm e \hh 0}}{\partial u_i(\vec{x}_{\vec{n}0}) \h \partial u_j(\vec{x}_{\vec{m}0})} \h \bigg|_{\vec u = \vec 0} = 
\bigg\langle\!\Psi_0 \h \bigg| \h \frac{\partial^2\hat V_{\rm ext}^{\rm n}}{\partial u_i(\vec{x}_{\vec{n}0}) \h \partial u_j(\vec{x}_{\vec{m}0})} \h \bigg|\Psi_0\!\bigg\rangle \h \bigg|_{\vec u = 0} \label{eq_snd_der_1} \\[8pt] \nonumber
& + 2\, \mathfrak{Re} \h \bigg( \sum_{s\neq 0} \frac{1}{E_0 - E_s} \, \bigg\langle\!\Psi_0 \h \bigg| \h \frac{\partial\hat V_{\rm ext}^{\rm n}}{\partial u_i(\vec{x}_{\vec{n}0})} \h \bigg|\Psi_s\!
\bigg\rangle\bigg\langle\!\Psi_s \h \bigg| \h \frac{\partial\hat V_{\rm ext}^{\rm n}}{\partial u_j(\vec{x}_{\vec{m}0})} \h \bigg|\Psi_0\!\bigg\rangle \bigg) \, \bigg|_{\vec u = \vec 0} \,. \label{eq_snd_der_2}
\end{align}
Here, we have used that the Hamiltonian $\hat H_{\rm e \hh 0}$ as defined by Eq.~\eqref{eq_elecHam} 
depends on $\vec u$ only through the external potential $\hat V_{\rm ext}^{\rm n}[\vec u]$. The latter can be expanded to second order in $\vec u$ as 
\begin{align}
 \hat V_{\rm ext}^{\rm n}[\vec u] & = Ze \h\sum_{\vec n} \int \! \de^3 \vec x' \, \hat \rho_{\rm e}(\vec x') \, \bigg( v(\vec x' - \vec x_{\vec n 0}) - \sum_i u_i(\vec x_{\vec n 0}) \h \frac{\partial v(\vec x' - \vec x_{\vec n 0})}{\partial x'_i} \nonumber \\[2pt]
 & \hspace{2.9cm} + \frac 1 2 \h \sum_{i, \h j} u_i(\vec x_{\vec n 0}) \h u_j(\vec x_{\vec n 0}) \, \frac{\partial^2 v(\vec x' - \vec x_{\vec n 0})}{\partial x'_i \h \partial x'_j} \h \bigg) \,.
\end{align}
Thus, we get for the first derivative (as in Sec.~\ref{sec_reconsider}),
\begin{align}
 \frac{\partial \hat V_{\rm ext}^{\rm n}}{\partial u_i(\vec x_{\vec n 0})} \h \bigg|_{\vec u = \vec 0} 
 = Z e \, \frac{\partial}{\partial x_i} \int \! \de^3 \vec x' \, v(\vec x - \vec x') \, \hat \rho_{\rm e}(\vec x') \h \bigg|_{\vec x \hh = \hh \vec x_{\vec n 0}} \,, \label{eq_put_1}
\end{align}
and similarly, for the second derivative,
\begin{equation}
\begin{aligned}
 & \frac{\partial^2 \hat V_{\rm ext}^{\rm n}}{\partial u_i(\vec x_{\vec n 0}) \h \partial u_j(\vec x_{\vec m 0})} \h \bigg|_{\vec u = \vec 0} \\[5pt]
 & = Ze \, \delta_{\vec n \vec m} \, \frac{\partial^2}{\partial x_i \h \partial x_j} \int \! \de^3 \vec x' \, v(\vec x - \vec x')
 \h \hat \rho_{\rm e}(\vec x') \h \bigg|_{\vec x \hh = \hh \vec x_{\vec n 0}} \,. \label{eq_put_2}
\end{aligned}
\end{equation}
By putting these results into Eq.~\eqref{eq_snd_der_1}, the first term yields
\begin{align}
 & \left. Ze \, \delta_{\vec n \vec m} \, \frac{\partial^2}{\partial x_i \h \partial x_j} \int \! \de^3 \vec x' \, v(\vec x - \vec x')  \h
 \langle \Psi_0 \mid \hat \rho_{\rm e}(\vec x') \mid \Psi_0 \rangle \h \right|_{\vec x \hh = \hh \vec x_{\vec n 0}} \\[5pt]
 & = Ze \, \delta_{\vec n \vec m} \, \frac{\partial^2 \varphi_{\rm ext}^{\rm e}(\vec x)}{\partial x_i \h \partial x_j} \h \bigg|_{\vec x \hh = \hh \vec x_{\vec n 0}} \,,
\end{align}
while the second term yields
\begin{align}
& Z^2 e^2 \frac{\partial^2}{\partial x_i \h \partial x'_j} \int \! \de^3 \vec y \int \! \de^3 \vec y' \, v(\vec x - \vec y) \label{above} \\[5pt]
& \hspace{0.75cm} \times 2 \, \mathfrak{Re} \,\bigg( \sum_{s \not = 0} \frac{\langle \Psi_0 \mid \hat \rho_{\rm e}(\vec y) \mid \Psi_s \rangle \langle \Psi_s \mid \hat \rho_{\rm e}(\vec y') \mid \Psi_0 \rangle}{E_0 - E_s} \bigg) \h v(\vec y' - \vec x') \h \bigg|_{\substack{ \vec x \, = \, \vec x_{\vec n 0} \\ \vec x' =\,\vec x_{\vec m 0}}} \,. \nonumber
\end{align}
The term in brackets coincides with the zero-frequency limit of the electronic density response function $\upchi_{\rm e \hh 0}$ as given by the Kubo formula 
in the spectral representation (see Eq.~\eqref{eq_statRF}). Hence, the second term in Eq.~\eqref{eq_snd_der_1} equals
\begin{align}
 & Z^2 e^2 \frac{\partial^2}{\partial x_i \h \partial x'_j} \int \! \de^3 \vec y \int \! \de^3 \vec y' \, v(\vec x - \vec y) \label{eq_below} \\ \nonumber
 & \hspace{2.2cm} \times \upchi_{\rm e \hh 0}(\vec y, \vec y' ; \h \omega = 0) \, v(\vec y' - \vec x') \h \bigg|_{\substack{ \vec x \, = \, \vec x_{\vec n 0} \\ \vec x' =\,\vec x_{\vec m 0}}} \\[3pt]
 & = \left. Z^2 e^2 \, \frac{\partial^2 v_{\rm n-n}^{\rm eff}(\vec x, \vec x' ; \h \omega = 0)}{\partial x_i \h \partial x'_j} \h \right|_{\substack{ \vec x \, = \, \vec x_{\vec n 0} \\ \vec x' =\,\vec x_{\vec m 0}}} \,,
\end{align}
where we have used again our central result \eqref{eq_veffnn}.
Thus, we have shown that Eq.~\eqref{eq_snd_der_1} is equivalent to
\begin{align}
 & \frac{\partial^2 E_{\rm e \hh 0}}{\partial u_i(\vec x_{\vec n 0}) \h u_j(\vec x_{\vec m 0})} \h \bigg|_{\vec u = \vec 0} \\[8pt] \nonumber 
 & = \left. Ze \, \delta_{\vec n \vec m} \, \frac{\partial^2 \varphi_{\rm ext}^{\rm e}(\vec x)}{\partial x_i \h \partial x_j} \h \right|_{\vec x \hh = \hh \vec x_{\vec n 0}} + \left. Z^2 e^2 \, \frac{\partial^2 v_{\rm n-n}^{\rm eff}(\vec x, \vec x' \h ; \h \omega = 0)}{\partial x_i \h \partial x'_j} \h \right|_{\substack{ \vec x \, = \, \vec x_{\vec n 0} \\ \vec x' =\,\vec x_{\vec m 0}}} \,.
\end{align}
This expression coincides precisely with Eq.~\eqref{eq_defM} for $M_{ij}(\omega = 0)$ and hence yields Eq.~\eqref{eq_matM}. This concludes our proof of the theorem. $\Box$

\bigskip \noindent
Finally, we remark that the possibility of expressing the dynamical matrix in terms of the effective core interaction ultimately hinges
on the connection of the former to the density response function. We therefore particularly acknowledge the excellent textbook \cite{SchafWegener} by W.\,Sch\"{a}fer and M.\,Wegener, where this crucial insight can already be found (see \cite[Eq.~(11.25)]{SchafWegener}).

\section{Conclusion}

We have proposed a general method to derive effective phonon-mediated and electron-mediated interactions, which consists in the elimination
of the phononic or electronic degrees of freedom by means of linear response theory. This approach starts directly from the fundamental
Hamiltonian of electrons and nuclei and is hence suitable for first-principles calculations.
In parti\-cular, it does not necessitate any approximation other than the decoupling procedure itself. 
A key advantage of the Response Theory of effective interactions lies in its conceptual simplicity:
a plethora of different motivations for effective interactions, ranging from functional integrals to second-order perturbation theory,
can be condensed into the simple, easily comprehensible ansatz \eqref{eq_veffee}--\eqref{eq_veffnn}.
Apart from this, the most important conceptual results of this approach can be summarized as follows:

\medskip \noindent
\begin{enumerate}
 \setlength{\itemsep}{1em}
\item Effective interactions serve as a heuristic tool to decouple a compound system interacting through the Coulomb potential into two formally independent subsystems. 
This is achieved by screening the interaction of each subsystem with the respective response function of the {\it other} subsystem.
Thus, the total effective interactions can be expressed in terms of screened interaction kernels (Secs.~\ref{sec_RFA}--\ref{subsec_screening}).
\item Consequently, in addition to the usual phonon-mediated electron interaction, one necessarily
has to introduce an electron-mediated core interaction (Sec.~\ref{sec_RFA}).
\item In their most abstract form, the expressions for these effective interactions are completely symmetric under the exchange of electrons and nuclei (Eqs.~\eqref{eq_veffee}--\eqref{eq_veffnn}).
\item The effective electron-electron and effective core interactions act on the respective density fluctuations, not on the densities themselves. 
Hence, they can be combined with the ordinary Coulomb interaction only if a Hartree potential is split off from the latter (Sec.~\ref{subsec_total}).
\item From the most abstract expressions, all well-known concrete exressions for the phonon-mediated electron interaction 
can be recovered in suitable limiting cases, in 
particular the expressions in terms of the elastic Green function and the phonon dispersion relation (Sec.~\ref{subsec_Rederivation}).
\item The phonon-mediated interaction is consistent with the notion of effective interactions used in the functional integral approach (Sec.~\ref{sec_JustFI}).
\item The standard expression of the dynamical matrix as used in electronic structure theory corresponds to an 
electron-mediated effective core interaction in the instantaneous limit (Sec.~\ref{sec_dyn}).
\end{enumerate}

\medskip \noindent
Our general formulae \eqref{eq_veffee}--\eqref{eq_veffnn} for the phonon-mediated and electron-mediated interactions in terms of the respective density response functions are valid for any material. In particular, they include all possible effects of inhomogeneity and aniso\-{}tropy. If combined with the Kubo formalism for the density response functions, they would therefore lend themselves to the {\itshape ab initio} calculation of effective interactions. These may in turn be used as initial interactions in subsequent fRG studies, and hence also allow for an unbiased, first-principles prediction of superconducting order parameters and transition temperatures in realistic material models.

\medskip
\section*{Acknowledgements}

This article grew out of the doctoral dissertation of one of the authors, R.\,S., at the University of Vienna in 2012. 
Special thanks therefore go to the Viennese Center for Computational Materials Science and the SFB ViCoM. 
R.\,S. further thanks the Institute for Theoretical Physics at the TU Bergakademie Freiberg for its hospitality. G.\,S.~was supported by the DFG Research Unit FOR 723.

\begin{appendices}

\bigskip
\section{Notations and Conventions} \label{app_notconv}

In this appendix, we assemble a number of conventions mainly concerning the Fourier transform taken to different limites. In the initial Sec.~\ref{app_FT}, we fix the formulae for the Fourier transform on a {\itshape finite lattice.} The latter
directly corresponds to the nuclear equilibrium positions of a real, i.e.~finite crystal.
Next, in Sec.~\ref{app_thLim} we consider the {\itshape thermodynamic limit} of infinitely many lattice points.
Its importance stems from theoretical materials physics: Strictly speaking, no quantity calculated for a finite system corresponds
to a material property ``as such''. Instead, it always depends on the concrete system characterized by a finite particle number $N$
(or, alternatively, a certain macroscopic length like $L=Na$, etc.). By contrast, a pure material property such as the dielectric constant $\varepsilon_{\rm r}=81$ or the refractive index
$n=1.3$ \cite[Table 3.2]{Hecht} of water does not refer to the number of molecules in the water probe or the latter's linear dimension. Therefore, in order to describe pure material properties,
one always has to take the results to the thermodynamic limit. Intuitively, in this limit the material would ``homogeneously'' fill out the whole space.
In the closing Secs.~\ref{app_cont} and \ref{app_comblim}, we then consider the {\itshape continuum limit}
both in combination with and without the thermodynamic limit. This limit is not so much of a conceptual, but of a practical importance.
In fact, in the continuum limit the lattice spacing goes to zero, and 
any function originally defined on the lattice becomes an ordinary function of a continuous variable.
Thus, this limit corresponds to the treatment of the displacement field as in ``macroscopic'' continuum mechanics.
However, we remark that the continuum limit is actually counterfactual, because there is a cutoff wavelength for the displacement field in the Fourier domain, which is determined by the Brillouin zone boundary and which may not simply be removed.
So, the importance of this limit lies rather in the fact that it allows to employ the usual differential calculus,
which simplifies certain calculations that would otherwise become rather cumbersome.
From this we have taken benefit especially in Secs.~\ref{sec_JustFI} and \ref{subsec_emMatProp}.

\subsection{Lattice Fourier transformation} \label{app_FT}

We first assemble the formulary of the lattice Fourier transformation, which is essentially the discrete Fourier transformation in disguise. 
For this purpose, we define the {\itshape dual lattice} $\Gamma^*$ as the set of possible discrete wavevectors associated with the direct lattice $\Gamma$ defined in Sec.~\ref{el_sector}. Per definitionem, dual lattice vectors (or {\itshape Bloch vectors}) $\vec k\equiv\vec k_{\vec m}$ obey
\begin{equation}
\vec x_{\vec n0}\mh\cdot\vec k_{\vec m}=2\pi\left(\frac{n_1m_1}{N_1}+\frac{n_2m_2}{N_2}+\frac{n_3m_3}{N_3}\right)
\end{equation}
for any direct lattice vector $\vec x_{\vec n0}$\hh.
The dual lattice vectors can be constructed by means of the base vectors of the {\it reciprocal lattice}, which are defined by the condition
\begin{equation}
 \vec a_i \cdot \vec b_j = 2\pi \, \delta_{ij} \,,
\end{equation}
and given explicitly by
\begin{equation}
\vec b_i=\pi \h \epsilon_{ijk} \, \frac{\vec a_j\times\vec a_k}{|\vec a_1\!\cdot(\vec a_2\times\vec a_3)|}\,,
\end{equation}
where the $\vec a_i$ ($i = 1, 2, 3$) are the base vectors of the direct lattice. Any dual lattice vector is a linear combination of the form
\begin{equation} \label{dual_vector}
\vec k_{\vec m}=\frac{m_1}{N_1} \h \vec b_1+\frac{m_2}{N_2}\h \vec b_2+\frac{m_3}{N_3}\h \vec b_3\,,
\end{equation}
where the $m_i$ are integers obeying
\begin{equation} \label{int_range}
0\leq m_i<N_i\,.
\end{equation}
Hence, the dual lattice is defined as
\begin{equation}
 \Gamma^* = \big\{\vec k_{\vec m} \h ; \, \vec m=(m_1,m_2,m_3), \, 0 \leq m_i < N_i \ (i = 1, 2, 3) \big\}\,.
\end{equation}
In particular, there are $N = N_1 N_2 N_2$ dual wavevectors, exactly as many as there are direct lattice points.
By contrast, the reciprocal lattice vectors themselves are linear combinations of the form
\begin{equation}
\vec G_{\vec m}=m_1\vec b_1 + m_2 \vec b_2 + m_3 \vec b_3\,,
\end{equation}
with arbitrary integers $m_i \in \mathbb Z$. These span the reciprocal lattice, which hence comprises infinitely many points.

With these prerequisites, we now define the Fourier transform of a field quantity 
$\vec u \equiv \vec u(\vec x_{\vec n0})$, $\vec x_{\vec n0} \in \Gamma$, by the unitary transformation
\begin{equation} \label{ft_1}
 \vec u(\vec k_{\vec m}) = \frac{1}{\sqrt{N}} \sum_{\vec x_{\vec n0} \hh \in \h \Gamma} \vec u(\vec x_{\vec n0}) \, \e^{-\i\vec k_{\vec m} \cdot\h \vec x_{\vec n0}}\,,
\end{equation}
where $\vec k_{\vec m} \in \Gamma^*$. The inverse transformation reads
\begin{equation} \label{ft_2}
\vec u(\vec x_{\vec n0}) = \frac{1}{\sqrt{N}} \sum_{\vec k_{\vec m} \in \h \Gamma^*} \vec u(\vec k_{\vec m}) \, \e^{\i\vec k_{\vec m} \cdot\h \vec x_{\vec n0}} \,, 
\end{equation}
which can be shown using the identity
\begin{equation}
 \delta_{\vec n {\vec m}} \equiv \h \delta_{\vec x_{\vec n0},\h \vec x_{\vec m0}} = \frac{1}{N} \sum_{\vec k \h \in\h \Gamma^*} \e^{\i \vec k \h \cdot \hh (\vec x_{\vec n0} - \vec x_{{\vec m0}})}\,.
\end{equation}
Similarly, the Fourier transformation of lattice integral kernels (such as the dynamical matrix or the elastic Green function)
is defined such that a relation
\begin{equation}
\vec f(\vec x_{\vec n0}) = \sum_{\vec m} \tsr K(\vec x_{\vec n0}, \vec x_{\vec m0}) \, \vec u(\vec x_{{\vec m0}})
\end{equation}
in the direct lattice is equivalent to
\begin{equation}
\vec f(\vec k_{\vec n}) = \sum_{\vec m} \tsr K(\vec k_{\vec n}, \vec k_{\vec m}) \, \vec u(\vec k_{\vec m})
\end{equation}
in the dual lattice. Explicitly, we thus obtain
\begin{align}
 \tsr K(\vec k_{\vec n}, \vec k_{\vec m}) & = \frac{1}{N} \sum_{\vec x\h\in\h\Gamma} \sum_{\vec x'\h\in\h\Gamma} \, \e^{-\i\vec k_{\vec n} \cdot \h \vec x} \h \tsr K(\vec x, \vec x') \, \e^{\i\vec k_{\vec m} \cdot \h \vec x'} \label{kft_1}\,, \\[5pt]
 \tsr K(\vec x_{\vec n0}, \vec x_{\vec m0}) & = \frac{1}{N} \sum_{\vec k\h\in\h\Gamma^*} \sum_{\vec k'\h\in\h\Gamma^*} \, \e^{\i\vec k \h \cdot \h \vec x_{\vec n0}} \h \tsr K(\vec k, \vec k') \, \e^{-\i\vec k' \cdot \h \vec x_{\vec m0}} \label{kft_2}\,.
\end{align}
Concretely, for a homogeneous kernel,
\begin{equation} \label{hom_ker}
 \tsr K(\vec x_{\vec n0}, \h \vec x_{\vec m0}) = \tsr K(\vec x_{\vec n0} - \vec x_{\vec m0})\,,
\end{equation}
these definitions imply in Fourier space that
\begin{equation} \label{hom_ker_ft}
 \tsr K(\vec k_{\vec n}, \h \vec k_{\vec m}) = \tsr K(\vec k_{\vec n}) \, \delta_{\vec n \vec m} \,.
\end{equation}
The Fourier transformation then simplifies as follows:
\begin{align}
 \tsr K(\vec k_{\vec m}) & = \sum_{\vec n} \tsr K(\vec r_{\vec n}) \, \e^{-\i\vec k_{\vec m} \cdot \h \vec r_{\vec n}}  \,, \label{analogous_1} \\[5pt]
 \tsr K(\vec r_{\vec n}) & = \frac{1}{N} \sum_{\vec m} \tsr K(\vec k_{\vec m}) \, \e^{\i\vec k_{\vec m} \cdot \h \vec r_{\vec n}} \,, \label{analogous_2}
\end{align}
where $\vec r_{\vec n}$ denotes the difference between any two direct lattice vectors and can itself be regarded as an element of $\Gamma$. The above relations are analogous to Eqs.~\eqref{ft_1}--\eqref{ft_2} up to the prefactors (cf.~\cite[Sec.~2.1]{ED1}). Finally, we remark that in the context of the lattice Fourier transformation, any dual lattice vector $\vec k_{\vec m}$ 
can be arbitrarily altered by a reciprocal lattice vector, $\vec k_{\vec m}\mapsto \vec k_{\vec m}+\vec G$. Thus, the dual lattice can also be chosen to lie more symmetric around the origin. For example, we may replace Eq.~\eqref{int_range} by
\begin{equation}
 -\frac{N_i}{2} \leq m_i < \frac{N_i}{2} \,,
\end{equation}
if $N_i$ is assumed even. This choice has the advantage that with any wavevector $\vec k_{\vec m}$, its reflected counterpart $-\vec k_{\vec m}$ lies also in the dual lattice, a fact which has often been used in the main text.

\subsection{Thermodynamic limit}\label{app_thLim}

Per definitionem, in the thermodynamic limit the number of lattice points goes to infinity, $N \to \infty$, such that the direct lattice becomes
\begin{equation}
 \Gamma \to \big \{\vec x_{\vec n0} \h ; \, \vec n=(n_1,n_2,n_3), \, n_i \in \mathbb Z \ (i = 1, 2, 3) \big\}\,.
\end{equation}
In Fourier space, the thermodynamic limit corresponds to a continuum limit in the sense  that the dual lattice turns into the {\itshape primitive unit cell} of the reciprocal lattice (see \cite[p.~73]{Ashcroft}),
\begin{equation} \label{gammastern}
 \Gamma^* \to \big \{ \vec k = \mu_1 \vec b_1 + \mu_2 \vec b_2 + \mu_3 \vec b_3 \h ; \, 0 \leq \mu_i < 1 \ (i = 1, 2, 3) \big\}\,.
\end{equation}
Again, by suitably adding reciprocal lattice vectors, the primitive unit cell of the reciprocal lattice can be chosen to be more symmetric with respect to the origin. For example, one may take
\h$-0.5 \leq \mu_i < 0.5$
\,in Eq.~\eqref{gammastern}. Here, we will make yet another choice (which is standard in solid-state physics): we  replace $\Gamma^*$ by the {\itshape (first) Brillouin zone}~$\mathcal B$, which is the Wigner-Seitz cell of the reciprocal lattice (see e.g.~\cite[p.~73]{Ashcroft}), i.e., the region in $\vec k$ space which is closer to the origin than to any other point of the reciprocal lattice. The volume of the Brillouin zone is given by
\begin{equation}
 |{\mathcal B}| = |\vec b_1 \cdot (\vec b_2 \times \vec b_3)| = \frac{(2\pi)^3}{V_0} \,,
\end{equation}
where
\begin{equation}
 V_0 = \frac V N = |\vec a_1 \cdot (\vec a_2 \times \vec a_3)| \smallskip \vspace{2pt}
\end{equation}
is the volume of the primitive unit cell of the direct Bravais lattice. 

Any summation over dual lattice vectors can now be interpreted as a Riemann sum, which in the thermodynamic limit approaches an integral over the Brillouin zone, i.e.,
\begin{equation}
 \sum_{\vec k_{\vec m} \in \h \Gamma^*} f(\vec k_{\vec m}) \, \Delta^3 \vec k_{\vec m}  \, \to \, \int_{\mathcal B} f(\vec k) \, \de^3 \vec k \,.
\end{equation}
Here, the differential $\Delta^3 \vec k_{\vec m}$ corresponds to the Brillouin zone volume per dual lattice vector, 
\begin{equation}
 \Delta^3 \vec k_{\vec m} = \frac{|{\mathcal B}|}{N}  = \frac{(2\pi)^3}{V} \,. \smallskip
\end{equation}
We thus find the replacement rule for summations over dual lattice vectors,
\begin{equation} \label{repl_sum}
 \sum_{\vec k_{\vec m} \in \h \Gamma^*} \mapsto \frac{V}{(2\pi)^3} \int_{\mathcal B}  \de^3 \vec k \,.
\end{equation}
Next, we consider the Kronecker delta for dual lattice vectors, which has the property that for an arbitrary function $f$,
\begin{equation} \label{kro}
 f(\vec k_{\vec m}) = \sum_{\vec k_{\vec n} \in \h \Gamma^*} \delta_{{\vec m}{\vec n}} \h f(\vec k_{\vec n}) \,.
\end{equation}
On the other hand, the Dirac delta distribution in momentum space is defined such that \smallskip
\begin{equation} \label{dir}
 f(\vec k) = \int_{\mathcal B} \de^3 \vec k' \, \delta^3(\vec k - \vec k') \h f(\vec k') \,. \smallskip \vspace{2pt}
\end{equation}
By taking the thermodynamic limit of Eq.~\eqref{kro}, \mbox{$f(\vec k_{\vec m}) \to f(\vec k)$,} using Eq. \eqref{repl_sum} and comparing the result with Eq.~\eqref{dir}, we obtain the replace\-{}ment rule for the Kronecker delta,
\begin{equation} \label{repl_delta}
 \delta_{\vec k_{\vec m} \vec k_{\vec n}} \mapsto \frac{(2\pi)^3}{V} \, \delta^3(\vec k - \vec k') \,. \smallskip
\end{equation}
We now come to the Fourier transformation. In the thermodynamic limit, the Fourier transform of the displacement field is defined by
\begin{equation} \label{th_ft_1}
\vec u(\vec k) = \frac{1}{\sqrt{|{\mathcal B}|}} \h \sum_{\vec x_{\vec n0} \in \hh \Gamma} \vec u(\vec x_{\vec n0}) \, \e^{-\i\vec k \h \cdot \h \vec x_{\vec n0}} \,,
\end{equation}
while the inverse transformation reads
\begin{equation} \label{th_ft_2}
 \vec u(\vec x_{\vec n0}) = \frac{1}{\sqrt{|{\mathcal B}|}} \int_{\mathcal B} \de^3 \vec k \,\h \vec u(\vec k) \, \e^{\i\vec k \h \cdot \h \vec x_{\vec n0}} \,,
\end{equation}
as can be shown by the identity
\begin{equation}
 \delta_{\vec n {\vec m}} = \frac{1}{|{\mathcal B}|} \int_{\mathcal B} \de^3 \vec k \,\h \e^{\i \vec k \h \cdot \hh (\vec x_{\vec n0} - \vec x_{{\vec m0}})}\,.
\end{equation}
By taking the thermodynamic limit of Eqs.~\eqref{ft_1}--\eqref{ft_2}---keeping $\vec u(\vec x_{\vec n0})$ in direct space fixed---and comparing the result with Eqs.~\eqref{th_ft_1}--\eqref{th_ft_2}, we find that the Fourier transform of any field quantity should be replaced in the thermodynamic limit as follows:
\begin{equation} \label{repl_uk}
 \vec u(\vec k_{\vec m}) \mapsto \sqrt{\frac{|{\mathcal B}|}{N}} \, \vec u(\vec k) = \sqrt{\frac{(2\pi)^3}{V}} \, \vec u(\vec k) \,.
\end{equation}
Similarly, by defining the Fourier transformation of the dynamical matrix in the thermodynamic limit as
\begin{align}
 \tsr K(\vec k, \vec k') & = \frac 1 {|\mathcal B|} \sum_{\vec x_{\vec n0}\in\hh \Gamma} \sum_{\vec x_{\vec m0}\in\hh \Gamma} \e^{-\i\vec k \h \cdot \h \vec x_{\vec n0}} \h \tsr K(\vec x_{\vec n0}, \vec x_{\vec m0}) \, \e^{\i\vec k' \cdot \h \vec x_{\vec m0}} \,, \label{ft_dyn_1} \\[5pt]
 \tsr K(\vec x_{\vec n0}, \vec x_{\vec m0}) & = \frac 1 {|\mathcal B|} \int_{\mathcal B} \de^3 \vec k \int_{\mathcal B} \de^3 \vec k' \, \e^{\i\vec k \h \cdot \h \vec x_{\vec n0}} \h \tsr K(\vec k , \vec k') \, \e^{-\i\vec k' \cdot \h \vec x_{\vec m0}} \,, \label{ft_dyn_2}
\end{align}
and comparing these equations with Eqs.~\eqref{kft_1}--\eqref{kft_2}, we obtain the replacement rule for the dynamical matrix,
\begin{equation} \label{repl_K}
 \tsr K(\vec k_{\vec m}, \vec k_{\vec n}) \mapsto \frac{(2\pi)^3}{V} \, \tsr K(\vec k, \vec k') \,.
\end{equation}
By using the lattice translation invariance of the dynamical matrix, which implies Eq.~\eqref{hom_ker_ft} and respectively, \begin{equation}
 \tsr K(\vec k, \vec k') = \tsr K(\vec k) \, \delta^3(\vec k - \vec k') \,,
\end{equation}
we further obtain (using Eqs.~\eqref{repl_delta} and  \eqref{repl_K}) the rule
\begin{equation}
 \tsr K(\vec k_{\vec m}) \mapsto \tsr K(\vec k)\,.
\end{equation}
This means, the reduced kernel of the dynamical matrix (depending on one wavevector only) remains invariant in the thermodynamic limit. The same applies to the elastic Green function defined by
\begin{equation}
 \big({-M} \omega^2 + \tsr K(\vec k)\big) \h \tsr D(\vec k, \omega) = \tsr 1\,,
\end{equation}
i.e., in the thermodynamic limit,
\begin{equation} \label{repl_dk}
 \tsr D(\vec k_{\vec m}, \omega) \mapsto \tsr D(\vec k, \omega) \,.
\end{equation}
Finally, we consider the mode expansion coefficients. In the discrete case they fulfill the Poisson bracket  relations \eqref{eq_Poi1a}, i.e.,
\begin{equation}
\big\{a_{\vec k_{\vec m} \lambda}, \h a^*_{\vec k_{\vec n} \lambda'}\big\}=(\i\hbar)^{-1} \h \delta_{\vec k_{\vec m}\vec k_{\vec n}} \h \delta_{\lambda \lambda'}\,,
\end{equation}
while in the thermodynamic limit they are supposed to obey
\begin{equation}
\big\{a_{\vec k \lambda}, \h a^*_{\vec k'\lambda'}\big\}=(\i\hbar)^{-1} \h \delta^3(\vec k-\vec k') \h \delta_{\lambda \lambda'} \,.
\end{equation}
Taking into account Eq.~\eqref{repl_delta}, we therefore find the relation
\begin{equation} \label{repl_ak}
a_{\vec k_{\vec m}\lambda} \mapsto \sqrt{\frac{(2\pi)^3}{V}} \, a_{\vec k\lambda}\,.
\end{equation}
With all these replacement rules, the mode expansion of the displacement field, Eq.~\eqref{eq_genSol}, can be written in the thermodynamic limit as follows:
\begin{equation} \label{u_th}
\begin{aligned}
	\vec{u}(\vec x_{\vec n0},t) & = \frac{1}{\sqrt{|\mathcal B|}} \h \int_{\mathcal B} \de^3 \vec k \, \sum_{\lambda=1}^3 \sqrt{\frac{\hbar}{2M\omega_{\vec{k}\lambda}}} \\[6pt]
	& \quad \, \times \big(a_{\vec k\lambda}(t) \, \vec{e}_{\vec{k}\lambda} \, \e^{{\rm i}\vec{k}\cdot\vec{x_{\vec n0}}}+
	a^*_{\vec k\lambda}(t) \, \vec{e}^{*}_{\vec{k}\lambda} \, \e^{-{\rm i}\vec{k}\cdot\vec{x_{\vec n0}}}\h\big)\,.
\end{aligned}
\end{equation}
Similarly, from Eq.~\eqref{eq_genSol1} one obtains the mode expansion of the conjugate momentum $\vec \pi(\vec x_{\vec n0}, t) = M \hh \partial_t \vec u(\vec x_{\vec n0}, t)$. The displacement field and its conjugate momentum obey the same Poisson bracket relations in real space, Eqs.~\eqref{eq_Poi1}--\eqref{eq_Poi3}, as in the case of a finite lattice.

\subsection{Continuum limit}\label{app_cont}

We first describe the continuum limit for a finite sample with volume~$V$. For this purpose, we define the new unit vectors $(i = 1, 2, 3)$
\begin{equation}
 \vec a_i^r  := \frac{1}{r} \, \vec a_i
\end{equation}
for any positive integer $r \in \mathbb N$, and let
\begin{equation}
 \vec x_{\vec n}^r := n_1 \hh \vec a_1^r  + n_2 \h \vec a_2^r  + n_3 \h \vec a_3^r  \,.
\end{equation}
Thus, we define a new Bravais lattice with $N^r  = r^3 N$ points as
\begin{equation}
 \Gamma^r := \big\{\vec x_{\vec n}^r  \h ; \, \vec n = (n_1, n_2, n_3), \, 0 \leq n_i < r  N_i \ (i = 1, 2, 3) \big\} \,.
\end{equation}
Since the volume of the primitive cell becomes
\begin{equation}
 V_0^r  = |\vec a_1^r  \cdot (\vec a_2^r  \times \vec a_3^r )| = \frac 1 {r^3} \h V_0 \,,
\end{equation}
the volume of the whole sample remains constant for each $r$,
\begin{equation}
 N^r \hh V_0^r  = N V_0 \h \equiv  V \,.
\end{equation}
The continuum limit is now defined by letting $r \to \infty$. Then the discrete Bravais lattice turns into a parallelepiped,
\begin{equation}
\Gamma^r  \h \to \h \mathcal V := \h \big\{ \vec x = \ell_1 \hh \vec a_1 + \ell_2 \h \vec a_2 + \ell_3 \h \vec a_3 \h ; \,\h \ell_i \in [0, N_i) \h \textnormal{ for } \h i = 1, 2, 3 \h \big\} \,,
\end{equation}
with sidelengths $L_i = N_i \h \vec a_i$ and volume $|\mathcal V| = V$. Furthermore, the unit vectors of the reciprocal lattice are given for any finite $r$ by
\begin{equation}
 \vec b_i^r  = r \h \vec b_i \,,
\end{equation}
and hence the dual lattice vectors are given by
\begin{align} \label{dual_lattice_given}
 \vec k_{\vec m} \h = \h \frac{m_1}{r N_1} \, \vec b_1^r  + \frac{m_2}{r N_2} \, \vec b_2^r  + \frac{m_3}{r N_3} \, \vec b_3^r  \h = \h \frac{m_1}{N_1} \, \vec b_1 + \frac{m_2}{N_2} \, \vec b_2 + \frac{m_3}{N_3} \, \vec b_3 \,,
\end{align}
with \smallskip
\begin{equation}
 0 \leq m_i < r  N_i \,. \smallskip
\end{equation}
This shows that the dual lattice becomes larger with increasing $r$, but the distances between the dual lattice vectors remain invariant. For $r \to \infty$, the dual lattice becomes infinite,
\begin{equation}
 (\Gamma^r )^* \to \big\{\vec k_{\vec m} \h ; \, \vec m = (m_1, m_2, m_3), \, m_i \in \mathbb Z \ (i = 1, 2, 3) \big\} \,.
\end{equation}
In the continuum limit, the displacement field becomes an ordinary continuous vector field defined on the volume $\mathcal V$, i.e.,
\begin{equation}
\vec u(\vec x_{\vec n0},t) \mapsto \vec u(\vec x,t)\,.
\end{equation}
Analogous to the thermodynamic limit, any summation over direct lattice vectors can now be interpreted as a Riemann sum which in the continuum limit approaches an integral over $\mathcal V$, i.e.,
\begin{equation}
 \sum_{\vec x_{\vec n0} \hh \in \h \Gamma} f(\vec x_{\vec n0}) \, \Delta^3 \vec x_{\vec n0} \to \int_{\mathcal V} \de^3 \vec x \, f(\vec x) \,.
\end{equation}
Here, the differential $\Delta^3 \vec x_{\vec n0}$ corresponds to the volume per lattice vector,
\begin{equation}
 \Delta^3 \vec x_{\vec n0} = \frac V N = V_0 \,.
\end{equation}
We thus find the replacement rule for summations over direct lattice vectors,
\begin{equation} \label{repl_cont}
 \sum_{\vec x_{\vec n0} \in \hh \Gamma} \mapsto \frac 1 {V_0} \h \int_{\mathcal V} \de^3 \vec x \,.
\end{equation}
Similarly, the Kronecker delta for direct lattice vectors turns into the Dirac delta distribution in real space,
\begin{equation} \label{Kron_dir}
 \delta_{\vec x_{\vec n0}, \h \vec x_{\vec m0}} \mapsto V_0 \, \delta^3(\vec x - \vec x') \,.
\end{equation}
Next, we come to the Fourier transformation. In the continuum limit, the Fourier transform of a field quantity is defined by
\begin{equation}
 \vec u(\vec k_{\vec m}) = \frac{1}{\sqrt V} \int_{\mathcal V} \de^3 \vec x \,\h \vec u(\vec x) \, \e^{-\i \vec k_{\vec m} \cdot \h \vec x} \,,
\end{equation}
while the inverse transformation reads
\begin{equation}
 \vec u(\vec x) = \frac{1}{\sqrt V} \h \sum_{\vec k_{\vec m} \in \h \Gamma^*} \vec u(\vec k_{\vec m}) \, \e^{\i \vec k_{\vec m} \cdot \h \vec x} \,,
\end{equation}
as can be shown by the identity
\begin{equation}
 \delta_{\vec k_{\vec m} \vec k_{\vec n}} = \frac{1}{V} \int_{\mathcal V} \de^3 \vec x \,\h \e^{\i (\vec k_{\vec m} - \vec k_{\vec n}) \h \cdot \h \vec x} \,.
\end{equation}
By comparing these equations with the corresponding discrete transformations, Eqs.~\eqref{ft_1}--\eqref{ft_2}, we see that the Fourier transform of the displacement field should be replaced in the continuum limit as
\begin{equation}
 \vec u(\vec k_{\vec m}) \mapsto \frac{1}{\sqrt{V_0}} \, \vec u(\vec k_{\vec m}) \,.
\end{equation}
We now turn to the dynamical matrix, which is defined in the continuum limit such that
\begin{equation}
V_{\rm phon} = \frac 1 2 \, \int_{\mathcal V} \de^3 \vec x \int_{\mathcal V} \de^3 \vec x' \, \vec u^{\rm T}(\vec x) \h\hh \tsr K(\vec x, \vec x') \, \vec u(\vec x') \,.
\end{equation}
Comparing this with Eq.~\eqref{eq_phonoHamClass} in the discrete case and using the replace\-{}ment rule \eqref{repl_cont} for the summations over lattice vectors, we find that
\begin{equation}
\tsr K(\vec x_{\vec n0}, \vec x_{\vec m0}) \mapsto V_0^2 \, \tsr K(\vec x, \vec x') \,.
\end{equation}
In Fourier space, this implies the replacement rule
\begin{equation}
\tsr  K(\vec k_{\vec m}, \vec k_{\vec n}) \mapsto V_0 \, \tsr K(\vec k_{\vec m}, \vec k_{\vec n}) \,,
\end{equation}
or in the homogeneous case,
\begin{equation}
\tsr K(\vec k_{\vec m}) \mapsto V_0 \, \tsr K(\vec k_{\vec m}) \,.
\end{equation}
The elastic Green function defined by
\begin{equation}
 \big({-M} \omega^2 + \tsr K(\vec k_{\vec m}) \big) \h \tsr D(\vec k_{\vec m}, \omega) = \tsr 1 \,,
\end{equation}
therefore has to be replaced in the continuum limit by
\begin{equation}
\tsr D(\vec k_{\vec m}, \omega) \mapsto \frac 1 {V_0} \h \tsr D(\vec k_{\vec m}, \omega) \,.
\end{equation}
Finally, the Poisson bracket relations remain invariant in the continuum limit,
\begin{equation}
 \big\{a_{\vec k_{\vec m} \lambda}, \h a_{\vec k_{\vec n}\lambda'}^* \big\} = (\i\hbar)^{-1} \h \delta_{\vec k_{\vec m} \vec k_{\vec n}} \h \delta_{\lambda \lambda'} \,,
\end{equation}
and so do the mode expansion coefficients. Therefore, the mode expansion \eqref{eq_genSol} can be written in the continuum limit as follows:
\begin{align}
 & \vec u(\vec x, t) = \\ \nonumber
 & \frac{1}{\sqrt V} \, \sum_{\vec k \h \in \h \Gamma^*} \sum_{\lambda = 1}^3 \h \sqrt{\frac{\hbar}{2 \hh \rho_0 \h \omega_{\vec k\lambda}}}  \left( a_{\vec k\lambda}(t) \, \vec e_{\vec k\lambda} \, \e^{\i \vec k \cdot \vec x} + a_{\vec k\lambda}^*(t) \, \vec e_{\vec k\lambda}^* \, \e^{-\i\vec k\cdot \vec x} \h \right) ,
\end{align}
where the reference mass density $\rho_0$ has been defined in Eq.~\eqref{mass_dens}. 
The continuous displacement field now obeys the Poisson bracket relation
\begin{equation} \label{cont_Poisson_1}
 \big\{ u_i(\vec x, t), \h \pi_j(\vec x', t) \big\} = \delta_{ij} \, \delta^3 (\vec x - \vec x') \,.
\end{equation}
By our convention that the displacement field in real space remains invariant, $\vec u(\vec x_{\vec m}, t) \mapsto \vec u(\vec x, t)$, and by Eq.~\eqref{Kron_dir}, it follows that the conjugate momentum in real space has to be replaced by
\begin{equation} \label{cont_Poisson_2}
 \vec \pi(\vec x_{\vec m}, t) \mapsto V_0 \, \vec \pi(\vec x, t) \,,
\end{equation}
and is therefore given by
\begin{equation} \label{cont_Poisson_3}
 \vec \pi(\vec x, t) = \frac{M}{V_0} \h \frac{\partial}{\partial t} \h \vec u(\vec x, t) = \rho_0 \h \frac{\partial}{\partial t} \h \vec u(\vec x, t) \,. \smallskip
\end{equation}
The mode expansion of $\vec \pi(\vec x, t)$ can then be directly obtained.

\subsection{Combined limites and synopsis} \label{app_comblim}

By taking the thermodynamic and the continuum limit at the same time, both the direct and the dual lattice vectors turn into continuous variables,
\begin{align}
 \vec x_{\vec n} & \to \vec x \,, \\[5pt]
 \vec k_{\vec m} & \to \vec k \,,
\end{align}
ranging in the whole three-dimensional space $\mathbb R^3$. Correspondingly, all summations should be replaced by integrals according to
\begin{align}
 \sum_{\vec x_{\vec n0} \in \Gamma} & \mapsto \frac{1}{V_0} \h \int\! \de^3 \vec x\,, \\[3pt]
 \sum_{\vec k_{\vec m0} \in \Gamma^*} & \mapsto \frac{V}{(2\pi)^3} \int \! \de^3 \vec k \,,
\end{align}
where $V$ is the volume of the probe and \h$V_0 = V/N$ \h the volume of the primi\-tive~cell. These two factors should in the end disappear from any relation between physical quantities. 
Similarly, the Kronecker deltas are replaced by Dirac delta distributions as in Eqs.~\eqref{repl_delta} and \eqref{Kron_dir}. The Fourier transformation is defined in the combined limit by
\begin{equation}
 \vec u(\vec k) = \frac 1 {(2\pi)^{\nicefrac 3 2}} \int \! \de^3 \vec x \,\h \vec u(\vec x) \, \e^{-\i \vec k \cdot \vec x} \,,
\end{equation}
while the inverse transformation reads
\begin{equation}
 \vec u(\vec x) = \frac 1 {(2\pi)^{\nicefrac 3 2}} \int \! \de^3 \vec k \,\h \vec u(\vec x) \, \e^{\i \vec k \cdot \vec x} \,,
\end{equation}
as can be shown by the identity
\begin{equation}
 \delta^3(\vec x - \vec x') = \frac 1 {(2\pi)^3} \int \! \de^3 \vec k \,\h \e^{\i \vec k \cdot (\vec x - \vec x')} \,.
\end{equation}
In particular, by requiring again that the displacement field in real space remains invariant, $\vec u(\vec x_{\vec m}) \mapsto \vec u(\vec x)$, this implies the replacement rule of its Fourier transform,
\begin{equation}
 \vec u(\vec k_{\vec m}) \h \mapsto \,\sqrt{\frac{1}{V_0}}\, \sqrt{\frac{(2\pi)^3}{V}} \, \vec u(\vec k) \h = \h \sqrt{\frac{(2\pi)^3 N}{V^2}} \, \vec u(\vec k) \,.
\end{equation}
Similar replacement rules can again be derived for the dynamical matrix and the elastic Green function.
Finally, the mode expansion coefficients transform according to Eq.~\eqref{repl_ak}, 
and the mode expansion can be expressed in the combined limit as
\begin{align}
& \vec{u}(\vec x, t) = \label{cont_u}  \\ \nonumber
 & \frac{1}{(2\pi)^{\nicefrac{3}{2}}} \int \! \de^3 \vec k \, \sum_{\lambda = 1}^3 \h \sqrt{\frac{\hbar}{2 \hh \rho_0 \h \omega_{\vec{k}\lambda}}}
\left(a_{\vec k\lambda}(t)\,\vec{e}_{\vec{k}\lambda} \, \e^{{\rm i}\vec{k}\cdot\vec{x}}
+a^*_{\vec k\lambda}(t) \, \vec{e}^{*}_{\vec{k}\lambda} \, \e^{-{\rm i}\vec{k}\cdot\vec{x}}\h \right),
\end{align}
with the mass density $\rho_0$ given by Eq.~\eqref{mass_dens}. As expected, this formula does not explicitly depend on the parameters $V$, $N$ or $M$ anymore (only through $\rho_0$). Finally, the continuous displacement field obeys again the Poisson bracket relation \eqref{cont_Poisson_1} with the conjugate momentum defined as in Eq.~\eqref{cont_Poisson_3}. 

Table \ref{synopsis_1} summarizes all the 
replacement rules derived in this appendix for the thermodynamic limit, the continuum limit and the combined limites. These rules can be used to unequivocally convert the exact microscopic formulae for a finite lattice to the heuristic field theoretical expressions in the thermodynamic and/or continuum limites, and vice versa. To illustrate this once more, consider e.g.~the equation of motion \eqref{eq_EoMPhonoGF} for the elastic Green function. Explicitly, this reads
\begin{equation} 
\begin{aligned}
 & \sum_{\vec x_{\vec \ell 0} \hh \in \h \Gamma} \Big( \delta_{\vec x_{\vec n0}, \h \vec x_{\vec \ell 0}} \, M \hh\partial_t^2 + \tsr K(\vec x_{\vec n 0}, \vec x_{\vec \ell 0}) \Big) \h \tsr D(\vec x_{\vec \ell 0}, t; \h \vec x_{\vec m0}, t') \\[3pt]
 & = \delta(c \h t - c \h t') \, \delta_{\vec x_{\vec n0}, \h \vec x_{\vec m0}} \tsr 1 \,.
\end{aligned}
\end{equation}
By replacing each quantity appearing in this equation according to Table \ref{synopsis_1}, we obtain the corresponding equation of motion in the combined thermodynamic and continuum limit,
\begin{equation}
\begin{aligned}
 & \int \! \de^3 \vec x \, \Big( \delta^3(\vec x - \vec y ) \h \rho_0 \h \partial_t^2 + \tsr K(\vec x, \vec y ) \Big) \h \tsr D(\vec y , t; \h \vec x', t') \\[3pt]
 & = \delta(c \h t - c \h t') \, \delta^3(\vec x - \vec x') \h \tsr 1 \,,
\end{aligned}
\end{equation}
which again depends on $V$, $N$ or $M$ only through the mass density $\rho_0$\hh.
Similarly, Table \ref{synopsis_2} summarizes the different versions of the mode expansion in the thermodynamic and/or continuum limites, which can be deduced from the exact Eqs.~\eqref{eq_genSol}--\eqref{eq_genSol1} by means of the replacement rules of Table~\ref{synopsis_1}.

\newpage
\begin{table}[t]
\begin{center}
\caption{Replacement rules. \label{synopsis_1}} \vspace{8pt}
\renewcommand{\arraystretch}{1.38}
\begin{tabular}{p{3.0cm}p{3.0cm}p{3.0cm}p{3.0cm}}
\toprule[1pt]
Finite crystal & Thermodynamic & Continuum & Combined \\[-8pt]
& limit & limit & limites \\
\midrule[1pt]
$\vec x_{\vec n0}$ & $\vec x_{\vec n0}$ & $\vec  x$ & $\vec x$ \\[2pt]
\midrule
$\vec k_{\vec m}$ & $\vec k$ & $\vec k_{\vec m}$ & $\vec k$ \\[2pt]
\midrule
$\sum_{\vec x_{\vec n0}}$ & $\sum_{\vec x_{\vec n0}}$ & $\frac{N}{V}  \, \mbox{\large $\int$}_{\!\!\mathcal V} \, \de^3 \vec x$ & $\frac{N}{V}  \, \mbox{\large $\int$} \de^3 \vec x$ \\[2pt]
\midrule
$\delta_{\vec x_{\vec n0}, \h \vec x_{\vec m0}}$ & $\delta_{\vec x_{\vec n0}, \h \vec x_{\vec m0}}$ & $\frac V N \, \delta^3(\vec x - \vec x')$ & $\frac V N \, \delta^3(\vec x - \vec x')$ \\[2pt]
\midrule
$\sum_{\vec k_{\vec m}}$ & $\frac{V}{(2\pi)^3}  \, \mbox{\large $\int$}_{\!\!\mathcal B} \, \de^3 \vec k$ & $\sum_{\vec k_{\vec m}}$ & $\frac{V}{(2\pi)^3}  \, \mbox{\large $\int$} \de^3 \vec k$ \\[2pt]
\midrule
$\delta_{\vec k_{\vec m}, \h \vec k_{\vec n}}$ & $\frac{(2\pi)^3}{V} \, \delta^3(\vec k - \vec k')$ & $\delta_{\vec k_{\vec m}, \h \vec k_{\vec n}}$ & $\frac{(2\pi)^3}{V} \, \delta^3(\vec k - \vec k')$ \\[2pt]
\midrule
$\vec u(\vec x_{\vec n0})$ & $\vec u(\vec x_{\vec n0})$ & $\vec u(\vec x)$ & $\vec u(\vec x)$ \\[2pt]
\midrule
$\vec u(\vec k_{\vec m})$ & $\sqrt{\frac{(2\pi)^3}{V}} \, \vec u(\vec k)$ & $\sqrt{\frac{N}{V}} \, \vec u(\vec k_{\vec m})$ & $\sqrt{\frac{(2\pi)^3 N}{V^2}} \, \vec u(\vec k)$ \\[2pt]
\midrule
$\vec \pi(\vec x_{\vec n0})$ & $\vec \pi(\vec x_{\vec n0})$ & $\frac V N \, \vec \pi(\vec x)$ & $\frac V N \, \vec\pi(\vec x)$  \\[2pt]
\midrule
$\vec \pi(\vec k_{\vec m})$ & $\sqrt{\frac{(2\pi)^3}{V}} \, \vec \pi(\vec k)$ &  $\sqrt{\frac{V}{N}} \, \vec \pi(\vec k_{\vec m})$ & $\sqrt{\frac{(2\pi)^3}{N}} \, \vec \pi(\vec k)$ \\[2pt]
\midrule
$ D(\vec x_{\vec n0}, \vec x_{\vec m0})$ & $ D(\vec x_{\vec n0}, \vec x_{\vec m0})$ & $ D(\vec x, \vec x')$ &  $ D(\vec x, \vec x')$ \\[2pt]
\midrule
$ D(\vec k_{\vec m}, \vec k_{\vec n})$ & $\frac{(2\pi)^3}{V} \,  D(\vec k, \vec k')$ & $\frac N V \,  D(\vec k_{\vec m}, \vec k_{\vec n})$ & $\frac{(2\pi)^3 N}{V^2} \,  D(\vec k, \vec k')$ \\[2pt]
\midrule
$ D(\vec k_{\vec m})$ & $ D(\vec k)$ & $\frac N V \,  D(\vec k_{\vec m})$ & $\frac N V \,  D(\vec k)$ \\[2pt]
\midrule
$ K(\vec x_{\vec n0}, \vec x_{\vec m0})$ & $ K(\vec x_{\vec n0}, \vec x_{\vec m0})$ & $\frac{V^2}{N^2} \h  K(\vec x, \vec x')$ &  $\frac{V^2}{N^2} \h  K(\vec x, \vec x')$ \\[2pt]
\midrule
$ K(\vec k_{\vec m}, \vec k_{\vec n})$ & $\frac{(2\pi)^3}{V} \,  K(\vec k, \vec k')$ & $\frac V N \,  K(\vec k_{\vec m}, \vec k_{\vec n})$ & $\frac{(2\pi)^3}{N} \,  K(\vec k, \vec k')$ \\[2pt]
\midrule
$ K(\vec k_{\vec m})$ & $ K(\vec k)$ & $\frac V N \,  K(\vec k_{\vec m})$ & $\frac V N \,  K(\vec k)$ \\[2pt]
\midrule
$a_{\vec k_{\vec m} \lambda}$ & $\sqrt{\frac{(2\pi)^3}{V}} \, a_{\vec k\lambda}$ & $a_{\vec k_{\vec m} \lambda}$ & $\sqrt{\frac{(2\pi)^3}{V}} \, a_{\vec k\lambda}$ \\[2pt]
\bottomrule[1pt]
\end{tabular}
\end{center}
\end{table}

\FloatBarrier

\begin{landscape}
\begin{table}[t]
\begin{center}
\caption{Mode expansions (with $\rho_0 = N \mhh M/V$). \label{synopsis_2}} \vspace{8pt}
\renewcommand{\arraystretch}{1.8}
\begin{tabular}{p{4.5cm}p{12.3cm}}
\toprule[1pt] \\[-20pt]
Finite crystal & $\vec{u}(\vec{x}_{\vec{n}0},t)
=\frac 1 {\sqrt N} \sum\limits_{\vec{k}} \sum\limits_\lambda \sqrt{\frac{\hbar}{2M\omega_{\vec{k}\lambda}}}
\left(a_{\vec{k}\lambda}(t) \, \vec{e}_{\vec{k}\lambda} \, \e^{{\rm i}\vec{k}\cdot\vec{x}_{\vec{n}0}}
\h + \h \mathrm{c.c.} \h \right)$  \\[8pt]
&  $\vec \pi(\vec{x}_{\vec{n}0},t)
=\frac 1 {\sqrt N} \sum\limits_{\vec{k}} \sum\limits_\lambda \h \sqrt{\frac{M\hbar \h \omega_{\vec k \lambda}}{2}}
\left(-\i \h a_{\vec{k}\lambda}(t) \, \vec{e}_{\vec{k}\lambda} \, \e^{{\rm i}\vec{k}\cdot\vec{x}_{\vec{n}0}}
\h + \h \mathrm{c.c.} \h \right)$ \\[8pt]
\midrule \\[-20pt]
Thermodynamic limit & $\vec{u}(\vec x_{\vec n0},t) = \sqrt{\frac{V}{(2\pi)^3 N}}  \,\h \mbox{\large $\int$}_{\!\!\mathcal B} \, \de^3 \vec k \, \sum\limits_{\lambda} \sqrt{\frac{\hbar}{2M\omega_{\vec{k}\lambda}}} \left(a_{\vec k\lambda}(t) \, \vec{e}_{\vec{k}\lambda} \, \e^{{\rm i}\vec{k}\cdot\vec{x_{\vec n0}}}\h + \h \mathrm{c.c.} \h \right)$ \\[8pt]
& $\vec{\pi}(\vec x_{\vec n0},t) =
	\sqrt{\frac{V}{(2\pi)^3 N}} \,\h \mbox{\large $\int$}_{\!\!\mathcal B} \, \de^3 \vec k \,  \sum\limits_{\lambda} \sqrt{\frac{M \hbar \h \omega_{\vec{k}\lambda}}{2}}\left({-\i} \h a_{\vec k\lambda}(t) \, \vec{e}_{\vec{k}\lambda} \, \e^{{\rm i}\vec{k}\cdot\vec{x_{\vec n0}}}\h + \h \mathrm{c.c.} \h \right)$ \\[11pt]
\midrule \\[-20pt]
Continuum limit & $\vec u(\vec x, t) =
 \frac{1}{\sqrt V} \sum\limits_{\vec k} \sum\limits_{\lambda} \sqrt{\frac{\hbar}{2 \hh \rho_0 \h \omega_{\vec k\lambda}}} \left( a_{\vec k\lambda}(t) \h \vec e_{\vec k\lambda} \, \e^{\i \vec k \cdot \vec x} \h + \h \mathrm{c.c.} \h \right)$ \\[8pt]
& $\vec \pi(\vec x, t) =
 \frac{1}{\sqrt V} \sum\limits_{\vec k} \sum\limits_{\lambda} \sqrt{\frac{\rho_0 \hh \hbar \hh \omega_{\vec k\lambda}}{2}} \left( -\i\h a_{\vec k\lambda}(t) \h \vec e_{\vec k\lambda} \, \e^{\i \vec k \cdot \vec x} \h + \h \mathrm{c.c.} \h \right)$ \\[8pt]
\midrule \\[-20pt]
Combined limites & $\vec u(\vec x, t) =
 \frac{1}{\sqrt{(2\pi)^3}} \, \mbox{\large $\int$} \de^3 \vec k \, \sum\limits_\lambda \sqrt{\frac{\hbar}{2 \hh \rho_0 \h \omega_{\vec k\lambda}}} \left( a_{\vec k\lambda}(t) \, \vec e_{\vec k\lambda} \, \e^{\i \vec k \cdot \vec x} \h + \h \mathrm{c.c.} \h \right)$ \\[8pt]
& $\vec \pi(\vec x, t) =
 \frac{1}{\sqrt{(2\pi)^3}} \, \mbox{\large $\int$} \de^3 \vec k \, \sum\limits_\lambda \sqrt{\frac{\rho_0 \hh \hbar \hh \omega_{\vec k\lambda}}{2}} \left( -\i\h a_{\vec k\lambda}(t) \, \vec e_{\vec k\lambda} \, \e^{\i \vec k \cdot \vec x} \h + \h \mathrm{c.c.} \h \right)$ \\[11pt]
\bottomrule[1pt]
\end{tabular}
\end{center}
\end{table}
\end{landscape}

\subsection{Born--von-Karman boundary conditions} \label{app_bvk}

As explained already in the main text (Sec.~\ref{el_sector}), the strict periodicity of the external nuclear potential exerted on the electrons together with the finite particle number require
the stipulation of the Born--von-Karman boundary conditions. Strictly speaking, however, this is not consistent with the fundamental Hamiltonian
of electrons and nuclei, because the application of the Coulomb potential operator on a many-body wave function does in general not respect these boundary conditions.
This is particularly transparent in Fourier space, where the Born--von-Karman boundary conditions restrict the possible wavevectors of a one-particle wave function $\psi(\vec k)$
to the form $\vec k = \vec k_{\vec m}$ (defined as in Eq.~\eqref{dual_vector}) with $m_i \in \mathbb Z$ ($i = 1, 2, 3$). By contrast, the Fourier transform of the Coulomb potential, $v(\vec k)=1/(\varepsilon_0|\vec k|^2)$, is defined for all wavevectors $\vec k \in \mathbb R^3$, which is also necessary for recovering the expression $v(\vec x - \vec x') = 1/(4\pi\varepsilon_0\h|\vec x - \vec x'|)$ in real space. This problem can be dealt with
by keeping the formal expression of the Coulomb potential in Fourier space fixed, but restricting it to the admissible wave\-{}vectors. We note, however,
that in real space the resulting ``Born--von-Karman Coulomb potential'' is not (and, in fact, must not be) of the usual form any more.

Let us examine more carefully the substitutions required by the Born--von-Karman boundary conditions. For field quantities defined on $\mathbb R^3$, the Fourier transformation is defined by
\begin{align}
\rho(\vec k)&=\frac{1}{(2\pi)^{\nicefrac 3 2}}\int\!\de^3\vec x\,\h\e^{-{\rm i}\vec k\cdot\vec x} \h \rho(\vec x)\,,\label{int_vol} \\[5pt]
\rho(\vec x)&=\frac{1}{(2\pi)^{\nicefrac 3 2}}\int\!\de^3\vec k\,\h\e^{{\rm i}\vec k\cdot\vec x} \h \rho(\vec k)\,,
\end{align}
while for translation-invariant interaction kernels (cf.~\cite[Sec.~2.1]{ED1})
\begin{align}
v(\vec k)&=\int \! \de^3\vec r\,\h\e^{-{\rm i}\vec k\cdot\vec r}\h v(\vec r)\,,\\[5pt]
v(\vec r)&=\frac{1}{(2\pi)^3} \int\!\de^3\vec k\,\h\e^{{\rm i}\vec k\cdot\vec r} \h v(\vec k)\,,
\end{align}
where $\vec r=\vec x-\vec x'$. On the other hand, for the corresponding quantities defined on a finite volume $V$ with Born--von-Karman boundary conditions, the Fourier transformations read
\begin{align}
\rho_V(\vec k)&=\frac{1}{\sqrt{V}}\int_V\de^3\vec x\,\h\e^{-{\rm i}\vec k\cdot\vec x} \h \rho_V(\vec x)\,, \label{int_fin} \\[5pt]
\rho_V(\vec x)&=\frac{1}{\sqrt{V}} \h \sum_{\vec k}\e^{{\rm i}\vec k\cdot\vec x} \h \rho_V(\vec k)\,, \label{bvk_Fourier}
\end{align}
and respectively,
\begin{align}
v_V(\vec k)&=\int_V\de^3\vec r\,\h\e^{-{\rm i}\vec k\cdot\vec r} \h v_V(\vec r)\,,\\[5pt]
v_V(\vec r)&=\frac{1}{V} \h \sum_{\vec k}\e^{{\rm i}\vec k\cdot\vec r} \h v_V(\vec k)\,. \label{BvCCoulomb}
\end{align}
In particular, these equations imply that $v_V(\vec r)\neq v(\vec r)$, even if $v_V(\vec k)$ coincides with $v(\vec k)$ on the admissible wavevectors.
However, if a real material is localized within the macroscopic volume $V$, then we expect the real-space density $\rho(\vec x)$ within $V$ to roughly coincide with the density $\rho_V(\vec x)$ of a (hypothetical) system satisfying the Born--von-Karman boun\-{}dary conditions. More precisely, this means \smallskip
\begin{equation}
\rho(\vec x)\approx\rho_V(\vec x) \smallskip
\end{equation}
in the bulk, but not in the vicinity of the boundaries of $V$. Furthermore, we have $\rho(\vec x) \approx 0$ outside the volume $V$ where the material is localized, which implies that the real-space integral in Eq.~\eqref{int_vol} is effectively restricted to the volume $V$ just as in Eq.~\eqref{int_fin}. Consequently, for the Fourier transforms of the densities we find the approximate equality
\begin{equation} \label{bvk_1}
\rho(\vec k) \approx \sqrt{\frac{V}{(2\pi)^3}}\,\h\rho_V(\vec k) \,,
\end{equation}
at the admissible wavevectors. By a similar argument, the potential energy of the real material localized in the volume $V$ should roughly coincide with the potential energy of the corresponding system with Born--von-Karman boundary condi\-{}tions, hence,
\begin{equation} \label{bvk_2}
\int \! \de^3\vec k\,\h\rho^*(\vec k)\h v(\vec k)\h \rho(\vec k) \h \approx \h \sum_{\vec k}\rho^*_V(\vec k) \h v_V(\vec k)\h \rho_V(\vec k) \,.
\end{equation}
As we already know that we can set approximately
\begin{equation}
\int \!\frac{\de^3\vec k}{(2\pi)^3} \h \approx \h \frac 1 V \sum_{\vec k} \ ,
\end{equation}
Eqs.~\eqref{bvk_1} and \eqref{bvk_2} together imply that the interaction kernel scales trivially in Fourier space, i.e., \smallskip
\begin{equation}
v_V(\vec k) \approx v(\vec k) \,, \smallskip
\end{equation}
for the admissible wavevectors. Thus, no prefactor appears in this equation, and we can indeed simply restrict the Coulomb potential in Fourier space to the admissible wavevectors in order to account for the Born--von-Karman boundary conditions. 
We stress again, however, that the resulting Hamiltonian differs from the fundamental Hamiltonian \eqref{eq_fund}, because 
the ``Born--von-Karman Coulomb potential'' $v_V(\vec x - \vec x')$ given in real space by the discrete sum \eqref{BvCCoulomb} does not coincide with the usual Coulomb potential $v(\vec x - \vec x')$ anymore (compare \cite[Eq.~(17.17)]{Ashcroft}). Strictly speaking, one therefore has to take the thermodynamic limit $V \to \infty$ in the end of all calculations. In this limit, Eq.~\eqref{BvCCoulomb} indeed approaches the usual Coulomb potential.

\section{Quantized Schr\"{o}dinger field} \label{sec:quantSF}

In this appendix, we give a short introduction to quantum field theory as applied to many-body theory.
We shall be brief as details can be found in the standard references like \cite{Giuliani,Hedin69,MartinRothen,Huang,Schweber,Thirring,Salmhofer}.
While the first Subsec.~\ref{subsecFockSpace} introduces the Fock space formalism, the second Subsec.~\ref{subsecSecondQuant} 
deals with the all-important second quantization.

\subsection{Fock space formalism}\label{subsecFockSpace}

{\it Many-particle quantum states.}---Formally, a one-particle wave function $\psi(\vec x,t)$ can be interpreted as a classical time-dependent field 
on the ordinary three-dimensional space $\mathbb R^3$. By contrast, already a two-particle wave function $\psi(\vec x_1,\vec x_2 \hh;\hh t)$ 
is a time-dependent function of {\itshape two} spatial variables $\vec x_1$ and $\vec x_2$, and hence cannot be interpreted as a classical field on $\mathbb R^3$.
The reason for this lies in the quantization procedure for multiparticle systems.
The quantization of a system of one particle with position degree of freedom $\vec{x}=(x_1,x_2,x_3)^{\rm T}$ and conjugate momentum $\vec{p}=(p_1,p_2,p_3)^{\rm T}$
is implemented by postulating the canonical commutator relations $(i, j = 1, 2, 3)$
\begin{align}
[\h\hat x_i\hh, \h \hat p_j\h] &= \i\hbar \, \delta_{ij}\,, \\[3pt]
[\h\hat x_i\hh, \h \hat x_j\h] &= 0\,,  \\[3pt]
[\h\hh\hat p_i\hh, \h \hat p_j\h] &= 0\,.
\end{align}
These commutator relations can be represented on wave functions on $\mathbb R^3$ by
\begin{align}
(\hat x_i \h \psi)(\vec x)&=x_i \h\hh \psi(\vec x)\,,\\[5pt]
(\hat p_i \h \psi)(\vec x)&=\frac{\hbar}{\i} \, \frac{\partial \psi(\vec x)}{\partial x_i} \,.
\end{align}
If one deals instead with two particles, one gets two sets of such commutator relations, one for each particle. 
Furthermore, the position and momentum operators corresponding to different particles commute with each other.
This enforces the two-particle state to depend on two spatial variables.

Although a generic function of two variables is not a product of two functions (each of one variable),
a basis of one-particle orbitals $\{\varphi_{i}(\vec x)\}$ still induces a basis in the two-particle state space given by
\begin{equation}
\langle \vec{x}_1\otimes\vec{x}_2 \h |\h \varphi_i\otimes\varphi_j\rangle\equiv
\langle \vec{x}_1, \vec{x}_2 \mid  \varphi_i,\varphi_j\rangle=\varphi_i(\vec{x}_1) \h \varphi_j(\vec{x}_2)\,.
\end{equation}
Using complex expansion coefficients $c_{ij}$\hh, an arbitrary two-particle state can now be written as \smallskip
\begin{equation}
|\Psi^2\rangle=\sum_{i, \, j} c_{ij} \h |\varphi_i\otimes\varphi_j\rangle\,, \smallskip
\end{equation}
or more concretely, in the position-state basis, as
\begin{equation}
\langle \vec x_1, \vec x_2 \mid \Psi^2 \rangle \equiv \Psi^2(\vec x_1,\vec x_2)=\sum_{i, \, j} c_{ij}\,\varphi_i(\vec x_1)\h\varphi_j(\vec x_2)\,.
\end{equation}
The corresponding state space of the two-particle system is referred to as the {\itshape tensor product} ``$\otimes$'' of the respective one-particle state spaces.
Formally, if there are two systems with state spaces $\mathcal{H}_1$ and $\mathcal{H}_2$ separately, 
then the combined system has the state space  $\mathcal{H}_1\otimes\mathcal{H}_2$\hh, i.e., the complex vector space
spanned by the formal basis $\varphi_i\otimes\psi_j$, where the $\varphi_i$ and $\psi_j$ are respectively bases of $\mathcal H_1$ and $\mathcal H_2$\hh.
Furthermore, the scalar product in the tensor product space $\mathcal{H}_1\otimes\mathcal{H}_2$ is induced by
\begin{equation}
\langle\varphi_{i_1}\otimes\psi_{j_1} \h |\h \varphi_{i_2}\otimes\psi_{j_2}\rangle_{\mathcal H_1\otimes\mathcal H_2}=
\langle\varphi_{i_1}\h |\h \varphi_{i_2}\rangle_{\mathcal H_1}\langle\psi_{j_1}\h |\h \psi_{j_2}\rangle_{\mathcal H_2}\,.
\end{equation}
In particular, if both particles have the same state space of square-integra-\linebreak ble, complex-valued functions 
on the three-dimensional space---which is denoted by $\mathcal{H}=L^2(\mathbb{R}^3,\mathbb{C})$---then the two-particle system has the state space
\begin{equation}
L^2(\mathbb{R}^3,\mathbb{C})\otimes L^2(\mathbb{R}^3,\mathbb{C})\cong
L^2(\mathbb{R}^6,\mathbb{C})\,,
\end{equation}
which comprises the square-integrable, complex-valued functions on the six-dimensional configuration space.
This result is generalized to a system of $N$ particles (each with state space $\mathcal H$) by defining the corresponding state space as \smallskip
\begin{equation}
\mathcal{H}^N = \h \underbrace{\mathcal H \otimes \ldots \otimes \mathcal H}_{N} \h \equiv \otimes^N \mathcal{H}\,. \smallskip
\end{equation}
This means, the $N$-particle state space is given by the square-integrable, complex-valued functions of $N$ position variables $\vec x_1, \ldots, \vec x_N$\hh.
For technical purposes, one also introduces a zero-particle space $\mathcal{H}^0$, which is identified with the complex numbers $\mathbb{C}$ (formally: complex-valued functions without any argument).

All particle numbers can be treated simultaneously by introducing the {\it Fock space} over the one-particle space $\mathcal{H}$, 
denoted by $\mathcal F=\mathcal F(\mathcal H)$. This is defined as the orthogonal sum of all $N$-particle spaces, i.e.,
\begin{equation}
\mathcal{F} \h = \h \bigoplus_{N=0}^\infty \, \mathcal{H}^N \,.
\end{equation}
Elements of $\mathcal{F}$ are infinite-dimensional vectors $|\Psi\rangle=(\Psi^0,\Psi^1,\Psi^2,\ldots)^{\rm T}$, \linebreak
where each entry $\Psi^N$ denotes an $N$-particle wave function. In the position-state basis, a Fock space vector can be written as
\begin{equation}
\Psi(\vec x_1;\vec y_1,\vec y_2;\vec z_1,\vec z_2, \vec z_3; \ldots )=
\begin{pmatrix}
\Psi^0\\[2pt]
\Psi^1(\vec x_1)\\[2pt]
\Psi^2(\vec y_1,\vec y_2)\\[2pt]
\Psi^3(\vec z_1,\vec z_2,\vec z_3)\\[2pt]
\ldots
\end{pmatrix} .
\end{equation}
The scalar product of two Fock space vectors is given by the infinite sum 
\begin{equation}
\langle\Psi\mid\Phi\rangle_\mathcal{F}\equiv\sum_{N=0}^\infty \langle\Psi^N\mid\Phi^N\rangle_{\mathcal H^N}\,,
\end{equation}
over the scalar products of the respective $N$-particle components,
\begin{equation}\label{scalarprod}
\langle\Psi^N|\Phi^N\rangle_{\mathcal H^N}=\int\!\de^3\vec{x}_1\ldots\de^3\vec{x}_N\,\Psi^{N*}(\vec{x}_1,\ldots,\vec{x}_N) \h \Phi^N(\vec{x}_1,\ldots,\vec{x}_N)\,.
\end{equation}
Finally, the Fock space vector
\begin{equation}
|0\rangle =(1,0,0,\ldots)^{\rm T}\,, \smallskip
\end{equation}
which has only a zero-particle component of unity, is called the {\it vacuum}. Obviously, the zero-particle state space
is given by all complex multiples of the above vacuum vector.

\bigskip\noindent
{\it Bosons and fermions.}---Up to now, we did not pay attention to the statistics of the particles under consideration.
By the {\it Bose-Fermi alternative} \cite[Sec.~I.3.4]{Haag}, wave functions of indistinguishable particles have to be either completely symmetric or completely antisymmetric 
(such particles being called {\itshape bosons} or {\itshape fermions,} respectively). Therefore, the states of $N$ indistinguishable particles are actually restricted to a subspace of $\otimes^N\mathcal{H}$. 
These completely symmetric or antisymmetric subspaces of $\otimes^N\mathcal{H}$ will be denoted by $\vee^N\mathcal{H}$ 
and by $\wedge^N\mathcal{H}$, respectively. The corresponding subspaces of $\mathcal{F}$ are called the {\it bosonic Fock space} $\mathcal F_+$ and the {\it fermionic Fock space} $\mathcal F_-$\h.
For technical purposes, we introduce the symmetrizing operator
\begin{equation}
 \mathbb A_+ : \otimes^N \mathcal H \rightarrow \vee^N \mathcal H \,,
\end{equation}
and the antisymmetrizing operator
\begin{equation}
 \mathbb A_- : \otimes^N \mathcal H \rightarrow \wedge^N \mathcal H \,,
\end{equation}
which act on $N$-particle product states as follows:
\begin{align}
\mathbb A_+(\varphi_1\otimes \ldots \otimes \varphi_N) & \h = \h \varphi_{1}\vee\ldots\vee\varphi_{N} \\[3pt]
 & \h \equiv \frac{1}{N!}\sum_{\pi \hh \in \hh S^N} \varphi_{\pi(1)}\otimes \ldots \otimes \varphi_{\pi(N)}\,, \\[6pt]
\mathbb{A}_{-}(\varphi_1\otimes \ldots \otimes \varphi_N) & \h = \h \varphi_{1}\wedge\ldots\wedge\varphi_{N} \\[3pt]
& \h \equiv \frac{1}{N!}\sum_{\pi \hh \in \hh S^N} (-1)^\pi \h \varphi_{\pi(1)}\otimes \ldots \otimes \varphi_{\pi(N)}\,. \label{antisym}
\end{align}
Here, $S^N$ denotes the group of permutations of $N$ elements, and
$(-1)^\pi$ the character of the respective permutation. In the position-state basis, the action of the (anti)symmetrizing operator reads
\begin{align}
& \langle\vec{x}_1,\ldots,\vec{x}_N \mid  \mathbb{A}_{\pm}(\varphi_1\otimes \ldots \otimes \varphi_N)\rangle \\[5pt] \nonumber
& =\frac{1}{N!}\sum_{\pi \hh \in \hh S^N} (\pm 1)^\pi \h \varphi_{1}(\vec{x}_{\pi(1)}) \ldots  \varphi_{N}(\vec{x}_{\pi(N)})\,.
\end{align}
For example, for a product of two orbitals, we get the well-known expressions
\begin{align}
\mathbb{A}_\pm(\varphi_i(\vec x_1) \h \varphi_j(\vec x_2))=\frac{1}{2}\left(\varphi_i(\vec x_1) \h \varphi_j(\vec x_2)\pm\varphi_i(\vec x_2) \h \varphi_j(\vec x_1)\right)\,.
\end{align}
Mathematically, the operators $\mathbb{A}_{\pm}$ are {\it projectors} on $\vee^N\mathcal{H}$ or $\wedge^N\mathcal{H}$\hh, respectively. By linearity, they can be extended to projectors on $\mathcal F_+$ and $\mathcal F_-$\h. 
In particular, in their quality as projectors, these operators are hermitean and idempotent, i.e.,
\begin{equation}
\mathbb{A}_{\pm}^\dagger = \mathbb{A}_{\pm}=\mathbb{A}_{\pm}^2\,.
\end{equation}
The second property implies that $\mathbb A_{\pm}$ has no effect on (anti)symmetric wave functions, i.e.,
$\mathbb{A}_{\pm}|\Psi\rangle=|\Psi\rangle$ \h if $|\Psi\rangle$ is already (anti)symmetric.

Scalar products of completely antisymmetrized products of orbitals can be evaluated as
\begin{equation}
\langle\psi_1\mh\wedge\ldots\wedge\psi_N \mid \varphi_1\mh\wedge\ldots\wedge\varphi_N\rangle=\frac{1}{N!}\h \det (\langle\psi_i\mid\varphi_j\rangle)\,, \smallskip
\end{equation}
where ``$\det$'' denotes the {\itshape determinant} of the $(N \times N)$ matrix. Similarly, for symmetrized products we find
\begin{equation}
\langle\psi_1\hspace{-1.5pt}\vee\ldots\vee\psi_N \mid \varphi_1\hspace{-1.5pt}\vee\ldots\vee\varphi_N\rangle
=\frac{1}{N!}\, {\rm per} (\langle\psi_i\mid\varphi_j\rangle) \,, \vspace{1pt}
\end{equation}
where ``${\rm per}$'' denotes the {\itshape permanent} (which is defined analogous to the determinant, but  with all minus signs replaced by plus signs). Now, if $\{\varphi_i\}$ denotes a complete, {\it orthonormal} system in the one-particle state space \mbox{$\mathcal{H}$\hh,}
then $\{\varphi_{i_1}\vee\ldots\vee\varphi_{i_N}\}$ is a complete, {\it orthogonal} system in 
the symmetric $N$-particle space $\vee^N\mathcal{H}$\hh, and $\{\varphi_{i_1}\wedge\ldots\wedge\varphi_{i_N}\}$ 
is a complete, {\it orthogonal} system in the antisymmetric $N$-particle space $\wedge^N\mathcal{H}$\hh. Concretely, one can calculate in the fermionic case,
\begin{align} \label{sp_1}
& \langle\varphi_{i_1}\mh\wedge\ldots\wedge\varphi_{i_N} \mid \varphi_{j_1}\mh\wedge\ldots\wedge\varphi_{j_N}\rangle\\[5pt] \nonumber
&=\left\{
\begin{array}{ll}
	(-1)^\pi/N! \,, & \textnormal{\rm if } \{i_1,\ldots,i_N\}=\{j_1,\ldots,j_N\} \,, \\[5pt]
	0 \,, & \textnormal{\rm otherwise} \,,
	\end{array} \right.
\end{align}
where $(-1)^\pi$ denotes the character of the permutation which translates the tupel $(i_1,\ldots,i_N)$ into $(j_1,\ldots,j_N)$. On the other hand, in the bosonic case,
\begin{align} \label{sp_2}
& \langle\varphi_{i_1}\!\vee\ldots\vee\varphi_{i_N} \mid \varphi_{j_1}\!\vee\ldots\vee\varphi_{j_N}\rangle\\[5pt] \nonumber
&=\left\{
\begin{array}{ll}
	n_1!\ldots n_k!\h /N! \,, & \textnormal{\rm if }\{i_1,\ldots,i_N\}=\{j_1,\ldots,j_N\} \,, \\[5pt]
	0 \,, & \textnormal{\rm otherwise} \,.
	\end{array} \right.\nonumber
\end{align}
In this equation, $n_1, \ldots, n_k$ are the {\itshape occupation numbers} of the respective orbitals $\varphi_1, \ldots, \varphi_k$ 
entering into the symmetrized tensor product \mbox{$\varphi_{i_1}\!\vee\ldots \vee \mh\varphi_{i_N}$\hh.} These are all positive and sum up to $n_1+\ldots+n_k=N$. By contrast, 
in the fermionic case all occupation numbers are either zero or one, hence each orbital may appear at most once in the corresponding antisymmetrized product.

The above Eqs.~\eqref{sp_1} and \eqref{sp_2} show that  $\varphi_{i_1}\mh\wedge\ldots\wedge\varphi_{i_N}$ and $\varphi_{i_1}\mh\vee\ldots\linebreak\vee\varphi_{i_N}$ 
are not normalized (which is clear because $\mathbb A_{\pm}$ is a projector). One therefore introduces the {\it Slater determinants}
\begin{equation} \label{Slater_det}
{\rm SL}_-(\varphi_{i_1},\ldots,\varphi_{i_N}) :=\sqrt{N!}\ \varphi_{i_1}\wedge\ldots\wedge\varphi_{i_N}\,,
\end{equation}
and the {\it Slater permanents}
\begin{equation}
{\rm SL}_+(\varphi_{i_1},\ldots,\varphi_{i_N}) :=\sqrt{\frac{N!}{n_1!\ldots n_k!}}\ \varphi_{i_1}\mh\vee\ldots\vee\varphi_{i_N}\,.
\end{equation}
In the position-state basis, they read as
\begin{align}
& \langle\vec{x}_1,\ldots,\vec{x}_N \h \mid \h {\rm SL}_-(\varphi_{i_1},\ldots,\varphi_{i_N})\rangle \\[5pt] \nonumber
& =\frac{1}{\sqrt{N!}}\, \det\mh\begin{pmatrix} \varphi_{i_1}(\vec x_1) & \ldots & \varphi_{i_1}(\vec x_N) \\
\vdots & & \vdots \\
\varphi_{i_N}(\vec x_1) & \ldots & \varphi_{i_N}(\vec x_N) \end{pmatrix} ,
\end{align}
and respectively,
\begin{align}
& \langle\vec{x}_1,\ldots,\vec{x}_N \h \mid \h {\rm SL}_+(\varphi_{i_1},\ldots,\varphi_{i_N})\rangle \\[5pt] \nonumber
& = \frac{1}{\sqrt{N!\,n_1!\ldots n_k!}}\,\h {\rm per} \mh \begin{pmatrix} \varphi_{i_1}(\vec x_1) & \ldots & \varphi_{i_1}(\vec x_N) \\
\vdots & & \vdots \\
\varphi_{i_N}(\vec x_1) & \ldots & \varphi_{i_N}(\vec x_N) \end{pmatrix} .
\end{align}
By construction, the Slater determinants and permanents over a complete, {\itshape orthonormal} system in the one-particle state space $\mathcal H$ also
form a complete, {\itshape orthonormal} system in the corresponding $N$-particle (anti)symmetric state space $\vee^N\mathcal{H}$ \h or \h$\wedge^N\mathcal{H}$\hh.

\bigskip\noindent
{\it Creators and annihilators.}---For any one-particle state $\varphi\in\mathcal{H}$, we introduce the {\itshape creation and annihilation operators,}
\begin{align}
{\hat a}^\dagger(\varphi)&:\mathcal{H}^N\rightarrow\mathcal{H}^{N+1}\,, \label{cr}\\[2pt]
\hat a(\varphi)&:\mathcal{H}^N\rightarrow\mathcal{H}^{N-1}\,. \label{an}
\end{align}
For $N \geq 1$, they are defined by
\begin{align}
{\hat a}^\dagger (\varphi) \, (\varphi_1\otimes \ldots \otimes \varphi_N)&=\sqrt{N+1}\,\h (\varphi\otimes \varphi_1\otimes \varphi_2\otimes \ldots \otimes \varphi_N)\,,\label{eq_defCrea} \\[5pt]
\hat a(\varphi) \, (\varphi_1\otimes \ldots \otimes \varphi_N)&=\sqrt{N} \, \langle \varphi \mid  \varphi_1\rangle\,(\varphi_2\otimes \ldots \otimes \varphi_N)\,, \label{eq_defAnn}
\end{align}
whereas their action on the vacuum $(N = 0)$ is defined as
\begin{align}
 {\hat a}^\dagger (\varphi) \h\hh \ket{0} & = \ket{\varphi} \,, \\[5pt]
 {\hat a}(\varphi) \h\hh \ket{0} & = 0 \,.
\end{align}
By linearity, these operators can be extended to the whole Fock space $\mathcal F$. As the notation suggests, creation and annihilation operators are adjoint to each other. Furthermore, the maps $\varphi \mapsto \hat a^\dagger(\varphi)$ and $\varphi \mapsto \hat a(\varphi)$ are linear and antilinear, respectively, meaning that for $\lambda_1, \lambda_2 \in \mathbb C$,
\begin{align}
\hat a^\dagger(\lambda_1 \h \varphi_1+\lambda_2\h \varphi_2)&=\lambda_1\,\hat a^\dagger(\varphi_1)+\lambda_2\,\hat a^\dagger(\varphi_2)\,,\label{eq_CreaLin} \\[5pt]
\hat a(\lambda_1\h\varphi_1+\lambda_2\h\varphi_2)&=\lambda_1^*\,\hat a(\varphi_1)+\lambda_2^*\,\hat a(\varphi_2)\,.\label{eq_AnnAnti}
\end{align}
{\it Bosonic} and {\it fermionic} creation and annihilation operators are respectively defined by restricting the operators \eqref{cr}--\eqref{an} to the bosonic and fermionic Fock spaces, i.e.,
\begin{align}
{\hat a}_\pm^\dagger(\varphi)&=\mathbb{A}_\pm\, {\hat a}^\dagger(\varphi)\, \mathbb{A}_\pm\,,\\[5pt]
\hat a_\pm(\varphi)&=\mathbb{A}_\pm\, \hat a(\varphi)\, \mathbb{A}_\pm\,.
\end{align}
In the following, we will explicitly write the subscripts ``$\pm$'' of the creators and annihilators only if misunderstandings are to be avoided. Let $\{\varphi_i\}$ denote again an orthonormal basis in the one-particle space $\mathcal H$. The action of the fermionic creator on a Slater determinant is given by
\begin{equation} \label{zzw_1}
 \hat a_-^\dagger(\varphi_1) \, {\rm SL}_-(\varphi_{i_1}, \ldots, \varphi_{i_N}) = \left\{ \begin{array}{ll} \mathrm{SL}_-(\varphi_1, \varphi_{i_1}, \ldots, \varphi_{i_N}) \,, & \textnormal{if } \, n_1 = 0 \,, \\[5pt]
 0 \,, & \textnormal{if } \, n_1 = 1 \,, \end{array} \right.
\end{equation}
where $n_1$ counts how often the orbital $\varphi_1$ appears in the Slater determinant $\mathrm{SL}_-(\varphi_{i_1}, \ldots, \varphi_{i_N})$.
The corresponding action of the fermionic annihilator reads
\begin{equation}
 \left\{ \begin{array}{ll} \hat a_-(\varphi_1) \, {\rm SL}_-(\varphi_{i_1}, \ldots, \varphi_{i_N}) = 0 \,, & \textnormal{if } \, n_1 = 0 \,, \\[5pt]
 \hat a_-(\varphi_1) \, {\rm SL}_-(\varphi_1, \varphi_{i_2}, \ldots, \varphi_{i_N}) = \mathrm{SL}_-( \varphi_{i_2}, \ldots, \varphi_{i_N}) \,, & \textnormal{if } \, n_1 = 1 \,. \end{array} \right.
\end{equation}
Here, we have assumed without loss of generality that if $n_1 = 1$, the orbital $\varphi_1$ appears as the first argument of the Slater determinant. (If it appears instead at another position, one may simply reorder the arguments using the antisymmetry of the Slater determinant.) In the bosonic case, the creator acts on a Slater permanent as
\begin{equation}
 \hat a_+^\dagger(\varphi_1) \, {\rm SL}_+(\varphi_{i_1}, \ldots, \varphi_{i_N}) = \sqrt{n_1 + 1} \ \mathrm{SL}_+(\varphi_1, \varphi_{i_1}, \ldots, \varphi_{i_N}) \,,
\end{equation}
while the bosonic annihilator yields
\begin{equation}
\hat a_+(\varphi_1) \, {\rm SL}_+(\varphi_{i_1}, \ldots, \varphi_{i_N}) = 0 \,, \quad \textnormal{if } \, n_1 = 0 \,,
\end{equation}
and \smallskip
\begin{align} \label{zzw_2}
 & \hat a_+(\varphi_1) \, {\rm SL}_+(\underbrace{\varphi_1, \varphi_1, \ldots, \varphi_1}_{n_1}, \varphi_{i_2}, \ldots, \varphi_{i_N}) \\[2pt] \nonumber
 & = \sqrt{n_1} \ \mathrm{SL}_+( \underbrace{\varphi_1, \ldots, \varphi_1}_{n_1 - 1}, \varphi_{i_2}, \ldots, \varphi_{i_N}) \,, \quad \textnormal{if } \, n_1 \geq 1 \,.
\end{align}
Again, we have assumed here that the $n_1$ orbitals $\varphi_1$ appear as the first arguments of the Slater permanent. (By the symmetry of the Slater permanent, this can always be achieved by simply permuting the arguments.) 
In particular, the above results imply the following representations of the Slater determinants and permanents:
\begin{align}
 \mathrm{SL}_-(\varphi_{i_1}, \ldots, \varphi_{i_N}) & = \hat a^\dagger_-(\varphi_{i_1}) \ldots \hat a^\dagger_-(\varphi_{i_N}) \, \ket{0} \,, \label{reprdet} \\[5pt]
 \mathrm{SL}_+(\varphi_{i_1}, \ldots, \varphi_{i_N}) & = \frac{1}{\sqrt{n_1! \ldots n_k!}} \, \hat a^\dagger_+(\varphi_{i_1}) \ldots \hat a^\dagger_+(\varphi_{i_N}) \, \ket{0} \,. \label{reprper}
\end{align}
Finally, a direct calculation using the explicit expressions Eqs.~\eqref{zzw_1}--\eqref{zzw_2} and the (anti)linearity of the maps $\varphi \mapsto \hat a(\varphi)$ and $\varphi \mapsto \hat a^\dagger(\varphi)$ shows that the bosonic or fermionic creation and annihilation operators
respectively fulfill the {\it canonical (anti)commutation relations}:
\begin{align}
\big[\hat a_\pm(\psi),\h{\hat a}^\dagger_\pm(\varphi)\big]_\mp&=\langle \psi \h |\h \varphi\rangle\,,\\[2pt]
\big[\hat a_\pm(\psi),\h\hat a_\pm(\varphi)\big]_\mp&=0\,, \\[2pt]
\big[{\hat a}^\dagger_\pm(\psi),\h{\hat a}^\dagger_\pm(\varphi)\big]_\mp&=0\,,
\end{align}
where \smallskip
\begin{equation}
[\hat A, \hat B\hh]_- \h \equiv \h [\hat A,\hat B\hh] \h = \h \hat A\hat B-\hat B\hat A \smallskip
\end{equation}
denotes the commutator, and
\begin{equation}
 [\hat A,\hat B\hh]_+= \hat A\hat B+\hat B\hat A
\end{equation}
the anticommutator of two operators $\hat A$ and $\hat B$.

\bigskip\noindent
{\it Field operators.}---After the introduction of general creators and annihilators, we can finally introduce 
(bosonic or fermionic) {\it field operators} \h$\hat\psi^\dagger(\vec x)$ and $\hat\psi(\vec x)$ (cf.~Sec.~\ref{subsec_secQelfistQnuc} in the main text). 
For this purpose, we consider the generalized basis of position eigenvectors $|\vec x\rangle$ in the one-particle state space,
and define the (bosonic or fermionic) field operators as the corresponding creators and annihilators:
\begin{align}
\hat\psi_\pm^\dagger(\vec x) & = \hat a_\pm^\dagger(\ket{\vec x}) \,, \\[3pt]
\hat\psi_\pm(\vec x) & = \hat a_\pm(\ket{\vec x}) \,.
\end{align}
Using the relation
\begin{equation}
 \langle \vec x \mid  \vec x' \rangle = \int \! \de^3 \vec y \,\h \delta^3(\vec y - \vec x) \h \delta^3(\vec y - \vec x') = \delta^3(\vec x - \vec x') \,,
\end{equation}
we obtain the canonical (anti)commutation relations for the field operators:
\begin{align}
\big[\hat\psi_\pm(\vec x),\h\hat\psi_\pm^\dagger(\vec{x}')\big]_\mp&=\delta^3(\vec{x}-\vec{x}')\,, \\[2pt]
\big[\hat\psi_\pm(\vec x),\h\hat\psi_\pm(\vec{x}')\big]_\mp&=0\,,\\[2pt]
\big[\hat\psi_\pm^\dagger(\vec x),\h\hat\psi_\pm^\dagger(\vec{x}')\big]_\mp&=0\,.
\end{align}
In the Fock space formalism, these (anti)commutation relations are straightforward deductions from the underlying definitions of the field operators.

The action of the annihilator $\hat a_{\pm}(\varphi)$ on the $N$-particle state space $\mathcal{H}^N$ reads in the position basis,
\begin{equation}
\big(\hat a_{\pm}(\varphi)\h\Psi^N\hh\big)(\vec{x}_2,\ldots,\vec{x}_N)=\sqrt{N}\int\!\de^3\vec{x}_1\,\varphi^*(\vec{x}_1)\h\hh\Psi^N(\vec{x}_1,\vec{x}_2,\ldots,\vec{x}_N) \,. \label{eq_annBasis}
\end{equation}
From this we immediately recover the important formula \eqref{eq_fieldOpExpl} in the main text,
which holds both in the bosonic and in the fermionic case. By contrast, the action of the creation operators depends 
on the statistics. In the position-state basis, it reads
\begin{align}
& \big({\hat a}_\pm^\dagger(\varphi) \h \Psi^{N-1}\hh\big)(\vec{x}_1,\ldots,\vec{x}_N) \\[2pt] \nonumber
& =\frac{1}{\sqrt{N}}
\sum_{j=1}^N(\pm 1)^{j-1} \h \varphi(\vec x_j) \h\hh\Psi^{N-1}(\vec x_1,\ldots,\widecheck{\vec{x}}_j,\ldots,\vec x_N)\,,
\end{align}
where the notation $\widecheck{\vec x}_j$ means that the argument $\vec x_j$ is omitted.
The creators and annihilators of one-particle states can be reconstructed from the field operators by means of the intuitive formulae
\begin{align}
{\hat a}^\dagger(\varphi)&=\int\!\de^3\vec x\,\h \varphi(\vec x)\h \hat\psi^\dagger(\vec x)\,,\\[2pt]
\hat a(\varphi)&=\int\!\de^3\vec x\,\h \varphi^*(\vec x)\h \hat\psi(\vec x)\,.
\end{align}
Conversely, we can express the field operators through the creators and annihilators
of a complete orbital system $\{\varphi_i\}$ in the one-particle space as
\begin{align}
\hat\psi^\dagger(\vec x)&=\sum_i \varphi^*_{i}(\vec x) \, {\hat a}^\dagger_{i}\,,\label{eq_fieldOpCrei} \\[2pt]
\hat\psi(\vec x)&=\sum_i \varphi_{i}(\vec x) \, {\hat a}_{i}\,,\label{eq_fieldOpAnni}
\end{align}
where $\hat a_i^\dagger \equiv \hat a^\dagger(\varphi_i)$ and $\hat a_i \equiv \hat a(\varphi_i)$\hh.
We further note that by the help of the field operators, 
a general (bosonic or fermionic) $N$-particle state $|\Psi^N\rangle$ can be expressed 
through its real-space wave function as
\begin{equation}
\begin{aligned}
&|\Psi^N\rangle=\frac{1}{\sqrt{N!}} \h \int\!\de^3\vec x_1\ldots\int\!\de^3\vec x_N\,\Psi^N(\vec x_1,\ldots,\vec x_N) \, \hat\psi^\dagger(\vec x_1)\ldots\hat\psi^\dagger(\vec x_N) \h |0\rangle\,,
\end{aligned}
\end{equation}
which can be shown from Eqs.~\eqref{reprdet} or \eqref{reprper}. In particular, $|\Psi^N\rangle$ is normalized if the first-quantized
real-space wave function $\Psi^N(\vec x_1,\ldots,\vec x_N)$ is. Furthermore, for a one-particle state $|\Psi^1\rangle$ we retrieve the well-known formula \smallskip
\begin{equation}
|\Psi^1\rangle=\int\!\de^3\vec x'\,\Psi^1(\vec x')\h \hat\psi^\dagger(\vec x') \h |0\rangle=
\int\!\de^3\vec x'\,\Psi^1(\vec x') \h |\vec x'\rangle\,. \smallskip
\end{equation}
Using that $\langle \vec x \mid \vec x' \rangle = \delta^3(\vec x - \vec x')$, this further implies
\begin{equation}
 \Psi^1(\vec x) = \int \! \de^3 \vec x' \, \Psi^1(\vec x') \, \langle \vec x \mid \vec x' \rangle =
\langle 0 \mid \hat \psi(\vec x) \mid \Psi^1 \rangle\,.
\end{equation}
Similarly, from a general (bosonic or fermionic) Fock space vector $|\Psi\rangle=(\ldots,\Psi^N,\ldots)\in\mathcal{F}_{\pm}$, one can gain back the $N$-particle component in the position-space basis by means of
\begin{equation}
\Psi^N(\vec x_1,\ldots,\vec x_N)=\frac{1}{\sqrt{N!}} \, \langle 0 \mid  \hat\psi(\vec x_N)\ldots\hat\psi(\vec x_1) \h |\h \Psi^N\rangle\,.
\end{equation}
We note that this equation connects the (anti)symmetry of the wave function to the (anti)commutativity of the field operators.

\subsection{Second quantization}\label{subsecSecondQuant}

{\it Introduction.}---There are two different notions of `second quantization': (i) The so-called second quantization of 
classical fields (such as the displacement field or the Schr\"{o}dinger field). In this sense, 
second quantization has actually the same meaning as {\itshape field quantization.}
In cases where the field had been known prior to the advent of quantum mechanics, the second quantization is actually
the first (field) quantization. By contrast, since the Schr\"odinger field had originally been discovered by quantizing a classical point-particle theory, 
the (first) field quantization of the (classical) Schr\"odinger field was then called a ``second'' quantization procedure. (ii) The promotion 
of one-particle operators to many-particle operators, or---equivalently---their expression in terms of field operators. It is this second aspect which concerns us in this subsection.

As a heuristic introduction, we first consider a system of two non-inter\-{}acting particles in a factorizing state, 
\begin{equation}
|\Phi\rangle=|\varphi_1, \varphi_2\rangle\equiv|\varphi_1\otimes\varphi_2\rangle\,. 
\end{equation}
As the particles do not interact, the total Hamiltonian is given by the sum of two one-particle Hamiltonians
(assumed to be identical), and the total energy therefore reads
\begin{equation}
E=\langle\varphi_1 \h |\h \hat H\h |\h \varphi_1\rangle+\langle\varphi_2\h |\h \hat H\h |\h \varphi_2\rangle\,.
\end{equation}
We want to write this in the form
\begin{equation}
E=\langle\Phi\h |\h \mathcal Q(\hat H)\h |\h \Phi\rangle \,,
\end{equation}
with an appropriate two-particle Hamiltonian $\mathcal Q(\hat H)$ constructed from the one-particle Hamiltonian $\hat H$.
This can be achieved by setting
\begin{equation}
\mathcal Q(\hat H)=\hat H\otimes\hat{\mathrm I}+\hat{\mathrm I}\otimes\hat H\,,
\end{equation}
where $\hat{\mathrm I}$ denotes the identity operator on each subspace. Equivalently, one can write
\begin{equation}
\mathcal Q(\hat H)=\hat H_1+\hat H_2\,, \smallskip
\end{equation}
where it is understood that $\hat H_i$ acts on the $i$-th coordinate of the 
two-particle wave function. These formulae already encapsulate the essence of the second quantization of operators, 
in that they show how many-particle observables are to be constructed
from their one-particle counterparts. We just have to generalize this now to arbitrary particle numbers.

\bigskip\noindent
{\it Second quantization of hermitean and unitary operators.}---First, for hermitean operators $\hat A$ on the one-particle state space $\mathcal{H}$,
the second quantization $\mathcal Q(\hat A)$ is defined as a hermitean operator on the Fock space $\mathcal{F}(\mathcal{H})$ given by
\begin{equation}
 \mathcal Q(\hat A) \left( \! \begin{array}{c} \Psi^0 \\[2pt] \Psi^1 \\[2pt] \Psi^2 \\[2pt] \Psi^3 \\[2pt] \ldots \end{array} \! \right)
= \left( \! \begin{array}{c} \!\! 0 \ \\[2pt] \!\! \hat A \, \Psi^1 \ \\[2pt] \!\! (\hat A \otimes \hat{\mathrm I} + \hat{\mathrm I} \otimes \hat A) \, \Psi^2 \ \\[2pt] \!\! ( \hat A \otimes \hat{\mathrm I} \otimes \hat{\mathrm I} + \hat{\mathrm I} \otimes \hat A \otimes \hat{\mathrm I} + \hat{\mathrm I} \otimes \hat{\mathrm I} \otimes \hat A) \, \Psi^3 \ \\[2pt] \!\! \ldots \ \end{array} \!\! \right)\,.
\end{equation}
That means, the second quantization of $\hat A$ is an infinite-dimensional diagonal matrix,
\begin{equation}
\mathcal Q(\hat A)={\rm diag}\h\big(0\,;\h\hat A_1\,;\h\hat A_1+\hat A_2\,;\h\hat A_1+\hat A_2+\hat A_3\,;\h\ldots\big)\,,
\end{equation}
where $\hat A_i$ is given by $\hat A$ in its action on the $i$-th coordinate of the $N$-particle wave function
$\Psi^N(\vec x_1,\ldots,\vec x_i,\ldots,\vec x_N)$. By contrast, for unitary operators $\hat U$ we define:
\begin{equation}
 \mathcal Q(\hat U) \left( \! \begin{array}{c} \Psi^0 \\[2pt] \Psi^1 \\[2pt] \Psi^2 \\[2pt] \Psi^3 \\[2pt] \ldots \end{array} \! \right)
= \left( \! \begin{array}{c} \!\! \Psi^0 \ \\[2pt] \!\! \hat U \, \Psi^1 \ \\[2pt] \!\! (\hat U \otimes \hat U ) \, \Psi^2 \ \\[2pt] \!\! ( \hat U \otimes \hat U \otimes \hat U ) \, \Psi^3 \ \\[2pt] \!\! \ldots \ \end{array} \!\! \right)\,,
\end{equation}
i.e., the second quantization of $\hat U$ is the infinite-dimensional diagonal matrix
\begin{equation}
\mathcal Q(\hat U)={\rm diag}\h\big(\h\hat{\mathrm I}\,; \h \hat U_1\,; \h \hat U_1\h \hat U_2\,;\h \hat U_1\h \hat U_2\h \hat U_3\,;\h\ldots\big)\,.
\end{equation}
The second quantization of hermitean and unitary operators is defined differently in order to recover
the relation between the Hamiltonian operator (which is hermitean) and its corresponding
time-evolution operator (which is unitary). In fact, if $\hat H$ generates the time evolution 
\begin{equation}
\hat U(t-t')=\exp\mh\left(-\frac{\i}{\hbar}\h (t-t') \hh \hat H\right), \label{eq_timeEvo}
\end{equation}
then the second-quantized time evolution $\mathcal Q(\hat U)$ is generated by $\mathcal Q(\hat H)$
through the analogous formula, i.e.,
\begin{equation}
\mathcal Q(\hat U(t-t'))=\exp\mh\left(-\frac{\i}{\hbar}\h (t-t') \h\mathcal Q(\hat H)\right).
\end{equation}
For hermitean operators, the second quantization is a linear mapping,
\begin{equation}
\mathcal Q(\lambda \h \hat A+\mu \h \hat B)=\lambda \h \mathcal Q(\hat A) +\mu \h \mathcal Q(\hat B)\,,
\end{equation}
which preserves the (anti)commutator relations
\begin{equation}
[\hat A,\hat B\hh]_{\mp}=\hat C
\end{equation}
in the sense that \smallskip
\begin{equation}
[\mathcal Q(\hat A),\mathcal Q(\hat B)]_{\mp}=\mathcal Q(\hat C)\,. \smallskip
\end{equation}
On the other hand, for unitary operators the general identity holds,
\begin{equation}
 \mathcal Q(\hat U \hh \hat V) = \mathcal Q(\hat U) \h \mathcal Q(\hat V) \,.
\end{equation}
For {\itshape both} hermitean and unitary operators, the original one-particle operator can be identified 
with its second quantization restricted to the one-particle subspace, \smallskip
\begin{equation}
\mathcal Q(\hat U)\h \big|_\mathcal{H}=\hat U\,. \smallskip
\end{equation}
Moreover, for both hermitean and unitary operators we have
\begin{equation}
(\mathcal Q(\hat U))^\dagger=\mathcal Q(\hat U^\dagger)\,,
\end{equation}
implying that the second quantization of a hermitean operator is again hermitean, and the second quantization of a unitary operator is again unitary.
The restriction of $\mathcal Q(\hat U)$ to the bosonic/fermionic Fock space is defined as
\begin{equation}
\mathcal Q_{\pm}(\hat U)=\mathbb A_\pm \h \mathcal Q(\hat U) \h \mathbb A_\pm\,.
\end{equation}
This is an identity in the sense of
\begin{equation}
  \mathcal Q_{\pm}(\hat U) = \mathcal Q(\hat U) \big|_{\mathcal F_{\pm}} \,, 
\end{equation}
because $\mathcal Q(\hat U)$ respects the (anti)symmetry of the wave function (on each subspace with fixed particle number).
For example, the standard expression for an $N$-particle kinetic energy operator,
\begin{equation}
\hat T_N=-\frac{\hbar^2}{2m} \h \sum_{i=1}^N\left(\frac{\partial}{\partial\vec x_i}\right)^{\!\!2}\,,
\end{equation}
if applied to an (anti)symmetric wave function, automatically yields another (anti)symmetric wave function.

\bigskip\noindent
{\it Normal form.}---The defining formulae for the second quantization of an operator are rather clumsy
and therefore primarily of a conceptual value. Fortunately though, it is possible to rewrite
the second quantization of {\itshape hermitean} operators in terms of the creation and annihilation operators introduced in Sec.~\ref{subsecFockSpace}.
This is called the {\it normal form} of a second-quan\-{}tized hermitean operator. Concretely, given a complete orthonormal system $\{\varphi_i\}$ of the one-particle state space $\mathcal{H}$,
the normal form of $\mathcal Q_{\pm}(\hat A)$ is just the generalization of the formula \smallskip \vspace{2pt}
\begin{equation}
\hat A=\sum_{i,\, j}|\varphi_i\rangle\langle\varphi_i \mid \hat A \mid \varphi_j\rangle\langle\varphi_j|\,, \smallskip
\end{equation}
valid on the one-particle space, to
\begin{equation}
\mathcal Q_{\pm}(\hat A)=\sum_{i, \, j} {\hat a}^\dagger_i\h \langle\varphi_i \mid \hat A \mid \varphi_j\rangle \h \hat a_j\,,\label{eq_normalform}
\end{equation}
valid on the bosonic or fermionic Fock space. Hence, in this 
last equation $\hat a_i^\dagger \equiv \hat a^\dagger_{\pm}(\varphi_i)$ and $\hat a_i \equiv \hat a_{\pm}(\varphi_i)$ denote 
the (anti)symmetrized creation and annihilation operators, respectively.
This analogy between creators and {\itshape ket} vectors, as well as anni\-{}hilators and {\itshape bra} vectors,
can be pushed further. For example, let $\{\varphi_i\}$ and $\{\psi_j\}$ denote 
different orthogonal bases in the one-particle space, which are related by
\begin{align}
 | \varphi_i \rangle & = \sum_j | \psi_j \rangle \h \langle \psi_j \mid  \varphi_i \rangle \,, \\[5pt]
 \langle\varphi_i| & = \sum_j \langle \varphi_i \h |\h \psi_j \rangle \h \langle \psi_j |  \,.
\end{align}
The operators $\hat a_{i}\equiv \hat a(\varphi_{i})$ can then be expressed analogously through the operators $\hat b_{j}\equiv \hat a(\psi_j)$, and vice versa, by means of
\begin{align}
{\hat a}^\dagger_{i}&=\sum_j \hat b^\dagger_j \, \langle\psi_j \mid  \varphi_i\rangle \,, \\[5pt]
\hat a_i&=\sum_j\langle\varphi_i \h |\h \psi_j \rangle \, \hat b_j\,.
\end{align}
A particularly important operator in the Fock space is the {\it particle number operator}. 
It can be defined as the second quantization of the identity operator on the one-particle state space, i.e.,
\begin{equation}
\hat N = \mathcal Q\big(\h\hat{\mathrm I}_{\mathcal H}\hh\big) \,.
\end{equation}
This gives the infinite-dimensional diagonal matrix
\begin{equation}
 \hat N={\rm diag}\h(\h 0 \h ; \h 1 \h ; \h 2 \h ; \h 3\h ; \h \ldots \h) \,.
\end{equation}
Its eigenstates are given
by those Fock space vectors which have only one non-vanishing component. The $N$-particle state space can therefore be gained back
from the Fock space as the eigenspace of the particle number operator with eigenvalue $N$. The normal form of the particle number operator reads
\begin{equation}
\hat N=\int\!\de^3\vec x\,\h\hat n(\vec x)=\int\!\de^3\vec x\,\h\hat\psi^\dagger(\vec x)\h\hat\psi(\vec x)=\sum_{m}{\hat a}^\dagger_m \h {\hat a}_m\,,\label{eq_numberdens}
\end{equation}
where the subscript $m$ may index {\it any} complete, orthonormal system in the one-particle state space.

Similarly, we can also express the second quantization of the {\itshape free} (one-particle) Hamiltonian $\hat H_0$ in terms of the field operators, whereby we retrieve the formula
\eqref{eq_kinEnSecQu} from the main text. Furthermore, in terms of a complete orthonormal system of energy eigenstates of the one-partical Hamiltonian,
\begin{equation}
\hat H_0 \h \varphi_m=\varepsilon_{0m} \h \varphi_m\,,
\end{equation}
the normal form of the second-quantized Hamiltonian reads
\begin{equation}
\mathcal Q(\hat H_0)=\sum_{m} \varepsilon_{0m} \, {\hat a}^\dagger_m \h \hat a_m \label{eq_secHamNormal}
\end{equation}
with $\hat a_m = \hat a(\varphi_m)$ and $\hat a^\dagger_m = \hat a^\dagger(\varphi_m)$\hh, hence it is given by a sum of harmonic oscillator Hamiltonians. This follows directly from Eq.~\eqref{eq_normalform}. By contrast, the second-quantized Coulomb
potential (Eq.~\eqref{sq_Coulomb} in the main text) is not the second quantization of any one-particle operator. Instead, it vanishes identically on
the one-particle state space. 

\bigskip\noindent
{\it Time-dependent field operators.}---We now finally come to the time-dependent field operators $\hat\psi(\vec x,t)$.
In principle, the time dependence of any operator is defined (in the Heisenberg picture) by
\begin{equation}
\hat A(t)=\e^{\i t \hat H/\hbar}\hat A\,\e^{-\i t \hat H / \hbar}\,.
\end{equation}
In the case of a field operator whose time dependence is induced by the second-quantized one-particle Hamiltonian \eqref{eq_secHamNormal},
this can be calculated explicitly: First, from the defining equations \eqref{eq_defCrea}--\eqref{eq_defAnn}, 
one shows easily that for any one-particle unitary operator $\hat U$ we have
\begin{align}
\mathcal Q(\hat U)^{-1}\,{\hat a}^\dagger(\varphi)\,\mathcal Q(\hat U)&={\hat a}^\dagger(\hat U^{-1}\varphi)\,, \\[5pt]
\mathcal Q(\hat U)^{-1}\,\hat a(\varphi)\,\mathcal Q(\hat U)&=a(\hat U^{-1}\varphi)\,.
\end{align}
In particular, by identifying $\hat U =\hat U_0(t) \equiv \e^{-\i t \hat H_0 / \hbar}$ with the time-evolution operator, we find for the time-dependent creators and annihilators,
\begin{align}
{\hat a}^\dagger(\varphi;\h t)&={\hat a}^\dagger(\hat U_0^{-1}(t) \h \varphi)\,, \\[5pt]
{\hat a}(\varphi;\h t)&={\hat a}(\hat U_0^{-1}(t) \h \varphi)\,.
\end{align}
For an energy eigenstate $\varphi_m$, the (reverse) time evolution reads
\begin{equation}
\hat U^{-1}_0(t) \h \varphi_m \h =\h\hat U_0(-t)\h \varphi_m\h =\h \e^{{\rm i}  t \h \varepsilon_{0m} / \hbar} \h \varphi_m \h = \h  \e^{{\rm i} \omega_{m} t} \h \varphi_m \,,
\end{equation}
where we have defined $\omega_m \equiv \varepsilon_{0m}/\hbar$.
Using Eqs.~\eqref{eq_CreaLin}--\eqref{eq_AnnAnti}, we obtain  for the time dependence of the corresponding creators and annihilators,
\begin{align}
{\hat a}^\dagger_m(t)&={\hat a}^\dagger_m\,\e^{{\rm i}\omega_{m} t}\,, \\[5pt]
\hat a_m(t)&=\hat a_m\,\e^{-{\rm i}\omega_{m} t}\,.
\end{align} 
With this, using the expressions \eqref{eq_fieldOpCrei}--\eqref{eq_fieldOpAnni}, we find the explicit formulae for the time-dependent field operators,
\begin{align}
\hat\psi^\dagger(\vec x,t)&=\sum_{m} {\hat a}^\dagger_m \,\e^{{\rm i}\omega_m t} \h\hh \varphi_m^*(\vec x)\,, \\[5pt]
\hat\psi(\vec x,t)&=\sum_{m} \hat a_m\, \e^{-{\rm i}\omega_m t} \h\hh \varphi_m(\vec x)\,. \label{eq_fieldOpTimeDep}
\end{align}
Their (anti)commutator can then be written as
\begin{equation}
\big[\hat\psi(\vec x,t),\hat\psi^\dagger(\vec x',t')\big]_{\mp}=U_0(\vec x,t;\vec x',t')
\end{equation}
in terms of the {\itshape propagator,}
\begin{equation}
 U_0(\vec x,t;\vec x',t'):=\big\langle\vec x \h \big| \h \e^{-\i (t - t') \hat H_0 / \hbar} \h \big| \hh \vec x' \big\rangle\,.
\end{equation}
Remarkably, Eq.~\eqref{eq_fieldOpTimeDep} is analogous to the {\itshape mode expansion,}
\begin{align} \label{schr_mode_exp}
\psi(\vec x,t)&=\sum_{m} a_m\, \e^{-{\rm i}\omega_m t} \h \varphi_m(\vec x)\,,
\end{align}
corresponding to the {\it classical} field equation
\begin{equation}
\big( \i\hbar \h \partial_t-\hat H_0 \big)\h\psi(\vec x,t)=0\,.
\end{equation}
Therefore, Eq.~\eqref{eq_fieldOpTimeDep} implies that the quantization of the Schr\"odinger field can be interpreted as the substitution of the mode expansion
coefficients with operators obeying the creation and annihilation algebra. This is in fact completely analogous to the quantization of the classical displacement field (see Eqs.~\eqref{CCRphon1}--\eqref{CCRphon3}). 
Correspondingly, it is straightforward to show that for any wave function
\begin{equation}
\psi(\vec x)=\sum_m a_m \h \varphi_m(\vec x)\,, 
\end{equation}
we have $\langle 0 \mid \hat a_m \mid \psi \rangle = a_m$\hh, and hence, by Eqs.~\eqref{eq_fieldOpTimeDep},
\begin{equation} \label{rech}
\langle 0 \mid \hat\psi(\vec x,t) \mid \psi\rangle
=\sum_m a_m \, \e^{-{\rm i}\omega_m t} \h \varphi_m(\vec x)
=\psi(\vec x,t)\,.
\end{equation}
With the usual identification
\begin{equation}
\hat\psi^\dagger(\vec x,t) \mid 0\rangle=|\vec x,t\rangle\,,
\end{equation}
the result \eqref{rech} obviously generalizes $\langle\vec x \mid\psi\rangle=\psi(\vec x)$ to
\begin{equation}
\langle\vec x,t \mid \psi\rangle=\psi(\vec x,t)\,. 
\end{equation}
Finally, we comment on the interpretation of the time-dependent field operator $\hat \psi^\dagger(\vec x, t)$. Using that
\begin{equation}
\hat\psi^{\dagger}(\vec x,t)=\hat U^{-1}(t) \, \hat a^{\dagger}(|\vec x\rangle) \, \hat U(t)=\hat a^{\dagger}(\hat U(-t) \h |\vec x \rangle) \,,
\end{equation}
we see that the operator $\hat\psi^{\dagger}(\vec x,t)$ creates the state vector
\begin{equation}
 |\varphi\rangle \h := \h \hat U(-t) \h \ket{\vec x} \h = \h \e^{\i t \hat H / \hbar} \h |\vec x\rangle \,,
\end{equation}
with the corresponding wave function in position space
\begin{equation}
 \varphi(\vec x') = ( \e^{\i t \hat H / \hbar} \h \delta^3)(\vec x'-\vec{x})\,.
\end{equation}
After a time lapse $t$, this wave function will take the form of a Dirac delta concentrated at the point $\vec x$, i.e.,
\begin{equation}
 (\e^{-\i t \hat H / \hbar} \h \varphi)(\vec x') = \delta^3(\vec x' - \vec x) \,.
\end{equation}
Sloppily, one therefore says that $\hat\psi^\dagger(\vec x,t)$ creates a particle at the space-time point $(\vec x,t)$. 
This, however, does not mean that $|\vec x,t\rangle$ is localized in time. Instead, for all times $t' \not = t$, 
the wave function will in general smear out, and its concrete form then depends on the Hamiltonian $\hat H$ under consideration.

\bigskip\noindent
{\it Spin in second quantization.}---So far, we have never treated the spin expli\-citly, although it actually determines the statistics of the particles: 
half-integer spin leads to fermions, integer spin to bosons (see e.g.~\cite[Sec.~3.3.3]{Itzykson}).
On a non-relativistic level, the fact that, say, electrons carry spin one-half, implies 
that the classical field equation is not the ``scalar'' Schr\"{o}dinger
equation but the {\itshape Pauli equation.} In the free case, this reads (see e.g.~\cite[Chap.~XX, \S\,29]{Messiah}, \cite[Sec.~1.4]{Bjorken}, or \cite[App.~C.2]{ED2})
\begin{equation}
\i \hh \hbar \, \partial_t \h \psi(\vec x,t)=\frac{\left(\boldsymbol\sigma\cdot(-\i\hbar\nabla)\right)^2}{2m} \, \psi(\vec x,t)\,,
\end{equation}
where $\boldsymbol\sigma=(\sigma_1,\sigma_2,\sigma_3)^{\rm T}$ is the vector of Pauli matrices.
Correspondingly, the wave function $\psi(\vec x) \equiv (\psi_1(\vec x),\psi_2(\vec x))^{\rm T}$
now takes values in the two-dimensional complex space $\mathbb C^2$,
and the one-particle state space is given by
\begin{equation}
\mathbb C^2\otimes L^2(\mathbb R^3, \mathbb C) \, \cong \,
L^2(\mathbb R^3,\mathbb C^2) \, \cong \, L^2(\mathbb R^3,\mathbb C)\oplus L^2(\mathbb R^3,\mathbb C) \,,
\end{equation}
with the scalar product
\begin{equation}
\langle\varphi \h |\h \psi\rangle=\int\!\de^3\vec x\,\h \varphi^\dagger(\vec x) \h \psi(\vec x)=
\int\!\de^3\vec x\,\h\big(\varphi_1^*(\vec x)\h \psi_1(\vec x)+\varphi^*_2(\vec x)\h \psi_2(\vec x)\big)\,.
\end{equation}
Therefore, the formulary worked out so far automatically carries over to sys\-{}tems with spin 
if we implicitly understand by the orbitals $\varphi_i$ elements of $L^2(\mathbb{R}^3,\mathbb{C}^2)$. 
When dealing with spin systems, however, it comes in handy to write a spin index explicitly, i.e.,
\begin{equation}
\hat\psi_{s}^\dagger(\vec x) \equiv \hat \psi^\dagger( \hh \ket{\vec x, s} )\,.\label{app_spin2quant}
\end{equation}
For $s\in\{\uparrow,\downarrow\}$, this field operator creates the respective states
\begin{equation}
 \ket{\vec x, \uparrow} = \left( \!\! \begin{array}{l} \ket{\vec x} \\[2pt] 0 \end{array} \!\! \right) , \qquad 
 \ket{\vec x, \downarrow} = \left( \!\! \begin{array}{l} 0 \\[2pt] \ket{\vec x} \end{array} \!\! \right) .
\end{equation}
The first-quantized spin observable,
\begin{equation}
\hat{\!\boldsymbol S}=\frac{\hbar}{2} \h \boldsymbol{\sigma} \,,
\end{equation}
then reads in second quantization,
\begin{align}
\mathcal Q(\,\hat{\!\vec S})
=\frac{\hbar}{2}\h\int\!\de^3\vec x\h \sum_{s, \h s'} \hat\psi^\dagger_s(\vec x) \, \boldsymbol{\sigma}_{ss'} \, \hat\psi_{s'}(\vec x)\,.
\end{align}
Correspondingly, we introduce the {\it spin density} by the formula
\begin{equation}
\hat{\! \vec S}(\vec x)=\frac{\hbar}{2} \h \sum_{s, \h s'} \hat\psi^\dagger_s(\vec x) \, \boldsymbol{\sigma}_{ss'} \, \hat\psi_{s'}(\vec x)\,,
\end{equation}
such that the total spin operator in second quantization is given by the spatial integral of the corresponding spin density operator. 
This expression is particularly important for the definition of a spin contribution to the electromagnetic current density (see e.g.~\cite[Sec.~3.2.5]{ED2}).

\section{Aspects of linear response theory} \label{app_AspectsResp}

In the first part of this appendix (Sec.~\ref{app_Response}), we assemble
for the convenience of the reader a number of important results in linear response
theory, which have already been used in the main text, namely: (i) the definition of the density response function,
(ii) the connection between its {\it direct} and its {\it proper} version, (iii) their respective relation to the dielectric function, 
and (iv) the spectral representation of the density response function as derived from the Kubo formula.
In the second part (Secs.~\ref{subsec_emMatProp}--\ref{app_Kubo}), we will investigate concretely 
the electromagnetic response generated by the displacement field of the charged nuclei.

\subsection{Response function and Kubo formalism} \label{app_Response}

{\itshape Definition.---}Typical experiments in solid state physics investigate the response of a material probe to an external (e.g.~electromagnetic)
perturbation \cite[Chap.~I, Sec.~1]{Madelung}. With any material probe, we can associate a microscopic, {\it internal} charge density, \smallskip
\begin{equation}
\rho\equiv\rho_{\rm int}(\vec x,t)\,, \smallskip
\end{equation}
which is the charge density produced by the degrees of freedom which constitute the material (typically: electrons and nuclei).
In the presence of a time-dependent {\itshape external} perturbation, such as an external potential,
\begin{equation}
\varphi\equiv\varphi_{\rm ext}(\vec x,t)\,,
\end{equation}
the internal charge density becomes a functional of the perturbation,
\begin{equation}
\rho_{\rm int}\equiv\rho_{\rm int}[\h\varphi\ext\h]\,.
\end{equation}
The corresponding {\it functional derivative},
\begin{equation}
\upchi(\vec x,t;\vec x',t')=\frac{\delta\rho_{\rm int}(\vec x,t)}{\delta\varphi\ext(\vec x',t')}\,,
\end{equation}
is called the {\it density response function}. With this, we can express the time evolution of the 
internal charge density to first order in the perturbation~as
\begin{equation}
\rho_{\rm int}(\vec x,t)=\rho_0(\vec x)+\int\!\de^3\vec x'\int\! c\,\de t'\,\upchi(\vec x,t;\vec x',t') \, \varphi_{\rm ext}(\vec x',t')\,, \smallskip
\end{equation}
where we have assumed the charge density in the absence of the external perturbation, $\rho_0 = \rho_{\rm int}[\h \varphi_{\rm ext} \hspace{-1.5pt}\equiv \mh 0\h]$, to be time independent. As in the main text, $c$ denotes the speed of light, such that $\de^3 \vec x \,\hh c \, \de t = \de^4 x$ coincides with the relativistic volume element. The difference,
\begin{equation}
 \rho_{\rm ind}(\vec x, t) = \rho_{\rm int}(\vec x, t) - \rho_0(\vec x)\,, \smallskip
\end{equation}
is called the {\it induced} charge density. This is the charge density produced in the system under the action of the external perturbation.
To be concrete, in the case of the electron-nuclear many-body 
system with the Hamiltonian $\hat H_0$ given by Eq.~\eqref{eq_fund}, the internal charge density is given by
\begin{equation}
\rho_{\rm int}(\vec x,t)=\langle\vec{\Psi}_0 | \h \hat\rho_{\rm n}(\vec x,t)+\hat\rho_{\rm e}(\vec x,t) \h | \vec{\Psi}_0\rangle\,,
\end{equation}
where $|\vec{\Psi}_0\rangle$ is the initial state of the many-body system under consideration (e.g.~the ground state of $\hat H_0$), while
the charge density operators evolve with the perturbed, time-dependent Hamiltonian
\begin{equation}
\hat H(t)= \hat H_0+\int\!\de^3\vec x\,\h(\h \hat\rho_{\rm n}(\vec x)+\hat\rho_{\rm e}(\vec x)) \, \varphi\ext(\vec x,t)\,.
\end{equation}
By contrast, the reference density $\rho_0(\vec x)$ is simply given by
\begin{equation}
\rho_{0}(\vec x)=\langle\vec{\Psi}_0|\h \hat\rho_{\rm n}(\vec x)+\hat\rho_{\rm e}(\vec x)\h |\vec{\Psi}_0\rangle\,.
\end{equation}
Together with the induced charge density, we introduce an induced potential
\begin{equation}
\varphi_{\rm ind}(\vec x,t)=\int\!\de^3\vec x' \mh \int \! c\,\de t' \, v(\vec x, t; \vec x', t') \h \rho\ind(\vec x',t')\,,
\end{equation}
where $v$ denotes the instantaneous Coulomb interaction kernel as given by Eq.~\eqref{eq_Coul}. In addition, we define a {\itshape total} potential by
\begin{equation}
\varphi\tot=\varphi\ext+\varphi\ind\,.
\end{equation}
With these definitions, we introduce the {\it proper density response function} as (cf.~\cite[Sec.~5.2]{ED1}) 
\begin{equation}
\widetilde\upchi(\vec x,t;\vec x',t')=\frac{\delta\rho_{\rm ind}(\vec x,t)}{\delta\varphi\tot(\vec x',t')}\,. \smallskip
\end{equation}
By the functional chain rule,
\begin{equation}\label{eq_FD}
\frac{\delta\rho_{\rm ind}}{\delta\varphi\ext}=
\frac{\delta\rho_{\rm ind}}{\delta\varphi\tot} \h \frac{\delta\varphi\tot}{\delta\varphi\ext}=
\frac{\delta\rho_{\rm ind}}{\delta\varphi\tot}+
\frac{\delta\rho_{\rm ind}}{\delta\varphi\tot}\h \frac{\delta\varphi_{\rm ind}}{\delta\rho_{\rm ind}}\h 
\frac{\delta\rho_{\rm ind}}{\delta\varphi\ext}\,,
\end{equation}
we find its relation to the (ordinary or {\itshape direct}) density response function, which is given by Eq.~\eqref{prop_dir} in the main text.
Note that the products in the above relations refer to spatial and temporal integrations of integral kernels.
With these results, one shows easily the standard relations
\begin{align}
\varepsilon_{\rm r}&=1-v \h\hh \widetilde\upchi\,,\label{eq_eps1}\\[5pt]
\varepsilon^{-1}_{\rm r}&=1+v\h\hh \upchi\,,\label{eq_eps2}
\end{align}
for the longitudinal dielectric function $\varepsilon_{\rm r}$\hh, which in turn is defined (by its inverse) as
\begin{equation}
\varepsilon^{-1}_{\rm r}\equiv\varepsilon^{-1}_{\rm r, \hh L}(\vec x,t;\vec x',t')\equiv\frac{\delta\varphi\tot(\vec x,t)}{\delta\varphi\ext(\vec x',t')}\,. \smallskip
\end{equation}
Finally, we note that the description of the electromagnetic response by means of the density function alone is actually insufficient,
because a time-dependent induced density is necessarily accompanied by an induced spatial current fulfilling the continuity equation,
\begin{equation}
\partial_\mu \h j^\mu\ind(x)= \frac{\partial}{\partial t} \h \rho\ind(\vec x,t)+\nabla\mh\cdot\vec j\ind(\vec x,t)=0\,,
\end{equation}
where $x \equiv x^\mu=(c\hh t,\vec x)$, $\partial_\mu=\partial/\partial x^\mu$, 
and $j^\mu=(c\rho,\h \vec j)$ is the four-current density. Similarly, as a consequence of the {\it Maxwell equations},
time-dependent external {\itshape electric} fields are usually accompanied by external {\itshape magnetic} fields.
To take into account these effects and to go beyond the description by means of the density response function alone, 
one has to introduce the {\it fundamental response tensor} defined by \cite{Melrose, Melrose1Book, Altland, ED1}
\begin{equation}
\chi\indices{^\mu_\nu}(\vec x,t;\vec x',t')=\frac{\delta j^\mu\ind(\vec x,t)}{\delta A^\nu\ext(\vec x',t')}\,, \label{eq_FRT}
\end{equation}
where $A^\nu(\vec x,t)=(\varphi/c,\vec A)$ is the electromagnetic four-potential. In particular, we then recover the density response function as the $00$-component of this $(4 \times 4)$ tensor, 
or more precisely, the relation
\begin{equation}
 \upchi(x, x') = \frac{\delta \rho_{\rm ind}(x)}{\delta \varphi_{\rm ext}(x')} = \frac 1 {c^2} \h \chi\indices{^0_0}(x, x') \,.
\end{equation}
As a matter of principle, it can be shown \cite{ED1} that all
linear electromagnetic response properties can be calculated analytically from the fundamental response tensor.
Inter alia, this also leads to a new interpretation of electrodynamics in media, the so-called Functional Approach,
which has been expounded by the authors of this article in Refs.~\cite{ED1,ED2,EDOhm,Refr,EDLor}.
In Sec.~\ref{subsec_emMatProp}, we will calculate the fundamental response tensor of the oscillating nuclei in terms of
the elastic Green function.

\bigskip \noindent
{\itshape Kubo formalism.---}We now come to the problem of how the density response function can, at least in principle, be calculated 
once we are given a microscopic state of the matter degrees of freedom (e.g.~an electron-nuclear ground state).
This problem is solved by the {\it Kubo formalism}. In general, this works as follows (cf.~\cite[App.~C.1]{ED2} for an even more general framework): 
We consider a Hamiltonian of the form
\begin{equation}
\hat H(t)=\hat H_0+\sum_j f_j(t) \h \hat B_j\,,\label{eq_pertHam} \vspace{-5pt}
\end{equation}
where $\hat H_0$ is the unperturbed Hamiltonian of the system under consideration (e.g.~the fundamental Hamiltonian
\eqref{eq_fund}), while the operators $\hat B_j$ represent the {\itshape external} \h perturbations and the functions $f_j(t)$ describe their respective time dependencies.
The system is assumed to be initially in some reference state (say, again, the ground state $\ket{\Psi_0}$), and the perturbation is assumed to set in at the time $t_0$, such that
\begin{equation}
 f_j(t)=0 \ \, \forall j \quad \textnormal{for} \ \, t<t_0 \,. \smallskip
\end{equation}
The Kubo formalism considers the ensuing time dependence of the expectation values,
\begin{equation} 
B_i(t)=\langle\Psi(t)|\h \hat B_i\h |\Psi(t)\rangle\,,\label{eq_expVal} \smallskip
\end{equation}
where $|\Psi(t)\rangle$ is determined by the {\it initial value problem} defined by
\begin{equation}
\i\hbar \, \partial_t \h |\Psi(t)\rangle=\hat H(t) \h |\Psi(t)\rangle
\end{equation}
and
\begin{equation}
|\Psi(t_0)\rangle=|\Psi_0\rangle\,. \smallskip
\end{equation}
The solution of this initial value problem can formally be written as
\begin{equation}
|\Psi(t)\rangle=\hat U(t,t_0)\h |\Psi_0\rangle\equiv\bigg(\mathcal T\exp\!\bigg(\!-\frac{\rm i}{\hbar}\int_{t_0}^t\de t'\,\hat H(t')\bigg)\mh\bigg) \h |\Psi_0\rangle\,,
\end{equation}
where the time-ordered exponential is a short-hand notation for the so-called {\it Dyson series} given by
\begin{equation} \label{eq_sol}
 \hat U(t,t_0)=\hat{\mathrm I}+
 \sum_{n=1}^\infty \frac{1}{n!}\left(-\frac{\rm i}{\hbar}\right)^{\!\!n}\int_{t_0}^t\de t_1\ldots\int_{t_0}^t\de t_n \, \mathcal T\big( \hat H(t_1)\ldots\hat H(t_n) \big)\,.
\end{equation}
In particular, this implies that the time-dependent vector $\ket{\Psi(t)}$ as well as the expectation values $B_i(t)$ defined in Eq.~\eqref{eq_expVal} depend on the functions $f_j(t')$ of the external perturbations, and hence they become {\it functionals},
\begin{equation}
B_i(t) \equiv B_i(t)\h [\, \ldots,f_j(t'),\ldots\, ]  \,.
\end{equation}
By Eq.~\eqref{eq_sol}, however, $B_i(t)$ depends on $f_j(t')$ only for earlier times $t'<t$.
The {\it Kubo formula} now expresses the corresponding functional derivatives of $B_i(t)$ with respect to  $f_j(t')$ as follows:
\begin{align}
\chi_{ij}(t,t') & = \left. \frac{\delta B_i(t)}{\delta f_j(t')} \h \right|_{f_k \h \equiv \h 0 \,\h \forall k} \\[2pt]
 & = -\frac{\i}{\hbar \h c} \, \varTheta(t-t')\, \langle\Psi_0 \mid [\hat B_i(t),\hat B_j(t')] \mid \Psi_0\rangle\,,
\end{align}
where $\varTheta$ denotes the Heaviside step function, and where the time-dependen\-{}cies of the operators $\hat B_i(t)$ are determined in the interaction picture, i.e., by the unperturbed Hamiltonian $\hat H_0$\hh:
\begin{equation}
 \hat B_j(t) \h \equiv \h \e^{\i t \hat H_0 /\hbar} \, \hat B_j \, \e^{-\i t \hat H_0 /\hbar} \,.
\end{equation}
Within {\it linear response theory}, the perturbed expectation value \eqref{eq_expVal} is then given by
\begin{equation}
B_i(t)=B_{i \hh 0}+\int\! c\,\de t' \, \sum_j\chi_{ij}(t,t') \h f_j(t') \,,
\end{equation}
where $B_{i0} \equiv B_i(t = t_0)$ is the unperturbed expectation value.

We now apply the Kubo formalism to the density response function. Here, the r\^{o}le of the index $j$ is played by the spatial variable $\vec x$, and the 
perturbing operator is the density $\hat \rho(\vec x)$ itself:
\begin{equation}
\sum_j f_j(t) \h \hat B_j \h \mapsto\int\!\de^3\vec x\,\h\varphi_{\rm ext}(\vec x,t)\h \hat\rho(\vec x)\,.
\end{equation}
We thus find for the (retarded) density response function in the reference state $|\Psi_0\rangle$ the expression (see \cite[Eq.~(3.84)]{ED2})
\begin{equation} \label{eq_zwischen}
\upchi(\vec x,t;\vec x',t')=-\frac{\i}{\hbar \h c}\, \varTheta(t-t')\, \langle\Psi_0 \mid [\h\hat\rho(\vec x,t),\hat\rho(\vec x',t')] \mid \Psi_0\rangle\,. \smallskip
\end{equation}
This is the fundamental {\it Kubo formula for the density response function}. In this form, however, it is not always 
particularly useful. Rather, one uses its {\it spectral representation}, which is obtained from the representation of the identity operator
\begin{equation} \label{eq_id_op}
\hat{\mathrm I}= \sum_s |\Psi_s\rangle\langle\Psi_s| \,,
\end{equation}
where the $\ket{\Psi_s}$ form an orthonormal basis of eigenvectors of $\hat H_0$ with corresponding energies $E_s$\h, i.e.,
\begin{equation}
 \hat H_0 \h \ket{\Psi_s} = E_s \h \ket{\Psi_s} \,.
\end{equation}
By inserting the identity operator \eqref{eq_id_op} between the density operators in the Kubo formula \eqref{eq_zwischen}, 
one obtains after a short calculation 
\begin{align}
&\upchi(\vec x, \vec x',\tau)= \\[5pt]
& -\frac{\i}{\hbar \h c} \, \varTheta(\tau) \, \sum^\infty_{s=0}\Big(\e^{-{\rm i}\tau(E_s-E_0)/\hbar} \, 
\langle\Psi_0\mid \hat\rho(\vec x)\mid \Psi_s\rangle\h\langle\Psi_s\mid \hat\rho(\vec x')\mid \Psi_0\rangle\nonumber\\[-3pt] \nonumber
&\hspace{2.95cm}-\e^{{\rm i}\tau(E_s-E_0)/\hbar} \,
\langle\Psi_0\mid \hat\rho(\vec x')\mid \Psi_s\rangle\h\langle\Psi_s\mid \hat\rho(\vec x)\mid \Psi_0\rangle\Big)\,,
\end{align}
where $\tau = t - t'$. In the frequency domain, this is equivalent to
\begin{align}
& \upchi(\vec x, \vec x';\omega)=\label{eq_densRespSpecRepr}\\[5pt] \nonumber
& \sum^\infty_{s=0}\left(
\frac{\langle\Psi_0\mid \hat\rho(\vec x)\mid \Psi_s\rangle\h\langle\Psi_s\mid \hat\rho(\vec x')\mid \Psi_0\rangle}{\hbar\hh\omega-(E_s-E_0)+\i\eta}-
\frac{\langle\Psi_0\mid \hat\rho(\vec x')\mid \Psi_s\rangle\h\langle\Psi_s\mid \hat\rho(\vec x)\mid \Psi_0\rangle}{\hbar\hh\omega+(E_s-E_0)+\i\eta}\right)\,,
\end{align}
where $\eta$ denotes a positive infinitesimal. We note that in the Kubo formula \eqref{eq_zwischen}, the density operators can be replaced without any effect by density fluctuation operators,
\begin{equation} \label{repl_kubo}
\delta\hat\rho(\vec x,t)=\hat\rho(\vec x,t)-\langle\Psi_0\mid \hat\rho(\vec x)\mid  \Psi_0\rangle\,,
\end{equation}
because the contribution from the reference densities would vanish in the commutator.
Correspondingly, the summation in the spectral representation \eqref{eq_densRespSpecRepr} can be restricted to $s>0$.
In particular, in the limit $\omega\rightarrow 0$ we get for the {\itshape instantaneous} response function
\begin{equation}\label{eq_statRF}
\upchi(\vec x,\vec x')=2\,\h\mathfrak{Re}\h \bigg(\h\sum_{s>0}\frac{\langle\Psi_0\mid \delta \hat \rho(\vec x)\mid \Psi_s\rangle\langle\Psi_s\mid \delta \hat\rho(\vec x')\mid \Psi_0\rangle}
{E_0-E_s} \h\bigg)\,.
\end{equation}
This expression is reminiscent of second-order perturbation theory, and indeed a connection to it is established in Sec.~\ref{subsec_2OrdPert}.

\subsection{Displacement field and electromagnetic response}\label{subsec_emMatProp}

We now consider concretely the electromagnetic response generated
by the displacement field of the charged nuclei.
We will first define the electromagnetic four-current corresponding to an oscillating elastic solid, and then consider the ensuing equation of motion
in an external four-potential. This will allow for the calculation of the entire 
fundamental electromagnetic response tensor \cite{Adler,Altland,Melrose,Strinati}, which is
a $(4\times 4)$ integral kernel containing the complete information about all linear electromagnetic response functions \cite{ED1}. Finally, this will lead to analytical formulae expressing the electromagnetic response of the oscillating solid in terms of the elastic Green function.

In the following, we consider the displacement field in the combined thermodynamic and continuum limit as explained in Sec.~\ref{disp_cont}.
With the continuous displacement field we associate the induced electromagnetic four-current given by
\begin{align}
\rho\ind(\vec x,t)&=-\rho_{\rm n0} \h \nabla\cdot\vec u(\vec x,t)\,,\label{eq_indfourCurr1}\\[5pt]
\vec j\ind(\vec x,t)&=\rho_{\rm n0}\, \partial_t \vec u(\vec x,t)\,.\label{eq_indfourCurr2}
\end{align}
While the first equation has already been derived (see Eq.~\eqref{eq_densContVers}) from the classical expression \eqref{eq_cldens} for the density of point particles, the second equation can be derived analogously from the corresponding expression for the classical current of point particles,
\begin{align}
\vec j_{\rm n}(\vec{x},t) & = Ze \h \sum_{\vec{n}} \hh (\partial_t \h \vec x_{\vec n})(t) \, \delta^3(\vec{x}-\vec{x}_{\vec{n}}(t)) \\[5pt]
& = Ze \h \sum_{\vec n} \hh (\partial_t \h \vec u)(\vec x_{\vec n0}, t) \, \delta^3(\vec x - \vec x_{\vec n0}) + \mathcal O(\vec u^2) \,,
\end{align}
where in the last step we have performed a first-order expansion in the displacement field analogous to Eq.~\eqref{nucdensdispl}. In the thermodynamic and continnum limit, 
this expression leads precisely to Eq.~\eqref{eq_indfourCurr2}.
 Per constructionem, the above four-current \eqref{eq_indfourCurr1}--\eqref{eq_indfourCurr2} fulfills the continuity equation:
\begin{equation}
\partial_t \h \rho\ind(\vec x,t)+\nabla\cdot\vec j\ind(\vec x,t)=0\,.
\end{equation}
With this definition of the electromagnetic four-current, 
we couple the elastic solid to an external four-potential by means of the standard interaction Lagrangean,
\begin{equation}
L_{\rm int}(t)=\int \! \de^3\vec x\,\h\big({-\rho\ind}(\vec x, t) \h \varphi\ext(\vec x, t)+\vec j\ind(\vec x, t)\cdot\vec A\ext(\vec x, t) \h \big)\,.
\end{equation}
Inserting the expressions \eqref{eq_indfourCurr1}--\eqref{eq_indfourCurr2} into this equation and performing a partial integration in the first term yields
\begin{align} \label{zwisc_1}
 L_{\rm int}(t) & = \rho_{\rm n0} \int \! \de^3 \vec x \,\h \big({-\vec u}(\vec x, t) \cdot \nabla \varphi_{\rm ext}(\vec x, t) + \dot{\vec u}(\vec x, t) \cdot \vec A_{\rm ext}(\vec x, t) \h \big) \\[5pt]
& \equiv L_{\rm int}[\vec u, \dot{\vec u}, \h t \h ]
\,,
\end{align}
where in the last step we have interpreted the Lagrangean as a functional in analogy to classical point-particle mechanics (see e.g.~\cite[Sec.~1.4]{Goldstein}).
We now use the fact that one can add a total time derivative to the Lagrangean without changing the equations of motion (see \cite[Eq.~(1.57$^\prime$)]{Goldstein}). Thus, we may replace
\begin{equation}
 L_{\rm int}[\vec u , \dot{\vec u} , \h t\h] \mapsto L_{\rm int}[\vec u , \dot{\vec u} , \h t\h] + \frac{\de}{\de t} F[\vec u , t \h] \,, \smallskip
\end{equation}
where the total time derivative is defined as
\begin{equation}
 \frac{\de}{\de t} F[\vec u , t \h ] = \frac{\partial F[\vec u , t \h ]}{\partial t} + \int \! \de^3 \vec x \,\h \dot{\vec u}(\vec x,t) \cdot \frac{\delta F[\vec u , t\h]}{\delta \vec u(\vec x,t)} \,.
\end{equation}
The proof that this replacement does not change the equations of motion is precisely analogous to the corresponding proof in point-particle mechanics. In our case, we choose the functional
\begin{equation}
 F[\vec u , t \h] = -\rho_{\rm n0} \int \! \de^3 \vec x\,\h \vec u(\vec x,t) \cdot \vec A_{\rm ext}(\vec x, t) \,,
\end{equation}
such that
\begin{equation}
 \frac{\de}{\de t} \h F[\vec u , \h t \h] = -\rho_{\rm n0} \int \! \de^3 \vec x \, \big( \vec u(\vec x, t) \cdot \hspace{3.5pt} \dot{\hspace{-3.5pt}\vec A}_{\rm ext}(\vec x, t) + \dot{\vec u}(\vec x, t) \cdot \vec A_{\rm ext}(\vec x, t) \big) \,.
\end{equation}
Adding this term to the Lagrangean \eqref{zwisc_1} yields the equivalent Lagrangean
\begin{align}
 L_{\rm int}(t) & = \rho_{\rm n0} \int \! \de^3 \vec x \,\h \vec u(\vec x, t) \cdot ({-\nabla} \varphi_{\rm ext} - \partial_t \vec A_{\rm ext} \h )(\vec x, t) \\[5pt]
 & = \rho_{\rm n0} \int \! \de^3 \vec x \,\h \vec u(\vec x, t) \cdot \vec E_{\rm ext}(\vec x ,t) \,. \label{eq_uE}
\end{align}
This interaction Lagrangean is of the standard form \eqref{lag_int} provided that we identify the force with
\begin{equation} \label{idfe}
 \vec f(\vec x, t) = \rho_{\rm n0} \h \vec E_{\rm ext}(\vec x, t) \,.
\end{equation}
In particular, the Euler--Lagrange equations then lead to the inhomogeneous equation of motion,
\begin{equation}
\rho_0 \h \partial^2_t \h \vec u(\vec x,t)+\int\! \de^3\vec x'\,\tsr K(\vec x-\vec x') \h \vec u(\vec x',t)=\rho_{\rm n0} \h \vec E_{\rm ext}(\vec x,t)\,.
\end{equation}
This shows that within our approximation \eqref{eq_indfourCurr1}--\eqref{eq_indfourCurr2} of the four-current, the magnetic field does not couple explicitly to the displacement field
(although it does so implicitly, because a transverse electric field is in general also accompanied by a magnetic field).

By combining the electromagnetic four-current \eqref{eq_indfourCurr1}--\eqref{eq_indfourCurr2} with the formalism of the Functional Approach to
electrodynamics of media \cite{ED1,ED2,EDOhm,Refr,EDLor}, we are now in a position to express the core electromagnetic response 
of the solid in terms of the elastic Green function. In fact, from the very definition of the elastic Green function (see Eq.~\eqref{uresp}) 
we obtain its characterization as the functional derivative,
\begin{equation}
D_{k\ell}(\vec x,t; \h \vec x',t')=\frac{\delta u_k(\vec x,t)}{\delta f_\ell(\vec x',t')}\,,
\end{equation}
of the displacement field with respect to the externally applied force. Interpreting this external force to be electromagnetic in its origin, 
such that $\vec f$ can be identified with the external electric field as in Eq.~\eqref{idfe}, 
the fundamental electromagnetic response tensor \eqref{eq_FRT} can be obtained by means of the functional chain rule,
\begin{align}
\chi\indices{^\mu_\nu}(x,x')=\sum_{k, \h \ell} \int \! \de^4 y\int \! \de^4 y'\, \frac{\delta j^\mu\ind(x)}{\delta u_k(y)} \, \frac{\delta u_k(y)}{\delta f_\ell(y')} \,\frac{\delta f_\ell(y')}{\delta A^\nu\ext(x')}\,. \label{fcr}
\end{align}
Now, from Eqs.~\eqref{eq_indfourCurr1}--\eqref{eq_indfourCurr2} we obtain
\begin{align}
 \frac{\delta \rho(\vec x, t)}{\delta u_k(\vec y, s)} & = -\rho_{\rm n0} \, \frac{\partial}{\partial x_k} \, \delta^3(\vec x - \vec y) \, \delta(c \h t - c \hh s) \,, \\[3pt]
 \frac{\delta j_i(\vec x, t)}{\delta u_k(\vec y, s)} & = \rho_{\rm n0} \, \delta_{ik} \, \delta^3(\vec x - \vec y) \, \frac{\partial}{\partial t} \h\hh \delta(c \h t - c\hh s) \,.
\end{align}
Furthermore, the defining equation $\vec E = -\nabla \varphi -\partial_t \h \vec A$ implies that
\begin{align}
\frac{\delta E_\ell(\vec y',s')}{\delta\varphi(\vec x',t')}&=-\frac{\partial}{\partial y'_\ell} \, \delta^3(\vec y'-\vec x') \, \delta(c\hh s'-c\h t')\,,\\[3pt]
\frac{\delta E_\ell(\vec y',s')}{\delta A_j(\vec x',t')}&=-\delta_{\ell j} \, \delta^3(\vec y'-\vec x') \, \frac{\partial}{\partial s'} \, \delta(c\hh s'-c \h t')\,.
\end{align}
By putting these results into Eq.~\eqref{fcr} and performing partial integrations, we obtain the desired explicit expressions 
for the fundamental electromagnetic response tensor in terms of the elastic Green function:
\begin{align}
\chi\indices{^0_0}(\vec x,t;\vec x',t')&=-c^2 \rho_{\rm n0}^2 \, \sum_{k, \h \ell} \frac{\partial^2}{\partial x^k \h \partial x'^{\ell}} \h D_{k\ell}(\vec x,t;\vec x',t')\,, \\[3pt]
\chi\indices{^0_j}(\vec x,t;\vec x',t')&=-c^2 \rho_{\rm n0}^2 \, \sum_k \frac{\partial^2}{\partial x^k \h \partial x'^{0}} \h D_{k j}(\vec x,t;\vec x',t')\,,\\[3pt]
\chi_{i \hh 0}(\vec x,t;\vec x',t')&=c^2 \rho_{\rm n0}^2 \, \sum_\ell \frac{\partial^2}{\partial x^0 \h \partial x'^{\ell}} \h D_{i \ell}(\vec x,t;\vec x',t')\,,\\[3pt]
\chi_{ij}(\vec x,t;\vec x',t')&=c^2 \rho_{\rm n0}^2 \,\h \frac{\partial^2}{\partial x^0 \h \partial x'^{0}} \h D_{ij}(\vec x,t;\vec x',t')\,.\label{eq_CurrentRespGF}
\end{align}
In Fourier space and for a homogeneous system, these equations simplify to
\begin{align}
\chi\indices{^0_0}(\vec k,\omega)&=-c^2 \rho_{\rm n0}^2\, \sum_{k, \h \ell} k_k \h D_{k\ell}(\vec k,\omega) \h\hh k_\ell\,, \\[3pt]
\chi\indices{^0_j}(\vec k,\omega)&=c^2 \rho_{\rm n0}^2\,\sum_k k_k \h D_{kj}(\vec k,\omega) \, \frac{\omega}{c} \,,\\[5pt]
\chi_{i \hh 0}(\vec k,\omega)&=-c^2 \rho_{\rm n0}^2\,\sum_\ell \frac{\omega}{c}\,D_{i\ell}(\vec k,\omega) \h\hh k_\ell \,,\\[1pt]
\chi_{ij}(\vec k,\omega)&=c^2 \rho_{\rm n0}^2\,\h\frac{\omega^2}{c^2}\, D_{ij}(\vec k,\omega)\,.
\end{align}
Consequently, the fundamental response tensor can be written more compactly in matrix notation as
\begin{equation}
\chi\indices{^\mu_\nu}(\vec k,\omega)=\rho_{\rm n0}^2\left( \!
\begin{array}{rr}
-c\hh \vec k^{\rm T}\tsr D(\vec k,\omega) \h c\hh \vec k & \omega\h c \hh \vec k^{\rm T}\tsr D(\vec k,\omega)\\[5pt]
-\omega \h \tsr D(\vec k,\omega) \h c\hh \vec k & \omega^2 \h \tsr D(\vec k,\omega)
\end{array}
\right)\,.
\end{equation}
In fact, this fundamental response tensor is in accord with the most general form of 
a fundamental response tensor in the homogeneous limit (see \cite[Eq.~(2.29)]{Refr}), provided that the current response tensor (which is the spatial part of the fundamental response tensor) is identified with
\begin{equation}
\tsr\chi(\vec k,\omega)  = \rho_{\rm n0}^2 \, \omega^2 \h \tsr D(\vec k,\omega)\,. \label{eq_currentResp}
\end{equation}
In particular, we remark that by the concrete form \eqref{D_concr} of the elastic Green function, 
the above equality implies that the zero-frequency limit of the current response tensor vanishes. 

Given Eq.~\eqref{eq_currentResp}, we can now derive concrete expressions for all linear electromagnetic response functions. 
For example, the standard relation between the current response tensor and the conductivity tensor (cf.~\cite[Eq.~(6.16)]{ED1}), \smallskip
\begin{equation}
\tsr \chi(\vec k,\omega)=\i\omega \h \tsr\sigma(\vec k,\omega)\,, \medskip
\end{equation}
implies that the conductivity of the cores can be expressed in terms of the elastic Green function as \smallskip
\begin{equation}
 \tsr \sigma(\vec k, \omega) = - \rho_{\rm n0}^2 \,\h \i\omega \h \tsr D(\vec k, \omega) \,, \medskip
\end{equation}
and hence also the zero-frequency limit of the conductivity vanishes.  
This is physically reasonable because an elastic solid cannot support a constant current.
We stress, however, that this statement refers to the elastically bound nuclei only, not to a possible electronic current.
Furthermore, concrete expressions for the dielectric tensor and the magnetic susceptibility will be derived in the {\itshape isotropic limit} in Sec.~\ref{em_resp}. 
Finally, for the density response function we obtain the simple expression in terms of the elastic Green function,
\begin{equation}\label{eq_densRespFuncGF}
\upchi(\vec k,\omega)\equiv\frac{\delta\rho\ind(\vec k,\omega)}{\delta\varphi\ext(\vec k,\omega)}
=\frac{1}{c^2} \h \chi\indices{^0_0}(\vec k,\omega)
=-\rho_{\rm n0}^2\, \vec k^{\rm T} \hh \tsr D(\vec k,\omega) \h\hh \vec k\,.
\end{equation}
This result is precisely the continuum analogon of Eq.~\eqref{eq_genConnection}. It is used in Sec.~\ref{sec_JustFI} for relating the Response Theory of effective interactions to the functional integral approach (see
Eqs.~\eqref{eq_fundEq_equiv}--\eqref{eq_fundEq}).

\subsection{Kubo formula for electromagnetic response of the cores} \label{app_Kubo}

In this last subsection, we discuss the consistency of the relation \eqref{eq_currentResp}, which we have derived by purely classical considerations, 
with the quantum field theoretical Kubo formula for the fundamental response tensor given by (see e.g.~\cite{Giuliani, Kubo, Altland, Bruus} or \cite[Sec.~3.2.4 and App.~C]{ED2}) 
\begin{align} \label{eq_kubo}
\chi\indices{^\mu_\nu}(\vec x,t;\vec x',t')=&\,\frac{\rm i}{\hbar \h c } \, \varTheta(t-t') \h \langle \Phi_0 \mid [\,\delta\,\hat{\! j}^\mu(\vec x,t), \h \delta\,\hat{\! j}_\nu(\vec x',t')] \mid \Phi_0 \rangle \\[3pt] \nonumber
&-\frac{Ze}{M} \, \sum_{i=1}^3\delta\indices{^\mu_i}\h \delta_{i\nu}\, \delta(c\h t - c\h t') \, \delta^3(\vec x- \vec x') \, \langle \Phi_0 \mid \hat\rho(\vec x) \mid \Phi_0 \rangle\,.
\end{align}
Here, the expectation values refer to the unperturbed state, e.g., the ground state $|\Phi_0\rangle$ of the nuclear subsystem. Similarly, the 
time evolution refers to the unperturbed core Hamiltonian. 
The second term on the right hand side corresponds to the local response of the so-called diamagnetic current (see e.g.~\cite[Eq.~(1.98c)]{Bruus}). Its prefactor involves the nuclear charge $Ze$ and mass $M$. 
Further note that in the first term, we have replaced the current operators by current fluctuation operators defined by
\begin{equation} \label{fluc_curr}
\delta\,\hat{\! j}^\mu(\vec x, t)=\hat{\! j}^\mu(\vec x, t) - \langle \Phi_0 \mid \, \hat{\! j}^\mu(\vec x) \mid \Phi_0 \rangle\,,
\end{equation}
which is analogous to the replacement \eqref{repl_kubo}.
In particular, the $00$-compo\-{}nent of Eq.~\eqref{eq_kubo} reproduces the Kubo formula for the density response function, Eq.~\eqref{eq_zwischen}, while
the spatial part reads
\begin{equation}
\begin{aligned}
\tsr\chi(\vec x,t;\vec x',t')=&\,\frac{\rm i}{\hbar \h c } \, \varTheta(t-t') \h \langle \Phi_0 \mid [\h\delta\,\hat{\!\vec j}(\vec x,t),\h\delta\,\hat{\!\vec j}(\vec x',t')\hh ] \mid \Phi_0 \rangle \\[3pt]
&-\frac{\rho_{\rm n0}^2}{\rho_0 \h c}\,\delta(t-t') \, \delta^3(\vec x-\vec x') \h \tsr 1\,,\label{eq_tbc}
\end{aligned}
\end{equation}
where we have used the fact that the expectation value of the charge density in the unperturbed state simply yields the constant reference density, $\langle \Phi_0 \mid \hat \rho(\vec x) \mid \Phi_0\rangle \equiv \rho_{\rm n0}$\h.
We also note that the contribution proportional to the four-dimensional Dirac delta ensures the constraint relations (see \cite{Altland, Fradkin} and \cite[Sec.~5.1]{ED1})
\begin{equation}
\partial_\mu \h \chi\indices{^\mu_\nu}(x,x')=0\,, \qquad \partial'^\nu\chi\indices{^\mu_\nu}(x,x')=0\,,
\end{equation}
respectively implying charge conservation and gauge invariance.

We now compare Eq.~\eqref{eq_tbc} with the central result \eqref{eq_currentResp} of the preceding subsection. For this purpose, we first identify the fluctuation current operators \eqref{fluc_curr} with the operators
for the {\itshape induced} current. Hence, with 
Eqs.~\eqref{eq_indfourCurr1}--\eqref{eq_indfourCurr2} we obtain the expressions
\begin{align}
\delta\hat\rho(\vec x,t)&=-\rho_{\rm n0} \h \nabla\cdot\hat{\vec u}(\vec x,t)\,,\label{eq_indfourCurr1op}\\[3pt]
\delta\,\hat{\!\vec j}(\vec x,t)&=\rho_{\rm n0}\, \partial_t \hat{\vec u}(\vec x,t)\,.\label{eq_indfourCurr2op}
\end{align}
We now show that Eq.~\eqref{eq_tbc} is equivalent to Eq.~\eqref{eq_currentResp} under these identifications. More precisely, we show that in real space,
\begin{equation}
\tsr\chi(\vec x,t;\vec x',t')=\rho_{\rm n0}^2 \, \frac{\partial^2}{\partial t\, \partial t'} \h \tsr D_{\rm ret}(\vec x,t;\vec x',t')\,,
\end{equation}
where
\begin{equation}
{-\i\hh \hbar \h c}  \, \tsr D_{\rm ret}(\vec x,t;\vec x',t')= \varTheta(t-t') \h \langle\Phi_0\mid [\hat{\vec u}(\vec x,t),\hat{\vec u}(\vec x',t')]\mid \Phi_0\rangle
\end{equation}
is the quantum field theoretical expression for the {\itshape retarded} Green function. In fact, acting twice with the time derivative on this expression gives the following four contributions:
\begin{align} \label{four_terms}
& {-\i\hh \hbar \h c}  \, \partial_t \h \partial_{t'} \hh \tsr D_{\rm ret}(\vec x, t; \vec x', t') =  \\[5pt] \nonumber
& \quad \quad \, \varTheta(t-t') \h \langle\Phi_0\mid [\partial_t\hat{\vec u}(\vec x,t),\partial_{t'}\hat{\vec u}(\vec x',t')]\mid \Phi_0\rangle\\[5pt] \nonumber
& \quad +\delta(t-t') \h \langle\Phi_0\mid [\hat{\vec u}(\vec x,t),\partial_{t'}\hh\hat{\vec u}(\vec x',t')]\mid \Phi_0\rangle\\[5pt] \nonumber
& \quad -\delta(t-t') \h \langle\Phi_0\mid [\partial_t \hh \hat{\vec u}(\vec x,t),\hat{\vec u}(\vec x',t')]\mid \Phi_0\rangle\\[5pt] \nonumber
& \quad -(\partial_t \hh \delta)(t-t') \h \langle\Phi_0\mid [\hat{\vec u}(\vec x,t),\hat{\vec u}(\vec x',t')]\mid \Phi_0\rangle\,.
\end{align}
Using the connection between the time-derivative of the displacement field and the electromagnetic current density operator, Eq.~\eqref{eq_indfourCurr2op},
we see directly that the first contribution reproduces the first term in Eq.~\eqref{eq_tbc}.
Then, the second and the third contribution can be evaluated using the equal-time commutator \eqref{CCR1}. The result yields two times the second contribution in Eq.~\eqref{eq_tbc}. Finally, by applying the fourth term to a test function $\varphi(t')$, using partial integration and the fact that the displacement field commutes with itself at equal times, we find the distributional identity
\begin{align} \label{distr_id}
 & (\partial_t \hh \delta)(t-t') \h \langle\Phi_0\mid [\hat{\vec u}(\vec x,t),\hat{\vec u}(\vec x',t')]\mid \Phi_0\rangle \\[5pt] \nonumber
 & = \delta(t - t') \h \langle\Phi_0\mid [\hat{\vec u}(\vec x,t),\partial_{t'} \hh \hat{\vec u}(\vec x',t')]\mid \Phi_0\rangle \,.
\end{align}
This shows that the fourth term in Eq.~\eqref{four_terms} equals the second or third one up to a sign, and thus restores the correct prefactor of the second term in Eq.~\eqref{eq_tbc}. This concludes our proof of the equivalence between Eq.~\eqref{eq_currentResp} and Eq.~\eqref{eq_tbc}.

\section{Connection to elasticity theory}\label{app_Iso}

In this appendix, we take the previously obtained results to the isotropic limit.
This limit is of both practical and conceptual importance for the description of the displacement field. Conceptually, 
it describes the simplest dynamical matrix which still has some realistic features to it. We have therefore used this limit already
in the main text (see Sec.~\ref{subsec_ExprDispRel}). Futhermore, it allows for a particularly transparent
connection to macroscopic elasticity theory. Practically, the isotropic limit is appropriate for most materials
anyway because the microscopic anisotropy stemming from the crystal structure is mostly averaged out on a macroscopic scale.
In fact, in the macroscopic {\itshape and} \h isotropic limit, the elastic properties can be completely described by only two material constants.
On the other hand, in Sec.~\ref{subsec_emMatProp} we have shown that the electromagnetic material properties can be expressed
in terms of the elastic Green function. To put this into practice for the said macroscopic and isotropic limit, we will connect the microscopic description of the displacement field in terms of the dynamical matrix to the macroscopic description
in terms of the elasticity tensor. This will then allow for the expression of the elastic Green function---and thus finally also
of the nuclear contribution to the electromagnetic response---in terms of the elastic constants.

\subsection{Dynamical matrix in the isotropic limit}\label{app_ElTensDM}

Per definitionem, the dynamical matrix is isotropic if it preserves its form under arbitrary spatial rotations. 
As a matter of principle, under a rotation $R \in \mathrm{SO(3)}$ the dynamical matrix in Fourier space transforms as
\begin{equation} \label{iso_cond}
(\tsr K)'(\vec k')=\tsr R\, \tsr K(\vec k) \tsr R{}^{-1}\,,
\end{equation}
where
\begin{equation}
\vec k'=\tsr R\,\vec k\,. \smallskip
\end{equation}
The dynamical matrix being of the same form in all coordinate systems rotated relatively to each other means that
\begin{equation}
(\tsr K)'(\vec k')=\tsr K(\vec k')\,,
\end{equation}
i.e.~explicitly, \smallskip
\begin{equation} \label{iso_cond_expl}
 \tsr R \, \tsr K(\vec k) \tsr R{}^{-1} = \tsr K(\tsr R \h \vec k) \,. \smallskip
\end{equation}
From this condition one arrives at far reaching results about the general form of the dynamical matrix, which we formulate in the following lemma:

\bigskip\noindent
{\bfseries Lemma.} In the isotropic limit, the dynamical matrix is of the general form
\begin{equation}\label{eq_genformDM}
\tsr K(\vec k)=K\L(\vec k)\tsr P\L(\vec k) + K\T(\vec k)\tsr P\T(\vec k)\,,
\end{equation}
where the longitudinal and transverse parts $K_{\rm L}(\vec k)$ and $K_{\rm T}(\vec k)$ depend on the modulus $|\vec k|$ of the wavevector only.

\bigskip\noindent
{\bfseries Proof.} We first consider a fixed wavevector $\vec k$ and restrict the condition \eqref{iso_cond_expl} to the subgroup of rotations around the axis parallel to $\vec k$. For such rotations, we have
\begin{equation}
 \tsr R \, \vec k = \vec k \,,
\end{equation}
and hence the condition \eqref{iso_cond_expl} yields
\begin{equation} \label{cond_schur}
 \tsr K(\vec k) \h \tsr R = \tsr R \, \tsr K(\vec k) \,.
\end{equation}
This means, the matrix $K(\vec k)$ commutes with all matrices representing rotations around the axis $\vec k$. The latter form a subgroup $G_{\vec k} \subset \mathrm{SO}(3)$ of the three-dimensional rotation group. The natural representation of $G_{\vec k}$ on the three-dimensional vector space $\mathbb R^3$ has two irreducible subspaces, i.e., subspaces being invariant under all rotations around $\vec k$. These are spanned by all vectors being parallel and respectively orthogonal to $\vec k$, hence they are given by  the longitudinal and transverse subspaces $P_{\rm L}(\vec k) \h \mathbb R^3$ and $P_{\rm T}(\vec k) \h \mathbb R^3$. Now, let us define the following matrices:
\begin{align}
 \tsr K_{\rm LL}(\vec k) & := \tsr P_{\rm L}(\vec k) \h \tsr K(\vec k) \h \tsr P_{\rm L}(\vec k) \,, \\[3pt]
 \tsr K_{\rm TT}(\vec k) & := \tsr P_{\rm T}(\vec k) \h \tsr K(\vec k) \h \tsr P_{\rm T}(\vec k) \,, \\[3pt]
 \tsr K_{\rm LT}(\vec k) & := \tsr P_{\rm L}(\vec k) \h \tsr K(\vec k) \h \tsr P_{\rm T}(\vec k) \,, \\[3pt]
 \tsr K_{\rm TL}(\vec k) & := \tsr P_{\rm T}(\vec k) \h \tsr K(\vec k) \h \tsr P_{\rm L}(\vec k) \,,
\end{align}
and for $R \in G_{\vec k}$ also
\begin{align}
 \tsr R_{\rm LL}(\vec k) & := \tsr P_{\rm L}(\vec k) \h \tsr R \,\h \tsr P_{\rm L}(\vec k) \,, \\[3pt]
 \tsr R_{\rm TT}(\vec k) & := \tsr P_{\rm T}(\vec k) \h \tsr R \,\h \tsr P_{\rm T}(\vec k) \,.
\end{align}
Using that $R$ itself commutes with the projection operators $P_{\rm L}(\vec k)$ and $P_{\rm T}(\vec k)$, one can easily show that the condition \eqref{cond_schur} implies
\begin{align}
 \tsr K_{\rm LL}(\vec k) \h \tsr R_{\rm LL}(\vec k) & = \tsr R_{\rm LL}(\vec k) \h \tsr K_{\rm LL}(\vec k) \,, \label{schur_1} \\[3pt]
 \tsr K_{\rm TT}(\vec k) \h \tsr R_{\rm TT}(\vec k) & = \tsr R_{\rm TT}(\vec k) \h \tsr K_{\rm TT}(\vec k) \,, \label{schur_2} \\[3pt]
 \tsr K_{\rm LT}(\vec k) \h \tsr R_{\rm TT}(\vec k) & = \tsr R_{\rm LL}(\vec k) \h \tsr K_{\rm LT}(\vec k) \,, \label{schur_3} \\[3pt]
 \tsr K_{\rm TL}(\vec k) \h \tsr R_{\rm LL}(\vec k) & = \tsr R_{\rm TT}(\vec k) \h \tsr K_{\rm TL}(\vec k) \,. \label{schur_4}
\end{align}
The first two Eqs.~\eqref{schur_1} and \eqref{schur_2} mean that the projections of the matrix $K(\vec k)$ on the longitudinal or transverse subspaces commute with all matrices of the corresponding irreducible representations of $G_{\vec k}$\hh. By Schur's lemma (see e.g.~\cite[Appendix~D.~\S\,8]{Messiah}), it follows that these projections of $K(\vec k)$ are multiples of the identity matrices on the respective subspaces, i.e.,
\begin{align}
 \tsr K_{\rm LL}(\vec k) & = K_{\rm L}(\vec k) \h \tsr P_{\rm L}(\vec k) \,, \\[2pt]
 \tsr K_{\rm TT}(\vec k) & = K_{\rm T}(\vec k) \h \tsr P_{\rm T}(\vec k) \,, 
\end{align}
with $\vec k$-dependent scalars $K_{\rm L}(\vec k)$ and $K_{\rm T}(\vec k)$. On the other hand, the last two Eqs.~\eqref{schur_3}--\eqref{schur_4} mean that the matrices $K_{\rm LT}(\vec k)$ and $K_{\rm TL}(\vec k)$ define homomorphic mappings between the two irreducible representations. Since these two irreducible representations are inequivalent (as the corresponding subspaces have different dimensions one and two, respectively), Schur's lemma also implies that
\begin{equation}
 \tsr K_{\rm LT}(\vec k) = \tsr K_{\rm TL}(\vec k) = 0 \,.
\end{equation}
Since any matrix $\tsr K(\vec k)$ can be decomposed as
\begin{equation}
 \tsr K(\vec k) = \tsr K_{\rm LL}(\vec k) + \tsr K_{\rm TT}(\vec k) + \tsr K_{\rm LT}(\vec k) + \tsr K_{\rm TL}(\vec k) \,,
\end{equation}
we thus obtain the formula \eqref{eq_genformDM}. Finally, using this form of the matrix $K(\vec k)$, we can evaluate the condition \eqref{iso_cond_expl} for arbitrary matrices $R \in \mathrm{SO}(3)$. With the help of the identities
\begin{align}
 \tsr R \, \tsr P_{\rm L}(\vec k) \h \tsr R{}^{-1} & = \tsr P_{\rm L}(\tsr R \h \vec k) \,, \\[2pt]
 \tsr R \, \tsr P_{\rm T}(\vec k) \h \tsr R{}^{-1} & = \tsr P_{\rm T}(\tsr R \h \vec k) \,,
\end{align}
it follows that
\begin{align}
 K_{\rm L}(\tsr R \h \vec k) & = K_{\rm L}(\vec k) \,, \\[2pt]
 K_{\rm T}(\tsr R \h \vec k) & = K_{\rm T}(\vec k) \,,
\end{align}
hence these functions depend on the modulus $|\vec k|$ only. $\Box$

\subsection{Elasticity tensor and macroscopic limit}\label{subsec_ElasticTens}

In this subsection, we draw the connection between the microscopic description of the displacement field in terms of the dynamical matrix
and the macroscopic approach used in classical {\it elasticity theory} 
(see \cite{FungTong,Landau65,Sigrist}, or \cite[Sec.~1.4]{MartinRothen}, \cite[Chap.~1]{Kleinert892} and \cite[Chap.~3]{Roemer} for short introductions).

In fact, macroscopic elasticity theory assumes the potential energy to be given up to second order in the displacement field by (see e.g.~\cite[Eq. (1.180)]{MartinRothen}; compare this to the microscopic, non-local expression \eqref{pot_lag} in terms of the dynamical matrix),
\begin{equation} \label{el_pot}
V[\vec u]=\frac{1}{2} \h \int \! \de^3\vec x\, u_{ij}(\vec x,t) \h\hh \lambda_{ijk\ell} \,\hh u_{k\ell}(\vec x,t)\,,
\end{equation}
where $\lambda_{ijk\ell}$ is the {\it elasticity tensor}, while
\begin{equation}
u_{ij}(\vec x, t)=\frac{1}{2} \, \bigg( \frac{\partial u_j(\vec x, t)}{\partial x_i} + \frac{\partial u_i(\vec x, t)}{\partial x_j}\bigg) \label{eq_defstrain}
\end{equation}
are the so-called {\it strains}. Here and in the following, we sum over all doubly appearing indices. The general form \eqref{el_pot} of the potential energy implies that the elasticity tensor can be chosen
to have the symmetries
\begin{equation}\label{eq_symmElast}
 \lambda_{ijk\ell}=\lambda_{jik\ell}=\lambda_{ij\ell k}=\lambda_{k\ell ij}\,.
\end{equation}
Further introducing the so-called {\it stresses} by
\begin{align}
\upsigma_{ij}(\vec x,t) := \frac{\delta V[\vec u]}{\delta u_{ij}(\vec x,t)} = \lambda_{ijk\ell} \h\hh u_{k\ell}(\vec x, t) \,,
\end{align}
we find the equation of motion in the following form \cite[Eq.~(23.1)]{Landau65}
\begin{align} \label{el_eom}
\rho_0 \, \frac{\partial^2 u_i(\vec x,t)}{\partial t^2}=-\frac{\delta V[\vec u]}{\delta u_i(\vec x,t)}=\frac{\partial \upsigma_{ij}(\vec x,t)}{\partial x_j}\,.
\end{align}
Here, we have used the functional chain rule
\begin{equation}
\frac{\delta V[\vec u]}{\delta u_i(\vec x,t)}=
\int \! \de^3\vec x'\,\frac{\delta V[\vec u]}{\delta u_{jk}(\vec x',t)}\h\frac{\delta u_{jk}(\vec x',t)}{\delta u_i(\vec x,t)}
\end{equation}
and the identity
\begin{equation}
\frac{\delta u_{jk}(\vec x',t)}{\delta u_i(\vec x,t)}= \frac 1 2 \, \bigg( 
\delta_{ij} \, \frac{\partial}{\partial x_k} \h + \h \delta_{ik} \, \frac{\partial}{\partial x_j} \h \bigg) \h \delta^3(\vec x-\vec x') \,,
\end{equation}
which follows from the definition \eqref{eq_defstrain}.
By substituting back the definition of the stress tensor in Eq.~\eqref{el_eom}, we further obtain
\begin{equation}
\rho_0 \, \frac{\partial^2 u_i(\vec x,t)}{\partial t^2}-\lambda_{ijk\ell} \, \frac{\partial^2 u_\ell(\vec x, t)}{\partial x_j \h \partial x_k}=0\,.\label{eq_standardEoM}
\end{equation}
At face value, this macroscopic equation of motion differs grossly from the microscopic theory in terms of the dynamical matrix
(see~Eq.~\eqref{eq_EoMcontlimit}). On closer inspection, however, 
we see that the {\it macroscopic theory is a special case of its microscopic counterpart} (cf.~\cite[pp.~443f.]{Ashcroft}).
In fact, the former can be reproduced from the latter by choosing the dynamical matrix in real space to be of the form
\begin{equation}\label{eq_DynMatElastCoeff}
K_{ij}(\vec x,\vec x')=-\lambda_{ik\ell j} \, \frac{\partial^2}{\partial x_k \h \partial x_\ell} \, \delta^3(\vec x-\vec x') \,,
\end{equation}
which in Fourier space is equivalent to
\begin{equation}\label{eq_DynMatElastCoeffFour}
K_{ij}(\vec k)=\lambda_{ik\ell j} \h k_k \h\hh k_\ell \,.
\end{equation}
This relation can be inverted as
\begin{equation}\label{eq_DynMatElastCoeffFT}
 \lambda_{ijk\ell}=\frac 1 8 \, \bigg( 
 \frac{\partial^2 K_{i\ell}(\vec k)}{\partial k_j \h \partial k_k}
 + \frac{\partial^2 K_{j\ell}(\vec k)}{\partial k_i \h \partial k_k}
 + \frac{\partial^2 K_{ik}(\vec k)}{\partial k_j \h \partial k_\ell}
 + \frac{\partial^2 K_{jk}(\vec k)}{\partial k_i \h \partial k_\ell}
 \bigg) \hh \bigg|_{\vec k \h = \h 0} \,,
\end{equation}
or equivalently (see \cite[Eq.~(22.79)]{Ashcroft}),
\begin{align}
&\lambda_{ijk\ell}=\\[2pt] \nonumber
& -\frac 1 8 \int \! \de^3\vec r\, \bigg( r_j \h K_{i\ell}(\vec r) \h r_k + r_i \h K_{j\ell}(\vec r) \h r_k + r_j \h K_{ik}(\vec r) \h r_\ell + r_i \h K_{jk}(\vec r) \h r_\ell \bigg) \,,
\end{align}
where $\vec r = \vec x - \vec x'$\h. In particular, if $K_{ij}(\vec k)$ is an arbitrary dynamical matrix of a homogeneous system (not necessarily of the form \eqref{eq_DynMatElastCoeffFour}), 
one can use Eq.~\eqref{eq_DynMatElastCoeffFT} as a definition of the associated elasticity tensor. The symmetry \eqref{sym_inv_k} of the dynamical matrix then guarantees that the symmetries \eqref{eq_symmElast} of the elasticity tensor are fulfilled.

\subsection{Macroscopic and isotropic limit}\label{subsec_Macr&Iso}

Using the results from the previous subsection, we now further express the dynamical matrix in terms of the elastic constants of an isotropic system: In the isotropic limit, 
the elasticity tensor is given by the standard formula (see e.g.~\cite[Eq.~(1.23)]{Kleinert892}),
\begin{equation}\label{eq_ElasTensorIso}
	\lambda_{ijk\ell}=\mu \h (\delta_{ik} \hh \delta_{j\ell}+\delta_{i\ell}\hh \delta_{jk})+\lambda\,\delta_{ij}\hh \delta_{k\ell}\,. \medskip
\end{equation}
with the {\itshape Lam\'{e} coefficient} $\lambda$ and the {\itshape shear modulus} $\mu$. 
They are related to the {\it Poisson ratio} $\nu$, to {\it Young's modulus} $E$ and to the {\it bulk modulus}~$B$ 
(also {\it modulus of compression}) via \cite[Table 1.3]{Kleinert892}
\begin{align}
\nu&=\frac{\lambda}{2 \h (\lambda+\mu)} \,,\\[6pt]
E&=2\mu \h (1+\nu)\,,\\[4pt]
B&=\lambda+\frac {2\mu} {3} \,.
\end{align}
One verifies directly that the isotropic elasticity tensor \eqref{eq_ElasTensorIso} corresponds via Eq.~\eqref{eq_DynMatElastCoeffFour} to the dynamical matrix in Fourier space
\begin{equation}
K_{ij}(\vec k)=(\lambda+\mu) \h k_i \h k_j +\mu \h\hh \delta_{ij} \h |\vec k|^2 \,,
\end{equation}
or in real space
\begin{equation}
K_{ij}(\vec x,\vec x')=-(\lambda+\mu) \h (\partial_i \h \partial_j \h \delta^3)(\vec x-\vec x')-\mu\h\hh \delta_{ij}\h (\Delta\delta^3)(\vec x-\vec x') \,.
\end{equation}
Comparing this with the general form \eqref{eq_genformDM} in the isotropic limit, we find 
for the longitudinal and transverse parts of the dynamical matrix,
\begin{align}
K\L(\vec k)&=(\lambda +2\mu) \h |\vec k|^2  \,,\\[5pt]
K\T(\vec k)&=\mu \h |\vec k|^2\,.
\end{align}
If we further compare these equations with the continuum version (see Table~\ref{synopsis_1}) of Eqs.~\eqref{readoff1}--\eqref{readoff2}, i.e.,
\begin{align}
 K\L(\vec k) & = \rho_0 \, \omega_{\vec k\rm L}^2 \,, \label{dyn_iso_1} \\[5pt]
K\T(\vec k) & = \rho_0 \, \omega_{\vec k\rm T}^2 \,, \label{dyn_iso_2}
\end{align}
we can read off the longitudinal and transverse dispersion relations as
\begin{align}
 \omega_{\vec k\rm L} & = v_{\rm L} \h |\vec k| \,, \label{el_disp_L} \\[5pt]
 \omega_{\vec k\rm T} & = v_{\rm T} \h |\vec k| \,, \label{el_disp_T}
\end{align}
where the longitudinal and transverse {\itshape speeds of sound} are given in terms of the elastic constants by \begin{align}
v\L^2&=(\lambda+2\mu)/\rho_0\,, \label{eq_longGF} \\[5pt]
v\T^2&=\mu/\rho_0\,. \label{eq_transGF}
\end{align}
Furthermore, one can show that the elastic potential energy \eqref{el_pot} takes the particularly simple form
\begin{equation}
	V[\vec u]=\frac{1}{2} \h \int \! \de^3\vec x\,\big((\lambda+2\mu) \h |\nabla\cdot\vec u(\vec x, t)|^2+\mu \h\hh |\nabla\times\vec u(\vec x, t)|^2\h \big)\,,
\end{equation}
and the equation of motion \eqref{eq_standardEoM} reverts to
\begin{equation}
	\rho_0 \, \frac{\partial^2}{\partial t^2} \, \vec u(\vec x,t)-(\lambda+\mu)\h \nabla(\nabla\!\cdot\vec u)(\vec x,t)-\mu\h\hh \Delta\vec u(\vec x,t)=0\,.\label{eq_EoMisotrop}
\end{equation}
Using the expression of the Laplace operator in terms of the divergence and the rotation,
\begin{equation}
\Delta\vec u(\vec x) \h =\h \nabla(\nabla\cdot\vec u)(\vec x)-\nabla\times(\nabla\times\vec u)(\vec x) \h \equiv \h \Delta \vec u_{\rm L}(\vec x) + \Delta \vec u_{\rm T}(\vec x) \,,
\end{equation}
we can separate the longitudinal and transverse parts of the displacement field, for which we then obtain
the respective wave equations (cf.~\cite[Eqs. (1.193)--(1.194)]{MartinRothen}),
\begin{align}
\left(\frac{1}{v\L^2} \h \frac{\partial^2}{\partial t^2}-\Delta\right)\vec u\L(\vec x,t)&=0\,,\label{eq_EoMisotropLong}\\[5pt]
\left(\frac{1}{v\T^2} \h \frac{\partial^2}{\partial t^2}-\Delta\right)\vec u\T(\vec x,t)&=0\,.\label{eq_EoMisotropTrans}
\end{align}
These equations can also be derived directly from the decomposition \eqref{dyn_iso_1}--\eqref{dyn_iso_2} of the dynamical matrix. Thus, in the macroscopic and isotropic limit,
the longitudinal and transverse oscillations of the elastic solid decouple completely.
Finally, in this limit the calculation of the elastic Green function becomes particularly simple. Its defining equation then reads
\begin{equation}
\bigg( \rho_0 \, \frac{\partial^2}{\partial t^2} -(\lambda+\mu)\h \nabla \nabla^{\rm T}-\mu\h \Delta\bigg)\h \tsr D(\vec x,t;\vec x',t')=\tsr 1 \h \delta^4(x-x')\,.\label{eq_GFiso}
\end{equation}
Using the continuum version of Eq.~\eqref{expl_DL} and substituting Eqs.~\eqref{el_disp_L}--\eqref{el_disp_T}, we obtain the explicit expression
\begin{equation} \label{gen_kleinert}
 \tsr D(\vec k, \omega) = -\frac{1}{\rho_0} \h \frac{1}{\omega^2 - v_{\rm L}^2 \h |\vec k|^2}\, \tsr P\L(\vec k)
-\frac{1}{\rho_0}\h \frac{1}{\omega^2 - v_{\rm T}^2 \h |\vec k|^2} \, \tsr P\T(\vec k) \,.
\end{equation}
This is precisely the frequency-dependent generalization of the result obtained by H.~Kleinert \cite[Eq.~(1.77)]{Kleinert892}. 
Furthermore, this shows that the ``elastic Green function'' is indeed a Green function in the sense of linear elasticity theory.

\subsection{Electromagnetic response in terms of elastic constants}\label{em_resp}

We finally come back to the electromagnetic response as produced by the nuclear displacement field.
In the previous subsections we have shown that the macroscopic equation of motion \eqref{eq_standardEoM} in terms of the elasticity tensor, or Eq.~\eqref{eq_EoMisotrop} in terms of the Lam\'e coefficient and the shear modulus, are only special cases of the more general microscopic equation of motion \eqref{eq_EoMcontlimit} in terms of the dynamical matrix.
Consequently, the Green function of the macroscopic equation of motion can be interpreted as an approximation for the elastic Green function
defined in Eq.~\eqref{eq_GFdef}. On the other hand, we have shown in Sec.~\ref{subsec_emMatProp} that all  linear electromagnetic response properties
can be expressed in terms of this elastic Green function. Using for the latter the expression \eqref{gen_kleinert} and Eqs.~\eqref{eq_longGF}--\eqref{eq_transGF}, thus describing the displacement
field in the macroscopic and isotropic limit, we see that all response properties can be expressed in terms of the elastic constants.
In this subsection, we shortly discuss the corresponding explicit formulae.

Starting with the connection between the current response tensor and the elastic Green function, Eq.~\eqref{eq_currentResp}, we find 
with the explicit expression of the latter in the macroscopic and isotropic limit, Eq.~\eqref{gen_kleinert}, the following formulae for the longitudinal and transverse current response functions:
\begin{align}
\chi\L(\omega,\vec k)&=-\frac{\rho_{\rm n0}^2}{\rho_0} \h \frac{\omega^2}{\omega^2-v\L^2 \h |\vec k|^2}\,,\label{eq_longCurrentResp}\\[5pt]
\chi\T(\omega,\vec k)&=-\frac{\rho_{\rm n0}^2}{\rho_0}\h \frac{\omega^2}{\omega^2-v\T^2 \h |\vec k|^2}\,.\label{eq_transCurrentResp}
\end{align}
From these equations we obtain directly
\begin{equation}
\frac{v\L^2}{v\T^2}=\lim_{\omega\rightarrow 0}\frac{\chi\T(\omega,\vec k)}{\chi\L(\omega,\vec k)}\,.
\end{equation}
This equality, which replaces the Lyddane-Sachs-Teller relation (see \cite[Eq.~(27.67)]{Ashcroft} or \cite[Chap.~10, Eq.~(62)]{Kittel}), allows one to retrieve the longitudinal and transverse speeds of sound from the respective response functions.

Next, for the generalized (wavevector- and frequency-dependent) magnetic susceptibility, we use the response relation \cite[Eq.~(7.38)]{ED1} to obtain
\begin{equation}
\chi_{\rm m}(\vec k,\omega)\hh =\h \mu_0 \h \frac{\chi\T(\vec k,\omega)}{|\vec k|^2}\h =\h -\frac{\mu_0 \h \rho_{\rm n0}^2}{\rho_0} \, \frac{1}{|\vec k|^2}\, \frac{\omega^2}{\omega^2-v\T^2\h |\vec k|^2}\,.
\end{equation}
In particular, this yields zero in the instantaneous limit $\omega\rightarrow 0$, showing that the elastic solid is not susceptible to static magnetic fields.
Furthermore, from the standard relation between the density response function $\upchi$ and the longitudinal 
current response function $\chi_{\rm L}$ (see \cite[Eq.~(7.11)]{ED1} or \cite[Eq.~(3.175)]{Giuliani}), we obtain
\begin{equation}
\upchi(\vec k,\omega) \hh =\h -\frac{|\vec k|^2}{\omega^2} \h \chi\L(\vec k,\omega)\h =\h \frac{\rho_{\rm n0}^2}{\rho_0}\h \frac{|\vec k|^2}{\omega^2-v\L^2|\vec k|^2}\,.\label{eq_densRespIsotrop}
\end{equation}
Alternatively, this relation can be derived directly from Eqs.~\eqref{eq_densRespFuncGF} and \eqref{gen_kleinert}.
Finally, the dielectric tensor can be expressed in the isotropic limit as \cite[Eq.~(2.79)]{Refr}
\begin{equation}
(\tsr\varepsilon_{\rm r})^{-1}(\vec k,\omega)=
\tsr 1-\frac{1}{\varepsilon_0 \h \omega^2}\,\chi\L(\vec k,\omega) \h \tsr P\L(\vec k)+\mathbbmsl D_0(\vec k, \omega) \h\hh \chi\T(\vec k,\omega) \h \tsr P(\vec k)\,,
\end{equation}
where \smallskip
\begin{equation} \label{again_GF}
\mathbbmsl D_0(\vec k,\omega)=\frac{1}{\varepsilon_0 \h (-\omega^2 + c^2|\vec k|^2)} \smallskip
\end{equation}
is the Green function of the d'Alembert operator \cite[Eq.~(3.9)]{ED1}. In particular, for the longitudinal dielectric function we thus obtain the concise result
\begin{equation}\label{eq_nuclDF}
\varepsilon_{\rm r, \hh L}^{-1}(\vec k,\omega)=1+\frac{\rho_{\rm n0}^2}{\rho_0 \h \varepsilon_0} \h \frac{1}{\omega^2-v\L^2|\vec k|^2}\,, \smallskip
\end{equation}
which coincides precisely with Eq.~\eqref{eq_dielectricFctLimit} in the main text. The above formulae
express the electromagnetic material properties in terms of the elastic properties of the oscillating nuclear lattice. Correspondingly, calculating with the above result for the nuclear dielectric function, Eq.~\eqref{eq_nuclDF}, the
total effective electron-electron interaction via the fundamental ansatz \eqref{eq_effint1},
one recovers precisely the result \eqref{eq_simplify}, provided one re-expresses the ``microscopic'' parameters
$M$ and $e$ by their ``macroscopic'' counterparts \mbox{$\rho_0$ and $\rho_{\rm n0}$\hh.}

In concluding this appendix, we stress again that none of the above formulae actually allows for a straightforward deduction of the real electromagnetic material response from the elastic constants only. The reason for this is that our formulae take into account only the contributions from the oscillating nuclei, whereas the real electromagnetic response is more often than not dominated by the electronic contributions. Furthermore, the whole formalism hinges on the association of an electromagnetic four-current with the displacement field, which applies only in case that the nuclei are at least partially ionized.

\section{Analogy to photon-mediated interactions}\label{app_QED}

In this appendix, we discuss in how far the effective phonon-mediated interaction treated in this article is analogous to the so-called photon-mediated interaction used in (quantum)
electrodynamics. Beginning with a short review of electrodynamics as a classical field theory (Sec.~\ref{sec_review}), 
we will perform the transition to the corresponding quantum field theory by taking recourse to the {\itshape functional integral formalism} \cite{Rivasseau, Salmhofer, Nagaosa}
to quantum electrodynamics \cite{Itzykson, Peskin, Ryder}. Thereby, we will point out the precise analogy between 
the continuous displacement field (as treated in Sec.~\ref{fi_nucl}) and the electromagnetic four-potential. In particular, in Sec.~\ref{rev_qed} we will rederive the photon-mediated interaction within the functional integral approach, and subsequently show that the ensuing standard result can also be obtained classically by a straightforward elimination procedure, which is completely analogous to the derivation of 
phonon-mediated interactions as proposed in this article. We will then shortly explain the problems of the effective electronic theory (Sec.~\ref{sec_nonlocal}) and, moreover, show that it coincides with the famous Feynman--Wheeler electrodynamics if applied to a classical four-current generated by point-particles (Sec.~\ref{FWED}). 
Finally, we will prove that the instantaneous limit of the effective photon-mediated electron-electron interaction  reproduces the well-known Darwin Lagrangean, which can thus also be interpreted as ``static Feynman--Wheeler electrodynamics'' (Sec.~\ref{Darwin}).
A similar result has been derived apparently for the first time by T.\,C.~Scott and R.\,A.~Moore \cite{Scott}
using an expansion of the Feynman--Wheeler action in powers of the inverse speed of light.

\subsection{Short review of classical electrodynamics} \label{sec_review}

We begin by reviewing some well-known concepts of electrodynamics, using the same conventions as in Ref.~\cite{ED1}. In particular, we choose the Minkowski metric \smallskip
\begin{equation}
 \eta_{\mu\nu} = \eta^{\mu\nu} = \mathrm{diag}(-1, 1, 1, 1) \,, \label{metric_convention} \smallskip
\end{equation}
which is particularly useful for the comparison with non-relativistic results. The classical {\itshape Lagrangean density} $\mathcal L \equiv \mathcal L(x)$ of electrodynamics has the gen-\linebreak eral structure
\begin{align}
\mathcal{L}[A,\psi]&=\mathcal L_{\rm em \hh 0}[A]+\mathcal L_{\rm int}[A,\psi]+\mathcal L_{\rm e \hh 0}[\psi] \,, \smallskip \label{eq_classaction}
\end{align}
meaning that it is composed of a free part for the electromagnetic fields, a free part for the ``matter'' degrees of freedom and an interaction term.  
More concretely, the free electromagnetic part reads
\begin{equation}
\mathcal L_{\rm em \hh 0}(x) =-\frac{1}{4\mu_0}F^{\mu\nu}(x) \h F_{\mu\nu}(x) \,,
\end{equation}
and the interaction part reads
\begin{equation}
\mathcal L_{\rm int}(x) = j^\mu(x) \h A_{\mu}(x) \,.
\end{equation}
Here, we have defined the electromagnetic field-strength tensor
\begin{equation} \label{fmn}
 F_{\mu\nu}(x)=\partial_\mu A_\nu(x)-\partial_\nu A_\mu(x) \smallskip
\end{equation}
in terms of the four-potential $A^\mu=(\varphi/c, \h \vec A)$, while $j^\mu=(c\rho, \h \vec j)$ denotes the four-current density.
The term $\mathcal{L}_{\rm e \hh 0}[\psi]$ in Eq.~\eqref{eq_classaction} generally refers to the charged matter. 
For concreteness, we identify it with the Lagrangean density for the Dirac field,
\begin{equation}
\mathcal L_{\rm e \hh 0}(x)=c\,\bar\psi(x)\h\hh \big( \hh \i \hh \hbar \h \gamma^\mu\partial_\mu-mc \hh \big)\h \psi(x) \,,
\end{equation}
in which case the electromagnetic four-current can be expressed in terms of the electronic field as 
\begin{equation}\label{eq_4currDirac}
j^\mu(x)=(-e)c\, \bar\psi(x) \h \gamma^\mu \h \psi(x) \,.
\end{equation}
Here, the gamma matrices $\gamma^\mu$ are defined such that they satisfy the anticommutation relations (cf.~our metric convention \eqref{metric_convention}),
\begin{equation}
 \big[ \hh \gamma^\mu, \h \gamma^\nu \h \big]_+ = -2 \h\eta^{\mu\nu} \,,
\end{equation}
and $\bar\psi(x)=\psi^\dagger(x) \h \gamma^0$ is the conjugate spinor field  (see e.g.~\cite{Itzykson,Ryder,Wachter}). 
The classical action of the coupled electromagnetic and electronic system is given by the integral
\begin{equation} \label{action_total}
 S[A, \psi] = \frac 1 c \h \int \! \de^4 x \,\h \mathcal L[A, \psi] = S_{\rm em \hh 0}[A] + S_{\rm int}[A, \psi] + S_{\rm e \hh 0}[\psi] \,.
\end{equation}
The Euler--Lagrange equations (cf.~Sec.~\ref{disp_cont}) for the electromagnetic field,
\begin{equation}
c \,\h \frac{\de S}{\de A^\mu(x)} \h \equiv \h 
\frac{\partial\mathcal L(x)}{\partial A^\mu(x)}-\frac{\partial}{\partial x^\nu}\frac{\partial\mathcal L(x)}{\partial(\partial_\nu A^\mu(x))} \h = \h 0\,,
\end{equation}
then lead to the classical equation of motion,
\begin{equation}
(\eta\indices{^\mu_\nu}\Box+\partial^\mu\partial_\nu) \h A^\nu(x)=\mu_0 \h j^\mu(x)\,,\label{eq_EoM4pot}
\end{equation}
with the d'Alembert operator
\begin{equation} \label{dAlembert}
\Box= -\partial^\mu \partial_\mu = \frac{1}{c^2}\frac{\partial^2}{\partial t^2}-\Delta\,. 
\end{equation}
Equation~\eqref{eq_EoM4pot} has a particular solution in terms of the electromagnetic Green function,
\begin{equation}
A^\mu(x)=\int \! \de^4 x' \, (D_0)^\mu_{~\nu}(x,x') \, j^\nu(x') + (\partial^\mu \mh f)(x) \,,\label{eq_EoM4potsol}
\end{equation}
where $\partial^\mu \mh f$ is a pure gauge which can be chosen arbitrarily. The tensorial Green function $D_0$ (see \cite[Secs.~3.3--3.4]{ED1}) 
can therefore be characterized as the functional derivative
\begin{equation}
(D_0)^\mu_{~\nu}(x,x')=\frac{\delta A^\mu(x)}{\delta j^\nu(x')}\,,
\end{equation}
and hence corresponds to the response function of the four-potential. Finally, the Euler--Lagrange equation for the Dirac~field,
\begin{equation}
c \,\h \frac{\de S}{\de \bar\psi(x)} \h = \h
\frac{\partial\mathcal L(x)}{\partial \bar\psi(x)}-\frac{\partial}{\partial x^\mu}\frac{\partial\mathcal L(x)}{\partial(\partial_\mu \bar\psi(x))} \h =\h 0\,,
\end{equation}
reproduces the Dirac equation,
\begin{equation}\label{eq_DiracEq}
\big(\gamma^\mu(\i\hbar \h \partial_\mu-eA_\mu(x))-mc \h\big) \h \psi(x)=0 \,,
\end{equation}
in the presence of the electromagnetic four-potential.

\subsection{Effective interaction in the functional integral formalism} \label{rev_qed}

In the main text, we have proposed a simple method for the derivation of effective interactions,
which consists in an elimination procedure for the degrees of freedom whose influence is to be described by an effective interaction. 
This elimination procedure is based on linear response theory (see Sec.~\ref{sec_RFA}). Furthermore, 
we have shown in Sec.~\ref{sec_JustFI} that this is consistent with the functional integral approach, where the notion of effective interactions
is well established. On the other hand, in quantum electrodynamics the idea of a photon-mediated electron-electron interaction is also well known.
The main goal of this subsection is therefore to show that these different notions of effective phonon- and photon-mediated interactions are in fact completely analogous. 

\bigskip \noindent
{\itshape Electromagnetic Green function in the functional integral formalism.}---First, we note that---apart from the problems associated with the gauge-freedom---the 
electromagnetic Green function has an analogous representation in the functional integral (or ``path integral'') formalism as the elastic Green function (see Sec.~\ref{fi_nucl}). 
In fact, the (time-ordered) Green function $D_0$ for the electromagnetic four-potential in the, say, Lorenz gauge is defined in the quantum field theoretical setting as (cf.~\cite[Chap.~5]{MandlShaw}) 
\begin{equation}\label{eq_EDGreenFct}
{-\i} \hh \hbar \h c \, (D_0)^{\mu\nu}(x, x')=\langle \h\mathcal T\hat A^\mu(x) \h \hat A^\nu(x') \rangle\,,
\end{equation}
which is analogous to Eq.~\eqref{eq_DefPhonoGF}. Here, $\hat A_\mu(x)$ is the free quantized four-potential in the Lorenz gauge obeying the equation of motion (cf.~Eq.~\eqref{eq_EoM4pot})
\begin{equation}
\Box \h \hat A^\mu(x)=0\,,\label{eq_EoM4potMod}
\end{equation}
and the equal-time commutator
\begin{equation}
\big[\hat A^\mu(\vec x,t), \h \partial_t \hat A^\nu(\vec x',t)\big]=\i\hbar \h c^2 \mu_0 \, \eta^{\mu\nu} \h \delta^3(\vec x-\vec x')\,.
\end{equation}
The expectation value in Eq.~\eqref{eq_EDGreenFct} is usually taken with respect to the non-interacting ground state 
(i.e., the non-interacting {\itshape vacuum} of quantum electrodynamics). 
Analogously to Eq.~\eqref{eq_EoMPhonoGF}, one shows that the Green function defined in Eq.~\eqref{eq_EDGreenFct} obeys the equation of motion (in the Lorenz gauge)
\begin{equation}
\Box  (D_0)^{\mu\nu}(x, x')=\mu_0 \, \eta^{\mu\nu} \, \delta^4(x-x')\,.
\end{equation}
Generally, time-ordered electromagnetic Green functions can be represented from the functional integral with sources (see e.g.~\cite{Mosel, Zee}),
\begin{equation} \label{fi_em}
 Z[J] = \int \! \mathcal D A \, \exp \mh \bigg( \frac 1 \hbar \h S_{\rm em \hh 0}\hh[A] + \frac{1}{\hbar \h c} \int \! \de^4 x \, A_\mu(x) \h J^\mu(x) \bigg) \,,
\end{equation}
as the second-order functional derivative
\begin{equation} \label{green_em}
 \frac{1}{\hbar \h c} \, (D_0)_{\mu\nu}(x, x') = \frac 1 {Z_0} \h \frac{\delta^2 Z[J]}{\delta J^\mu(x) \h \delta J^\nu(x')} \h \bigg|_{J \h \equiv \h 0} \,,
\end{equation}
where $Z_0 = Z[J \equiv 0]$. This representation is analogous to the corresponding representation \eqref{ZJ}--\eqref{fder} of the elastic Green function. We remark, 
however, that the above electromagnetic path integral is actually ill-defined due to the gauge freedom. This is a standard problem in gauge theory, which is cured by the 
introduction of the Faddeev-Popov determinant 
(see e.g.~\cite{Bertlmann, Itzykson, Peskin, Ryder}). Here, we do not go into these problems because we use the path integral only as a heuristic tool, with the only relevant property being the standard result for Gaussian integrals (i.e., the equivalence of Eqs.~\eqref{zw_equiv_5} and \eqref{zw_5} below). We also note that in the literature, the path integral is often introduced with an additional factor~$\i$ (imaginary unit) in the exponent, which, however, is not necessary for our purposes.

To show Eq.~\eqref{green_em}, we proceed analogously as in Sec.~\ref{fi_nucl} (see also \cite[Chap.~12]{Mosel}. We first rewrite the action as
\begin{align}
 c \h\hh S_{\rm em \hh 0} & = -\frac 1 {4 \mu_0} \h \int \! \de^4 x \,\h F^{\mu\nu}(x) \h F_{\mu\nu}(x) \\[5pt]
 & = -\frac 1 {2 \mu_0} \int \! \de^4 x \, \h A_\mu(x) \, (\eta^{\mu\nu} \h \Box + \partial^\mu \partial^\nu) \h A_\nu(x) \label{notinv}\\[5pt]
 & = - \frac 1 2 \h \int \! \de^4 x \int \! \de^4 x'\h\hh A_\mu(x) \, (D_0^{-1})^{\mu\nu}(x, x') \h A_\nu(x') \,, \label{use_again}
\end{align}
where in the second step we have performed partial integrations after using Eq.~\eqref{fmn}. We remark again that, strictly speaking, the differential operator
in Eq.~\eqref{notinv} is not invertible (cf.~\cite[p.~157]{Bertlmann}), and hence
the precise definition of the tensorial Green function (see the discussion in \cite[Sec.~3.3]{ED1}) and the meaning of the corresponding Gaussian integral require some care.
This is precisely the aforementioned problem in the definition of the electromagnetic functional integral (see also \cite[Sec.~3.1]{Kleinert891} for a straightforward solution).
In the context of this article, we may also identify the classical Green function $D_0$ in the above expression \eqref{use_again} with the gauge-fixed Green function resulting from the quantum field theoretical definition, Eq.~\eqref{eq_EDGreenFct}, which in fact {\it is} invertible. (In this case, the spoiler
term $\partial^\mu\partial_\nu A^\nu$ in the free action can be dropped on grounds of the Lorenz gauge.)
Ignoring---for the sake of argument---these technical difficulties, the functional integral can be written~as
\begin{equation} \label{zw_equiv_5}
 Z[J] = \int \! \mathcal D A \, \exp \mh \bigg( \frac{1}{\hbar \h c} \h \bigg( {-\frac 1 2 \h A_\mu \h (D_0^{-1})^{\mu\nu} A_\nu  + A_\mu \h J^\mu} \bigg) \bigg) \,.
\end{equation}
Formally, this is a Gaussian functional integral, which yields explicitly (cf. the calculation for the displacement field in Eqs.~\eqref{zw_3}--\eqref{zw_4})
\begin{equation} \label{zw_5}
 Z[J] = Z_0 \h \exp \mh \bigg( \frac 1 {2 \h \hbar \h c} \, J_\mu \h\hh D_0^{\mu\nu} J_\nu \bigg) \,.
\end{equation}
Here, we have used again the symmetry of the Green function,
\begin{equation}
 (D_0)^{\mu\nu}(x, x') = (D_0)^{\nu\mu}(x', x) \,,
\end{equation}
which follows from the symmetry of the corresponding differential operator in Eq.~\eqref{eq_EoM4pot} provided one chooses a time-ordered Green function.
Finally, applying the functional derivatives to Eq.~\eqref{zw_5} and evaluating the result at vanishing sources yields the desired relation \eqref{green_em}.

\bigskip \noindent
{\itshape Photon-mediated interaction in the functional integral formalism.}---Within the functional integral approach, the {\itshape effective action} $S_{\rm eff}[\psi]$ 
for the electronic subsystem is defined by {\itshape integrating out} the electromagnetic degrees of freedom (cf.~\cite[Eq.~(9.80)]{Lawrie}), i.e.~through
\begin{equation}
 \exp \mh \bigg( \frac 1 \hbar \h S_{\rm eff}[\psi] \bigg) = \frac{1}{Z_0} \h \int \! \mathcal D A \, \exp \mh \bigg( \frac 1 \hbar \h S[A, \psi] \bigg) \,.
\end{equation}
This definition is again analogous to the corresponding definition \eqref{def_eff_int} of 
the phonon-mediated interaction.
Writing the total action formally as
\begin{equation}
 S[A, \psi] = -\frac 1 {2 \h c} \h\hh A_\mu \h (D_0^{-1})^{\mu\nu}  A_\nu + \frac 1 c \h\hh A_\mu \h j^\mu +  S_{\rm e \hh 0}[\psi] \,,
\end{equation}
and performing the Gaussian functional integral over the electromagnetic degrees of freedom as in Eqs.~\eqref{zw_equiv_5}--\eqref{zw_5}, 
we obtain the well-known result (see e.g.~\cite{Travisanutto, Olevano}, or \cite[Sec.~I.5]{Zee} and \cite[Eq.~(9.81)]{Lawrie}),
\begin{equation} \label{dzv}
 S_{\rm eff}[\psi] \h = \h S_{\rm e \hh 0}[\psi] + \frac 1 {2 \h c} \, j_\mu \h D_0^{\mu\nu}  j_\nu \h \equiv \h S_{\rm e \hh 0}[\psi] + S_{\rm e-e}^{\rm eff}[\psi] \,.
\end{equation}
Thus, in addition to the free part $S_{\rm e \hh 0}$ of the electronic action, we obtain an effective, photon-mediated electron-electron interaction part given by
\begin{equation} \label{seffee}
 S_{\rm e-e}^{\rm eff} = \frac 1 c \h \int \! \de^4 x \int \! \de^4 x' \, \mathcal L_{\rm e-e}^{\rm eff}(x, x') \,, \vspace{-3pt}
\end{equation}
with the interaction Lagrangean \smallskip
\begin{equation}
\mathcal L_{\rm e-e}^{\rm eff}(x,x')= \frac{1}{2} \,\hh j_\mu(x) \, (D_0)^{\mu\nu}(x,x') \, j_\nu(x')\,. \label{eq_effinteraction}  \smallskip
\end{equation}
This is the photonic analogon of the expression \eqref{effelintprop} (or \eqref{veffcont} in the continuum limit).
Thus, the effective photon-mediated electron-electron interaction introduced in quantum electrodynamics is indeed fully analogous
to the effective phonon-mediated electron-electron interaction as introduced in the Response Theory of the electron-phonon coupling.

\bigskip \noindent
{\itshape Classical rederivation of photon-mediated interaction.}---Finally, we will now prove that the above effective photon-mediated interaction 
can also be obtained directly from the classical Lagrangean field theory, which is again analogous to our classical derivation of the 
phonon-mediated interaction at the end of Sec.~\ref{fi_nucl}.
Concretely, we will show that the effective action \eqref{dzv} for the electronic field can be written as
\begin{equation} \label{action_id}
S_{\rm eff}[\psi] = S[A[\psi],\psi]\,,
\end{equation}
where $S[A, \psi]$ is the fundamental action given by Eq.~\eqref{action_total}, and $A^\mu[\psi]$ is given by Eq.~\eqref{eq_EoM4potsol} with the four-current being defined through Eq.~\eqref{eq_4currDirac}.
In other words, the effective photon-mediated interaction can also be derived in purely classical terms by eliminating the electromagnetic four-potential $A^\mu$ from the fundamental action \eqref{action_total}
using its equation of motion \eqref{eq_EoM4pot}.

To prove Eq.~\eqref{action_id}, we first note that both sides of the equation contain the purely electronic term $S_{\rm e \hh 0}[\psi]$, and hence it suffices to show that
\begin{equation}
 S^{\rm eff}_{\rm e-e}[\psi] = S_{\rm int}[A[\psi], \psi] + S_{\rm em \hh 0}[A[\psi]] \,.
\end{equation}
For the interaction term $S_{\rm int}$ the result is straightforward: expressing $A^\mu$ via Eq.~\eqref{eq_EoM4potsol} through its response function $D_0$ yields
\begin{align}
S_{\rm int}&=\frac 1 c \h \int\!\de^4 x\,\h j_\mu(x) \h A^\mu(x) \\[3pt]
&=\frac 1 c \h \int\!\de^4 x\int\!\de^4 x'\,j_\mu(x) \, (D_0)^{\mu\nu}(x,x')\, j_\nu(x')\,.\label{eq_contr1}
\end{align}
Next, we eliminate the four-potential from the purely electromagnetic term $S_{\rm em \hh 0}$\hh. For this purpose, we use again the expression \eqref{use_again}, which upon insertion of $A^\mu=(D_0)^{\mu\nu} \h j_\nu$ turns into
\begin{equation}
S_{\rm em \hh 0}=-\frac{1}{2 \h c} \h \int \!\de^4x\int \!\de^4x'\,j_\mu(x) \, (D_0)^{\mu\nu}(x,x') \, j_\nu(x')\,.\label{eq_contr2}
\end{equation}
Now the contributions \eqref{eq_contr1} and \eqref{eq_contr2} together yield
\begin{equation}
 S_{\rm em \hh 0} + S_{\rm int} \h = \h \frac{1}{2 \h c} \h \int \!\de^4x\int \!\de^4x'\,j_\mu(x) \, (D_0)^{\mu\nu}(x,x') \, j_\nu(x') \h \equiv \h S_{\rm e-e}^{\rm eff}\,,
\end{equation}
which exactly reproduces
the effective action \eqref{seffee}--\eqref{eq_effinteraction} and thus proves the identity \eqref{action_id}. We conclude that the well-known result \eqref{seffee}--\eqref{eq_effinteraction} for the effective action---originally obtained 
within the path integral formalism---can be reproduced more directly by eliminating the electromagnetic four-potential 
from the fundamental Lagrangean \eqref{eq_classaction} via its Green function (i.e.~its response function).
The effective photon-mediated electron interaction is therefore exactly analogous to the effective phonon-mediated electron interaction,
which was obtained by eliminating the nuclear degrees of freedom by their response functions. 
This corroborates again that the Response Theory of effective interactions is in accord with standard knowledge in quantum electrodynamics.

\subsection{Non-local Lagrangeans and initial value problems} \label{sec_nonlocal}

As explained already, effective interactions are in general non-local in {\it both} space {\it and} time.
In principle, the formalism of Lagrangean field theory carries over to non-local actions of the general form 
\begin{equation}
S=\frac 1 c \h \int \! \de^4x\int \! \de^4x'\,\mathcal{L}(x,x')\,,
\end{equation}
where the Lagrangean $\mathcal{L}(x, x')$ depends on the space-time arguments through the field, say $\psi(x)$, provided that $\mathcal L(x, x')$ is symmetric in $x$ and $x'$.
In this case, we can calculate a derivative of the action with respect to the field as
\begin{equation}
 c \,\h \frac{\delta S}{\delta \psi(y)} \h  = \h \int \! \de^4 x' \, \frac{\partial \mathcal L(y, x')}{\partial \psi(y)} + \int \! \de^4 x \, \frac{\partial \mathcal L(x, y)}{\partial \psi(y)} \h = \h 2 \int \! \de^4 x' \, \frac{\partial \mathcal L(y, x')}{\partial \psi(y)} \,.
\end{equation}
Correspondingly, we then obtain the {\it formal} Euler--Lagrange equations
\begin{equation}
c \,\h \frac{\de S}{\de \psi(x)} \h = \h 2 \int \! \de^4x' \, \bigg(\frac{\partial\mathcal L(x,x')}{\partial\psi(x)}-
\frac{\partial}{\partial x^\mu} \h \frac{\partial\mathcal L(x,x')}{\partial(\partial_\mu\psi(x))}\bigg) \h = \h 0\,.
\end{equation}
Concretely, for the electronic effective action \eqref{dzv}--\eqref{eq_effinteraction}, these equations lead to the Dirac
equation \eqref{eq_DiracEq} but with the four-potential expressed in terms of the Dirac spinor through the explicit equation \eqref{eq_EoM4potsol} and the definition of the four-current, Eq.~\eqref{eq_4currDirac}. Thus, in this case the formal equation
of motion reads
\begin{equation} \label{formal_eom}
\begin{aligned}
0 & = \gamma^\mu \h (\i\hbar \h \partial_\mu - mc) \h \psi(x) \\[3pt]
 & \quad \, + e^2 \h \gamma^\mu \h \psi(x)\int \! \de^4 x' \, (D_0)_{\mu\nu}(x,x') \h \psi^\dagger(x') \h \gamma^0 \h \gamma^\nu \h \psi(x')=0 \,.
\end{aligned}
\end{equation}
This shows that the effective interaction is {\it not} a true interaction, because for a non-local action the Euler--Lagrange equations only hold on a formal level, 
but do not lead to an {\itshape initial value problem.}

In fact, the above formal Euler--Lagrange equations involve the field at all times 
through an integration $\de^4 x'$ over the whole space-time. Concretely, 
this can be seen from Eq.~\eqref{formal_eom}, where  the time-derivative of the Dirac field $\partial_t \hh \psi(\vec x, t)$
cannot be expressed exclusively by the field $\psi(\vec x', t)$ at the {\itshape same} time $t$, but involves the field $\psi(\vec x', t')$ at {\itshape all} times $t'$. 
Hence, in the case of a non-local action the Euler--Lagrange equations yield only a complicated integro-differential equation. A further problem lies 
in the fact that the time-ordered Green functions are in general not real valued (see \cite[p.~35]{Itzykson}), which leads to problems for expressions assumed to be real
by their very nature (such as the interaction energy). These considerations show again that effective interactions in general do not yield interactions stricto sensu.

\subsection{Feynman--Wheeler electrodynamics} \label{FWED}

Broadly speaking, the impossibility of formulating an initial value problem for a non-local action is the reason why one
actually has to introduce electromagnetic fields in the first place. This has been stressed already in \cite[Sec.~5.8]{Davies}. In particular, even in the case of
finitely many interacting classical particles, one has to introduce infinitely
many degrees of freedom---namely the electromagnetic field---in order to get
rid of the retardation effects in the equations of motion: keeping the number
of degrees of freedom finite would necessitate self-consistent integro-differential equations of
motion which involve the particle trajectories at all times.

Surprisingly, however, exactly such an approach to classical electrodynamics has been put forward by R.\,P.~Feynman and
J.\,A.~Wheeler. Their famous {\it absorber theory of radiation} (see the original articles \cite{Feynman45,Feynman49}, or e.g.~the textbook \cite[Sec.~2.4]{Zeh}) has it that the dynamics of $N$ relativistic,
interacting point particles of mass $m$ and charge $q$ is described by a non-\linebreak local action \smallskip
\begin{equation}\label{eq_FeynmanWheeler}
S_{\rm FW} =S_{\rm e \hh 0}+S_{\rm e-e}^{\rm FW} \,, \smallskip
\end{equation}
composed of a free part \vspace{-7pt}
\begin{equation}
S_{\rm e \hh 0} = -mc \, \sum_{i=1}^N \h \int \! c \h\hh \de\tau_i\,, \label{eq_freePPaction}
\end{equation}
and an interaction part given by
\begin{equation}\label{eq_FeynmanWheelerEffWW}
S_{\rm e-e}^{\rm FW} =\frac{\mu_0 \h q^2}{8\pi c} \h \sum_{i\neq j} \h \int \! c\,\de\tau_i\int \! c\,\de\tau_j\,\h u^\mu_i(\tau_i) \,\hh \delta\big((x_i(\tau_i)-x_j(\tau_j))^2\h \big) \h\hh u_{j, \hh \mu}(\tau_j)\,.
\end{equation}
In these equations, $x_i^\mu=x_i^\mu(\tau_i)$ denotes the four-dimensional trajectory of the~$i$-th particle in Minkowski space, which is parametrized by its respective proper time \smallskip
\begin{equation}
 \de \tau_i = \de t \h \sqrt{1 - |\vec v_i(t)|^2/c^2} \,, \smallskip
\end{equation}
with $\vec v_i(t) = \de \vec x_i(t)/\de t$. Furthermore, 
\begin{equation}
(x_i-x_j)^2=-c^2(t_i-t_j)^2+|\vec x_i-\vec x_j|^2
\end{equation}
refers to the Minkowski scalar product. The relativistic four-velocity of the $i$-th particle is given by
\begin{equation}
u_i^\mu(\tau_i) \equiv\frac{\de x_i^\mu(\tau_i)}{\de\tau_i}= \gamma_i(t) \begin{pmatrix} c\\[2pt] \vec v_i(t) \end{pmatrix} ,
\end{equation}
where
\begin{equation}
\gamma_i(t) =\frac{1}{\sqrt{1-|\vec v_i(t)|^2/c^2}}\,. \smallskip \vspace{2pt}
\end{equation}
In particular, with these definitions, the relativistic action for free particles \eqref{eq_freePPaction} can be written explicitly as
\begin{equation}
S_{\rm e \hh 0} = -m c^2 \h \sum_{i = 1}^N \int \! \de t\,\sqrt{1-|\vec v_i(t)|^2/c^2} \,.
\end{equation}
By the Euler--Lagrange equations, the Feynman--Wheeler action leads to the relativistic standard equation of motion (for each particle $i = 1, \ldots, N$),
\begin{equation}
m \, \frac{\de}{\de\tau_i} \h u_i^\mu(\tau_i)=q \h F^{\mu\nu}_i(x_i(\tau_i)) \, u_{i, \hh \nu}(\tau_i)\,,
\end{equation}
where the field-strength tensor is expressed explicitly in terms of the corresponding four-potential by Eq.~\eqref{fmn}, 
while the four-potential is again given by Eq.~\eqref{eq_EoM4potsol} with the definition of the current of classical point-particles,
\begin{equation}
j_i^\mu(x)=q \h \sum_{j\neq i} \int \! c\,\de\tau_j\,u_j^\mu(\tau_j) \, \delta^4(x-x_j(\tau_j))\,.\label{eq_4currPoint}
\end{equation}
Explicitly, the equation of motion therefore reads
\begin{equation}
\begin{aligned}
m \h \frac{\de}{\de\tau_i} \h u_i^\mu(\tau_i) = q \, \big( \partial^\mu A_i^\nu - \partial^\nu A_i^\mu \h \big) (x_i(\tau_i))\, u_{i, \hh \nu}(\tau_i) \,,
\end{aligned}
\end{equation}
where
\begin{equation}
A_i^\mu(x)=\frac{\mu_0 \h q}{4\pi} \h \sum_{j \not = i}\int \! c \, \de\tau_j\,\hh \delta((x-x_j(\tau_j))^2\h) \, u_j^\mu(\tau_j)\,.
\end{equation}
In this last formula, we have used the explicit expression of the Feynman Green function (see \cite[Sec.~3.4]{ED1}, or Eq.~\eqref{eq_GFFeyn} below).
In particular, the above equation of motion implies  that the time derivative of the $i$-th particle posi\-{}tion, $\de^2 \vec x_i(t)/ \de t^2$, cannot be expressed 
exclusively by the positions $\vec x_j(t)$ of all other particles at the {\itshape same} time $t$, but instead involves their trajec\-{}tories at {\itshape all} times~$t'$ (or at least on the light cone of the respective $i$-th praticle).
Quite as its field theoretical counterpart described in the previous subsection, Feynman--Wheeler electrodynamics does therefore not allow for an initial value problem.

We will now show that the Feynman--Wheeler action \eqref{eq_FeynmanWheeler} is just the classical point-particle counterpart
of the well-known effective action \eqref{dzv}--\eqref{eq_effinteraction}. 
Remarks in this direction can already be found in the classic textbook \cite[pp.~120f.]{Misner} by C.\,W.~Misner, K.\,S.~Thorne and J.\,A.~Wheeler,
and in the equally classic textbook \cite[pp.~250f.]{FeynmanHibbs} by R.\,P.~Feynman and A.\,R.~Hibbs.
In fact, the proof for this is straightforward: one simply has to interpret the electromagnetic four-current in Eq.~\eqref{eq_effinteraction} as being generated by classical point-particles,
and to specify the Green function $D_0$ in the effective interaction. In other words, one has to show that the interaction term
$S_{\rm e-e}^{\rm FW}$ in the Feynman--Wheeler action is of the form of an effective interaction given by Eq.~\eqref{seffee}--\eqref{eq_effinteraction}.

To show this explicitly, we use the expression for  the current of classical relativistic point-particles given by Eq.~\eqref{eq_4currPoint}.
Furthermore, we introduce the 
retarded {\itshape scalar} Green function $\mathbbmsl D_0^+$ and its advanced counterpart $\mathbbmsl D_0^-$ (cf.~\cite[Sec.~3.1]{ED1}), which both fulfill the equation
\begin{equation}
\Box\mathbbmsl D_0^{\rm}(\vec x-\vec x', \h t-t')=\mu_0\, \delta^3(\vec x-\vec x') \, \delta(c \h t- c \h t')\,,
\end{equation}
with the d'Alembert operator defined in Eq.~\eqref{dAlembert}. These scalar Green functions are given explicitly by
\begin{equation}
\mathbbmsl D_0^{\pm}(\vec x-\vec x', \h t-t')=\frac{\mu_0}{4\pi c} \, \frac{1}{|\vec x - \vec x'|} \, \delta\hh\bigg(\mh t-t'\mp\frac{|\vec x-\vec x'|}{c}\h\bigg) \,,
\end{equation}
or equivalently by (cf.~\cite[Eq.~(1.170)]{Itzykson})
\begin{equation}
\mathbbmsl D_0^{\pm}(x-x')=\frac{\mu_0}{2\pi} \, \varTheta(\pm(t-t'))\,\delta((x-x')^2\hh)\,,
\end{equation}
where $\varTheta$ is the Heaviside step function. With these, we further introduce the 
(tensorial) electromagnetic {\itshape Feynman Green function} (cf.~\cite[Sec.~3.4]{ED1}) as
\begin{equation}
(D_0)_{\mu\nu} = \frac{1}{2} \, (\mathbbmsl D_0^{+}+\mathbbmsl D_0^-) \, \eta_{\mu\nu} = \frac{\mu_0}{4\pi} \, \delta((x-x')^2\hh) \, \eta_{\mu\nu} \,. \label{eq_GFFeyn}
\end{equation}
The latter is a Green for Eq.~\eqref{eq_EoM4pot} in the Lorenz gauge $\partial_\mu A^\mu = 0$.

Now, with Eqs.~\eqref{eq_4currPoint} and \eqref{eq_GFFeyn} one shows directly that the Feynman--Wheeler action \eqref{eq_FeynmanWheeler} 
can be written as an effective interaction in the form
\begin{equation}
 S_{\rm FW} \h = \h S_{\rm e \hh 0} + \frac{1}{2\h c} \h \int \! \de^4 x \int \! \de^4 x' \, j_\mu(x) \, (D_0)^{\mu\nu}(x,x') \, j_\nu(x') \h \equiv \h S_{\rm eff} \,,
\end{equation}
where singular particle self-interaction terms have to be discarded. We have thus shown that Feynman--Wheeler electrodynamics simply corresponds 
to ordinary electrodynamics, where the electromagnetic fields
have been eliminated formally through an appropriate symmetric Green function, and where the charge and current densities are assumed to 
be produced by classical relativistic point-particles. As the electromagnetic Green function can be interpreted as the response function of the electromagnetic
four-potential, all this is completely analogous to the Response Theory of the electron-phonon coupling proposed in this article.

\subsection{Darwin Lagrangean}\label{Darwin}

As in the case of phonon-mediated interactions (see Eq.~\eqref{instLim}), we now consider the effective photon-mediated interaction in the instantaneous limit,
\begin{equation}
D_0(\vec x,t;\vec x',t')\mapsto D_0(\vec x,\vec x'; \h \omega=0)\,\delta(c \h t - c \h t')\,.
\end{equation}
This implies for the interaction Lagrangean \eqref{eq_effinteraction},
\begin{equation}
 \mathcal L_{\rm e-e}^{\rm eff}(\vec x, t; \vec x', t') = \frac 1 2 \, j_\mu(\vec x, t) \, (D_0)^{\mu\nu}(\vec x, \vec x'; \h \omega = 0 ) \, j_\nu(\vec x', t) \, \delta(c\h t -c\h t') \h  \,,
\end{equation}
and hence for the corresponding action \eqref{seffee},
\begin{equation}
 S_{\rm e-e}^{\rm eff} \h \equiv \h \frac 1 c \h \int \! \de^3 \vec x \int \! c\,\de t \h \int \! \de^3 \vec x' \! \int \! c\,\de t' \, \mathcal L_{\rm e-e}^{\rm eff}(\vec x, t; \vec x', t') \h = \h \int \! \de t \, L_{\rm e-e}^{\rm eff}(t) \,,
\end{equation}
with
\begin{equation} \label{effintl}
 L_{\rm e-e}^{\rm eff}(t) = \frac 1 2 \h \int \! \de^3 \vec x \int \! \de^3 \vec x' \,  j_\mu(\vec x, t) \, (D_0)^{\mu\nu}(\vec x, \vec x'; \h \omega = 0 ) \, j_\nu(\vec x', t) \,.
\end{equation}
We are now going to show that this instantaneous interaction Lagrangean exactly reproduces the famous Darwin Lagrangean (see \cite[Eq.~(12.81)]{Jackson}, or the original article \cite{DarwinArticle}). This assertion ist most easily proven
in the Coulomb gauge, which leads to the concrete expression for the electromagnetic Green function (see \cite[Eq.~(3.52)]{ED1}),
\begin{equation}
(D_0)^{\mu}_{~\nu}(\vec k, \omega) = \left( \! \begin{array}{cc} \mu_0 / |\vec k|^2 & 0 \\[5pt]
0 & \mathbbmsl{D}_0(\vec k, \omega) \h \tsr P_{\rm T}(\vec k)
\end{array} \! \right) ,
\end{equation}
where
\begin{equation}
\mathbbmsl{D}_0(\vec k,\omega)=\frac{\mu_0}{-\omega^2/c^2 + |\vec k|^2} \smallskip
\end{equation}
is the scalar Green function in the Fourier domain (given also in Eq.~\eqref{again_GF}). With this, the effective interaction Lagrangean \eqref{effintl} simplifies to
\begin{equation}
\begin{aligned}
L_{\rm e-e}^{\rm eff} & =  \frac{c^2}{2} \int \! \de^3 \vec x \int \! \de^3 \vec x' \, \rho(\vec x) \, (D_0)^{00}(\vec x, \vec x'; \h \omega = 0) \, \rho(\vec x') \\[5pt]
& \quad \, + \frac 1 2 \h \int \! \de^3 \vec x \int \! \de^3 \vec x' \, \vec j^{\rm T}(\vec x) \, \tsr D_0(\vec x, \vec x'; \h \omega = 0) \, \vec j(\vec x') \,.\label{eq_effIntCoul}
\end{aligned}
\end{equation}
In particular, the $00$-component of the Coulomb Green function equals the Coulomb potential up to the prefactor. In real space, it is given by
\begin{equation}
(D_0)^{00}(\vec x,t;\vec x',t')=-\frac{\mu_0}{4\pi} \h \frac{\delta(c\h t-c\h t')}{|\vec x-\vec x'|} \,,
\end{equation}
and hence automatically yields an instantaneous interaction between the charge densities, which in fact coincides  with the ordinary Coulomb interaction (cf.~\cite{Travisanutto, Zee}).
Approximating also the spatial part of the electromagnetic Green function by its zero-frequency limit, we obtain
\begin{equation}
(D_0)_{ij}(\vec{x}-\vec{x}'; \h \omega = 0)=\int\!\frac{\de^3\vec{k}}{(2\pi)^3} \, \e^{{\rm i}\vec{k}\cdot(\vec x - \vec x')} \, \frac{\mu_0}{|\vec{k}|^2} \h \bigg(\delta_{ij}
-\frac{k_ik_j}{|\vec{k}|^2} \bigg)\,.
\end{equation}
The first term in this integral gives again the Coulomb potential up to the prefactor. Further using the regularization (see \cite[p.~765, Eq.~(1.89)]{Kleinert892})
\begin{equation}\label{usefulint}
\int\!\frac{\de^3\vec{k}}{(2\pi)^3}\,\h\e^{{\rm i}\vec{k}\cdot(\vec{x}-\vec x')}\,\frac{k_ik_j}{|\vec{k}|^4}=\frac{1}{8\pi} \, \frac{\partial^2}{\partial x_i \h \partial x_j} \h |\vec{x}-\vec{x}'| \,,
\end{equation}
we obtain after a short calculation the overall result
\begin{equation}
(D_0)_{ij}(\vec{x}-\vec{x}'; \h \omega = 0) = \frac{\mu_0}{8\pi} \, \bigg( \h \frac{\delta_{ij}}{|\vec{x}-\vec{x}'|}
+\frac{(x_i-x'_i)(x_j-x'_j)}{|\vec{x}-\vec{x}'|^3} \hh \bigg) \,.
\end{equation}
Collecting together all terms in Eq.~\eqref{eq_effIntCoul}, we arrive at the following expression for the effective interaction Lagrangean:
\begin{align}
 & L^{\rm eff}_{\rm e-e} = - \frac{1}{8\pi \varepsilon_0} \int \! \de^3 \vec x \int \! \de^3 \vec x' \,\h \frac{\rho(\vec x) \h \rho(\vec x')}{|\vec x - \vec x'|} \label{zw_6} \\[5pt] \nonumber
 & + \frac{\mu_0}{16 \pi} \int \! \de^3 \vec x \int \! \de^3 \vec x' \, \vec j^{\rm T}(\vec x) \, \bigg( \h \frac{1}{|\vec x - \vec x'|} \h \tsr 1 + \frac{(\vec x - \vec x')(\vec x - \vec x')^{\rm T}}{|\vec x- \vec x'|^3} \h \bigg) \, \vec j(\vec x') \,.
\end{align}
We now assume a point-particle current produced by only two particles ($i = 1, 2$) with charges $q_i$ and spatial currents $q_i \h \vec v_i$\h, such that 
\begin{equation}
 \rho(\vec x) = \sum_{i = 1}^2 q_i \, \delta^3(\vec x-\vec x_i)\,, \qquad
 \vec j(\vec x) = \sum_{i = 1}^2 q_i \h \vec v_i \, \delta^3(\vec x-\vec x_i)\,.
\end{equation}
These expressions can be derived directly from Eq.~\eqref{eq_4currPoint}. By putting them into Eq.~\eqref{zw_6} and discarding the particle self-interactions, we obtain
the interaction Lagrangean
\begin{equation}
\begin{aligned}
L_{\rm e-e}^{\rm eff}& =\frac{q_1q_2}{4\pi\varepsilon_0 \h |\vec x_1-\vec x_2|} \\[5pt]
& \quad \, \times \left(-1+
\frac{1}{2}\h \frac{\vec{v}^{\rm T}_1}{c}\h \bigg(\tsr 1+\frac{(\vec x_1-\vec x_2)(\vec x_1-\vec x_2)^{\rm T}}{|\vec x_1-\vec x_2|^2}\bigg) \h 
\frac{\vec v_2}{c}\right).
\end{aligned}
\end{equation}
This result indeed coincides exactly with the Darwin Lagrangean \cite[Eq. (12.81)]{Jackson}. 
Thus, we have shown that the Darwin interaction Lagrangean corresponds to ``static Feynman--Wheeler electrodynamics'',
i.e., an electron-electron interaction obtained from Feynman--Wheeler electrodynamics by taking the instantaneous limit as explained in Sec.~\ref{subsec_InstLim}.

Our final remark regards the gauge independence of the effective interaction. In fact, in this subsection we have chosen the Green function in the Coulomb gauge for our rederivation of the Darwin Lagrangean.
By contrast, in the previous subsection we used the Lorenz gauge.
On the other hand, in Ref.~\cite[Eq.~(3.41)]{ED1} we have shown that the most general form of the
electromagnetic Green function reads
\begin{equation} \label{eq_gen_GF}
 (D_0)\indices{^\mu_\nu}(k) = \mathbbmsl D_0(k) \left( \eta\indices{^\mu_\nu} + \frac{c k^\mu}{\omega} \, f_\nu(k) + g^\mu(k) \, \frac{c k_\nu}{\omega} + \frac{c k^\mu}{\omega} \, h(k) \, \frac{c k_\nu}{\omega} \right),
\end{equation}
where the functions $f_{\nu}$, $g^\mu$ and $h$ can be chosen arbitrarily up to the constraints of Minkowski-transversality,
\begin{equation}
f_\nu(k) \, k^\nu = k_\mu \, g^\mu(k) = 0 \,.
\end{equation}
Now, the effective action reads in Fourier space
\begin{equation}
S^{\rm eff}_{\rm e-e}=\frac 1 {2 \h c} \h \int\! \de^4 k\,(j^\mu)^*(k) \, (D_0)_{\mu\nu}(k) \, j^\nu(k) \,.
\end{equation}
Putting the general expression \eqref{eq_gen_GF} into this formula, performing partial integrations and using the continuity equation in the form
\begin{equation}
k_\mu \h\hh j^\mu(k) =0 \,,
\end{equation}
leads to the expression
\begin{equation}
S^{\rm eff}_{\rm e-e}=\frac 1 {2 \h c} \h \int\! \de^4 k\, (j^\mu)^*(k) \, \mathbbmsl D_0(k) \, j_\mu(k) \,,
\end{equation}
which does not depend on the arbitrary functions $f$, $g$ and $h$ anymore. This means that the effective interaction  is independent of the choice of the tensorial Green function $D_0$ (cf.~also \cite{Olevano}). In particular, it delivers the same
expressions no matter whether we use the Green function in the Coulomb gauge or in the Lorenz gauge.
As the former induces an instantaneous Coulomb interaction complemented by a retarded, tensorial
current-current interaction, while the latter implies a retarded Coulomb interaction for both charge and current densities,
the Response Theory also provides the simplest proof of the fact that {\itshape the instantaneous Coulomb interaction does not violate the principle \linebreak of causality.}
A direct proof of this fact can be rather complicated (see e.g. \cite[p.~66]{Tannoudji}).
At the same time, these considerations show that the above-mentioned technical difficulties with the definition
of the electromagnetic path integral can indeed by ignored in the derivation of the effective photon-mediated interaction, because the ambiguities in the definition of the corresponding Green function cancel
out in the final result.

\end{appendices}

\bibliographystyle{model1-num-names}
\bibliography{masterbib}

\end{document}